\documentclass[11pt, a4paper, twosides]{report}

\makeatletter
\newcommand\ackname{Acknowledgements}
\if@titlepage
  \newenvironment{acknowledgements}{%
      \titlepage
      \null\vfil
      \@beginparpenalty\@lowpenalty
      \begin{center}%
        \bfseries \ackname
        \@endparpenalty\@M 
      \end{center}}%
     {\par\vfil\null\endtitlepage}
\else
  
\fi
\makeatother

\usepackage[english]{babel}
\usepackage{amsfonts,amsmath,amssymb}
\usepackage{float,braket}
\usepackage{pslatex}
\usepackage[small,bf,hang]{caption2}
\usepackage{a4wide}
\usepackage{graphicx,color}
\usepackage{subfigure}
\usepackage{graphicx,wrapfig}

\renewcommand{\d}{\ensuremath{\mathrm{d}}}
\newcommand{\ii}{\ensuremath{\mathrm{i}}}
\newcommand{\p}{\partial}
\newcommand{\Tr}{\ensuremath{\mathrm{Tr}}}
\newcommand{\e}{\ensuremath{\mathrm{e}}}
\newcommand{\GZ}{\ensuremath{\mathrm{GZ}}}
\newcommand{\RGZ}{\ensuremath{\mathrm{RGZ}}}

\newcommand{\s}{\ensuremath{\mathrm{s}}}
\newcommand{\nl}{\ensuremath{\mathrm{nl}}}
\newcommand{\h}{\ensuremath{\mathrm{h}}}

\newcommand{\ext}{\ensuremath{\mathrm{ext}}}

\newcommand{\loc}{\ensuremath{\mathrm{loc}}}

\newcommand{\subs}{\ensuremath{\mathrm{subs}}}
\newcommand{\phys}{\ensuremath{\mathrm{phys}}}

\newcommand{\YM}{\ensuremath{\mathrm{YM}}}
\newcommand{\FP}{\ensuremath{\mathrm{FP}}}
\newcommand{\gf}{\ensuremath{\mathrm{gf}}}
\newcommand{\m}{\ensuremath{\mathrm{m}}}

\newcommand{\quadr}{\ensuremath{\mathrm{quadr}}}

\newcommand{\aux}{\ensuremath{\mathrm{aux}}}
\newcommand{\QCD}{\ensuremath{\mathrm{QCD}}}

\newcommand{\beqa}{\begin{eqnarray}}
\newcommand{\eeqa}{\end{eqnarray}}
\newcommand{\beq}{\begin{equation}}
\newcommand{\eeq}{\end{equation}}

\begin{document}
\title{{\bf The Gribov problem and QCD dynamics}}
\author{N.~Vandersickel \thanks{nele.vandersickel@ugent.be}\,\,$^a$, Daniel Zwanziger \thanks{daniel.zwanziger@nyu.edu}\,\,$^b$\\
\\
\small $^a$ \textnormal{Ghent University, Department of Mathematical Physics and Astronomy} \\
\small \textnormal{Krijgslaan 281-S9, 9000 Gent,Belgium}\\
\\
\small $^b$  \textnormal{New York University, New York, NY 10003, USA}\\
\normalsize}

\date{}
\maketitle

\abstract{In 1967, Faddeev and Popov were able to quantize the Yang-Mills theory by introducing new particles called ghost through the introduction of a gauge. Ever since, this quantization has become a standard textbook item. Some years later, Gribov discovered that the gauge fixing was not complete, gauge copies called Gribov copies were still present and could affect the infrared region of quantities like the gauge dependent gluon and ghost propagator. This feature was often in literature related to confinement. Some years later, the semi-classical approach of Gribov was generalized to all orders and the GZ action was born. Ever since, many related articles were published. This review tends to give a pedagogic review of the ideas of Gribov and the subsequent construction of the GZ action, including many other toipics related to the Gribov region. It is shown how the GZ action can be viewed as a non-perturbative tool which has relations with other approaches towards confinement. Many different features related to the GZ action shall be discussed in detail, such as BRST breaking, the KO criterion, the propagators, etc. We shall also compare with the lattice data and other non-perturbative approaches, including stochastic quantization.}

\tableofcontents

\newcommand{\club}{$\clubsuit$}

\chapter{Introduction}
\section{Patchwork quilt of QCD}
Quantum Chromodynamics (QCD) is the theory which describes the strong interaction, one of the four fundamental forces in our universe. This force describes the interactions between quarks and gluons, which are fundamental building blocks of our universe. At very high energies, QCD is asymptotically free, meaning that quarks and gluons behave like free particles\footnote{E.g.~in deep inelastic scattering (DIS) experiments, quarks can be treated as free particles.}. However at low energies, i.e~our daily world, due to the strong force, quarks and gluons interact and form bound states called hadrons. A well known example of these bound states are the proton and the neutron, but a whole zoo of hadrons has been observed in particle detectors. In fact, the only way to obtain information about the strong force is through bound states, as no free quark or gluon has even been detected. We call this phenomenon \textit{confinement}. Although 40 years of intensive research have passed since the formulation of the standard model (which includes QCD),  no good answer has been found to probably one of the most fundamental questions in QCD. Even the formulation of what confinement really is, is under discussion \cite{Greensite:2011zz}.\\
\\
The difficulty for solving confinement lies in the fact that the standard techniques which have been so successful in QED, are not applicable in QCD. In QED the coupling constant is small enough\footnote{This depends of course on the energy range of interest.}, so one can apply perturbation theory, which amounts to writing down a series in the coupling constant. In QCD, the coupling constant increases with decreasing energy, a phenomenon know as ``infrared slavery", and at low energy the coupling constant becomes too large and perturbation theory alone can never give a good description of the theory.  Therefore, other techniques are required which we call non-perturbative methods. There exists a wide range of non-perturbative methods, which all try to approach QCD from one way or another. One should not see these different techniques as competing, but as patchwork trying to cover all aspects of QCD. Many different approaches have been developed to describe confinement, e.g.~Abelian dominance \cite{Mandelstam:1974pi,thooft:1976,Ezawa:1982bf},  center-vortex dominance \cite{Greensite:2003bk}, light-cone dominance \cite{Brodsky:1995rn}, the Kugo-Ojima confinement mechanism \cite{Kugo:1979gm,Kugo:1995km}, Wilson's lattice gauge theory approach \cite{Wilson:1974sk}, and the approach initiated by Gribov \cite{Gribov:1977wm}. The last approach has been elaborated in widely scattered articles, but it has not been reviewed, and the relation of different approaches is not well known.\\
\\
Let us mention that even if one omits quarks in QCD, one can still call the remaining theory confining. Although there is no real experimental evidence, because the theory of gluons without quarks is a gedankentheorie, lattice simulations have shown that gluons form bound states which we call glueballs, and no free gluon can occur. Therefore, it is of interest to investigate pure QCD without quarks, and try to find out what happens. One could say that confinement is hidden in the behavior of the gluons.

\section{Gribov's gluon confinement scenario}\label{massgap}

The gauge concept was introduced into physics by Hermann Weyl to describe electromagnetism, by analogy with Einstein's geometrical theory of gravity.  According to Weyl, the vector potential $A_\mu(x)$ is the electromagnetic analog of the transporter or Christoffel symbol $\Gamma_{\lambda \mu}^\nu$ of general relativity, and serves to transport the phase factor of a charged field.  It is remarkable that the theories of the weak and strong interaction which have been developed since Weyl's time and which, with electromagnetism, form the standard model, have also proven to be gauge theories.  The group is extended from U(1) to SU(2) and, for QCD, SU(3), but these theories embody the geometrical gauge character introduced by Weyl.  In particular they respect the principle of local gauge invariance in the same way that Einstein's theory of gravity respects coordinate invariance.  Here we shall discuss heuristically how the gauge principle leads to Gribov's confinement scenario.\\
\\
According to the gauge principle, physical observables $O(A)$ and the action $S(A)$ are invariant under local gauge transformations $U(x)$
\beq
O({^U}A) = O(A)
\eeq
and $S({^U}A) = S(A)$, where the gauge transformation is defined by
\beq
{^U}A_\mu(x) \equiv U(x)A_\mu(x)U^\dag(x) - {\ii \over g} U(x)\p_\mu U^\dag(x).
\eeq
(Conventions are given below.)  We have written these statements for the continuum theory, but the corresponding statements hold in lattice gauge theory.\\
\\
Let us consider a toy example that illustrates how imposition of a gauge symmetry produces a mass gap.  The hamiltonian for a free non-relativistic particle in one dimension is given by $H = (-1/2m) {d^2 \over dx^2}$.  With no further symmetry condition, the spectrum is {\it continuous,} with energy $E =  k^2/2m$, where $k$ is any real number.  Now suppose that the group of translations, $x \to x + pL$,  which is a symmetry of the hamiltonian, is taken to be a ``gauge" symmetry, so the point $x$ and the point $x + pL$ are identified physically.  Here $p$ is any integer and $L$ is a fixed length.  The wave function $\psi(x)$ is required to be invariant under this gauge transformation, $\psi(x+pL) = \psi(x)$.  Then the spectrum becomes {\it discrete}, $E_n  = (2 \pi n)^2/2mL^2$, where $n$ is an integer.  We describe this situation by saying that the gauge symmetry has changed the physical configuration space --- that is, the space of $x$'s --- from the real line to the circle of circumference $L$, and this changes the continuum spectrum into a discrete spectrum.   This example illustrates a general phenomenon: a gauge theory must obey the constraint that the configurations $A$ and ${^U}A$ are identified physically, and such constraints have a dynamical consequences, making the spectrum more restrictive.  According to Gribov's scenario, this changes the spectrum in QCD and explains why gluons are absent from the physical spectrum.\footnote{A spectrum may of course develop a mass gap for other reasons such as the occurrence of a mass term in the action or by the operation of the Higgs mechanism.}  \\
\\
The key step in the development of this idea was taken by Gribov \cite{Gribov:1977wm} who showed that for a non-Abelian gauge theory, such as $SU(2)$ or $SU(3)$, the local gauge group is more powerful than for an Abelian theory, and imposes additional constraints.  In an Abelian gauge theory the gauge condition $\p \cdot A = 0$ fixes the gauge (essentially) uniquely, whereas in a non-Abelian gauge theory, as Gribov showed, there are distinct transverse configurations, $A \neq A'$ with $\p \cdot A' = \p \cdot A = 0$, that are related by a `large' gauge transformation $A' = {^U}A$.  These are known as Gribov copies, and a non-Abelian gauge theory is subject to the additional constraint that these Gribov copies must be identified physically.  Thus the physical configuration space is restricted to a region free of Gribov copies, sometimes called a fundamental modular region.  Singer \cite{Singer:1978dk} showed that this situation is unavoidable in a non-Abelian gauge theory, and that the physical configuration space, is topologically non-trivial, in contrast to the situation in Abelian theories where the space of transverse configurations, satisfying $\p \cdot A = 0$, is a linear vector space.  Gribov also understood the dynamical implications of the existence of Gribov copies, including a suppression of infrared modes due to the proximity of the Gribov horizon in infrared directions.  He made an approximate calculation of the gluon propagator and found that it did not have a physical pole.  Thus the gluon is expelled from the physical spectrum as a result of the constraints of non-Abelian gauge invariance \cite{Gribov:1977wm}.  It was subsequently shown that the fundamental modular region (a region free of Gribov copies, defined below in sect.\ \ref{discussFMR}) in the minimal Landau gauge for a non-Abelian gauge theory is bounded in every direction \cite{Zwanziger1982a}.  This means that when you walk along any straight line emerging from the origin in the space of transverse configurations, you inevitably arrive at points that are (physically identified with points that are) back inside the bounded region, whereas in Abelian gauge theory you keep going forever.  It was found in approximate calculations by Cutkosky \cite{Cutkosky:1983jd,Cutkosky:1987yi,Cutkosky:1988ti} and Koller and van Baal \cite{Koller:1987fq}, in which only a few low-energy modes were kept, that when gauge invariance of the wave functional is imposed, the spectrum changes  from the perturbative spectrum of gluons to one that approximates the physical spectrum of glueballs.\\
\\		
In the above remarks the physical configuration space was described in terms of a gauge fixing to the minimal Landau gauge.  However to make sense, the physical configuration should be defined gauge-invariantly.  We briefly indicate how this is done.  The physical configuration $A_\phys$ corresponding to a configuration $A$ is identified with the {\em gauge orbit} through $A$.  This consists of all configurations $A' = {^U}A$ that are gauge-equivalent to $A$ for all gauge transformations $U = U(x)$,\footnote{It is worth noting that, in addition to gauge transformations $U(x)$ that are continuously connected to the identity, the gauge orbit may also contain ``large" gauge transformations that cannot be so connected.  For example, if the gauge-structure group is the Lie group SU(2), whose group manifold is the 3-sphere $S_3$, and if the Euclidean base-manifold ($x$-space) is also the 3-sphere $S_3$, then a local gauge transformation $U(x)$ is a mapping $S_3 \to S_3$.  Such mappings, if they are continuous, are characterized by an integer winding number $n$, and the gauge orbit falls into disconnected pieces characterized by the winding number.}
\beq
A_\phys \equiv \{A': A' = {^U}A\}.
\eeq
The physical configuration is the space of gauge orbits, $\cal P = \{ A_\phys\}$.  This situation is described by the statement that the physical configuration space $\cal P$ is the {\it quotient space} of the space of configurations ${\cal A} = \{A\}$, modulo the group of local gauge transformations ${\cal G} = \{U\}$,
\beq
\label{quotient}
{\cal P} = {\cal A}/ {\cal G}.
\eeq
A complete gauge fixing is a parametrization of the quotient space by choosing a single representative from each equivalence class.  In practice the gauge must be defined in a convenient way.  Singer's theorem is the statement that a choice of unique representative on each gauge orbit by a linear gauge condition, such as $\p_\mu A_\mu = 0$, that is also continuous is impossible in a non-Abelian gauge theory.\\  
\\

\section{Approaches to the confinement problem}

There are many different approaches to the confinement problem.  We shall not attempt to survey them here, beyond mentioning them.  Lattice gauge theory, pioneered by Wilson, provides an unobjectionable and numerically powerful approach, which for practical, non-perturbative calculations of the hadron spectrum has out-distanced the competition.  However other approaches offer the possibility of analytic results, and possibly of exhibiting a simple intuitive confinement mechanism, the way electroweak interactions are understood by means of the Higgs mechanism.  Among these is the approach in the maximal Abelian gauge, with as a confinement mechanism the dual Meissner effect and condensation of color-magnetic monopoles, and a flux tube with a string tension.  This approach is closely related to the approach in the center-vortex gauge.  Other approaches make use of the light-cone gauge and the Coulomb gauge.  Schwinger-Dyson calculations have been done in the minimal Landau gauge.   Impressive results in 2+1 dimensions have been obtained by Nair and co-workers \cite{Nair:1990, Nair:2008}.

\section{Outline of the article}
The purpose of this review is to give a pedagogic overview of this approach by Gribov and the succeeding construction of the so-called GZ action, which is scattered in the literature. This shall be the topic of chapter \ref{sec1}. We shall first explain the Gribov problem and discuss in detail the existence of the Gribov horizon with the corresponding Gribov region, which shall be of great importance for the solution of the Gribov problem. Here, we can already show the relation of the Gribov horizon to Abelian and center-vortex dominance. We shall also cover many aspects of the even smaller fundamental modular region (FMR) and review all its properties. We shall then discuss the semi-classical solution of Gribov and its quantum generalization in which the cut-off at the Gribov horizon is implemented by a local and renormalizable action, which we shall call, as it has been called in the literature, the GZ action. We shall show the consequences on the gluon and ghost propagators, i.e.~they are modified in the infrared region.\\
 \\
Subsequently, in chapter \ref{sec2}, we shall elaborate on many different aspects of the GZ action.  Firstly, we shall comment on the transversality of the gluon propagator. We shall also elaborate on the breaking of the BRST symmetry of the GZ action, and revisit the Maggiore-Schaden construction \cite{Maggiore:1993wq}, which attempts to interpret the BRST breaking as a kind of spontaneous symmetry breaking. We shall also comment on the fact that it is possible to restore BRST symmetry by the introduction of additional fields. Secondly, we shall investigate the right form of the horizon condition, which was under discussion in the literature \cite{Kondo:2009wk,Dudal:2010fq} and show that the solution can be found with the help of renormalizability. Thirdly, we shall dwell upon the famous Kujo-Ojima criterion \cite{Kugo:1979gm}, often used as an explanation of confinement and we shall show its relation to the GZ action. The KO criterion gives a relation between an enhanced ghost propagator and confinement and which is often discussed in the literature \cite{Zwanziger:1992qr,Kondo:2009wk,Kondo:2009ug,Aguilar:2009nf,Boucaud:2009sd,Fischer:2008uz}. We shall also show one should be careful with imposing a kind of boundary condition on the ghost propagator. Indeed one can show that imposing an enhanced ghost propagator leads directly to the GZ action \cite{Dudal:2009xh}, and the KO criterion is not really fulfilled.
In section \ref{sec3} the comparison with the lattice data shall be scrutinized in detail.  We would also give a possible explanation for the discrepancy of the newest lattice data \cite{Cucchieri:2007rg,Cucchieri:2007md,Cucchieri:2008fc,Maas:2007uv,Sternbeck:2007ug,Bogolubsky:2007ud,Bogolubsky:2009dc,Bornyakov:2008yx,Gong:2008td,Cucchieri:2010xr} and the GZ action \cite{Dudal:2007cw,Dudal:2008sp,Dudal:2008rm,Dudal:2008xd,Dudal:2010tf}. We note other analytical approaches, which are in agreement with the latest lattice data \cite{Binosi:2009qm,Fischer:2008uz,Boucaud:2008ky}.  Other relevant analytic calculations of the gluon and ghost propagators are given in \cite{Alkofer:2000wg,Zwanziger:2001kw, Zwanziger:2002ea, Lerche:2002ep,Pawlowski:2003hq,Alkofer:2003jj}.
In sect.~\ref{reflection.positivity} we present the lattice data which show clear evidence of violation of reflection positivity in the gluon propagator.  It is noted that the Gribov form of the gluon propagator $k^2 / [(k^2)^2 + \hat\gamma^4]$ also violates reflection positivity because of the unphysical poles at $k^2 = \pm i \hat\gamma^2$, and it is proposed that the presence of unphysical singularities in the gluon propagator is the manifestation in the GZ approach of confinement of gluons.  In sect.~\ref{color.confinement} a plane-wave source $J_\mu^b(x) = H_\mu^b \cos(k \cdot x)$ is introduced, with free energy $W(J) = W_k(H)$.  A bound on the free energy per unit Euclidean volume, $w_k(H) = W_k(H)/V$, is reported that results from the proximity of the Gribov horizon in infrared directions.  It implies $\lim_{k \to 0} w_k(H) = 0$,  which is the statement that the system does not respond to a constant external color field $H$, no matter how strong.  In this precise sense, the color degree of freedom is absent from the system and may be said to be confined.  Finally in the concluding sect.~\ref{overview.of.approaches}, a critique of various approaches to the Gribov problem is presented, with a view toward lessons learned, open problems, and directions for future research.  It includes in particular a review of the merits and difficulties of stochastic quantization which by-passes the Gribov problem.

\section{Limitations of the present approach}

In the best of worlds, an alternative perturbation series based on the GZ action, and starting with the Gribov propagator, would give numerically accurate results.  Alas we do not live in the best of worlds and, as reported in sect.~\ref{lattice.data}, numerical simulation on large lattices gives a finite result $D(0) > 0$ for the gluon propagator $D(k)$ at $k = 0$ in Euclidean dimension $d = 3$ and 4 \cite{Cucchieri:2004mf, Bowman:2007hd}, whereas the alternative perturbation series gives $D(0) = 0$.  (Numerical simulation does however give $D(0) = 0$ in dimension $d = 2$.)  We conclude that the GZ action successfully describes confinement of gluons by unphysical singularities in the gluon propagator, but the perturbation theory generated from this action does not give quantitatively accurate results.  This deficiency of the alternative perturbation series may possibly be remedied by non-perturbative calculations with the GZ action \cite{Dudal:2007cw,Dudal:2008sp,Dudal:2008rm,Dudal:2008xd,Dudal:2010tf}.\\
 \\

\section{The Yang-Mills theory: definitions and conventions}
Before starting all this, let us introduce the Yang-Mills action and establish our conventions for they can easily differ in different books and articles. QCD is a gauge theory, as all theories for the fundamental forces. In fact, QCD is a special case of the more general $SU(N)$ Yang-Mills theory, with $N=3$. Therefore, we introduce the standard Yang-Mills theory and set the definitions and conventions used throughout this review. The derivation of the Yang-Mills action can be found in any standard textbook on quantum field theory \cite{Peskin}.\\
\\
We start with the compact group $SU(N)$ of $N \times N$ unitary matrices $U$ which have determinant one. We can write these matrices as
\begin{eqnarray}\label{U}
U &=& \e^{-\ii g \theta_a X_a} \;,
\end{eqnarray}
where $X^a$ represent the generators of the $SU(N)$ group.  For $SU(2)$ we have $X_a = \sigma_a/2$.  The index $a, b, c, \ldots$ is called the color index and runs from $\{1, \ldots, N^2 -1\}$. These generators obey the following commutation rule
\begin{eqnarray}
[X^a, X^b] &=& \ii f_{abc} X^c\;,
\end{eqnarray}
and the $SU(N)$ group corresponds to a simple Lie group. We can choose these generators to be hermitian, $X^\dagger = X$, and normalize them as follows
\begin{eqnarray}
\Tr[X_a X_b] &=& \frac{\delta_{ab}}{2}\;.
\end{eqnarray}
The generators $X_a$ belong to the adjoint representation of the group $SU(N)$, i.e.
\begin{eqnarray}
U X_a U^\dagger &=&  X_b (D^A)_{ba}  \;,
\end{eqnarray}
with $(D^A (X_a))_{bc}= - \ii f_{abc}$. The structure constants of the $SU(N)$ group have the following property,
\begin{eqnarray}\label{liestructure}
f^{abc}f^{dbc} &=& N \delta^{ad} \;.
\end{eqnarray}
Now we can construct a Lagrangian, which is symmetric under this group.
\\
\\
Firstly, we define the standard $SU(N)$ Yang-Mills action as
\begin{eqnarray}
S_{\YM} &=& \int \d^4 x \frac{1}{2} \Tr F_{\mu\nu}  F_{\mu\nu}\;,
\end{eqnarray}
whereby $F_{\mu\nu}$ is the field strength
\begin{eqnarray}\label{ff}
F_{\mu\nu}  &=& \p_\mu A_{\nu}  - \p_\nu A_\mu  - \ii g [A_{\mu}, A_{\nu}]\;,
\end{eqnarray}
and $A_\mu$ the gluon fields which belongs to the adjoint representation of the $SU(N)$ symmetry, i.e.
\begin{eqnarray}
A_\mu &=& A_\mu^a X^a \;.
\end{eqnarray}
The field strength can thus also be written as
\begin{eqnarray}
F_{\mu\nu} &=& F_{\mu\nu}^a X^a\;,
\end{eqnarray}
whereby
\begin{eqnarray}
F_{\mu\nu}^a \ =\ \p_\mu A_{\nu}^a  - \p_\nu A_\mu^a  +g  f_{akl} A_{\mu}^k A_{\nu}^l	\;.
\end{eqnarray}
Under the $SU(N)$ symmetry, we define $A_\mu$ to transform as
\begin{eqnarray}\label{notinf}
A_{\mu}' &=& U A_\mu U^{\dagger} - \frac{\ii}{g} (\p_\mu U) U^{\dagger}\;,
\end{eqnarray}
and one can check that this is compatible with the fact the $A_\mu$ belongs to the adjoint representation. Consequently, from \eqref{ff} we find
\begin{eqnarray}
F_{\mu\nu}' &=& U F_{\mu\nu} U^{\dagger}\;,
\end{eqnarray}
and therefore the Yang-Mills action is invariant under the $SU(N)$ symmetry. Infinitesimally, the transformation \eqref{notinf} becomes
\begin{eqnarray}\label{infinitesimal}
\delta A_{\mu}^a &=& -D_\mu^{ab} \theta^b \;,
\end{eqnarray}
with $D_\mu^{ab}$ the covariant derivative in the adjoint representation
\begin{eqnarray}\label{covariantderivativeadjoint}
D_\mu^{ab} &=& \p_\mu \delta^{ab} - g f^{abc} A_\mu^c\;.
\end{eqnarray}
Secondly, we can also include the following matter part in the action, when considering full QCD
\begin{eqnarray}
S_{\m} &=& \int \d^4 x \left( \overline \psi^i_\alpha (\gamma_\mu)_{\alpha \beta} D_\mu^{ij} \psi^j_\beta \right)\;,
\end{eqnarray}
where flavor indices and possible mass terms are suppressed, and which contains the matter fields $\overline \psi_i$ and $\psi_i$ belonging to the fundamental representation of the $SU(N)$ group, i.e.
\begin{eqnarray}
\overline \psi_i' &=& U_{ij} \overline \psi_j\;,
\end{eqnarray}
or infinitesimally
\begin{eqnarray}
\delta \overline \psi_i' &=& -\ii g \theta^a X^a_{ij} \overline \psi_j\;.
\end{eqnarray}
The index $i$ runs from $\{1, \ldots, N\}$. Every $\overline \psi_i$ and $\psi_i$ is in fact a spinor, which is indicated with the indices $\{\alpha, \beta, \ldots$\}. $D_\mu^{ij}$ is the covariant derivative in the fundamental representation
\begin{eqnarray}
    D_\mu^{ij} &=& \p_\mu \delta^{ij} - \ii g A_\mu^a (X^a)^{ij} \;,
\end{eqnarray}
and $\gamma_\mu$ are the Dirac gamma matrices. One can again check that also the matter part is invariant under the $SU(N)$ symmetry. The matter field $\overline \psi$ represents the quarks of our model. As is known from the standard model, there is more than one type of quark, which we call flavors. For each flavor, we would need to add a term like $S_\m$, but keeping the notation simple, we shall not introduce a flavor index here. The starting point of the Yang-Mills theory including quarks is thus given by
\begin{eqnarray}\label{ym}
S &=& S_\YM + S_\m \;.
\end{eqnarray}
Most of the time, we shall however omit the matter part, and work with pure Yang-Mills theory.

\chapter{From Gribov to the local action\label{sec1}}
In this section we shall give an overview of the literature concerning the Gribov problem. First, we shall uncover the Gribov problem in detail by reviewing the Faddeev-Popov quantization \cite{Faddeev:1967fc}. Next, we shall treat the Gribov problem semi-classically, as done in \cite{Gribov:1977wm}. Then, we shall translate the ideas of Gribov into a quantum field theory by formulating a local, renormalizable action \cite{Zwanziger:1989mf,Zwanziger:1992qr}.  This is often referred to in the literature as the GZ action and we shall use this designation.

\section{The Faddeev-Popov quantization}
In this section we recall the Faddeev-Popov method which solves the quantization problem of the Yang-Mills action at the perturbativev level \cite{DeWitt:1964yg,DeWitt:1967yk,Faddeev:1967fc}. Although this has now become a standard textbook item (see \cite{Weinberg:1996kr,Masujima,Ryder,Zinn-Justin2002,thooft2007} for some examples), we shall go into the details of the calculations to point out some subtleties.

\subsection{Zero modes}
We start from the Yang-Mills action given in the Introduction. Here we shall only consider the purely gluonic action, as the quantization problem arises in this sector. We recall that
\begin{eqnarray}
S_{\YM} &=& \int \d^d x \frac{1}{4}  F_{\mu\nu}^a  F_{\mu\nu}^a \;.
\end{eqnarray}
Naively, we would assume the generating functional $Z(J)$ (see \cite{Peskin}) to be defined by,
\begin{eqnarray}\label{genym}
Z(J) &=& \int [\d A] \e^{-S_\YM + \int \d x J_\mu^a A_\mu^a}\;.
\end{eqnarray}
Unfortunately, this functional is not well defined. Indeed, taking only the quadratic part of the action,
\begin{eqnarray}
Z(J)_\quadr &=& \int [\d A] \e^{- \frac{1}{4} \int \d x (\p_\mu A_\nu(x) - \p_\nu A_\mu(x) )^2 + \int \d x J_\mu^a(x) A_\mu^a(x)} \nonumber\\
            &=& \int [\d A] \e^{ \frac{1}{2} \int \d x \d y  A_\nu^a(x)\left[ \delta^{ab} \delta (x -y)  ( \p^2 \delta_{\mu\nu} - \p_\mu \p_\nu )\right] A_\mu^b(y) + \int \d x J_\mu^a(x) A_\mu^a (x)}\;,
\end{eqnarray}
and performing a Gaussian integration \eqref{gauss1}
\begin{eqnarray}
Z(J)_\quadr            &=& (\det A)^{-1/2} \int [\d A] \e^{ -\frac{1}{2} \int \d x \d y  J_\nu^a(x)  A_{\mu\nu}(x,y)^{-1} J_\mu^a(y)} \;,
\end{eqnarray}
with $A_{\mu\nu}(x,y) =  \delta (x-y)  ( \p^2 \delta_{\mu\nu} - \p_\mu \p_\nu )$, we see that this expression is ill-defined as the matrix $A_{\mu\nu}(x,y)$ is not invertible. This matrix has vectors with zero eigenvalues, e.g.~the vector $ Y_\mu(x) = \p_\mu \chi (x)$,
\begin{eqnarray}
\int \d y A_{\mu\nu}(x,y) Y_\nu(y)~=~ \int \d y [\delta (x-y)  ( \p^2 \delta_{\mu\nu} - \p_\mu \p_\nu )] \p_\nu \chi (y) &=& 0\;.
\end{eqnarray}
Therefore, something is wrong with the expression of the generating function \eqref{genym}. Notice that this problem is present for $SU(N)$ Yang-Mills action as well as for QED, i.e.~the abelian version of the Yang-Mills action.\\
 \\
The question is now where do these zero modes come from? Let us consider a gauge transformation \eqref{notinf} of $A_\mu = 0$, whereby we take $U = \exp (\ii g X^a \chi^a)$,
\begin{eqnarray}
A_{\mu}' &=& - \frac{\ii}{g} (\p_\mu U) U^{\dagger} ~=~  X_a \p_\mu \chi^a \;,
\end{eqnarray}
or thus $A_\mu^{a\prime} = \p_\mu \chi^a$. This means that our examples of zero modes $Y_\mu$ are in fact gauge transformations of $A_\mu =0$. As we are integrating over the complete space of all possible gluon fields $A_\mu$, we are also integrating over gauge equivalent fields. As these give rise to zero modes, we are taking too many configurations into account.

\subsection{A two dimensional example}
\begin{figure}[H]
\begin{center}
\includegraphics[width=8cm]{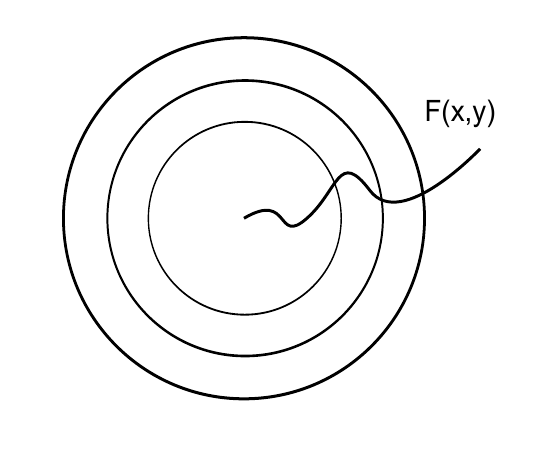}
\caption{Gauge orbits of a system with rotational symmetry in a plane and a function $\mathcal F$ which picks one representative from each gauge.}\label{2fig1}
\end{center}
\end{figure}

To fix our` thoughts, let us consider a two dimensional example. Consider an action, $S(r)$, invariant under a rotation in a two-dimensional space,
\begin{eqnarray}\label{start}
W &=& \int \d \vec r \e^{ -S(r)} ~=~ \int_0^{2\pi} \d \theta \int_0^{\infty} r \d r \e^{-S(r)} \;.
\end{eqnarray}
The ``gauge orbits'' of this example are concentric circles in a plane, see Figure \ref{2fig1}. All the points on the same orbit, give rise to the same value of the action $S(\vec r)$. Therefore to calculate $W$, we could also pick from each circle exactly one point, i.e.~the representative of the ``gauge orbit'', and multiply with the number of points on the circle (see Figure \ref{2fig1}). Now how exactly can we implement this?  Mathematically, we know that for each real-valued function, we have that \cite{Riley}
\begin{eqnarray}
\delta ( f (x)) &=& \sum_i \frac{\delta ( x - x_i)}{ | f' (x_i)|} \;,
\end{eqnarray}
with $x_i$ the solutions of $f ( x) = 0$ and provided that $f$ is a continuously differentiable function with $f'$ nowhere zero.
Integrating over $x$ yields
\begin{eqnarray}
\int \d x  \delta (f ( x)) &=& \sum_i \frac{1}{| f' (x_i)|}  \;,
\end{eqnarray}
or thus we find the following identity
\begin{eqnarray}
\frac{1}{\sum_i \frac{1}{ |f' (x_i)|}} \int \d x  \delta ( f ( x))  &=&  1 \;.
\end{eqnarray}
Applying this formula in our 2 dimensional plane, we can write
\begin{eqnarray}\label{gribovc}
\frac{1}{\sum_i \frac{1}{ \left. \left| \frac{\p \mathcal F (r, \phi) }{\p \phi} \right| \right|_{ \mathcal F( r,\phi) =0} }}   \int \d \phi  \delta ( \mathcal F (r, \phi))  &=&  1 \;,
\end{eqnarray}
whereby $\mathcal F$ represents the line which intersects each orbit. Now assuming \textit{that our function $\mathcal F$ intersects each orbit only once}, we can write
\begin{eqnarray}
 \left. \left| \frac{\p \mathcal F (r, \phi) }{\p \phi} \right| \right|_{ \mathcal F( r,\phi) =0}    \int \d \phi  \delta ( \mathcal F (r, \phi))  &=&  1 \;.
\end{eqnarray}
However, to make an analogy with the Yang-Mills gauge theory in the next section, we rewrite this,
\begin{equation} \label{unity}
\left. \left| \frac{\p \mathcal F (r, \theta + \phi ) }{\p \phi} \right| \right|_{ \mathcal F( r,\theta + \phi) =0}    \int \d \phi  \delta ( \mathcal F (r, \theta + \phi))  =  \left. \left| \frac{\p \mathcal F (\vec r^\phi ) }{\p \phi} \right| \right|_{ \mathcal F( \vec r^\phi) =0}    \int \d \phi  \delta ( \mathcal F (\vec r^\phi))  = 1 \;,
\end{equation}
and in this notation $\phi$ represents the rotation angle of the vector $\vec r$. This identity will allow us to pick on every orbit (a given $r$) only one representative, where $\mathcal F( \vec r^\phi) = 0$. Notice however that for every representative we have to multiply with a Jacobian, i.e.~the derivative of the function $\mathcal F$ at this point with respect to the symmetry parameter $\phi$. As this Jacobian only depends on the distance $r$, we denote this measure as follows
\begin{equation}\label{prefaddeev}
\Delta_{\mathcal F}(r) = \left. \left| \frac{\p \mathcal F (\vec r^\phi ) }{\p \phi} \right| \right|_{ \mathcal F( \vec r^\phi) =0} \;.
\end{equation}
Now inserting this identity into expression \eqref{start} gives
\begin{equation}
W = \int \d \theta \int r \d r  \Delta_{\mathcal F}(r)     \int \d \phi   \delta ( \mathcal F ( r, \theta + \phi )) \e^{\ii S(r)} \;,
\end{equation}
and transforming $\theta \to \theta -\phi$,
\begin{eqnarray} \label{2transf}
W  &= &\int \d \phi  \int \d \theta \int r \d r  \Delta_{\mathcal F}(r)     \delta ( \mathcal F ( r, \theta )) \e^{\ii S(r)}\;,
\end{eqnarray}
we are able to perform the integration over $\phi$, which gives a factor of $2 \pi$.
\begin{equation}
W = 2 \pi \int \d \theta \int r \d r  \Delta_{\mathcal F}(r)     \delta ( \mathcal F ( r, \theta )) \e^{\ii S(r)} \;.
\end{equation}
This factor represents the ``volume'' of each orbit. We shall see that we obtain something similar for the Yang-Mills action.\\
\\
Finally, let us remark that it is of uttermost importance that $\mathcal F$ intersects each orbit only once. Otherwise this derivation is not valid, and one should stick with formula \eqref{gribovc}.

\subsection{The Yang-Mills case}
We can repeat an analogous story for the Yang-Mills action \cite{Faddeev:1967fc,bach,Pokorski}, by keeping in mind the pictorial view of the previous section. Due to the gauge invariance of the Yang-Mills action, we can also divide the configuration space ${A_\mu (x)}$ into gauge orbits of equivalent classes. Two points of one equivalency class are always connected by a gauge transformation $U = \exp(-\ii g X^a \theta^a)$,
\begin{eqnarray}\label{AU}
A^U = U A_\mu U^{\dagger} - \frac{\ii}{g} (\p_\mu U) U^{\dagger} \;,
\end{eqnarray}
 see equations \eqref{notinf} and \eqref{U}. In analogy with the two dimensional example, we shall therefore also try to pick only one representative from each gauge orbit, which shall define a surface in gauge-field configuration space. As we are now working in a multivariable setting, i.e.~infinite dimensional space time coordinate system $x$, and the $N^2 - 1$ dimensional color coordinate system $a, b, \ldots$, the analogue of the identity \eqref{unity} becomes,
\begin{equation}\label{unity2}
\Delta_{\mathcal F}  \int  [\d U]  \delta ( \mathcal F (A^U ))  =  1 \;,
\end{equation}
whereby we have used a shorthand notation,
\begin{eqnarray}
\delta ( \mathcal F (A^U )) &=& \prod_{x } \prod_a \delta ( \mathcal F^a (A_\mu^U (x) ) \nonumber\\
\left[\d U\right] & \sim & \prod_x \prod_a \d \theta^a (x) \;.
\end{eqnarray}
Due to the multivariable system, the Jacobian \eqref{prefaddeev} needs to be replaced by the absolute value of the determinant,
\begin{equation}\label{absvalue}
\Delta_{\mathcal F} (A)=  |\det \mathcal  M_{ab} (x,y) | \qquad \text{with} \qquad  \mathcal M_{ab} (x,y) = \left. \frac{\delta \mathcal F^a (A_\mu^U (x))  }{\delta \theta^b(y)} \right|_{\mathcal F(A^U) =0} \;.
\end{equation}
This determinant is called the Faddeev-Popov determinant. Just as in the two dimensional example, this determinant is independent from the gauge parameter $\theta^a$. \\
\\
Inserting this identity into the generating function \eqref{genym} gives
\begin{eqnarray}\label{chap2Z2}
Z &=&  \int [ \d U] \int [\d A]  \Delta_{\mathcal F}(A)    \delta ( \mathcal F (A^U))   \e^{-S_\YM } \;,
\end{eqnarray}
where we temporarily omit the source term $J A$. Analogous to the two dimensional example (see equation \eqref{2transf}), we perform a gauge transformation of the field $A \to A^{U^\dagger}$, so that $A_\mu^U$ transforms back to $A_\mu$:
\begin{equation}
A^U_\mu = U A_\mu U^{\dagger} - \frac{\ii}{g} (\p_\mu U) U^{\dagger} \qquad \to\qquad  U A_\mu^{U^\dagger} U^{\dagger} - \frac{\ii}{g} (\p_\mu U) U^{\dagger} = A_\mu \;.
\end{equation}
Expression \eqref{chap2Z2} becomes
\begin{eqnarray}
Z &=&  \int [ \d U] \int [\d A]  \Delta_{\mathcal F}(A)    \delta ( \mathcal F (A))   \e^{-S_\YM } \;,
\end{eqnarray}
as the action, the measure $ [\d A] $ and the Faddeev-Popov determinant are invariant under gauge transformations. Now we have isolated the integration over the gauge group $U$, so we find
\begin{eqnarray}\label{genZZZ}
Z &=& V \int [\d A]  \Delta_{\mathcal F}(A)    \delta ( \mathcal F (A))   \e^{-S_\YM } \;,
\end{eqnarray}
with $V$ an infinite constant. As is common in QFT, one can always omit constant factors. It is exactly this infinite constant which made the path integral \eqref{genym} ill-defined.\\
\\
Let us now work out the Faddeev-Popov determinant. This determinant is gauge invariant, and does not depend on $\theta^a$, therefore we can choose $A$ so that if satisfies the gauge condition  $\mathcal F (A) = 0$. In this case, we can set $\theta^a =0$,
\begin{equation}
  \mathcal M_{ab} (x,y) = \left. \frac{\delta \mathcal F^a (A_\mu^U (x))  }{\delta \theta^b(y)} \right|_{\theta =0 \& \mathcal F(A) =0 } \;.
\end{equation}
Applying the chain rule yields,
\begin{eqnarray}
 \mathcal M_{ab} (x,y) &=& \int \d z \left. \frac{\delta \mathcal F^a (A_\mu (x))   }{\delta A_\mu^c (z)} \frac{\delta A^{c,U}_\mu (z)} {\delta \theta^b(y)}    \right|_{\theta =0 \& \mathcal F(A) =0 }\;.
\end{eqnarray}
First working out expression \eqref{AU} for small $\theta$
\begin{eqnarray}
A_\mu^U &=& A_\mu - (D_\mu \theta)^a X^a + O(\theta^2) \;,
\end{eqnarray}
and thus
\begin{eqnarray}
 \mathcal M_{ab} (x,y) &=& \int \d z \left. \frac{\delta \mathcal F^a (A_\mu (x))   }{\delta A_\mu^c (z)} (- D_\mu^{bc} \delta (y-z) )   \right|_{ \mathcal F(A) =0 }\;.
\end{eqnarray}
This is the most general expression one can obtain without actually choosing the gauge condition, $\mathcal F$. Let us now continue to work out the Faddeev-Popov determinant for the linear covariant gauges. For this, we start from the Lorentz condition, i.e.
\begin{eqnarray}\label{notideal}
\mathcal F^a (A_\mu (x)) &=& \p_\mu A^{\mu a} (x) - B^a (x) \;,
\end{eqnarray}
with $B^a (x)$ an arbitrary scalar field.  With this condition, $ \mathcal M_{ab} (x,y) $ becomes,
\begin{eqnarray}\label{mab}
 \mathcal M_{ab} (x,y) &=&  \left.    - \p_\mu  D_\mu^{ab} \delta (y-x)    \right|_{ \mathcal F(A) =0 } \;.
\end{eqnarray}
Because in the delta function in expression \eqref{genZZZ}, the condition $\mathcal F(A) =0$ is automatically fulfilled, so we find,
\begin{eqnarray}
Z&=& \int [\d A] [\det [ - \p_\mu  D_\mu^{ab} \delta (y-x) ]]    \delta (  \p A - B)   \e^{-S_\YM } \;.
\end{eqnarray}
Still, we cannot calculate with this form. However, luckily, there is a way to lift this determinant into the action. For this, we need to introduce Grasmann variables, $c$ and $\bar c$, known as Faddeev-Popov ghosts. As described in the appendix in expression \eqref{ghostapp}, we have that (by setting $\eta = \overline \eta =0$)
\begin{eqnarray}\label{absvalue2}
\det \mathcal M_{ab} (x,y) &=& \int [\d c] [\d \overline c] \exp \int \d x \d y \overline c^a (x) \mathcal M_{ab} (x,y) c^b (y)\;,
\end{eqnarray}
and thus
\begin{eqnarray}\label{genz5}
Z&=& \int [\d A][\d c] [\d \overline c]      \delta (\p A - B )   \exp \left[- S_\YM  -   \int \d x  \overline c^a (x)  \p_\mu  D_\mu^{ab}  c^b (x)  \right] \;,
\end{eqnarray}
and we have been able to express the determinant by means of a local term in the action.\\
\\
Finally, we would like to get rid of the dirac delta function. For this, we can perform a little trick: since gauge-invariant quantities should not be sensitive to changes of auxiliary conditions, we average over the arbitrary field $B^a (x)$ by multiplying with a Gaussian factor,
\begin{eqnarray}
  \int [\d B] \delta (\p A - B ) \exp \left( \frac{1}{2 \alpha} \int \d x B^2 \right)  &=& \exp \left( \frac{1}{2 \alpha} \int \d x (\p_\mu A_\mu^a )^2 \right)\;,
\end{eqnarray}
whereby $\alpha$ corresponds to the width of the Gaussian distribution. Taking all the results together, we obtain the following gauge fixed action:
\begin{eqnarray}\label{actiongauge}
S &=& S_\YM + \underbrace{\int \d x \left(   \overline c^a   \p_\mu  D_\mu^{ab}  c^b -  \frac{1}{2 \alpha} (\p_\mu A_\mu^a )^2   \right)}_{S_\gf}\;,
\end{eqnarray}
This gauge is called the \textit{linear covariant gauge}. Taking the limit $\alpha \to 0$ returns the \textit{Landau gauge}. In this case the width $\alpha$ vanishes and thus the Landau gauge is equivalent to the Lorentz gauge \eqref{notideal} with $B = 0$. Another widely known gauge is the \textit{Feynman gauge} whereby $\alpha =1$. The Landau gauge has the advantage of being a fixed point under renormalization, while in the Feynman gauge, the form of the gluon propagator has the most simple form.\\
\\
In conclusion, we have obtained the following well defined generating functional:
\begin{eqnarray}
Z(J) &=&   \int [\d A][\d c] [\d \overline c]         \exp \left[- S + \int \d x J_\mu^a A_\mu^a  \right]\;,
\end{eqnarray}
with the action $S$ given in equation \eqref{actiongauge}.

\subsection{Two important remarks concerning the Faddeev-Popov derivation}
We need to make two important remarks.
\begin{itemize}\label{twoimportantremarks}
\item First, notice that in fact, we need to take \textit{the absolute value} of the determinant, see expression \eqref{absvalue}. In many textbooks this absolute value is omitted, without mentioning that mathematically, it should be there. Subsequently, in equation \eqref{absvalue2}, we have neglected this absolute value in order to introduce the ghosts. It was thus implicitly assumed that this determinant is always positive. However, in the next section, we shall prove that this is not always the case. Only when considering infinitesimal fluctuations around $A_\mu =0$, i.e.~in perturbation theory, is this determinant a positive quantity (see section \ref{thegribovproblem}).
\item Secondly, closely related to the first remark, this derivation is done under the assumption of having a gauge condition which intersects with each orbit \textit{only once}. We call this an \textit{ideal} condition. If this is not the case, one should in fact stay with an analogous formula such as \eqref{gribovc}, namely,
    \begin{eqnarray}
    1 + N(A) &=& \Delta_{\mathcal F}  \int  [\d U]  \delta ( \mathcal F (A^U ))\;,
    \end{eqnarray}
    where $N(A)$ is the number of Gribov copies for a given orbit\footnote{For each copy, the Faddeev-Popov determinant is the same, therefore, the sum in equation \eqref{gribovc} can be replaced by $1 + N(A)$.}.  Again, in the next section, we shall show that the condition \eqref{notideal} is not ideal by demonstrating that the orbit can be intersected more than once. In fact, a mistake is made here.
\end{itemize}

\subsection{Other gauges}
Here we have worked out the Faddeev-Popov quantization for the \textit{Linear covariant gauges} which encloses the \textit{Landau} and \textit{Feynman gauge} as a special case. However, many other gauges are possible. We can divide the gauges in several classes. The first class are the \textit{covariant gauges}, which besides the linear covariant gauges also includes e.g.~the \textit{'t Hooft gauge} \cite{'tHooft:1971fh} and the \textit{background fields gauge} \cite{DeWitt:1967ub}. \\
\\
A second class of gauges are the noncovariant gauges, i.e.~gauges which break Lorentz invariance. The most famous example is probably the \textit{Coulomb gauge}, whereby $\mathcal F^a = \nabla_i A_i^a$ (see e.g.~\cite{Kleinert:2006} for a derivation).  This gauge condition has the same form as the Landau gauge condition but in $s = d-1$ dimensions.  Consequently our results concerning the Gribov region, the fundamental modular region etc.~that hold for configurations $A_\mu(x)$ in Landau gauge, hold also in Coulomb gauge for the {\it space} components of configurations $A_i(t, {\bf x})$ in Coulomb gauge  at a {\it fixed} time $t$, with the substitution $d \to s = d-1$.  A local action in Coulomb gauge that implements a cut-off at the Gribov horizon has been given in \cite{Zwanziger:2006sc}, similar to the action we shall present here shortly in Landau gauge. \\
\\
Some other examples are the \textit{axial gauge}, the \textit{planar gauge}, \textit{light-cone gauge} and the \textit{temporal gauge}. A nice overview on the second class can be found in \cite{Leibbrandt:1987qv}.
Finally, there are some other gauges like the \textit{Maximal Abelian gauge} \cite{Shinohara:2001cw}, which breaks color symmetry, and some more exotic gauges which break translation invariance. For a nice overview on different gauges, we refer to \cite{Klaus}.\\
\\
For this review, we shall mainly work in the Landau gauge.

\subsection{The BRST symmetry\label{globalmark}}
Now that we have fixed the gauge, the local gauge symmetry is obviously broken. Notice however that the global gauge symmetry is still present as one can check by performing a global gauge transformation on the action \eqref{actiongauge}. Fortunately, after fixing the gauge, a new symmetry appears that involves the ghosts, namely the BRST symmetry. This symmetry is most conveniently expressed by introducing the $b$-field,
\begin{eqnarray}\label{bS}
S &=& S_\YM +  \int \d^d x \left( b^a \p_\mu A_\mu^a + \alpha \frac{(b^a)^2}{2} + \overline{c}^{a}\partial _{\mu } D_{\mu}^{ab}c^b \right) \;,
\end{eqnarray}
whereby the path integral is now given by
\begin{eqnarray}
Z(J) &=&   \int [\d A][\d c] [\d \overline c][\d b]         \exp \left[- S + \int \d x J_\mu^a A_\mu^a  \right]\;,
\end{eqnarray}
with $S$ here the action \eqref{bS}, and $b$ is a bosonic field. $b^a$ is in fact an auxiliary field, sometimes referred to as the Nakanishi-Lautrup field \cite{Nakishi:1966di}.  Since it has no interaction vertices, one can easily integrate out this field to get back the action \eqref{actiongauge}. Therefore, the actions \eqref{bS} and \eqref{actiongauge} are equivalent.  It was found by Becchi, Rouet and Stora \cite{Becchi:1974xu} and independently by Tyutin \cite{Tyutin:1975qk}, that the action $S$ of equation \eqref{bS} enjoys a new symmetry, called the BRST symmetry \cite{Becchi:1996yh},
\begin{eqnarray}
s S &=& 0 \;,
\end{eqnarray}
with
\begin{align}\label{BRST}
sA_{\mu }^{a} &=-\left( D_{\mu }c\right) ^{a}\,, & sc^{a} &=\frac{1}{2}gf^{abc}c^{b}c^{c}\,,   \nonumber \\
s\overline{c}^{a} &=b^{a}\,,&   sb^{a}&=0\,, \nonumber\\
s \psi^i_\alpha &= - \ii g   c^a (X^a)^{ij} \psi^j_\alpha & s \overline \psi^i_\alpha &= - \ii g  \overline \psi^j_\alpha c^a (X^a)^{ji} \;.
\end{align}
One can check that $s$ is nilpotent,
\begin{eqnarray}
s^2 &=& 0\,.
\end{eqnarray}
a property which will turn out to be very important.\footnote{Without the introduction of the $b$-field, the BRST operator would be nilpotent only on-shell, i.e.~using the equation of motion.}  \label{physicalsubspace} This BRST symmetry is of utmost importance, as it is useful for several properties. Firstly, it is the key to the proof of the renormalizability of the Yang-Mills action. Secondly, the BRST symmetry is also the key to the proof that the Yang-Mills action is \textit{unitary} in perturbation theory . Let us explain what (perturbative) unitarity means. We define the physical state space $\mathcal H_\subs$, which is a subspace of the total Hilbert space, as the set of all physical states $\ket{\psi}_\phys$. A physical state is defined by the cohomology of the free BRST symmetry\footnote{This means, switch off interactions or set $g= 0$.} \cite{Slavnov:1989jh,Frolov:1989az}
\begin{equation}
s_0 \ket{\psi}_\phys = 0 \qquad \text{and} \qquad  \ket{\psi}_\phys \not= s_0 (\ldots) \;,
\end{equation}
where $s_0$ is the free BRST symmetry. Now a theory is unitary if
\begin{enumerate}
\item Starting from physical states belonging to $\mathcal H_\subs$, after these states have interacted, one ends up again with physical states $\in \mathcal H_\subs$.
\item All physical states have a positive norm.
\end{enumerate}
It is precisely the BRST symmetry which allows one to prove these properties\footnote{See \cite{Lehmann:1954rq,Henneaux:1992ig} for the original proofs, or \cite{Dudal:2007ch} for a more recent version of the proof.}. Also notice that by fixing the gauge, we have introduced extra particles, the ghost particles $c$ and $\overline c$. Because these particles are scalar and anticommuting, they violate the spin statistics theorem. For the theory to be physically acceptable, these ghost particles must not appear in the physical spectrum. This is of course related to issue of unitarity and one can show that the ghosts are indeed excluded from $\mathcal H_\subs$ by invoking the BRST symmetry.\\
\\
Finally, let us remark that in some approaches, BRST symmetry is regarded as a first principle of a gauge fixed Lagrangian \cite{Baulieu:1983tg,Nakanishi:1990qm}, rather than instead using Faddeev Popov quantization.

\section{The Gribov problem\label{thegribovproblem}}
Let us now explicitly show that in the Landau gauge\footnote{From now on, we shall work in the Landau gauge, unless explicitly mentioned.}, the gauge condition is not ideal. Gribov demonstrated this first in his famous article \cite{Gribov:1977wm} in 1977, which has been reworked pedagogically in \cite{Sobreiro:2004us}. For the gauge condition \eqref{notideal}, he explained that one can have three possibilities. A gauge orbit can intersect with the gauge condition only once ($L$), more than once ($L'$) or it can have no intersection ($L''$). In \cite{Gribov:1977wm} Gribov explains that no examples of the type $L''$ are known, however that many examples of the type $L'$ are possible.

\begin{figure}[H]
\begin{center}
\includegraphics[width=8cm]{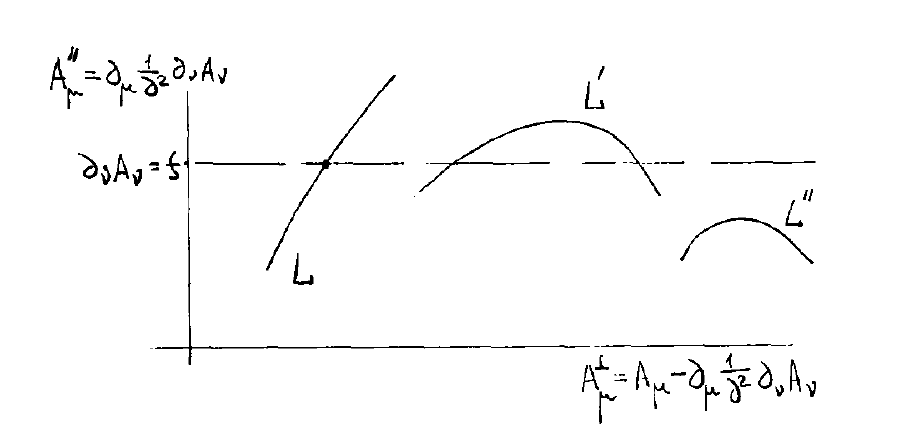}
\caption{The three possibilities for a gauge orbit w.r.t.~a gauge condition. Original figure from \cite{Gribov:1977wm}.}\label{2figgribov}
\end{center}
\end{figure}

\noindent Let us quantify this. Take two equivalent fields, $A_\mu$ and $A_\mu'$ which are connected by a gauge transformation \eqref{notinf}. If they both satisfy the same gauge condition, e.g.~the Landau gauge, we call $A_\mu$ and $A_\mu'$ \textit{Gribov copies}. We can work out this condition a bit further,
\begin{eqnarray}
&A_\mu' = U A_\mu U^{\dagger} - \frac{\ii}{g} (\p_\mu U) U^{\dagger}\;,  \qquad  \p_\mu A_\mu =0 \quad \& \quad  \p_\mu A_\mu' = 0 \;,&\nonumber\\
&\Downarrow& \nonumber\\
& \p_\mu U A_\mu U^{\dagger} +  U A_\mu \p_\mu U^{\dagger} - \frac{\ii}{g} (\p^2_\mu U) U^{\dagger} - \frac{\ii}{g} (\p_\mu U) (\p_\mu U^{\dagger}) =0 \;.&
\end{eqnarray}
Taking an infinitesimal transformation, $U = 1 + \alpha$, $U^\dagger = 1- \alpha$, with $\alpha = \alpha^a X^a$, this expression can be expanded to first order,
\begin{eqnarray}\label{zeromode}
-\p_\mu (\p_\mu \alpha + \ii g  [\alpha, A_\mu]) &=& 0\;,
\end{eqnarray}
or from equation \eqref{covariantderivativeadjoint} we see that this is equivalent to
\begin{eqnarray}
- \p_\mu D_\mu \alpha &=& 0\;.
\end{eqnarray}
When $A_\mu$ is transverse, $\p_\mu A_\mu = 0$, the Faddeev-Popov operator that appears here is hermitian,
\beq
- \p_\mu D_\mu = - D_\mu \p_\mu,
\eeq
It has a trivial null space consisting of constant eigenvectors, $\p_\mu \psi = 0$, that generate global gauge transformations which are unfixed by our gauge fixing.  The relevant infinitesimal Gribov copies are in the space orthogonal to the trivial null space.\\
\\
In conclusion, the existence of (infinitesimal) Gribov copies is connected to the existence of zero eigenvalues of the Faddeev-Popov operator. This is a very important insight as now we can understand the two remarks made in the previous section. \\
\\
Firstly, for small $A_\mu$, this equation reduces to $-\p_\mu^2 \alpha = 0$. However, it is obvious that the eigenvalue equation
\begin{equation}
-\p_\mu^2 \psi = \epsilon \psi \;,
\end{equation}
only has only positive eigenvalues $\epsilon = p^2 > 0$ (apart from the trivial null space). This means that also for small values of $A_\mu$ we can expect the eigenvalues $\epsilon (A)$ to be larger than zero. However, for larger $A_\mu$, this cannot be guaranteed anymore, so negative eigenvalues can appear (and will appear) for sufficiently large $A$, and thus the Faddeev-Popov operator shall also have zero eigenvalues. This means that our gauge condition is not ideal. Secondly, if the Faddeev-Popov operator has negative eigenvalues, the determinant of this operator can switch sign and the positivity of this determinant is no longer ensured. An explicit construction of a zero mode of the Faddeev Popov operator has been worked out in \cite{Henyey:1978qd,vanBaal:1991zw,Sobreiro:2004us}.\\
\\
Finally, let us also mention that in QED no Gribov copies are present. We can show this with a simple argument. In QED, the gauge transformations are given by
\begin{eqnarray}
A_\mu' &=& A_\mu - \p_\mu \chi \;,
\end{eqnarray}
and thus, for the Landau gauge, $\p_\mu A_\mu =0$, the condition for $A_\mu'$ to be a gauge copy of $A_\mu$ becomes
\begin{eqnarray}
 \p_\mu A_\mu' = 0 &\Rightarrow& \p_\mu^2 \chi = 0 \;,
\end{eqnarray}
which does not have any solutions besides plane waves. As a plane wave does not vanish at infinity, they cannot be used for constructing a gauge copy $A_\mu'$.

\subsection{The Gribov region: a possible solution to the Gribov problem?}
\subsubsection{Definition of the Gribov region}
\begin{figure}[H]
\begin{center}
\includegraphics[width=8cm]{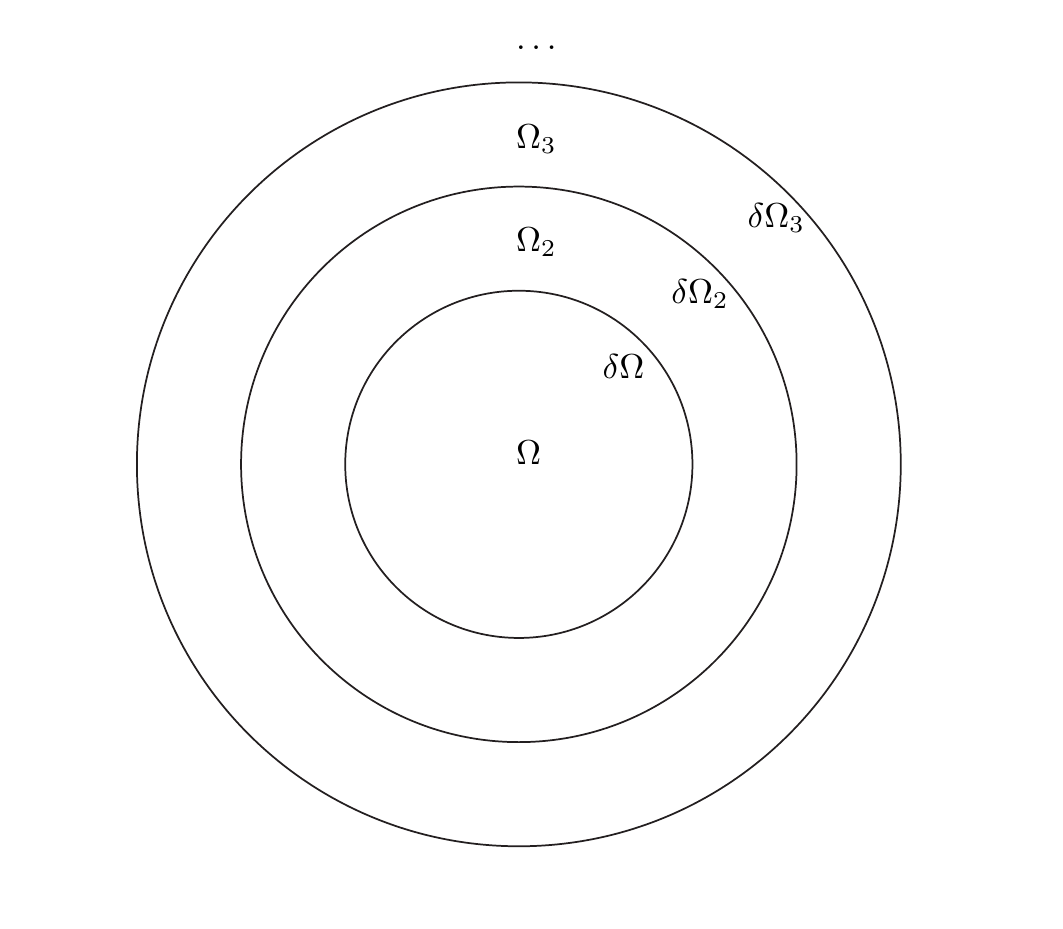}
\caption{The different regions in the hyperspace $\p A = 0$.}\label{2fighorizon}
\end{center}
\end{figure}

Now that we have shown that the Faddeev-Popov quantization is incomplete, we need to improve the gauge fixing. Gribov was the first to propose in 1977 \cite{Gribov:1977wm} to further restrict to a region of integration, the so-called Gribov region $\Omega$, which is defined as follows:
\begin{eqnarray}\label{defgribovregion}
\Omega &\equiv &\{ A^a_{\mu}, \, \p_{\mu} A^a_{\mu}=0, \, \mathcal{M}^{ab}  >0  \} \,,
\end{eqnarray}
whereby the Faddeev-Popov operator $\mathcal{M}^{ab}$ is given by equation \eqref{mab}
\begin{eqnarray}\label{mab2}
\mathcal M^{ab}(x,y) &=& - \p_\mu  D_\mu^{ab} \delta(x-y)  = \left( - \p_\mu^2 \delta^{ab} +  \p_\mu f_{abc} A_\mu^c \right) \delta(x-y)\;.
\end{eqnarray}
This is the region of gauge fields obeying the Landau gauge and for which the Faddeev-Popov operator is positive definite. We recall that a matrix is positive definite if for all vectors $\omega$,
\begin{eqnarray}\label{positivedef}
\int \d x \d y \omega^a(x)  \mathcal M^{ab} (x,y) \omega^b(y) > 0 \;.
\end{eqnarray}
In this way the problem of the absolute value of the determinant would already be solved (first remark on p.\pageref{twoimportantremarks}). The border of this region $\delta \Omega$ is called the first Gribov horizon and at this border the first (non-trivial) eigenvalue of the Faddeev-Popov operator becomes zero. Crossing this horizon, this eigenvalue becomes negative. This is depicted in Figure \ref{2fighorizon}. Consecutively, one can define the other horizons similarly, as drawn on the picture, where the second ($\delta \Omega_2$), the third ($\delta \Omega_3$), \ldots eigenvalue becomes zero. However, keep in mind that this picture is a very simplified pictorial view. In reality, the space of gauge fields is much more complicated.

\subsubsection{An alternative formulation of the Gribov region}
We can also define the Gribov region as the set of \textit{relative} minima of the following functional\footnote{The original derivation can be found in \cite{Semenov1986,Maskawa:1978rf,Zwanziger1982a}, a more recent version in the appendix of \cite{Capri:2005dy}. }
\begin{eqnarray}\label{functional}
F_A(U) \equiv ||A^U ||^2 &=&   \Tr \int \d x \ A^U_\mu (x) A^U_\mu (x)  = \frac{1}{2} \int \d x \ A^{U a}_\mu (x) A^{U a}_\mu (x) \;,
\end{eqnarray}
which corresponds to selecting on each gauge orbit the gauge configuration which minimizes $A^2$. Notice that there can be more than one minimum. It can be seen relatively easily  that this definition agrees with the Gribov region. Assume we have a gluon field $A_\mu$ for which the functional \eqref{functional} as a local minimum at $U = 1$. Firstly, in order to have an extremum, varying $||A||^2$ w.r.t.~an infinitesimal gauge transformation \eqref{infinitesimal} must be zero,
\begin{eqnarray}
\delta ||A||^2 &=& \delta  \left( \frac{1}{2} \int \d x A^{a }_\mu (x) A^{a }_\mu (x) \right) = \int \d x \delta A^{a }_\mu (x) A^{a }_\mu (x) = -\int \d x D_\mu^{ab} \theta^b(x) A^{a }_\mu (x) \nonumber\\
&=& - \int \d x \p_\mu \theta^a(x) A^{a }_\mu (x)  =  \int \d x \theta^a(x) \p_\mu  A^{a }_\mu (x) = 0\;.
\end{eqnarray}
As this equation must be zero for all $\theta^a$, we must have that $\p_\mu  A^{a }_\mu (x) = 0$. Secondly, this extremum must be a minimum, therefore, differentiating again,
\begin{eqnarray}
\delta^2 ||A||^2 &=&  - \int \d x \p_\mu  \theta^a (x) \delta A^{a }_\mu (x)  = \int \d x  \theta^a (x) (-\p_\mu  D_\mu^{ab}) \theta^b (x) > 0 \quad \forall\ \theta \;.
\end{eqnarray}
This implies that the operator $-\p_\mu  D_\mu^{ab} = \mathcal M^{ab}$ must be positive definite, see equation \eqref{positivedef}.

\subsubsection{Properties of the Gribov region}
How exactly one can implement this restriction, is the topic of the next sections. First, we shall address some profound geometrical questions and discuss some properties of the Gribov region.
\begin{itemize}
\item \label{propertiesofgribovregion} Firstly, does each orbit of gauge equivalent fields intersect with the Gribov region? This property is of course of paramount importance as it would not make any sense to integrate over an incomplete region of gauge fields. A first step towards establishing this property was made in \cite{Gribov:1977wm} where is was proved that for every field infinitesimally close to the horizon $\delta \Omega$, there exists an gauge copy at the other side of the horizon, infinitesimally close again. Later, it was then actually proven that every gauge orbit indeed intersects with the Gribov region \cite{Zwanziger1982a,Dell'Antonio:1991xt}. In \cite{Dell'Antonio:1991xt} is was mathematically rigorously proven that for every gauge orbit, the functional \eqref{functional} achieves its absolute minimum. Moreover, since every minimum belongs to the Gribov region, every gauge orbit intersects with the Gribov region.
\item Does $A_\mu =0$ belong to the Gribov region? This is important as this means that the perturbative region is also incorporated in the Gribov region. In fact, we can prove this very easily.  Taking $A_\mu =0$, the Faddeev-Popov operator becomes $\mathcal M^{ab} = -\p^2 \delta^{ab}$, which is positive definite.
\item We can also prove that the Gribov region is convex \cite{Zwanziger1982a}. This means that for two gluon fields $A_\mu^1$ and $A_\mu^2$ belonging to the Gribov region, also the gluon field $A_\mu = \alpha A_\mu^1 + \beta A_\mu^2$ with $\alpha, \beta \geq 0$ and $\alpha + \beta  =1$, is inside the Gribov region. To demonstrate this, we need to show that $\mathcal M^{ab} ( \alpha A_\mu^1 + \beta A_\mu^2)$ is positive definite. However, from expression \eqref{mab2}, we immediately see that
    \begin{eqnarray*}
    \mathcal M^{ab} ( \alpha A_\mu^1 + \beta A_\mu^2) &=& \alpha \mathcal M^{ab} (A_\mu^1) + \beta \mathcal M^{ab} (A_\mu^2) \;.
    \end{eqnarray*}
    As $\alpha, \beta \geq 0$, this sum of two positive definite matrices is again a positive definite matrix.
\item Finally we can show rather easily that the Gribov region is bounded in every direction \cite{Zwanziger1982a}.  This is the essential property that leads to a mass gap, as explained above in sec.~\ref{massgap}.  Assume we have a gluon field $A_\mu \neq 0$ located inside the Gribov region $\Omega$.  Then we can show that,  for sufficiently large $\lambda > 0$, the gluon field $\lambda A_\mu$ is located outside of $\Omega$. Firstly, as the matrix $ \mathcal M^{ab}_2(A_\mu) = \p_\mu f_{abc} A_\mu^c$ is traceless (it is already traceless on the color indices), the sum of all the eigenvalues of $\mathcal M^{ab}_2$ is zero. Therefore, for $A_\mu \not= 0$, there should     exist at least one eigenvector\footnote{We assume this eigenvector to have norm 1.} $\omega$ with negative eigenvalue $\kappa$, i.e.
     \begin{eqnarray}
      \int \d x \d y \omega^a(x)  \mathcal M_2^{ab} (x,y) \omega^b(y) = \kappa  < 0 \;.
     \end{eqnarray}
     Secondly, as $\mathcal M^{ab}_2(A_\mu)$ is linear in $A_\mu$,  $\mathcal M^{ab}_2( \lambda A_\mu) = \lambda \mathcal M^{ab}_2( A_\mu)$ has the same eigenvector $\omega$ with eigenvalue $\lambda \kappa$. Therefore,
     \begin{eqnarray}
       \int \d x \d y \ \omega^a(x) \mathcal M^{ab}( \lambda A_\mu) (x,y) \omega^b(y)  &=& \int \d x \  \omega^a(x) ( -  \p_\mu^2 )  \omega^a(x) +  \lambda \kappa \;,
     \end{eqnarray}
     which shall become negative for large enough $\lambda$. Consequently, $\mathcal M^{ab}( \lambda A_\mu)$ is no longer positive definite and $\lambda A_\mu$ is located outside the horizon. Therefore, $\Omega$ is bounded in every direction. Moreover, in \cite{Dell'Antonio:1989jn} it has been proven that the Gribov region is contained within a certain ellipsoid. One may also find the exact location of the Gribov horizon for $A$ which has a single fourier componant \cite{Zwanziger:2011},
\beq
\label{onefourier}
A_\mu^b(x) = c_\mu^b \cos(k \cdot x),
\eeq
where $c_\mu^b$ is orthogonal to $k_\mu$ for $A$ transverse, $k_\mu c_\mu^b = 0$.  Let  $t_{\mu \nu}(c)$ be the Lorentz tensor defined by $t_{\mu \nu}(c) \equiv c_\mu^b c_\nu^b$, and let  $n^2(c)$ be its largest eigenvalue, $t_{\mu \nu} V_\nu = n^2 V_\mu$.  Configurations of the form \eqref{onefourier} intersect the Gribov horizon at
\beq
n^2(c) = 2 k^2.
\eeq
Because $n^2(c)$ depends quadratically on $c_\mu^b$,  if we write $c_\mu^b = k C_\mu^b$, then the Gribov horizon occurs at $n^2(C) = 2$.  This exhibits the suppression of infrared modes by the Gribov horizon.
\end{itemize}
With all these properties, restricting the integration of gluon fields to the Gribov region looks like a very attractive option to improve the gauge fixing. Unfortunately, the Gribov region still contains Gribov copies. This was first discussed in \cite{Semenov1986}. Let us repeat their reasoning. Assume a gluon field $A_\mu$ belonging to the boundary of the Gribov region, then we have that
\begin{eqnarray}
\delta ||A||^2 &=& 0 \;,\nonumber\\
\delta^2 ||A||^2 &\not>& 0 \quad\Rightarrow \quad \exists \theta, \quad  \int \d x  \theta^a (x) \mathcal M^{ab}(x) \theta^b (x) = 0\;.
\end{eqnarray}
As the Faddeev-Popov determinant has zero modes, this means that it is inconclusive whether $||A||^2$ is a minimum\footnote{This is a consequence of the second derivative test as can be found in any textbook on basic mathematics.}. We have to consider the third variation $\delta^3 ||A||^2$,
\begin{eqnarray}
\delta^3 ||A||^2 &=& g f_{abc} \int \d x \p_\mu \theta^a(x) \theta^b(x) D_\mu^{cd}(x) \theta^d(x) \;,
\end{eqnarray}
which is, generally speaking, not zero\footnote{One can compare this with $x^3$ whose first and second derivatives are zero at $x=0$, while the third derivative is positive. }. Therefore, a gluon field on the boundary of the Gribov region is not a relative minimum of the functional \eqref{functional}, and thus there must exist a transformation $\tilde U$:
\begin{equation}\label{AUA}
||A||^2 > ||A^{\tilde U}||^2 \;.
\end{equation}
Since $||[(1-\epsilon) A]^{\tilde U}||$ and $||(1-\epsilon)A||$ are continuous in $\epsilon$, and the difference $B \equiv ||A||^2 - ||A^{\tilde U}||^2$ is a finite number, it follows that for $\epsilon > 0$ sufficiently small, the configuration $(1 - \epsilon)A$ satisfies the same inequality, $||(1-\epsilon)A||^2 > ||[(1-\epsilon) A]^{\tilde U}||^2$.  Thus the configuration $(1 - \epsilon)A$ is {\it not} an absolute minimum and moreover because the configuration $A = 0$ lies in the interior of $\Omega$ and because $\Omega$ is convex, the configuration $(1-\epsilon)A$ lies inside $\Omega$.  Thus $\Omega$ contains Gribov copies.
\\
\\
Also numerical results have confirmed this, see e.g.~\cite{Cucchieri:1997dx}.  In fact, it is not surprising that the Gribov region still contains copies. By looking at the functional \eqref{functional}, it seems obvious that on a gauge orbit, the functional \eqref{functional} can have more than one relative minima. Two relative minima of \eqref{functional} on the same orbit are Gribov copies which both belong to $\Omega$. Also Gribov was already aware of this possibility \cite{Gribov:1977wm}. 
Interesting explicit examples of Gribov copies are given in \cite{Lavelle:2007} and references found there.

\subsection{The fundamental modular region (FMR)}\label{discussFMR}
\subsubsection{Definition of the FMR}
What is then the configuration space free from Gribov copies? It is obvious that from the functional \eqref{functional} which defines the Gribov region, we can also define a more strict region, i.e.~the set of \textit{absolute} minima of the functional \eqref{functional}. As we take for each gauge orbit, the absolute minimum, we shall select, on a given orbit, only one gluon field, namely the gauge configuration closest to the origin. This region is then called \textit{the fundamental modular region} $\Lambda$. Restricting to this region of integration is also called \textit{the minimal Landau gauge}\footnote{Sometimes this is also called the \textit{absolute Landau gauge}, while the minimal Landau gauge can refer to taking one arbitrary minimum of the functional \eqref{functional}, depending on the author or article. }. $\Lambda$ is then a proper subset of $\Omega$, $\Lambda \subset \Omega$.
Notice that the absolute minimum of the functional \eqref{functional} can only determine the minimum up to a global gauge transformation. Indeed, as mentioned on p.\pageref{globalmark}, fixing the gauge does not break the global gauge symmetry and by performing a global gauge transformation $H$ independent from the space time coordinate $x$, expression \eqref{functional} does not change,
\begin{eqnarray}
||A^U ||_H^2 &=&   \Tr \int \d x H A^U_\mu (x) H^\dagger  H A^U_\mu (x) H^\dagger  =  \Tr \int \d x  A^U_\mu (x) A^U_\mu (x) = ||A^U ||^2  \;.
\end{eqnarray}
Therefore, saying that we picked out from a gauge orbit exactly one configuration always means modulo global gauge transformations.\\
\\
In fact, the FMR $\Lambda$ would be the exact gauge fixing if the global minima of the functional \eqref{functional} are non-degenerate. However, it is proven that degenerate minima can and do only occur on the boundary of the FMR, $\delta \Lambda$ \cite{vanBaal:1991zw}.  Therefore, if one would integrate over
\begin{eqnarray}
Z &=& \int_{\Lambda} [\d A] \e^{-S_\YM} \;,
\end{eqnarray}
these degenerate minima do not play any role, as they have zero measure. This agrees with endpoints of a function which do not play a role when integrating over a function.

\subsubsection{Properties of the FMR}
Let us discuss again some properties of the FMR, which resemble those of the Gribov region.
\begin{itemize}
\item Firstly, all gauge orbits intersect with the FMR. This is in fact already demonstrated in the first bullet point on p.\pageref{propertiesofgribovregion}.
\item $A_\mu = 0$ belongs to the FMR as 0 is the smallest possible norm.
\item $\Lambda$ is convex \cite{Semenov1986}. This is a bit more involved to prove than for the case of the Gribov region $\Omega$. We have to show that if $A^1_\mu, A^2_\mu \in \Lambda$, also $B_\mu = t A_\mu^1 + (1-t) A_\mu^2$, with $t \in [0,1]$. For this we work out the functional \eqref{functional},
\begin{eqnarray*}
||A^U ||^2 &=& \Tr \int \d x A^U_\mu (x) A^U_\mu (x)  \nonumber\\
&=& \Tr \int \d x \left( U A_{\mu} U^\dagger - \frac{\ii}{g} (\p_\mu U ) U^\dagger  \right) \left( U A_{\mu} U^\dagger - \frac{\ii}{g} (\p_\mu U ) U^\dagger  \right).	 \nonumber\\
&=& ||A||^2 - 2 \frac{\ii}{g} \Tr \int \d x \left( A_{\mu} U^\dagger \p_\mu U  \right) - \frac{1}{g^2} \Tr \int \d x \left( (\p_\mu U) U^\dagger (\p_\mu U ) U^\dagger  \right) \;.
\end{eqnarray*}
As $A_\mu^1$ and $A_\mu^2$ both belong to the FMR, we have that
\begin{eqnarray*}
||A^{1,U} ||^2 - ||A^1||^2 \geq 0 &\Leftrightarrow& - 2 \frac{\ii}{g} \Tr \int \d x \left( A^1_{\mu} U^\dagger \p_\mu U  \right) - \frac{1}{g^2} \Tr \int \d x \left( (\p_\mu U) U^\dagger (\p_\mu U ) U^\dagger  \right) \geq 0 \nonumber\\
||A^{2,U} ||^2 - ||A^2||^2 \geq 0 &\Leftrightarrow& - 2 \frac{\ii}{g} \Tr \int \d x \left( A^2_{\mu} U^\dagger \p_\mu U  \right) - \frac{1}{g^2} \Tr \int \d x \left( (\p_\mu U) U^\dagger (\p_\mu U ) U^\dagger  \right) \geq 0 \;.
\end{eqnarray*}
These two inequalities are linear in $A$, and they yield
\begin{eqnarray}
- 2 \frac{\ii}{g} \Tr \int \d x \left( (t A_\mu^1 + (1-t) A_\mu^2) U^\dagger \p_\mu U  \right) - \frac{1}{g^2} \Tr \int \d x \left( (\p_\mu U) U^\dagger (\p_\mu U ) U^\dagger  \right) \geq 0 \;,
\end{eqnarray}
from which follows
\begin{eqnarray}
||B^{U} ||^2 - ||B||^2 \geq 0 \;.
\end{eqnarray}
Therefore, $B$ belongs to the FMR and the FMR is convex.

\item $\Lambda$ is bounded in every direction. This is obvious as $\Lambda \subset \Omega$ with $\Omega$ bounded in every direction.
\item The boundary of $\Lambda$, $\delta \Lambda$ has some points in common with the Gribov horizon \cite{vanBaal:1991zw}.
\item Some points on the boundary $\delta \Lambda$ are Gribov copies of each other.
\end{itemize}

\subsection{Other solutions to the Gribov problem}
Firstly, a very important result has been proven by Singer in \cite{Singer:1978dk}, whereby it was shown that with suitable regularity conditions at infinity there is no gauge choice that is continuous, that is, there is no choice of (unique) representative of each gauge orbit that is continuous in the space of gauge orbits.  Thus a gauge free of Gribov copies is a singular gauge that is therefore very difficult to handle in calculations. These gauges do exist, e.g.~the space-like planar gauge \cite{Bassetto:1983rq} which has no Gribov copies, but which breaks Lorentz invariance. 
\\
\\
Many other attempts have been made, such as improving the Faddeev-Popov gauge fixing in \cite{Ghiotti:2005ih} whereby the absolute value of the Faddeev-Popov determinant was lifted into the action. However, the number of copies is not properly accounted for, and therefore, as far as we know, no further calculations have been done in their framework. Also in \cite{Slavnov:2008xz,Quadri:2010vt} an attempt to improve the gauge fixing has been done. However, as far as we know, the meaning of this model remains unclear in the infrared.\\
\\
It has been pointed out  \cite{Hirschfeld:1978yq, Friedberg:1995ty} that if one takes the Faddeev-Popov determinant, $\det \mathcal M(A)$, {\it without the absolute value,} and integrates over all configurations, including {\it all} Gribov copies, then one gets the correct result.  This happens because one is evaluating the signed intersection number of the intersection of the gauge fixing surface, for example $\p \cdot A = 0$, with each gauge orbit, and this is a topological invariant, the same for each orbit.  This has the feature that the Euclidean Landau-gauge or Coulomb-gauge functional weight changes sign, and the correct result is obtained by cancellation between different Gribov copies which give equal but opposite contribution.  This could lead to large errors if approximations are made. \\
\\
Finally we mention that stochastic quantization \cite{Parisi:1980ys} with stochastic gauge fixing \cite{Zwanziger:1981kg} provides a geometrically exact solution to the quantization of gauge fields whereby a gauge-fixing ``force" is introduced that is tangent to the gauge orbit.  The stochastic process can be represented by a local, perturbatively renormalizable \cite{ZinnJustin:1987ux, Chan:1985kf, Baulieu:1999wz, Baulieu:2000bgz} action in $d + 1$ dimensions.  The idea here is a continuum analog of the Monte Carlo method in lattice gauge theory whereby a stochastic process is invented whose ``time" average approaches the desired equilibrium average.  Here ``time" is machine time or the number of sweeps over the $d$-dimensional Euclidean lattice.  There is also a ``time"-independent $d$-dimensional version of stochastic quantization.  This method has perhaps been insufficiently explored although, in fact, the infrared critical exponent of the gluon propagator has been calculated in the time-independent stochastic quantization \cite{Zwanziger:2001kw, Zwanziger:2002ea}.  The merits of this approach are discussed in sect.~\ref{stochastic.quantization}.

\subsection{Summary}
In conclusion, to be absolutely sure that one has a correct quantization of the Yang-Mills theory, one should really restrict to the FMR in order to have a completely correct gauge fixing whereby only one gauge configuration is chosen per orbit. However, no practical implementation of this region has been found so far in the continuum. Some other attempts of improving gauge fixing are interesting, but not very convenient or too difficult to handle. However, if we restrict ourself to the Gribov region, it is possible to perform practical calculations. Gribov has done this semi-classically, and as we shall describe in detail below, it is possible to build an action which automatically restricts to the Gribov region. One can still object that the Gribov region still contains Gribov copies, but there has been a conjecture \cite{Zwanziger:1993dh,Greensite:2004ke, Greensite:2004ur}, that the important configurations lie on the common boundary $\delta \Lambda \cap \delta \Omega$ of the Gribov region $\Omega$ and the FMR $\Lambda$. Therefore, the extra copies inside the Gribov region would not play a significant role, and it would be sufficient to restrict to the Gribov region.

\section{The relation of the Gribov horizon to Abelian and center-vortex dominance}
\begin{figure}[H]
\begin{center}
\includegraphics[width=14cm]{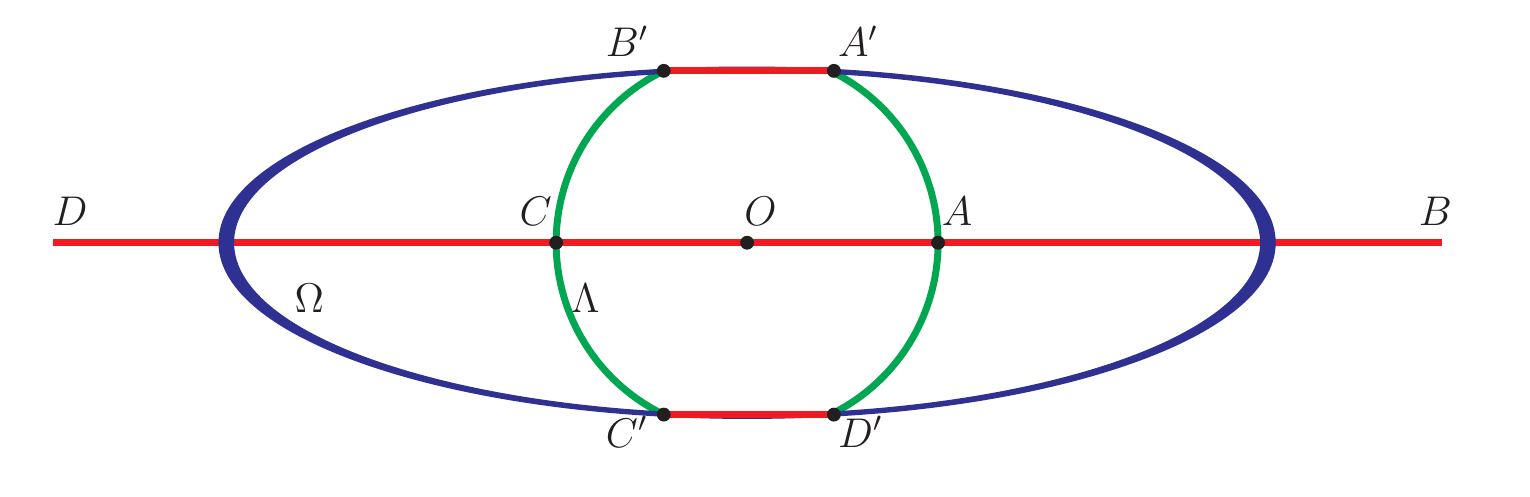}
\caption{The plane of the figure represents transverse configurations $\partial\cdot A = 0$.  $\Lambda$ is the fundamental modular region, with boundary in green, and $\Omega$ is the Gribov region with boundary in blue.  The straight line $DCOAB$, in red, represents the linear vector space of transverse, Abelian configurations, $A_\mu^b = \delta^{b3} a_\mu$.  The segments $AB$ and $CD$ that lie outside $\Lambda$ are gauge equivalent to the segments $A'B'$ and $C'D'$, also in red, that lie on the common boundary of $\Lambda$ and $\Omega$.}\label{commonboundary}
\end{center}
\end{figure}

Let us go a bit more into the details of the hypothesis that the important configurations lie on the common boundary $\delta \Lambda \cap \delta \Omega$, as a unified discussion is not available in literature. \\
\\
We start by considering whether or not, for a given configuration $A$, there exists a non-zero solution $\omega$ to the equation,
\begin{equation}\label{degenerate}
D_\mu(A)\omega = 0\;.
\end{equation}
This equation is gauge-invariant, as shown in the appendix \ref{appgaugeinv}.  So the property for a configuration $A$ to allow a non-zero solution to \eqref{degenerate} holds also for every gauge copy of $A$, $D_\mu({^U}A){^U}\omega = 0$.  This property, which distinguishes certain gauge orbits, is thus of geometrical significance, and we may suspect that these gauge orbits play a distinguished role in the dynamics of QCD.\\
\\
A gauge orbit with this property has a peculiar feature. In (at least) one direction, namely the direction defined by $\omega$, the gauge orbit is degenerate because the infinitesimal gauge transformation,
\begin{equation}
A_\mu' = A_\mu + D_\mu(A) \varepsilon \omega = A_\mu\;,
\end{equation}
with $\varepsilon$ infinitesimal, leaves $A$ invariant. Therefore, the gauge orbit through $A$ has (at least) one dimension less that a generic gauge orbit.  We call \eqref{degenerate} the ``degeneracy property''.\footnote{In the mathematical literature such an orbit is called {\it reducible}.}\\
\\
We now relate degenerate gauge orbits to the confinement scenario in maximal Abelian gauge.  According this scenario, the functional integral in the maximal abelian gauge is dominated by Abelian configurations (or more precisely those nearby) \cite{Ezawa:1982bf}.  In $SU(2)$ gauge theory, any Abelian configuration can be written as\footnote{This can be generalized to other $SU(N)$.}
\begin{equation}\label{abelian}
{A}_\mu^a(x) = \delta^{a3} a_\mu(x)\;,
\end{equation}
where $a_\mu(x)$ is an arbitrary Abelian configuration.  This configuration possesses the degeneracy property \eqref{degenerate}.  Indeed for the the $x$-independent global generator,
\beq
\label{global}
{\omega}^a = c \delta^{a3},
\eeq
where $c$ is a constant, we have $\p_\mu \omega = 0$, and we easily verify
\begin{equation}
D_\mu(A)\omega = \p_\mu \omega - \ii g [A_\mu, \omega] = 0\;,
\end{equation}
which is the condition for a gauge orbit to be degenerate.  Thus, {\it the hypothesis that in the maximal Abelian gauge the functional integral is dominated by Abelian configurations is compatible with the gauge-invariant hypothesis that the functional integral is dominated by degenerate gauge orbits.}\\
\\
The abelian form \eqref{abelian} is preserved by the group of local $U(1)$ gauge transformations about the 3-axis, that induces the local Abelian gauge transformations, $a_\mu \to a_\mu + \p_\mu \lambda$,  and we may suppose that, by such Abelian gauge transformations, the Abelian configurations are made transverse, $\p_\mu a_\mu = 0$.  Each transverse Abelian configuration corresponds to a unique distinct Abelian field tensor $f_{\mu \nu}(x) = \p_\mu a_\nu - \p_\nu a_\mu$, with inversion $a_\nu = (\p^2)^{-1} \p_\mu f_{\mu \nu}$ automatically satisfying $\p_\nu a_\nu = 0$, so different transverse Abelian configurations are gauge inequivalent.  On the other hand the FMR, $\Lambda$, in minimal Landau gauge is bounded in every direction.  So some transverse Abelian configurations lie inside $\Lambda$ and some lie outside.  This situation is pictured in Figure \ref{commonboundary}.\\
\\
\hspace{-0.6cm}We shall now show that {\it when a transverse Abelian configuration $A$ that lies outside the fundamental modular region $A \notin \Lambda$ is gauge transformed $A' = {^U}A$ (by a non-Abelian gauge transformation $U$) to the fundamental modular region, $A' \in \Lambda$, where $\Lambda$ is the set of absolute minimum of $\|{^U}A\|^2$ on every gauge orbit, then $A'$ lies on the common boundary of the fundamental modular region $\Lambda$ and the Gribov region $\Omega$, $A' \in \p\Lambda \cap \p\Omega$} \cite{Greensite:2004ke}.  This is also illustrated in Fig. \ref{commonboundary}.\\
\\
To prove the statement we observe first that the equation, $D_\mu(A') \omega' = 0$, follows from \eqref{degenerate} by gauge invariance, where $\omega' = U\omega U^\dag$ is now an $x$-dependent gauge transformation $\p_\mu \omega' \neq 0$.  From this we have immediately
\begin{equation}\label{2zhorizon}
\p_\mu D_\mu(A') \omega' = 0\;,
\end{equation}
where the operator on the left is recognized as the Faddeev-Popov operator.  The existence of an $x$-dependent solution $\omega'$ to \eqref{2zhorizon} is the defining condition\footnote{The condition that $\omega$ be $x$-dependent, $\p_\mu \omega \neq 0,$ is necessary because the minimal Landau gauge condition does not fix global gauge transformations $U = \exp\omega$, which have as infinitesimal generator $x$-independent $\omega$, with $\p_\mu \omega = 0$.  These satisfy \eqref{2zhorizon} for every transverse configuration $A$ (including those in the interior of $\Lambda$), when $\omega$ is $x$-independent, $\p_\mu\omega = 0$, for we have $\p_\mu D_\mu(A)\omega = D_\mu(A) \p_\mu \omega = 0$.} for a configuration $A' \in \Omega$ to lie on the Gribov horizon $A' \in \p\Omega$, and moreover we have $A' \in \Lambda \subset \Omega$, so we conclude that $A' \in \p\Omega$.  Furthermore since $A' \in \p\Omega$ and $A' \in \Lambda$ and $\Lambda$ is included in $\Omega$, $\Lambda \subset \Omega$, it follows that $A'$ necessarily also lies on the boundary of $\Lambda$, $A \in \p\Lambda$.  Thus it lies on the common boundary $A' \in \p\Lambda \cap \p\Omega$. QED.\footnote{Note that relative minima of the minimizing functional $F_A(U) = \|{^U}A\|^2$ for degenerate gauge orbits occur on the boundary $\p\Omega$ by the argument given above for absolute minima.}\\
\\
This result is interesting because the Gribov horizon $\p \Omega$ arises as an artifact of gauge fixing in the minimal Landau gauge, whereas the degenerate gauge orbits have a geometrical significance.  On the lattice there are also center vortex configurations where, for $SU(2)$, all link variables have the value $\pm 1$.  There is a confinement scenario in the maximal center gauge (as there is for the maximal Abelian gauge)  according to which the dominant configurations in the maximal center gauge are center vortex configurations.  Center vortex and Abelian configuration both lie on degenerate gauge orbits: for Abelian configurations the number of missing dimensions is the rank of the gauge group, whereas for center vortex configurations in $SU(N)$ there are $N^2-1$.  When the missing dimension is greater than 1, then degenerate configurations are singular points of the Gribov horizon of wedge or conical type \cite{Greensite:2004ur}.  The proof in \cite{Greensite:2004ur} was presented for lattice gauge theory, but the argument carries over to continuum gauge theory.  The argument given above that, in minimal Landau gauge, degenerate gauge orbits intersect the common boundary, $\p\Lambda \cap \p\Omega$, applies to both Abelian and center-vortex configurations.  Thus {\it the hypothesis that abelian configurations (center vortex configurations) dominate the functional integral in the maximal Abelian gauge (maximal center gauge) is compatible with the hypothesis that configurations on the common boundary of the fundamental modular region and of the Gribov region, $\p\Lambda \cap \p\Omega$, dominate the functional integral in the minimal or absolute Landau gauge.}  Since the present state of QCD is a patchwork of different confinement scenarios, it is gratifying that the confinement scenario in center vortex gauge or maximal Abelian gauge is compatible with the scenario of dominance of configurations on the Gribov horizon.  {\it All three scenarios are compatible with a gauge-invariant scenario of dominance by degenerate gauge orbits.}


\section{Semi classical solution of Gribov}
\subsection{The no-pole condition}
Gribov was the first one to try to restrict the region of integration to the Gribov region \cite{Gribov:1977wm,Sobreiro:2004us}, which was done in a semi-classical way. He restricted the generating functional to the Gribov region by introducing a factor $V(\Omega)$ in expression \eqref{genz5},
\begin{eqnarray}\label{ZJ}
Z(J) &=& \int_\Omega [\d A]   \exp \left[- S_\YM   \right] \nonumber\\
&=& \int [\d A][\d c] [\d \overline c]    V(\Omega)  \delta (\p A )   \exp \left[- S_\YM  -   \int \d x  \overline c^a (x)  \p_\mu  D_\mu^{ab}  c^b (x)  \right] \;,
\end{eqnarray}
whereby we are working in the Landau gauge,  $\delta (\p A )$. Now the question is how to determine this factor $V(\Omega)$. One can see that there is a close relationship between the ghost sector and the Faddeev-Popov determinant, which is clear from calculating the exact ghost propagator. For this, we start from expression \eqref{ghostapp}
\begin{eqnarray}
	I &=& \int [\d c][\d \overline{c}] \exp\left[ \int \d^d x \d^d y \ \overline{c}_a(x) A_{ab} (x,y)c_b(y) + \int \d^d x \ (J^{a}_c(x) c_a(x) + \overline{c}_a(x) J_{\overline{c}}^a (x)) \right]  \nonumber \\
	&=&C \det A \exp\left[  -\int \d^d x \d^d y \   J_c^a(x) A^{-1}_{ab}(x,y) J_{\overline{c}}^b(y)\right]\;,
\end{eqnarray}
where in our case:
\begin{eqnarray}
A_{ab}(x,y) &=& -\p_\mu D_\mu^{ab} \delta(x-y)\;.
\end{eqnarray}
From this we can calculate the ghost propagator,
\begin{eqnarray}
\braket{ \overline c_a(x)  c_b (y)}_c  &=&  \frac{\delta }{\delta \tilde J_{ c }^b (y)} \frac{\delta }{\delta \tilde J_{\overline c}^a(x)} Z\nonumber\\
&=&  \int [\d A]  V(\Omega) \delta(\partial_{\mu}A_{\mu}^{a})\det(-\partial_{\mu}D_{\mu}^{ab})  A_{ab}^{-1}(x,y) \e^{-S_{YM}}\;.
\end{eqnarray}
Taking the Fourier transform and keeping in mind that we have conservation of momentum
\begin{equation}\label{ghosthorbla}
\braket{ \overline c_a(p)  c_b (-p)}_c   =  \int [\d A]  V(\Omega) \delta(\partial_{\mu}A_{\mu}^{a})\det(-\partial_{\mu}D_{\mu}^{ab})\left( \int \d (x-y) \e^{\ii p (x-y)} A_{ab}^{-1}(x,y) \right)\e^{-S_{YM}}\;,
\end{equation}
we can compare this expression with the one loop renormalization improved ghost propagator starting from the Faddeev-Popov action,
\begin{eqnarray}\label{ghosthor}
\braket{\overline c_a(p)  c_b (k)}_c  &=&   \delta(p + k) \delta^{ab} \mathcal G(k^2) \nonumber\\
\mathcal G(k^2) &=& \underbrace{\frac{1}{k^2}}_{\mathcal P_1} \underbrace{ \frac{1}{ \left( 1 - \frac{11 g^2 N}{48 \pi^2} \ln \frac{\Lambda^2}{k^2} \right)^{\frac{9}{44}}}}_{\mathcal P_2}\;.
\end{eqnarray}
From this expression, we can make some interesting observation. Firstly, for large momentum $k^2$ we are within the Gribov region $\Omega$, as perturbation theory should work there. Indeed, for large $k^2$, $\mathcal G(k^2) \approx 1/ (k^2 \ln \frac{\Lambda}{k^2} )$, which is the perturbative result. Secondly, we notice this expression to have two poles: one pole at $k^2 = 0$ and one pole at $k^2 = \Lambda^2 \exp \left(- \frac{1}{g^2} \frac{48 \pi^2}{11 N} \right)$. The first pole indicates that for $k^2 \approx 0$, we are approaching a horizon, see expression \eqref{ghosthor}. As for all $k^2$, $\mathcal P_1$ is always positive, we stay inside the Gribov region. The second part of the ghost propagator $\mathcal P_2$ is not always positive for all $k^2$. For $k^2 < \Lambda^2 \exp \left(- \frac{1}{g^2} \frac{48 \pi^2}{11 N} \right)$, $\mathcal P_2$ becomes complex, indicating that we have left the Gribov region. Therefore, $V(\Omega)$ should make it impossible for a singularity to exist except at $k^2 = 0$.\\
\\
From these observations, we can construct the no-pole condition. For this, we shall calculate  $\mathcal G (k^2, A)_{ab}$ where the gluon field is considered as an external field. This comes down to calculating $\det(-\partial_{\mu}D_{\mu}^{ab})$  $ \times A^{-1}(x,y)$ from expression \eqref{ghosthorbla}, i.e. we shall calculate the following diagrams:

\begin{figure}[H]
\begin{center}
\includegraphics[width=16cm]{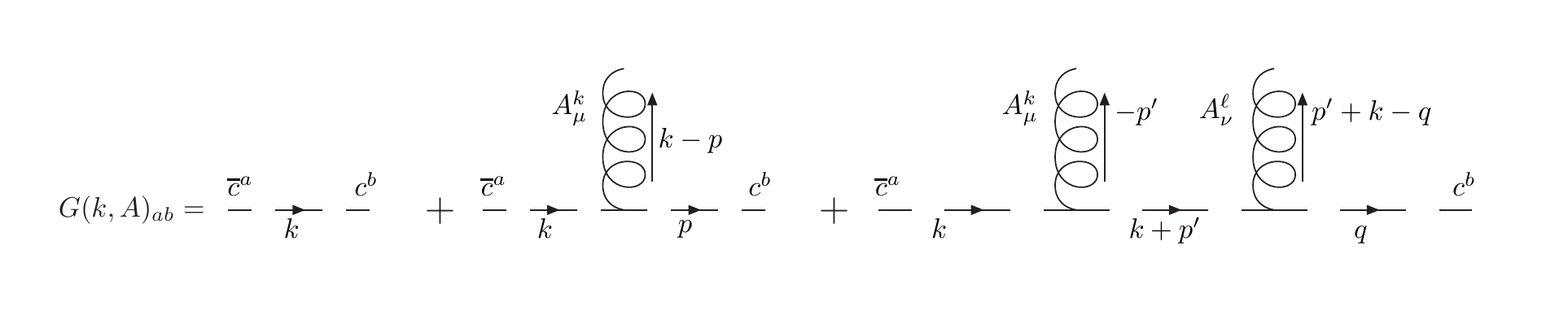}
\caption{The ghost propagator with external field to second order.}\label{ghost}
\end{center}
\end{figure}

\noindent In momentum space, these three diagrams are given by\footnote{The Feynman rule for the ghost-ghost gluon vertex is given by $\ii k_\mu f_{akb}$ with $akb$ resp. from $\overline c$, $A$, and $c$ whereby the outgoing momentum $k_\mu$ stems from $\overline c$.}
\begin{eqnarray}\label{2.66}
I_1 &=& \delta^{ab} (2\pi)^d \delta(k-q) \frac{1}{k^2}  \nonumber\\
I_2 &=& g\frac{1}{k^2}  \frac{1}{p^2} f_{a k b}\  \ii p_\mu    A_\mu^k (k -p) \nonumber\\
I_3 &=& g^2\int \frac{\d^d p'}{(2\pi)^d} \frac{1}{k^2} \frac{1}{(p' + k)^2} \frac{1}{q^2} f_{a k c}\ \ii ( p' + k)_\mu    A_\mu^k ( -p')  f_{c \ell b}\ \ii q_\nu A_\nu^\ell (p' + k - q)\;.
\end{eqnarray}
In fact, this is all we can say about these diagrams, unless we take into account that after determining $V(\Omega)$, which shall be a function of the external gluon field, we shall always need to integrate over $A$. This means that the gluon lines are connected, rendering the second diagram to be equal to zero. For the third diagram, the incoming momentum $k$ shall equal the outcoming momentum $q$. Formally, we can therefore rewrite the third diagram as
\begin{eqnarray}
I_3 &=& - g^2\frac{\delta(k-q) (2\pi)^d}{V}  \frac{1}{k^4} f_{a k c}  f_{c \ell b} \int \frac{\d^d p'}{(2\pi)^d} \frac{ k_\nu (p'+ k)_\mu }{(p' + k)^2}      A_\mu^k ( -p')  A_\nu^\ell (p')\;,
\end{eqnarray}
whereby we have introduced the infinite volume factor $V$, to maintain the right dimensionality. Moreover, we also know that the color indices  $k = \ell$. In order to calculate the correct prefactor, we take the sum over the color factors, see formula \eqref{liestructure}
\begin{eqnarray}\label{coloraway}
\mathcal G (k^2, A) &=& \frac{1}{N^2 -1}\delta_{ab} \mathcal G (k^2, A)_{ab}=  \frac{1}{k^2} + \frac{1}{V}\frac{1}{k^4} \frac{N g^2}{N^2 - 1} \int\frac{ \d^d q}{(2 \pi)^d} A_\mu^\ell(-q) A_\nu^{\ell} (q)  \frac{(k-q)_\mu  q_\nu}{(k-q)^2} \nonumber\\
&=& \frac{1}{k^2}\left( 1 + \sigma(k, A) \right)\;,
\end{eqnarray}
whereby
\begin{eqnarray}
\sigma(k,A) &=& \frac{1}{V}\frac{1}{k^2} \frac{Ng^2}{N^2 - 1} \int\frac{ \d^d q}{(2 \pi)^d} A_\mu^\ell(-q) A_\nu^{\ell} (q) \frac{ (k-q)_\mu  k_\nu }{(k-q)^2}\;.
\end{eqnarray}
Now we can rewrite this
\begin{eqnarray}
\mathcal G (k^2, A) &\approx&  \frac{1}{k^2}\frac{1}{ 1 - \sigma(k, A) }\;,
\end{eqnarray}
as in this way we are considering the inverse, or the 1PI diagram. This inverse contains more information as we are in fact resumming an infinite tower of Feynmandiagrams. The condition that the Faddeev-Popov operator has no zero modes, reduces to the requirement that
\begin{eqnarray}\label{condition}
\sigma(k,A) < 1\;.
\end{eqnarray}
We look into this requirement a bit more. As we are working in the Landau gauge, $q_\mu A_\mu(q) = 0$, and thus $A_\mu^\ell A_\nu^\ell$ is transverse,
\begin{eqnarray}
A_\mu^\ell(-q) A_\nu^{\ell}(q) &=& \omega(A) \left( \delta_{\mu\nu} - \frac{q_\mu q_\nu}{q^2} \right) = \omega (A) P_{\mu\nu}\;,
\end{eqnarray}
moreover, multiplying with $\delta_{\mu\nu}$, we find that $\omega(A) = \frac{1}{d-1}A_\mu^\ell A_\mu^\ell $, with $d$ the number of dimensions. Therefore, we can simplify $\sigma$,
\begin{eqnarray}\label{2.67}
\sigma(k,A) &=& \frac{1}{V}\frac{1}{d-1} \frac{Ng^2}{N^2 - 1} \frac{k_\mu  k_\nu}{k^2} \int\frac{ \d^d q}{(2 \pi)^d} A_\alpha^\ell(-q) A_\alpha^{\ell} (q) \frac{ 1 }{(k-q)^2} P_{\mu\nu} \;.
\end{eqnarray}
As it is possible to prove that $\sigma(k,A)$ decreases with increasing $k^2$, see appendix \ref{sigma}, where we have used the fact that $A_\alpha^\ell(-q) A_\alpha^{\ell}(q)$ is positive, condition \eqref{condition} becomes,
\begin{eqnarray}\label{nopolecondition}
\sigma(0, A) < 1\;.
\end{eqnarray}
Taking the limit $k^2 \to 0$ in $\sigma(k,A)$ yields,
\begin{eqnarray}\label{nopole}
\sigma(0,A) &=& \frac{1}{V}\frac{1}{d-1} \frac{Ng^2}{N^2 - 1} \lim_{k^2 \to 0} \frac{k_\mu  k_\nu}{k^2} \frac{d-1}{d} \delta_{\mu\nu}  \int\frac{ \d^d q}{(2 \pi)^d} A_\alpha^\ell(-q) A_\alpha^{\ell} (q) \frac{ 1 }{q^2} \nonumber\\
&=& \frac{1}{V} \frac{1}{d} \frac{Ng^2}{N^2 - 1} \int\frac{ \d^d q}{(2 \pi)^4} A_\alpha^\ell(-q) A_\alpha^{\ell} (q) \frac{ 1 }{q^2}\;,
\end{eqnarray}
whereby we used the fact that $\int \d^dq f(q^2) q_\mu q_\nu/q^2 = 1/d\  \delta_{\mu\nu} \int \d^d q f(q^2)$. \\
\\
In summary, the no-pole condition is given by
\begin{eqnarray}
V(\Omega) &=& \theta (1 - \sigma(0,A))\;,
\end{eqnarray}
with $\sigma(0,A)$ given by expression \eqref{nopole}, or thus, using the Heaviside function,
\begin{eqnarray}\label{thetareplace}
V(\Omega) &=&  \int_{- \ii \infty + \epsilon}^{+ \ii \infty + \epsilon}\frac{\d \beta}{2\pi \ii \beta} \e^{\beta (1 - \sigma(0,A))} \;,
\end{eqnarray}
we can insert this into the path integral \eqref{ZJ}.

\subsection{The gluon and the ghost propagator\label{gribovgluonghost}}
\subsubsection{The gluon propagator}
Our goal is calculate the gluon propagator in Fourier space
\begin{eqnarray}
\Braket{A_\mu^a (k) A_\nu^b(p)}\;,
\end{eqnarray}
at lowest order including the restricting to the Gribov region. We start from the path integral \eqref{ZJ}, while introducing appropriate sources for the gluons,
\begin{equation*}
Z(J) =  \mathcal N \int \frac{\d \beta}{2\pi \ii \beta} \int [\d A]    \e^{\beta (1 - \sigma(0,A))}    \exp -\left[ S_\YM^\quadr   + \int \d^d x \frac{1}{2 \alpha} (\p_\mu A_\mu)^2 + \int \d^d x A_\mu^a(x) J_\mu^a (x) \right]\;,
\end{equation*}
with $N= Z^{-1}(J=0)$. We do not need to take into account the integration over $[\d c][\d \overline c]$ as we are only calculating the free gluon propagator, also we only need the free part of $S_\YM$.  Translating this in Fourier space, we have that
\begin{multline*}
 S_\YM^\quadr   + \int \d^d x \frac{1}{2 \alpha} (\p_\mu A_\mu)^2 + \int \d^d x A_\mu^a(x) J_\mu^a (x) \\
 =  \int \frac{\d^d k}{(2\pi)^d} \left( \frac{1}{2} A_\mu^a(k)  \left(\delta_{\mu\nu} k^2 + \left(\frac{1}{\alpha} - 1 \right)k_\mu k_\nu  \right) A_\nu^a(-k)   -  A_\mu^a(k) J_\mu^a (-k) \right) \;,
\end{multline*}
or thus,
\begin{multline}
\Braket{A_\mu^a (k) A_\nu^b(p)} \\ \left. = \frac{\delta^2 }{\delta J_\mu^a(-k) \delta J_\nu^b(-p)}   \int \frac{\d \beta\e^{\beta}}{2\pi \ii \beta}   \int [\d A]  \e^{ -\int \frac{\d^d k}{(2\pi)^d} \frac{1}{2} A_\mu^a(k)   K_{\mu \nu}^{ab} (k) A_\nu^b(-k) +    \int \frac{\d^d k}{(2\pi)^d}  A_\mu^a(k) J_\mu^a (-k) } \right|_{J=0} \;,
\end{multline}
whereby
\begin{eqnarray}
K_{\mu \nu}^{ab} (k) &=&\delta^{ab} \left( \beta\frac{1}{V} \frac{2}{d} \frac{N g^2}{N^2 - 1} \delta_{\mu\nu} \frac{1}{k^2} + \delta_{\mu\nu} k^2 + \left(\frac{1}{\alpha} - 1 \right)k_\mu k_\nu \right)\;,
\end{eqnarray}
also includes the part stemming from $\sigma(0,A)$ in expression \eqref{nopole}. Now invoking the Fourier transform of \eqref{gauss1}, we find
\begin{equation}
\Braket{A_\mu^a (k) A_\nu^b(p)} = \delta(k+p) \mathcal N  \int \frac{\d \beta\e^{\beta}}{2\pi \ii \beta}  (\det  K_{\mu \nu}^{ab} )^{-1/2}  (K_{\mu \nu}^{ab})^{-1} (k)\;.
\end{equation}
This determinant has been worked out in appendix \ref{sigma2}, resulting in
\begin{eqnarray}
 (\det  K_{\mu \nu}^{ab} )^{-1/2}  &=&  \exp\left[ -\frac{d-1}{2} (N^2 - 1) V \int \frac{\d^d q}{(2\pi)^d} \ln \left( q^2 + \frac{\beta N g^2}{N^2 - 1} \frac{2}{d V}\frac{1}{q^2} \right)  \right]\;,
\end{eqnarray}
and thus
\begin{align}
\Braket{A_\mu^a (k) A_\nu^b(p)} =& \delta(k+p)\mathcal N \int \frac{\d \beta}{2\pi \ii }  \e^{f(\beta)} (K_{\mu \nu}^{ab})^{-1} (k)\;, \nonumber\\
f(\beta) =& \beta - \ln \beta  -\frac{d-1}{2} (N^2 - 1) V \int \frac{\d^d q}{(2\pi)^d} \ln \left( q^2 + \frac{\beta N g^2 }{N^2 - 1} \frac{2}{d V}\frac{1}{q^2} \right)\;.
\end{align}
As we assume $(K_{\mu \nu}^{ab})^{-1} (k)$ not to be oscillating too much, we apply the method of steepest descent\footnote{The infinite parameter to apply the method of steepest descent is the Euclidean volume $V$, as is illustrated by the explicit calculation \eqref{pr2}.} to evaluate the integral over $\beta$,
\begin{eqnarray}
\Braket{A_\mu^a (k) A_\nu^b(p)} &=& \delta(k+p)\mathcal N' \e^{f(\beta_0)} \left. (K_{\mu \nu}^{ab})^{-1} (k) \right|_{\beta = \beta_0}\;,
\end{eqnarray}
whereby we have absorbed $2\pi \ii$ into $\mathcal N$. $\beta_0$ is the minimum of $f(\beta)$, i.e.
\begin{eqnarray}\label{betaverwaar}
&&f'(\beta_0) =  0 \nonumber\\
&\Rightarrow& 1 =\frac{1}{\beta_0}  +\frac{d-1}{d}N g^2 \int \frac{\d^d q}{(2\pi)^d} \frac{1}{ \left( q^4 + \frac{\beta_0 N g^2}{N^2 - 1}  \frac{2}{d V} \right) } \;.
\end{eqnarray}
We define the Gribov mass,
\begin{eqnarray}
\gamma^4 &=& \frac{\beta_0 N }{N^2 - 1} \frac{2}{d V} g^2\;,
\end{eqnarray}
which serves as an infrared regulating parameter in the integral. As in fact, $V$ is equal to infinity, in order to have a finite $\gamma$, $\beta_0 \sim V$. Therefore, $1/\beta_0$ can be neglected and we obtain the following gap equation,
\begin{eqnarray}\label{gapequation}
1 = \frac{d-1}{d}N g^2 \int \frac{\d^d q}{(2\pi)^d} \frac{1}{ \left( q^4 + \gamma^4 \right) } \;,
\end{eqnarray}
which shall determine $\gamma^4$. Now, we only have to calculate the inverse of
\begin{eqnarray}
(K_{\mu \nu}^{ab})(k) &=&\delta^{ab} \left( \gamma^4 \delta_{\mu\nu} \frac{1}{k^2} + \delta_{\mu\nu} k^2 + \left(\frac{1}{\alpha} - 1 \right)k_\mu k_\nu \right)\;,
\end{eqnarray}
whereby we have set $\beta = \beta_0$ which yields,
\begin{eqnarray}
(K_{\mu \nu}^{ab})(k)^{-1} &=&\delta^{ab} \left( \frac{k^2}{k^4 + \gamma^4} P_{\mu\nu}(k) + \alpha \frac{k^2}{\alpha \gamma^4 + k^4} \frac{k_\mu k_\mu}{k^2} \right) \;,
\end{eqnarray}
as one can check by calculating $(K_{\mu \nu}^{ab})^{-1} (k) (K_{\nu\kappa}^{bc}) (k) = \delta^{ac} \delta^{\mu\kappa}$. For $\alpha = 0$, the inverse becomes tranverse and the gluon propagator is given by
\begin{eqnarray}\label{gluonprop}
\Braket{A_\mu^a (k) A_\nu^b(p)} &=& \delta(k+p) \delta^{ab} \frac{k^2}{k^4 + \gamma^4} P_{\mu\nu}(k) \;,
\end{eqnarray}
as $\mathcal N'$ shall cancel $\e^{f(\beta_0)}$ due to normalization.

\subsubsection{The ghost propagator}
Now that we have found the gluon propagator, we can calculate the ghost propagator. In fact, this comes down to connecting the gluon legs in expression \eqref{2.66}. We easily find, as in \eqref{2.67},
\begin{eqnarray}\label{ghostpropagator}
\mathcal G^{ab} (k^2) &=& \delta^{ab} \frac{1}{k^2}\frac{1}{ 1 - \sigma(k) }\;, \nonumber\\
\sigma(k) &=& Ng^2  \frac{k_\mu  k_\nu}{k^2} \int\frac{ \d^d q}{(2 \pi)^d} \frac{q^2}{q^4 + \gamma^4} \frac{ 1 }{(k-q)^2} \left( \delta_{\mu\nu} - \frac{q_\mu q_\nu}{q^2}\right)\;.
\end{eqnarray}
To calculate $1-\sigma(k)$, we rewrite the gap equation \eqref{gapequation} as
\begin{eqnarray}
1 = \frac{k_\mu k_\nu}{k^2} N g^2 \int \frac{\d^d q}{(2\pi)^d} \frac{1}{ \left( q^4 + \gamma^4 \right) } \left( \delta_{\mu\nu} - \frac{q_\mu q_\nu}{q^2}\right)\;,
\end{eqnarray}
and thus we write unity in a complicated way,
\begin{equation*}
1-\sigma(k) = \frac{k_\mu k_\nu}{k^2} N g^2 \int \frac{\d^d q}{(2\pi)^d} \frac{1}{ \left( q^4 + \gamma^4 \right) } \left( \delta_{\mu\nu} - \frac{q_\mu q_\nu}{q^2}\right) \left( 1- \frac{q^2}{(k-q)^2}\right) = \frac{k_\mu k_\nu}{k^2} N g^2 R_{\mu\nu} (k)\;.
\end{equation*}
To investigate the infrared behavior, we expand this integral for small $k^2$, whereby up to order $k^2$
\begin{eqnarray}
\left( 1- \frac{q^2}{(k-q)^2}\right) &=& 1 - \frac{1}{\frac{k^2}{q^2} - \frac{2 k_\mu q_\mu}{q^2} + 1} = \frac{k^2}{q^2} - \frac{2 k_\mu q_\mu}{q^2} - 4 \left( \frac{k_\mu q_\mu}{q^2}\right)^2\;,
\end{eqnarray}
and thus we can split $R_{\mu\nu}$ in three parts. The first part is given by
\begin{equation}
R_{\mu\nu}^1 (k) = k^2 \int \frac{\d^d q}{(2\pi)^d} \frac{1}{q^2 \left( q^4 + \gamma^4 \right) } \left( \delta_{\mu\nu} - \frac{q_\mu q_\nu}{q^2}\right) = \frac{d-1}{d}\delta_{\mu\nu} k^2   I_\gamma \;,
\end{equation}
whereby $I_\gamma = \int \frac{\d^d q}{(2\pi)^d} \frac{1}{q^2 \left( q^4 + \gamma^4 \right) } $ is a number depending on $\gamma$.  The second part is zero, at is it odd in $q$, and the third part is given by
\begin{equation}
R_{\mu\nu}^3 (k) = -4 \delta_{\mu\nu} k_\alpha k_\beta \int \frac{\d^d q}{(2\pi)^d} \frac{q_\alpha q_\beta}{q^4 \left( q^4 + \gamma^4 \right) } + 4 k_\alpha k_\beta \int \frac{\d^d q}{(2\pi)^d} \frac{q_\alpha q_\beta q_\mu q_\nu}{q^6 \left( q^4 + \gamma^4 \right) } \;.
\end{equation}
The first term of this expression is given by,
\begin{equation*}
 - 4 \delta_{\mu\nu} \frac{\delta^{\alpha \beta}}{d} k_\alpha k_\beta \int \frac{\d^d q}{(2\pi)^d} \frac{1}{q^2 \left( q^4 + \gamma^4 \right) } = -\frac{4}{d}  \delta_{\mu\nu} k^2 I_\gamma \;,
\end{equation*}
while the second term is given by
\begin{multline*}
 4 k_\alpha k_\beta (\delta_{\alpha \beta} \delta_{\mu\nu} +  \delta_{\alpha \mu} \delta_{\beta\nu} + \delta_{\alpha \nu} \delta_{\beta\mu}) \frac{1}{d^2 + 2d} \int \frac{\d^d q}{(2\pi)^d} \frac{1}{q^2 \left( q^4 + \gamma^4 \right) }  = 4 (k^2 \delta_{\mu\nu} + 2 k_\mu k_\nu)  \frac{1}{d^2 + 2d} I_\gamma\;.
\end{multline*}
Therefore,
\begin{eqnarray}
R_{\mu\nu}^3 (k) &=&  4\left( - \frac{1}{d}  \delta_{\mu\nu} k^2  + (k^2 \delta_{\mu\nu} + 2 k_\mu k_\nu)  \frac{1}{d^2 + 2d} \right)I_\gamma\;.
\end{eqnarray}
Taking all results together, we obtain,
\begin{eqnarray}
1-\sigma(k) &=& N g^2\frac{k_\mu k_\nu}{k^2}\left[ \frac{d-1}{d}\delta_{\mu\nu} k^2     +  4\left( - \frac{1}{d}  \delta_{\mu\nu} k^2  + (k^2 \delta_{\mu\nu} + 2 k_\mu k_\nu)  \frac{1}{d^2 + 2d} \right) \right]I_\gamma \nonumber\\
&=&  N g^2k^2 \left[ \frac{d-1}{d}    +  4\left( - \frac{1}{d}   +   \frac{3}{d^2 + 2d} \right) \right]I_\gamma \nonumber\\
&=& N g^2 k^2 \frac{d^2 - 3 d + 2}{d^2 +2d} I_\gamma \;,
\end{eqnarray}
or thus, the ghost propagator is enhanced,
\begin{eqnarray}
\mathcal G^{ab} (k^2) &=& \delta^{ab} \frac{1}{k^4}   \frac{d^2 +2d}{d^2 - 3 d + 2} \frac{1}{N g^2 I_\gamma}\;.
\end{eqnarray}
As an example, for $d = 4$, we find easily that $I_\gamma^{d=4} = 1/(32\pi^2 \gamma^2)$ and thus
\begin{eqnarray}
\left.\mathcal G^{ab} (k^2)\right|_{d=4} &=& \delta^{ab} \frac{1}{k^4}    \frac{128\pi^2 \gamma^2}{Ng^2 }\;.
\end{eqnarray}
Also in three dimension we find enhancement of the ghost. In two dimensions, the calculations are not so straightforward as there is a problem with switching the limit and the integration. We refer to \cite{Dudal:2008xd} for more details on this. The conclusion however remains the same, also in  $2$ dimensions the ghost propagator is enhanced.\\
\\
In fact, looking at the calculations, $\sigma = 1$ means that the $\theta$-function has become a $\delta$-function. This is due to the fact that we could neglect $1/\beta_0$ in expression \eqref{betaverwaar} as we are working in an infinite volume $V \to \infty$. In other words, by limiting to the Gribov region, the ghost propagator has an extra pole, which indicates that the region close to the boundary has an important effect on the ghost propagator.

\section{The local renormalizable action}
After the publication of Gribov, his result was generalized to all orders by constructing a local renormalizable action \cite{Zwanziger:1989mf,Zwanziger:1992qr} which implements the restriction to the Gribov region, and which, following custom, we shall call the GZ action. In this section we shall first analyze a toy model to demonstrate how the GZ action was obtained \cite{Zwanziger:1989}.

\subsection{A toy model}

We start with the simple quadratic action for a real scalar field in Euclidian dimension $d$,
\begin{eqnarray}
S &=& \int \d^d x\frac{1}{2} (\p A(x))^2\;,
\end{eqnarray}
whereby we omit color and Lorentz indices. We assume the Gribov region is contained within the ``ellipsoid" in A-space,
\beq
H(A) \equiv \int \d^d x\frac{1}{2} A(x) [(- \p^2)^{-1}A](x) = cV.
\eeq
It will be essential that the action $S(A)$ and the ``horizon function" $H(A)$ are both integrals over a density and are thus bulk quantities, of order of the Euclidean volume $V$.  Consequently the constant $c$ remains finite in the limit $V \to \infty$.
We work at finite but large volume $V$, and use the fourier transform
\beq
A(x) = V^{-1} \sum_k A_k \exp(i k \cdot x),
\eeq
\beq
J(x) = V^{-1} \sum_k J_k \exp(i k \cdot x),
\eeq
where $\int J(x)A(x) = V^{-1}\sum_k J_{-k} A_k $ will be a source term.  This gives
\beq
S(A) = \frac{1}{2V}\sum_k A_k k^2 A_{-k},
\eeq
and the Gribov region is bounded by
\begin{eqnarray}\label{ellips}
\underbrace{\frac{1}{2V}\sum_k \frac{A_k A_{-k}}{k^2}}_{H(A)} = cV  \;.
\end{eqnarray}
which is seen to be an ellipsoid in the infinite dimensional space of the $A_k$.  To restrict to the Gribov region, we need to consider the following generating functional
\begin{eqnarray}\label{gentheta}
Z &=& \int [\d A] \ \theta(cV - H(A)) \ \e^{-S}\;,
\end{eqnarray}
as the $\theta$-function assures that $H(A) < cV$.\\
\\
If we introduce $y_k = A_k/k$ the ellipsoid becomes a hypersphere and, as is known for hyperspheres, the volume gets more and more concentrated on the surface as the dimension grows.  (Indeed in a space of dimension $N$ the volume element in radius is given by $r^{N-1}dr$, and the (normalized) integral over the interior of a sphere $r \leq R$, given by $\int_0^\infty \theta(R - r) r^{N-1}dr$, approaches$\int_0^\infty \delta(R - r) r^{N-1}dr$, where the $\theta$-function is replaced by the $\delta$-function as the number of dimensions $N$ becomes infinite, $N \to \infty$.)   Therefore, we can replace the $\theta$-function with a $\delta$-function and the generating functional \eqref{gentheta} becomes
\begin{eqnarray}\label{gendelta}
Z &=& \int [\d A] \delta(cV - H(A)) \e^{-S}\;.
\end{eqnarray}
Let us remark that also Gribov already noticed this, see the end of the previous section.\\
\\
We use the formula
\begin{eqnarray}
\delta (x - y) &=& \int_{-\infty}^{\infty} \frac{\d t}{2\pi} \ \e^{\ii (x-y) t} \;,
\end{eqnarray}
so we find
\begin{eqnarray}\label{wick}
Z &=&\int_{-\infty}^\infty \frac{\d t}{2\pi} \int [\d A]  \e^{-S} \e^{(\ii t +\beta) (cV - H(A))} \nonumber\\
&=&  \int_{-\infty}^\infty \d t \e^{-G(\ii t +\beta)} \;,
\end{eqnarray}
where $G(\beta) \equiv -\ln \int [\d A] \e^{\beta(cV-H(A))} \e^{-S}$.
Here we have distorted the contour of integration $\ii t \to\ii t + \beta$ for real $\beta > 0$, which improves the convergence because $H(A) > 0$ is positive.  A saddle point approximation for the $t$-integration is now justified because $G(\beta) = O(V)$ is a large quantity
\begin{eqnarray}
Z &\approx&  \e^{-G(\beta^*)}\;,
\end{eqnarray}
whereby $\beta^*$ is the solution of
\begin{eqnarray}\label{determiningbeta}
G'(\beta) &=& 0 \nonumber\\
cV&=& \frac{\int [\d A] H(A) \e^{\beta(cV-H(A))} \e^{-S} }{\int [\d A] \e^{\beta(cV-H(A))} \e^{-S}}  \nonumber\\
cV&=& \frac{\int [\d A] H(A) \e^{-\beta H(A) - S} }{\int [\d A] \e^{-\beta H(A) - S} } \nonumber\\
cV &=& \Braket{H(A)}_{\beta} \;.
\end{eqnarray}
We thus obtain a generating functional of Boltzmann type,
\begin{eqnarray}\label{partZ}
Z &=& \int [\d A] \e^{- [ \beta^*H(A) + S]} \;,
\end{eqnarray}
with $\beta^*$ determined by \eqref{determiningbeta}.  As in statistical mechanics, the micro-canonical ensemble defined by the ``horizon condition" $H(A) = cV$ has been replaced by the canonical ensemble \cite{munster} of Boltzmann type.\footnote{Here the horizon function $H(A)$ and $\beta^*$ are mathematically {\it analogous} to a Hamiltonian and inverse temperature in statistical mechanics, and should not be confused with a mechanical Hamiltonian and physical inverse temperature.}\\
\\
We may verify by explicit calculation that with the partition function of canonical type \eqref{partZ}, the horizon function $H(A)$ has zero (relative) variance in the infinite volume limit, so that it is equivalent to the micro-canonical ensemble.  More precisely we now show that $\langle H \rangle =  O(V)$, $\langle H^2 \rangle = O(V^2)$, and the variance $\delta H^2 \equiv \langle H^2 \rangle - \langle H \rangle^2 = O(V)$, is of order $V$, which is smaller than the mean-square by a volume factor, so the relative variance $\delta H^2/\langle H^2 \rangle = O(V^{-1})$ vanishes for $V \to \infty$.  This is the behavior of a generic bulk quantity in statistical mechanics.  We have
\begin{eqnarray}\label{pr1}
\braket{H(A)} &=&  \frac{1}{2 V}\sum_k \frac{1}{ k^2} \int [\d A] A_k A_{-k}  \exp\left[- \frac{1}{2V} \sum_p A_p \left( p^2 +  \frac{\beta}{p^2} \right) A_{-p}\right]    \nonumber\\
&=& \left. \frac{1}{2 V}\sum_k \frac{V^2}{k^2} \frac{\p^2}{\p J_k \p J_{-k} }  \int [\d A]   \exp\left[-\frac{1}{2 V}\sum_p A_p \left( p^2 +  \frac{\beta}{p^2} \right) A(-p)  +  \frac{1}{V}\sum_p A_p J_{-p}\right] \right|_{J =0}\nonumber\\
&=& \left. \frac{V}{2}\sum_k \frac{1}{ k^2} \frac{\p^2}{\p J_k \p J_{-k} }    \exp \left[ \frac{1}{2 V}\sum_p J_p \left( p^2 +  \frac{\beta}{p^2} \right)^{-1} J_{-p} \right] \right|_{J =0}\nonumber\\
&=&\frac{1}{2} \sum_k \frac{1}{k^4 + \beta} \nonumber\\
& \Rightarrow & \frac{V}{2} \int \frac{\d^d k}{(2\pi)^d} \frac{1}{k^4 + \beta} \;,
\end{eqnarray}
where we have written $\beta$ for $\beta^*$.  Here we used the Fourier transform of expression \eqref{gauss1} in the appendix, and $\Rightarrow$ means to leading order in (large) $V$, so the sum over $k$ may be replaced by an integral.  This shows that $\langle H \rangle =  O(V)$, as stated.  We now evaluate $\langle H^2(A)\rangle$,
\begin{align}\label{pr2}
\braket{H^2(A)} &=\frac{1}{4V^2}\left. \sum_{k,p} \frac{1}{k^2}\frac{1}{p^2} \frac{V^4\p^4}{\p J_k \p J_{-k} \p J_p \p J_{-p} }   \exp \left[ \frac{1}{2V}\sum_q J_q \left( \frac{q^2}{q^4 + \beta} \right) J_{-q} \right] \right|_{J =0}\nonumber\\
&=\frac{1}{4}\left. \sum_{k,p} \frac{1}{k^2}\frac{1}{p^2} \frac{\p^2}{\p J_k \p J_{-k} } \left[ V \frac{p^2}{p^4 + \beta} + \left(\frac{p^2}{p^4 + \beta}\right)^2 J_p J_{-p} \right] \exp \left[ \frac{1}{2V}\sum_q J_q \left(\frac{q^2}{q^4 + \beta}\right) J_{-q} \right] \right|_{J =0}\nonumber\\
&= \frac{1}{2}\sum_k \frac{1}{k^4 + \beta} \frac{1}{2}\sum_p \frac{1}{p^4 + \beta} + \frac{1}{2}\sum_p  \frac{1}{(p^2)^2}  \left(\frac{p^2}{p^4 + \beta}\right)^2 \nonumber\\
&\Rightarrow \braket{H(A)}^2 + \frac{V}{2} \int \frac{\d^d p}{(2\pi)^d} \frac{1}{(p^4 + \beta)^2} \;,
\end{align}
Thus we have $\braket{H^2(A)} = \braket{H(A)}^2 +O(V)$, and the relative variance $\delta H^2 / \braket{H^2(A)} = O(V^{-1})$, vanishes like $V^{-1}$ as asserted.  We have verified that in the thermodynamic limit the Boltzmann distribution \eqref{partZ} does behave like a $\delta$-function distribution.\\
\\

\subsection{The non-local GZ action}
\label{non-localGZaction}
To restrict the region of integration of the Yang-Mills action to the Gribov region, we need to consider the path integral,\footnote{In this section we give a more complete derivation of the results of \cite{Zwanziger:1989mf}.}
\begin{eqnarray}\label{startder}
Z &=& \int [\d A] \det\mathcal M(A) \e^{S_{\YM}} \theta( \lambda (A))\;,
\end{eqnarray}
where $\lambda (A)$ is the lowest eigenvalue of the Faddeev-Popov operator,
\begin{equation}
\mathcal M^{ab} = \mathcal M_0^{ab} + \mathcal M_1^{ab} =  - \p^2 \delta^{ab} + g f_{abc} A^c_\mu \p_\mu\;,
\end{equation}
whereby we are working on-shell, $\p_\mu A_\mu = 0$. By introducing the $\theta$-function, $\theta(\lambda(A))$, we insure that the lowest eigenvalue $\lambda(A)$ is always greater than zero.  Note that all constant vectors $\p_\mu \omega = 0$ are eigenvectors of the Faddeev-Popov operator, with zero eigenvalue $\mathcal M \omega = 0$. As these eigenvalues never become negative, we shall not consider these trivial eigenvectors and work in the space orthogonal to this trivial null-space.

\subsubsection{Degenerate perturbation theory}
In order to find the lowest lying (non-trivial) eigenvalue $\lambda (A)$, we shall apply perturbation theory whereby $\mathcal M_0^{ab}$ is the unperturbed operator. For the moment, we work in a finite periodic box of edge $L$, which shall approach infinity in the infinite volume limit.  The momentum eigenstates $\ket{\Psi^{(0)}_{\vec n s}}$: $\vec n \in \mathbb Z^d / \{ 0\}$, while $s$ runs over all the colors, $s = 1 , \ldots, N^2 -1$, are given in configuration space by
\begin{equation}\label{conf1}
\Braket{x, a|\Psi^{(0)}_{\vec n s}} = \delta^{as} \left( \frac{1}{L} \right)^\frac{d}{2} \exp{\left(\ii \frac{2\pi}{L} \vec n \cdot \vec x \right)}\;.
\end{equation}
They form a convenient basis of eigenvectors of the operator $\mathcal M_0$,
\beq
\mathcal M_0 \ket{\Psi^{(0)}_{\vec n s}} = \left(\frac{2\pi}{L}\right)^2 \vec n^2 \ket{\Psi^{(0)}_{\vec n s}}.
\eeq
We designate by $\vec n_0$ the lowest non-zero momenta, i.e.~$\vec n_0 = (0,\ldots,0, \pm 1, 0, \ldots, 0)$, with $\vec n_0^2 = 1$. Thus $\ket{\Psi^{(0)}_{\vec n_0 s}}$ represent the vectors belonging to the lowest eigenvalue\footnote{The lowest lying eigenvalue is of course zero, belonging to the constant vectors, but we are not considering these constant vectors anymore.}  $\lambda_{\vec n_0 s}^{(0)}$ which is given by
\begin{equation}
\lambda_{\vec n_0}^{(0)} = \left( \frac{2 \pi}{L} \right)^2\;.
\end{equation}
Notice that the space spanned by the vectors $\ket{\Psi^{(0)}_{\vec n_0 s}}$ is $2d(N^2 - 1)=T$ dimensional, so we must apply degenerate perturbation theory. We call this space $\mathcal H_0$
\begin{equation}\label{hho}
\mathcal H_0 = \text{span}(\ket{\Psi^{(0)}_{\vec n_0 s}}, s=1, \ldots, T)\;.
\end{equation}
The projector onto this space $\mathcal H_0$ is given by
\begin{eqnarray}\label{projector}
P_0 = \sum_{\vec n_0 s}  \ket{\Psi^{(0)}_{\vec n_0s}} \bra{\Psi^{(0)}_{\vec n_0s}}\;.
\end{eqnarray}
The other eigenvectors of $\mathcal M_0$, $\ket{\Psi^{(0)}_{\vec n s}}$ have corresponding eigenvalues $\lambda_{\vec n s}^{(0)}$ given by
\begin{equation}
\lambda_{\vec n s}^{(0)} = \left( \frac{2 \pi}{L} \right)^2 \vec n^2, \ \ \   \ \text{with} \ \vec n^2 > 1 \;.
\end{equation}
(The entire space can be decomposed into $\mathcal H_0 + \mathcal H_1 + \mathcal H_2 + \ldots$, whereby the $\mathcal H_n$ are defined as the spaces spanned by the vectors belonging to the same eigenvalue\footnote{We can order the eigenvalues by size, $\mathcal H_n$ belongs to the $(n+1)$th eigenvalue.}).\\
\\
Let us now switch on the perturbation $\mathcal M_1$. The $T$ degenerate eigenvalues $\lambda_0$ of $\mathcal M_0$ split up into $T$ different eigenvalues, $\lambda_1, ... \lambda_T$, of $\mathcal M$ which are its $T$ lowest (non-trivial) eigenvalues.  Within the degenerate subspace $\mathcal H_0$, with eigenvalue $\lambda_0$, any linear combination $\ket{\psi_{a}^{\prime (0)}} = \sum_b U_{ab} \ket{\psi_{b}^{(0)}}$ of the eigenvectors, $\ket{\psi_{b}^{(0)}}$, is also an eigenvector, where $U_{ab}$ is an arbitrary $T$-dimensional unitary matrix, and $a = (\vec n_0, s)$.  Consequently, a very small perturbation $\mathcal M_1$ causes a (large) finite change from the (arbitrarily chosen) eigenvectors $\ket{\psi_{a}^{(0)}}$, given in \eqref{conf1}, into some new $T$-dimensional basis of eigenvectors, and this change is non-perturbative.  Fortunately however a small perturbation causes only a small change to the $T$-dimensional degenerate subspace $\mathcal H_0$ itself, because the projector $P_0$ onto $\mathcal H_0$, given in \eqref{projector} is basis independent, and is thus adapted to any basis.  Indeed, if we change basis in $\mathcal H_0$, $\ket{\psi_{a}^{\prime (0)}} = \sum_b U_{ab} \ket{\psi_{b}^{(0)}}$, we have
\beq
P_0^\prime = \sum_a \ket{\psi_{a}^{\prime (0)}} \bra{\psi_{a}^{\prime (0)}} = \sum_{abc}U_{ab} \ket{\psi_{b}^{(0)}} U_{ac}^*\bra{\psi_{c}^{(0)}} = \sum_b \ket{\psi_{b}^{(0)}} \bra{\psi_{b}^{(0)}} = P_0.
\eeq
For this reason degenerate perturbation theory is done in 2 steps.  In the first step we make a transformation $S$, which will be calculated perturbatively, that provides a diagonalization of the Faddeev-Popov operator $\mathcal M$ to within a finite $T \times T$ dimensional matrix $\kappa$,
\beq
\label{similarity}
S^{-1} \mathcal M S = \kappa.
\eeq
Here $\kappa$ acts within the subspace $\mathcal H_0$ and satisfies
\beq
\label{projectkappa}
P_0 \kappa = \kappa P_0 = \kappa,
\eeq
and $S$ satisfies
\beq
\label{SPzero}
SP_0 = S.
\eeq
In the second step the diagonalization of $\mathcal M$ is completed by a simple $T \times T$ matrix diagonalization
\beq
U^{-1} \kappa U = \lambda_{\rm diag},
\eeq
where $\lambda_{\rm diag}$ is a $T \times T$-dimensional diagonal matrix that consists of the $T$ lowest (non-trivial) eigenvalues, $\lambda_1$ to $\lambda_T$ of $\mathcal M$, $\lambda_{\rm diag} \equiv \text{diag}(\lambda_1 ... \lambda_T)$, and $U_{ab}$ is a finite $T \times T$ unitary matrix that is calculated non-perturbatively.  For our purposes it will not be necessary to perform step 2 because it will be sufficient to make use of the property
\beq
\label{tracekappa}
\sum_{n = 1}^T \lambda_n = \Tr (U^{-1} \kappa U) = \Tr\kappa.
\eeq\\
\\
To make the first  step, we seek a transformation $S$, and a $T \times T$-dimensional matrix $\kappa$ that satisfy \eqref{similarity}, or
\begin{equation}\label{ms}
\mathcal M S = S \kappa \;,
\end{equation}
where $S$ and $\kappa$ also satisfy \eqref{SPzero} and \eqref{projectkappa}.  To determine $S$ and $\kappa$ we write them as a perturbation series:
\begin{align}
S & = \sum_{n = 0}^\infty S_n & \kappa &= \sum_{n = 0}^\infty \kappa_n \;.
\end{align}
By substituting them into \eqref{ms}, and identifying equal orders, we find
\begin{subequations}
\begin{eqnarray}\label{uitgebreid}
\mathcal M_0 S_0 &=& S_0 \kappa_0 \;,\\
\mathcal M_0 S_1 + \mathcal M_1 S_0 &=& S_1 \kappa_0 + S_0 \kappa_1 \label{rxs} \;,\\
\mathcal M_0 S_2 + \mathcal M_1 S_1 &=& S_2 \kappa_0 + S_1 \kappa_1 + S_0 \kappa_2 \label{qrz} \;, \\
\mathcal M_0 S_3 + \mathcal M_1 S_2 &=&  S_3 \kappa_0 + S_2 \kappa_0 + S_1 \kappa_2 + S_0 \kappa_3 \;,\\
&\vdots& \nonumber
\end{eqnarray}
\end{subequations}\\
\\
The first equation is the free equation, which is solved by
\beq
S_0 = P_0, \ \ \ \ \ \ \kappa_0 = \lambda_0 P_0,
\eeq
where for simplicity we have written $ \lambda_0 \equiv\lambda_{\vec n_0}^{(0)} = (2\pi/L)^2$.
To solve the higher order equations, we use the normalization condition, that $S_n$ for $n \geq 1$ maps into the space orthogonal to $\mathcal H_0$,
\begin{equation}
P_0 S_n  = 0 \quad \forall n \geq 1\;,
\end{equation}
and we also have
\beq
S_n P_0 = S_n \ \ \ \ \  \forall n.
\eeq
We now multiply the remaining equations by $P_0$, and use
\beq
P_0 \mathcal M_0 = \mathcal M_0 P_0 = \lambda_0 P_0.
\eeq
Firstly, we find
\begin{eqnarray}
\label{kappa1}
P_0 \mathcal M_1 P_0 &=&  \kappa_1
\end{eqnarray}
In an analogous fashion, we find for the other equations
\begin{align}\label{kappa2}
\kappa_2 &= P_0 \mathcal M_1 S_1  & \kappa_3 &= P_0 \mathcal M_1 S_2 & \ldots \;,
\end{align}
etc.  To find $S_1$, we start from equation \eqref{rxs} which may be written, upon using equation \eqref{kappa1},
\begin{eqnarray}
\mathcal M_0 S_1 - S_1\kappa_0 &=& P_0  \mathcal M_1 P_0 - \mathcal M_1 P_0 \;.
\end{eqnarray}
As $\kappa_0 = \lambda_0 P_0$, we find
\begin{equation}\label{S12}
(\mathcal M_0 - \lambda_0) S_1 = (P_0 - I) \mathcal M_1 P_0 \Rightarrow S_1 =  \left( \mathcal M_0 -  \lambda_0  I  \right)^{-1}  (P_0 - I) \mathcal M_1 P_0 \;.
\end{equation}
In an analogous fashion, we can deduce $S_2$ from equation \eqref{qrz},
\begin{eqnarray}
 \mathcal M_0 S_2 - S_2 \kappa_0 &=& - \mathcal M_1 S_1 + S_1 \kappa_1 + S_0 \kappa_2 \nonumber\\
\Rightarrow (\mathcal M_0 - \lambda_0  I  ) S_2 &=&   (P_0 - I) \mathcal M_1 \left( \mathcal M_0 -  \lambda_0  I\right)^{-1} (P_0 - I) \mathcal M_1 P_0 \nonumber\\
&&+ \left( \mathcal M_0  -  \lambda_0  I   \right)^{-1} (P_0 - I) \mathcal M_1 P_0 \mathcal M_1 P_0\;,
\end{eqnarray}
or thus
\begin{equation}
S_2 = \left[ \left( \mathcal M_0 -  \lambda_0  I   \right)^{-1} (P_0 - I) \mathcal M_1  \right]^2 P_0 + \left( \mathcal M_0  -  \lambda_0  I   \right)^{-2} (P_0 - I) \mathcal M_1 P_0 \mathcal M_1 P_0 \;,
\end{equation}
whereby we made use of equations \eqref{kappa1}, \eqref{kappa2} and \eqref{S12}. With the expressions for $S_1$ and $S_2$, we obtain
\begin{eqnarray}
\kappa_0 &=& \lambda_0 P_0 \;,\nonumber\\
\kappa_1 &=& P_0 \mathcal M_1 P_0 \;, \nonumber\\
\kappa_2 &=& P_0 \mathcal M_1 \left( \mathcal M_0 -  \lambda_0  I   \right)^{-1}  (P_0 - I) \mathcal M_1 P_0 \;, \nonumber\\
\kappa_3 &=& P_0 \mathcal M_1 \left[ \left( \mathcal M_0 -  \lambda_0  I   \right)^{-1} (P_0 - I) \mathcal M_1  \right]^2 P_0 \;,\nonumber\\
 &&+ P_0 \mathcal M_1 \left( \mathcal M_0  -  \lambda_0  I   \right)^{-2} (P_0 - I) \mathcal M_1 P_0 \mathcal M_1 P_0 \;.
\end{eqnarray}
We can write these expressions in terms of the matrix elements,
\begin{eqnarray}
\label{kappamatixelements}
\kappa_0^{\vec n_0' u, \vec n_0 t} &=& \lambda_{\vec n_0}^{(0)} \delta^{ut} \delta_{\vec n_0, \vec n_0'} \;,\nonumber\\
\kappa_1^{\vec n_0' u, \vec n_0 t} &=& \bra{\Psi_{\vec n_0' u}^{(0)}} \mathcal M_1 \ket{\Psi_{\vec n_0 t}^{(0)}} \;, \nonumber\\
\kappa_2^{\vec n_0' u, \vec n_0 t} &=& - \bra{\Psi_{\vec n_0'  u}^{(0)}} \mathcal M_1 \left( \mathcal M_0  - \lambda_0  I      \right)^{-1}  (I - P_0 ) \mathcal M_1 \ket{\Psi_{\vec n_0  t}^{(0)}} \;, \nonumber\\
\kappa_3^{\vec n_0' u, \vec n_0 t} &=& \bra{\Psi_{\vec n_0' u}^{(0)}} \mathcal M_1 \left[ \left(\mathcal M_0  - \lambda_0  I   \right)^{-1} (I - P_0  ) \mathcal M_1  \right]^2 \ket{\Psi_{\vec n_0 t}^{(0)}} \nonumber\\
 &&- \bra{\Psi_{\vec n_0' u}^{(0)}} \mathcal M_1 \left( \mathcal M_0  -  \lambda_0  I   \right)^{-2} (I - P_0 ) \mathcal M_1 P_0 \mathcal M_1 \ket{\Psi_{\vec n_0 t}^{(0)}} \;.
\end{eqnarray}

\subsubsection{The infinite volume limit}

We shall now show that in the large-volume limit, we can do some simplifications.\\
\\
In this limit we let $L \to \infty$, while keeping a typical momentum $\vec k = 2\pi \vec n/L$ finite.  So it is advantageous to change notation, which we also simplify, and write
\beq
\ket{\vec k, s} \equiv \ket{\Psi_{\vec ns}^{(0)}}; \ \ \ \ \ \ \ \ \bra{\vec  k, s} = \bra{\Psi_{\vec ns}^{(0)}},
\eeq
where $\vec k \equiv (2\pi/L) \vec n$.  Compared to a typical momentum $\vec k$, the lowest non-zero momentum, $\vec k_0 = 2\pi \vec n_0/L$, with $|\vec n_0| = 1$ goes to zero, $\vec k_0 \to 0$, and the simplification comes from neglecting $|\vec k_0| = 2\pi/L$ compared to $|\vec k|$.  We shall use the notation $\vec k_0, \vec k_0'$ etc.\ for momentum vectors with lowest non-zero magnitude $|\vec k_0| = |\vec k_0'| = 2\pi/L$.

We start with the expression for $\kappa_2$. We rewrite $ (I - P_0) $ in terms of a bra-ket expansion,
\begin{equation}
 I - P_0 = \sum_{||\vec k ||>0} \sum_{s=1}^{N^2-1} \ket{\vec k, s} \bra{\vec k, s} - \sum_{||\vec k || = 2\pi/L} \sum_{s=1}^{N^2-1} \ket{\vec k, s} \bra{\vec k, s} =  \sum_{||\vec k||>2\pi/L} \sum_{s=1}^{N^2-1} \ket{\vec k, s} \bra{\vec k, s} \;.
\end{equation}
If we now let the operator $\left( \mathcal M_0  -  \lambda_0  I   \right)^{-1}$ act on this term, we obtain
\begin{align}
 \left( \mathcal M_0  -  \lambda_0  I   \right)^{-1} (I - P_0) &=  \sum_{||\vec k||>2\pi/L} \sum_{s=1}^{N^2-1}  (k^2 - k_0^2)^{-1}\ket{\vec k, s} \bra{\vec k, s}.
 \end{align}
In the infinite volume limit we neglect $k_0^2 = (2\pi/L)^2$ compared to $k^2$
\beq
(k^2 - k_0^2)^{-1} \to (k^2)^{-1}
\eeq
and the restriction on the summation becomes vacuous,
\beq
\sum_{||\vec k||>2\pi/L} \to V \int \frac{d^dk}{(2\pi)^d},
\eeq
where $V = L^d$ is the Euclidean volume.  Consequently we may make the substitution
\beq
\left( \mathcal M_0  -  \lambda_0  I   \right)^{-1} (I - P_0) \to  \sum_{\vec k} \sum_{s=1}^{N^2-1}  (k^2)^{-1}\ket{\vec k, s} \bra{\vec k, s} = \mathcal M_0^{-1}.
\eeq
Inserting this in $\kappa_2^{\vec n_0' u, \vec n_0 t}$ yields,
\begin{equation}
\kappa_2^{\vec k_0' u, \vec k_0 t} = - \sum_{\vec k} \sum_{s=1}^{N^2-1}   \bra{\vec k_0'  u} \mathcal M_1 \ket{\vec k  s} \left( \frac{1}{ k ^2 } \right) \bra{\vec k  s}      \mathcal M_1 \ket{\vec k_0  t} \;.
\end{equation}
We can work out the matrix elements with the help of equation \eqref{conf1}
\begin{eqnarray}\label{lalala}
\bra{\vec k s} \mathcal M_1 \ket{\vec k_0  t}&=& \ii k_{0,\mu}  \frac{1}{V}\int \d^d x \ \e^{-\ii  (\vec k - \vec k_0) \cdot \vec x} g f_{stc} A_\mu^c(x) \nonumber\\
& \to & \ii k_{0,\mu}  \frac{1}{V}\int \d^d x \ \e^{-\ii  \vec k \cdot \vec x} g f_{stc} A_\mu^c(x) \nonumber\\
\bra{\vec k_0'  u} \mathcal M_1 \ket{\vec k s}  &=& \ii k_{0,\mu}' \frac{1}{V} \int \d^d x \ g f_{usc} A_\mu^c(x)\e^{\ii (\vec k - \vec k_0') \cdot \vec x} \nonumber  \\
& \to & \ii k_{0,\mu}' \frac{1}{V} \int \d^d x \ g f_{usc} A_\mu^c(x)\e^{\ii \vec k \cdot \vec x} \;,
\end{eqnarray}
where we have made use of partial integration and the Landau gauge condition $\p_\mu A_\mu = 0$, and neglected $\vec k_0$ and $\vec k_0'$, which are of order $2\pi/L$, compared to $\vec k$.  We thus obtain at large $L$,
\begin{eqnarray}
\kappa_2^{\vec k_0' u, \vec k_0 t} & = & - \sum_{\vec k}  \ii k_{0,\mu}' \frac{1}{V} \int \d^d x \ g f_{usc} A_\mu^c(x)\e^{\ii \vec k \cdot \vec x} \left( \frac{1}{ k ^2 } \right) \ii k_{0,\nu}  \frac{1}{V}\int \d^d y \ \e^{-\ii  \vec k \cdot \vec y} g f_{std} A_\nu^d(y) \nonumber \\
& = &
 \frac{1}{V} (\ii k_{0,\mu}') \int \d^d x \d^dy \ g f_{usc} A_\mu^c(x) \ \frac{1}{V} \sum_{\vec k}\left( \frac{\e^{\ii \vec k \cdot (\vec x - \vec y) }}{ k ^2 } \right) \ g f_{std} A_\nu^d(y) (\ii k_{0,\nu}) \; ,
 \end{eqnarray}
 which gives, for large $V$, the simple expression,
 \beq
 \kappa_2^{\vec k_0' u, \vec k_0 t} =
- (\ii k_{0,\mu}')  \frac{1}{V}  \int \d^d x \d^dy \ g f_{usc} A_\mu^c(x) \ \left( \mathcal M_0^{-1}\right)_{\vec x \vec y} \ g f_{std} A_\nu^d(y) (\ii k_{0,\nu}) \; .
\eeq \\
\\
We now turn to $\kappa_3$, given in \eqref{kappamatixelements}.  It consists of two terms.  The first term contains only projectors $I - P_0$ orthogonal to the subspace $\mathcal H_0$ in intermediate positions and may be treated like $\kappa_2$.  The second term in \eqref{kappamatixelements} contains the projector $P_0$ in an intermediate position, instead of $I-P_0$ which appears in the expression,
\begin{eqnarray}
... (I - P_0 ) \mathcal M_1 P_0 \mathcal M_1 \ket{\vec k_0 t} & = & ... \sum_{\vec k, r, \vec k_0'', s}
\ket{\vec k, r} \bra{\vec k, r} \mathcal M_1
 \ket{\vec k_0'', s} \bra{\vec k_0'', s}M_1 \ket{\vec k_0 t}
 \nonumber  \\
 & = & ... \sum_{\vec k, r, \vec k_0'', s}
\ket{\vec k, r}  \ii k_{0,\mu}''  \frac{1}{V}\int \d^d x \ \e^{-\ii  (\vec k - \vec k_0'') \cdot \vec x} g f_{rsc} A_\mu^c(x)
 \nonumber  \\
 && \ \ \ \ \ \times \ii k_{0,\nu}  \frac{1}{V}\int \d^d y \ \e^{-\ii  (\vec k_0'' - \vec k_0) \cdot \vec y} g f_{std} A_\nu^d(y) \;,
\end{eqnarray}
where we have used \eqref{lalala}.  We observe that with $P_0$ in an intermediate position (instead of $I-P_0$), there is (1) an extra factor $\vec k_0''$ of small magnitude $|\vec k_0''| = 2\pi/L$ (instead of a finite momentum $\vec k$), and (2) an extra factor of $1/V$ that is associated with the finite sum $\sum_{|\vec k_0''| = 2\pi/L}$ (instead of the sum $\sum_{\vec k} \to V \int \d^d k / (2\pi)^d$ which cancels the $1/V$).  Consequently, in the large-volume limit, we may neglect the second term in \eqref{kappamatixelements} compared to the first term.  The first term in this equation is evaluated at large $V$ by the argument used for $\kappa_2$, with the result for $\kappa_3$ given in the next equation below.\\
\\
In conclusion, in the large volume limit, we obtain the following matrices
\begin{eqnarray}
\kappa_0^{\vec k_0' u, \vec k_0 t} &= & \left( \frac{2\pi}{L}\right)^2 \delta^{ut} \delta_{\vec k_0, \vec k_0'}  \;,\nonumber\\
\kappa_1^{\vec k_0' u, \vec k_0 t} &= & \ii k_{0,\mu}  \frac{1}{V}\int \d^d x \ \e^{\ii  (\vec k_0 - \vec k_0') \cdot \vec x} g f_{utb} A_\mu^b(x) \;, \nonumber\\
\kappa_2^{\vec k_0' u, \vec k_0 t} &= &
- (\ii k_{0,\mu}')  \frac{1}{V}  \int \d^d x \d^dy \ g f_{usc} A_\mu^c(x) \ \left( \mathcal M_0^{-1} \right)_{\vec x \vec y} \ g f_{std} A_\nu^d(y) (\ii k_{0,\nu}) \;, \nonumber\\
\kappa_3^{\vec k_0' u, \vec k_0 t} & = &
(\ii k_{0,\mu}')  \frac{1}{V}  \int \d^d x \d^dy \ g f_{urc} A_\mu^c(x) \ \left(  \mathcal M_0^{-1} \mathcal M_1  \mathcal M_0^{-1} \right)_{\vec x r. \vec y s} \ g f_{std} A_\nu^d(y) (\ii k_{0,\nu}) \; , \nonumber \\
 & & \text{etc.}
\end{eqnarray}
In the large volume limit, the higher order terms $\kappa_n$ may be evaluated like $\kappa_3$.  For each $\kappa_n$, all terms in which the projector $P_0$ appears in an intermediate position are negligible compared to the one remaining term which is evaluated like $\kappa_2$.
We notice that
\begin{equation}
\mathcal M^{-1} =    \mathcal M_0^{-1}  -\mathcal M_0^{-1} \mathcal M_1  \mathcal M_0^{-1}  + \mathcal M_0^{-1} \mathcal M_1  \mathcal M_0^{-1} \mathcal M_1  \mathcal M_0^{-1} - ... \;,
\end{equation}
so we can sum the whole series $\kappa_n$ starting from $n =2$,
\begin{equation}
\label{Tracekappar}
\kappa_r^{\vec k_0' u, \vec k_0 t} = \sum_{n\geq2} \kappa_n^{\vec k_0' u, \vec k_0 t} =- (\ii k_{0,\mu}')  \frac{1}{V}  \int \d^d x \d^dy \ g f_{urc} A_\mu^c(x) \ \left( \mathcal M^{-1}\right)_{\vec x r, \vec y s} \ g f_{std} A_\nu^d(y) (\ii k_{0,\nu}) \;.
\end{equation}


\subsubsection{Taking the trace of $\kappa$}

\begin{figure}[H]
\begin{center}
\includegraphics[width=14
cm]{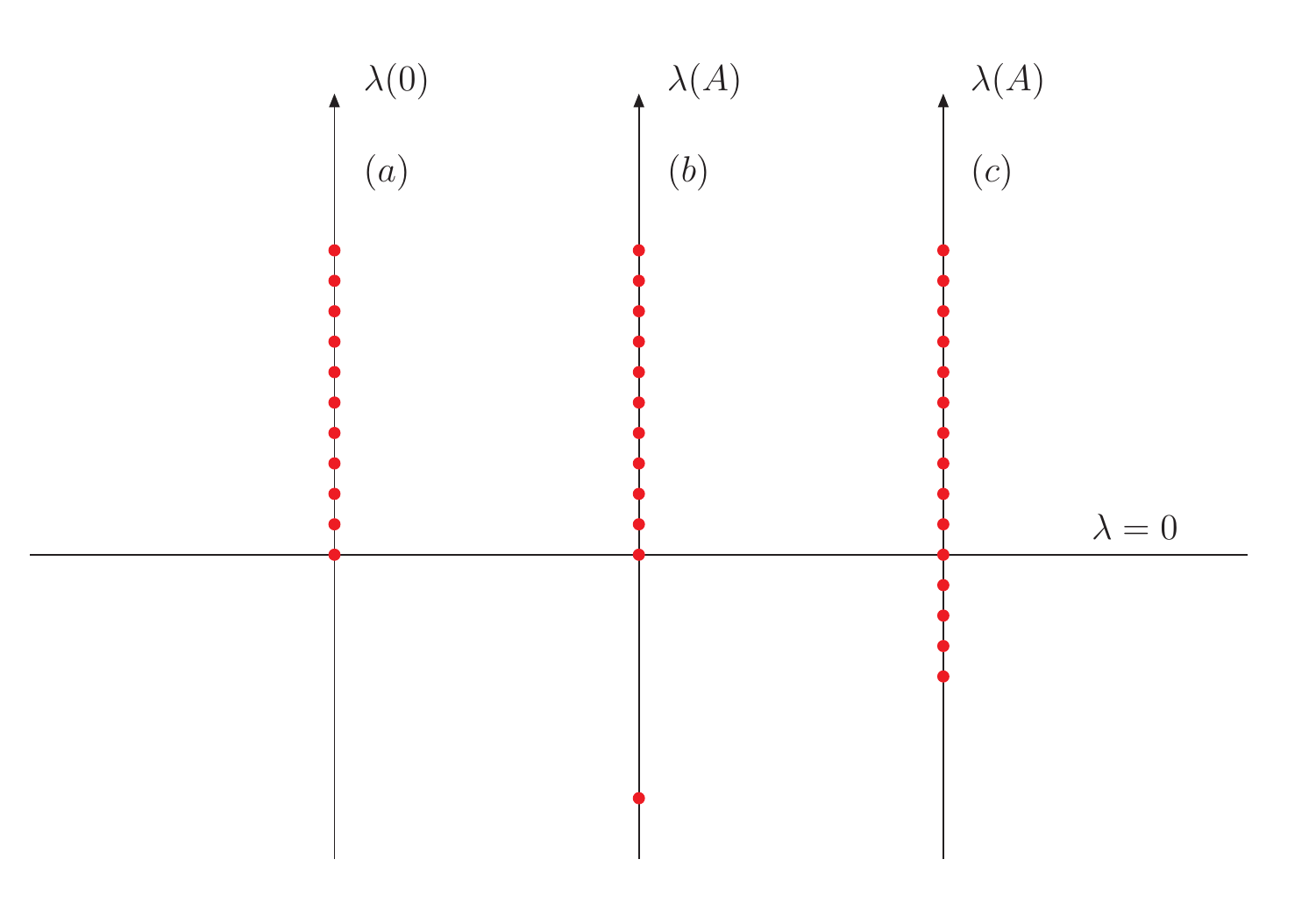}
\caption{Eigenvalues $\lambda_n(A)$ of the Faddeev-Popov operator $\mathcal M(A)$ at large Euclidean volume $V$.  (a) The unperturbed spectrum $A = 0$.  (b) A configuration $A$ slightly outside the Gribov horizon, with one bound state, at a finite negative value. (c)  A configuration $A$ slightly outside the Gribov horizon, for which the almost continuous spectrum begins at a negative value.}\label{spectrum}
\end{center}
\end{figure}

At large volume, the spectrum of (nontrivial) eigenvalues $k^2 = (2\pi \vec n^2 /L)^2 \geq (2\pi/L)^2$ of $\mathcal M_0 = - \p^2$ becomes dense on the positive real line $0 < k^2 < \infty$.  As the perturbation $\mathcal M_1(A)$ gets turned on, the spectrum at first remains dense on the positive real line.  However when $A$ crosses the Gribov horizon (which is bounded in every direction) the lowest eigenvalue becomes negative.  Let us consider how this can happen.\\
\\
One possibility is that the spectrum behaves as in ordinary non-relativistic potential theory.  One bound state develops with finite negative binding energy $\lambda_1 < 0$, while the rest of the spectrum $\lambda_n$ for $n > 1$ remains dense on the positive real axis.  This is illustrated in Figure \ref{spectrum}, case (b).  Among the $T$ lowest eigenvalues $\lambda_1 ... \lambda_T$ (which are the eigenvalues of the matrix $\kappa$ we have just calculated), $\lambda_1$ will be finitely negative, while $\lambda_2 ... \lambda_T$ are at the bottom of the almost dense positive continuum starting at 0,
$\lambda_n \approx 0$ for $n = 2, ... T$.  In this case the sum of the first $T$ eigenvalues becomes negative when the first one becomes negative
\beq
\lambda_1 < 0 \Leftrightarrow \sum_{n=1}^T \lambda_n < 0.
\eeq\\
\\
Another possibility is that the spectrum remains (almost) dense, but the bottom of the spectrum moves a finite distance into the negative region.  This is also illustrated in Figure \ref{spectrum}, case (c).  In this case, the lowest $T$ eigenvalues all become negative (almost) together.  These are the eigenvalues of the matrix $\kappa$ we have just evaluated and $\sum_{n = 1}^T \lambda_n = \Tr \kappa$.\\

Next, we next evaluate $\Tr\kappa = \Tr\kappa_0 + \Tr\kappa_1 + \Tr\kappa_r$.  Firstly, the trace of $\kappa_0$ is given by
\begin{equation}
\Tr \kappa_0 =\sum_{||\vec n|| =1} \sum_{u=1}^{N^2-1} \kappa_0^{\vec n u, \vec n u} =    \lambda_0 T \;.
\end{equation}
Secondly, the trace of $\kappa_1$ is zero,
\begin{eqnarray}
\Tr \kappa_1 & = & \sum_{||\vec n|| =1} \sum_{u=1}^{N^2-1}  \kappa_1^{\vec n u, \vec n u}   = \sum_{||\vec n|| =1} \sum_{u=1}^{N^2-1}  \bra{\Psi_{\vec n u}^{(0)}} \mathcal M_1 \ket{\Psi_{\vec n u}^{(0)}}\nonumber \\
& = & \ii n_{0,\mu} \frac{2 \pi}{L}  \frac{1}{V}\int \d^d x \ g f_{uuc} A_\mu^c(x) = 0\;,
\end{eqnarray}
which vanishes because $f_{uuc} = 0$, and we have used \eqref{lalala}. Finally, by \eqref{Tracekappar}, the trace of $\kappa_r$ at large volume $V$ is given by
\begin{equation}
\Tr \kappa_r = \sum_{||\vec n|| =1} \sum_{u=1}^{N^2-1}  \kappa_r^{\vec n u, \vec n u} = - \frac{2 \lambda_0}{V}  \int \d^d x \d^dy \ g f_{urc} A_\mu^c(x) \ \left( \mathcal M^{-1}\right)_{\vec x r, \vec y s} \ g f_{usd} A_\mu^d(y) \;.
\end{equation}
\\
Other, perhaps more pathological, cases may be considered, but we shall assume that for the configurations that dominate the Euclidean functional integral in the infinite-volume limit, it is justified, as it is in the two cases just considered. to replace the condition that all eigenvalues of $\kappa$ be positive by the condition that the sum of its eigenvalues be positive, $\Tr \kappa > 0$.  In any case, this condition is weaker than the one it replaces, just as the restriction to the Gribov region is weaker than the restriction to the FMR.  Consequently results obtained by this procedure will be an underestimate of the effect on QCD dynamics of the restriction to the FMR.\\
\\
In conclusion, the following quantity should be positive
\begin{multline*}
\Tr \kappa = 2 \left( \frac{2\pi}{L}  \right)^2  \left( d (N^2 -1) - \frac{1}{V} \int \d^d x \int \d^d y g f_{ba\ell} A_\mu^a (x) (\mathcal M^{-1})^{\ell m} \delta(x-y) g f_{bkm} A^k_\mu (y)\right) \\
> 0 \;.
\end{multline*}
Therefore, we can set
\begin{eqnarray}
\label{Zcutoff}
Z &=& \int [\d A] \e^{- S_{\text{eff}}} \theta( d(N^2 -1)V - H(A)) \;,
\end{eqnarray}
where the integral extends over transverse configurations, $S_{\text{eff}} \equiv S_{\YM} - \ln \det \mathcal M$, where the ``horizon function" is given by
\begin{eqnarray}
\label{horizonfunction}
H(A)& \equiv & \int \d^d x \int \d^d y g f_{ba\ell} A_\mu^a (x) (\mathcal M^{-1})^{\ell m} \delta(x-y) g f_{bkm} A^k_\mu (y) \;.
\end{eqnarray}
It takes its name because, at large volume, the Gribov horizon is given by
\beq
H(A) = d(N^2 -1)V.
\eeq
Note that the coefficient $d(N^2-1)$ is the number of components of the gluon field, $A_\mu^b$.

\subsubsection{The non-local GZ action}\label{sectnonlocal}

We start from \eqref{Zcutoff}, and we represent the $\theta$-function, $\theta[d(N^2-1) - H(A)]$ by its fourier decomposition,\footnote{As before, we have suppressed the source term $(J,A)$ in the action.  This is justified in the present section, as long as the source $J(x)$ remains of compact support in the infinite-volume limit, because the relevant terms will turn out to be bulk or extensive quantities, of the order of the Euclidean volume $V$.}
\beq
\theta[K(A)] = { 1 \over 2\pi i } \int_{-\infty}^\infty  \ { d\omega  \over  (\omega - i\epsilon) } \exp[i \omega K(A)],
\eeq	
where, for convenience, we have written
\beq
K(A) \equiv d(N^2-1) - H(A)\;.
\eeq
Upon interchanging order of integration we obtain from \eqref{Zcutoff}
\beq
Z = { 1 \over 2\pi i } \int_{-\infty}^\infty  \ { d\omega  \over  (\omega - i\epsilon) } \exp[\mathcal W(i\omega)],
\eeq
where the ``extended free energy'' $\mathcal W(z)$, depending on the complex variable $z$, is defined by
\beq
\label{extended}
\exp \mathcal W(z) \equiv \int [dA] \ \exp[z K(A) - S_{\rm eff}(A)].
\eeq
The position of the $i \epsilon$ allows us to continue the path of the $\omega$ integration into the lower half plane, $\omega \to \omega - i \lambda$, with $\lambda > 0$, where we will look for a saddle point.  The $\omega$ integration now reads
\beq
Z = { 1 \over 2\pi } \int_{-\infty}^\infty   d\omega \ \exp[ - \ln(\lambda + i\omega) + \mathcal W(\lambda + i\omega)],
\eeq
We look for a saddle point in $\lambda$ at $\omega = 0$.  If so, it is located at a stationary point of
\beq
\Phi(\lambda) \equiv - \ln \lambda  + \mathcal W(\lambda),
\eeq
namely at the solution $\lambda^*$ of
\beq
{\p \Phi(\lambda) \over \p \lambda} = { \p \mathcal W(\lambda) \over  \p \lambda } - {1 \over \lambda} = 0.
\eeq
We have
\beq
\label{lambdaensemble}
\exp \mathcal W(\lambda) = \int [dA] \exp[ \lambda K(A) - S_{\rm eff}(A)],
\eeq
so
\beq
{ \p \mathcal W(\lambda) \over  \p \lambda } = \langle K \rangle_{\lambda},
\eeq
where the subscript indicates that the expectation-value is in the ensemble (\ref{lambdaensemble}), and the saddle-point condition reads
\beq
\label{saddlehorizona}
\langle K \rangle_{\lambda} - \frac{1}{\lambda} = 0,
\eeq
or
\beq
\label{saddlehorizon}
\langle H \rangle_{\lambda} = d(N^2-1)V - \frac{1}{\lambda}.
\eeq
It determines a solution $\lambda = \lambda^*$.  Since the Gribov region is defined by $H(A) \leq d(N^2-1)V$, the sign is correct for $\lambda > 0$.\\
\\
We expand $\Phi(\lambda + i \omega)$ about the saddle point $\lambda^*$, so the  $\omega$ integration now reads, to leading order
\beqa
\label{LOomegaint}
Z & = & { 1 \over 2\pi } \exp[ - \ln \lambda^* + \mathcal W(\lambda^*)]
\\  \nonumber
&& \times \int_{-\infty}^\infty d\omega \ \exp\left[ - {\p^2 \Phi(\lambda) \over \p \lambda^2 }\Big|_{\lambda =\lambda^*} \omega^2/2) \right].
\eeqa
Whether or not the saddle point method is justified and accurate depends on the sign and magnitude of the second derivative,
\beqa
{ \p^2 \Phi(\lambda) \over  \p \lambda^2 } & = & { 1 \over \lambda^2} + { \p^2 \mathcal W(\lambda) \over  \p \lambda^2 }
\nonumber \\
& = & { 1 \over \lambda^2} + \langle K^2 \rangle_{\lambda} - \langle K \rangle_{\lambda}^2
\nonumber \\
& = & { 1 \over \lambda^2} + \langle H^2 \rangle_{\lambda} - \langle H \rangle_{\lambda}^2 > 0.
\eeqa
It is positive, because $1/\lambda^2$ is positive, as is the variance of any quantity, so the $\omega$ integration is splendidly convergent.\\
\\
We now come to an important point.   The horizon function $H(A)$ is a bulk or extensive quantity.  Indeed, from \eqref{horizonfunction} it may be written as the integral over a density,
\begin{eqnarray}
\label{horizonfunctiona}
H(A)& \equiv & \int \d^d x \ g f_{ba\ell} A_\mu^a (x) N_{bl}(x)\;.
\end{eqnarray}
where
\beq
N_{bl}(x) \equiv \left[(\mathcal M^{-1})^{\ell m} g f_{bkm} A^k_\mu \right] (x) =  \int \d^d y \ (\mathcal M^{-1})^{\ell m} \delta(x-y) g f_{bkm} A^k_\mu (y).
\eeq
Shortly we will write it as an integral over a density which is a product of local fields.  It is a general property in statistical mechanics that the variance of a bulk quantity is of order $V$,
\beq
\langle H^2 \rangle_{\lambda} - \langle H \rangle_{\lambda}^2 = O(V).
\eeq
as is illustrated by the explicit calculation in the toy model, (\ref{pr2}).  This makes the coefficient, of $\omega^2$, in the $\omega$ integration \eqref{LOomegaint} also of order $V$, ${\p^2 \Phi(\lambda) \over \p \lambda^2 } = O(V)$.  Thus the width of the peak in $\omega$ is of order $\Delta\omega = O(V^{-1/2}) \to 0$, so the saddle-point approximation becomes accurate in the limit $V \to \infty$, and the $\omega$ integration
contributes a factor of order $V^{-1/2}$.  Thus, we obtain for the partition function
\beq
Z \sim V^{-1/2} \exp \mathcal W(\lambda^*).
\eeq
Again, it is a general property that the (extended) free energy $\mathcal W(\lambda)$ is a bulk quantity of order $V$, so at large volume the coefficient may be neglected and from \eqref{lambdaensemble} we obtain the main result,
\beq
Z = \int [dA] \ \exp\{ \ \lambda[d(N^2-1)V - H(A)] - S_{\rm eff} \ \},
\eeq
where $\lambda$ has the value $\lambda^*$ determined by (\ref{saddlehorizon}).\\
\\
We have just seen that $H$ is a bulk quantity $H = O(V)$.  On the other hand $\lambda^*$ is a parameter, analogous to the Boltzmann factor $\beta = 1/kT$, that remains finite in the limit $V \to \infty$, as we shall verify by explicit calculation.  Therefore the term $1/\lambda^2$ in the saddle-point equation (\ref{saddlehorizon}) becomes negligible at large $V$, so at large $V$ we may write the saddle-point equation as
\beq
{\langle H(A) \rangle_{\lambda} \over V} = d(N^2-1) - 0_+,
\eeq
where $0_+$ is an arbitrarily small positive number.
Since the Gribov region is defined by $H(A) \leq d(N^2-1) V$, with $H(A) = d(N^2-1) V$ on the boundary, this equation shows that the probability distribution of $H(A)$ gets concentrated on the boundary $\p \Omega$ in this limit.  We call this equation the ``horizon condition."  It may also be written
\beq
\label{partiallnZ}
\frac{\p \ln Z}{\p \lambda} = 0_+
\eeq\\
\\
In the following, we change notation slightly, by setting $\lambda = \gamma^4$, and we write the last result as
\begin{eqnarray}
\label{SZhsub1}
Z  =  \int [\d A][\d c][\d \overline c]  \e^{-[S_\YM + S_\gf +   \int \d^d x h_1 (x) - \gamma^4 \int \d^d x \ d (N^2 -1 )]} \;,
\end{eqnarray}
where $h_1(x)$ is given by
\begin{eqnarray}\label{2h1}
h_1(x) &=&  \gamma^4 \int \d^d y g f_{ba\ell} A_\mu^a (x) (\mathcal M^{-1})^{\ell m} \delta(x-y) g f_{bkm} A^k_\mu (y) \;.
\end{eqnarray}
The parameter $\gamma$ is fixed by the gap equation
\begin{eqnarray}
\Braket{h_1(x)}_\gamma &=& \gamma^4 d (N^2 - 1) - 0_+ \;.
\end{eqnarray}

\subsubsection{Remarks}
One thing which needs to be pointed out, is what happens with the path integral at the boundary of the Gribov region. At the boundary, one of the eigenvalues $\lambda$ of the Faddeev-Popov operator $\mathcal M$ approaches zero. Because $h_1$ contains the inverse of the Faddeev-Popov operator, the probability in the path integral vanishes rapidly; roughly speaking, a factor $\e^{-\frac{1}{\lambda}}$ enters the path integral. On the other hand, we have argued that only the boundary of the Gribov region gives contributions for $V \to \infty$. This could sound contradictory. However, we can give an example which demonstrates what is going on. The path integral shall be a result of two competing functions. Firstly, we have a factor $r^{N-1}$ stemming from the integration, where $N$ approaches infinity in the thermodynamic limit, and we simple take $r$ to represents the fields, while secondly, a factor $\e^{-\frac{1}{R-r}}$, with $R$ the size of the boundary, represents the horizon function. The following function shows us what is going on in the path integral
\begin{equation}\label{expressionqq}
\lim_{N \to \infty} r^{N-1} \e^{-\frac{1}{R-r}} \;.
\end{equation}
In the figures below, one can see how for larger $N$, this function evolves into a delta function
\begin{figure}[H]
  \centering
     \includegraphics[width=5cm]{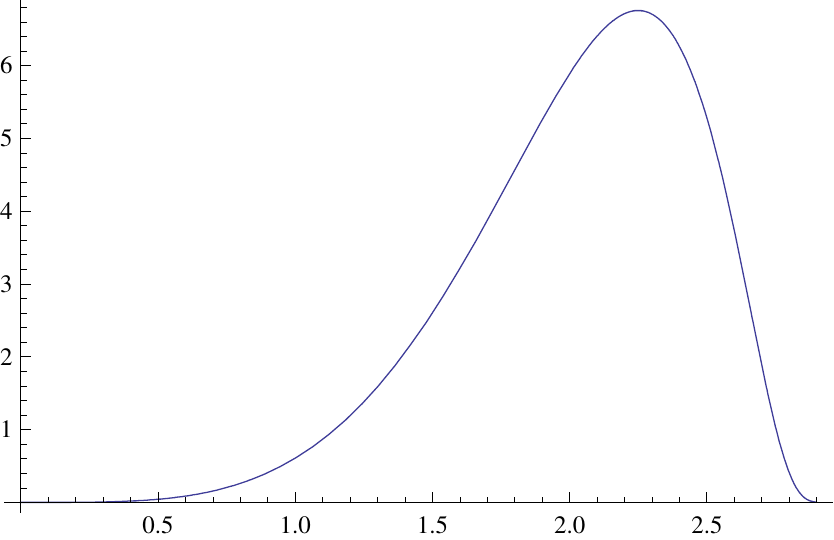}
     \includegraphics[width=5cm]{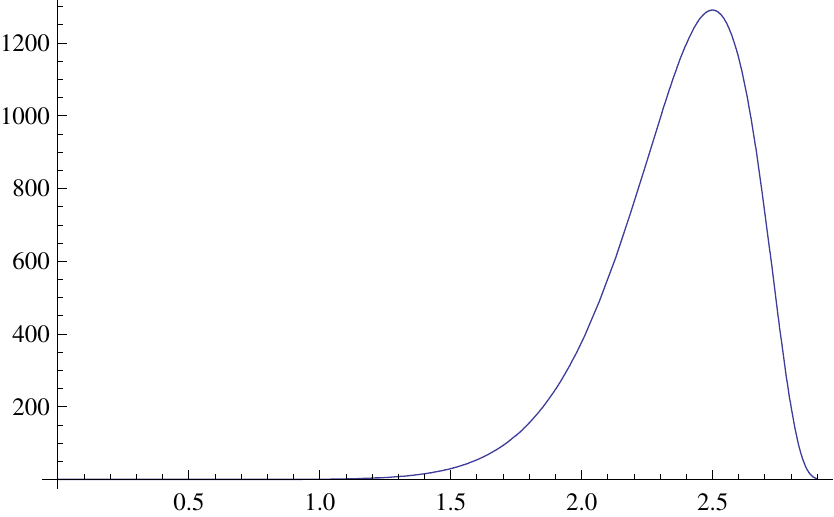}
     \includegraphics[width=5cm]{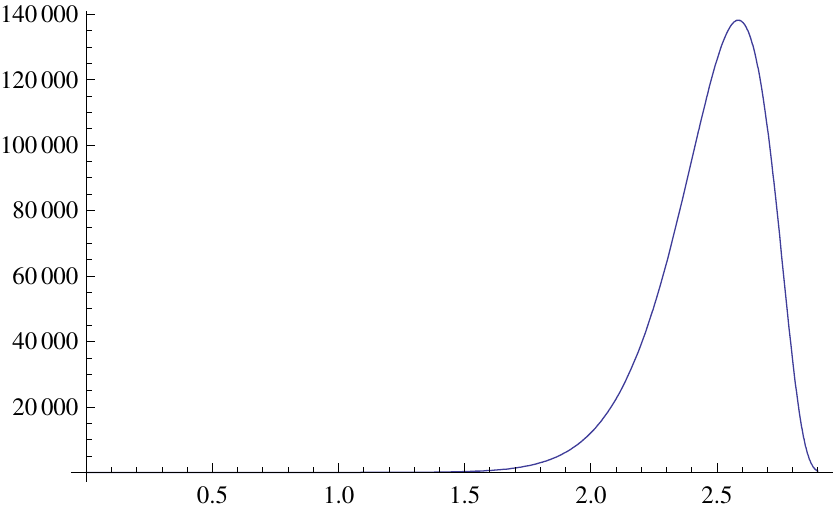}
     \includegraphics[width=5cm]{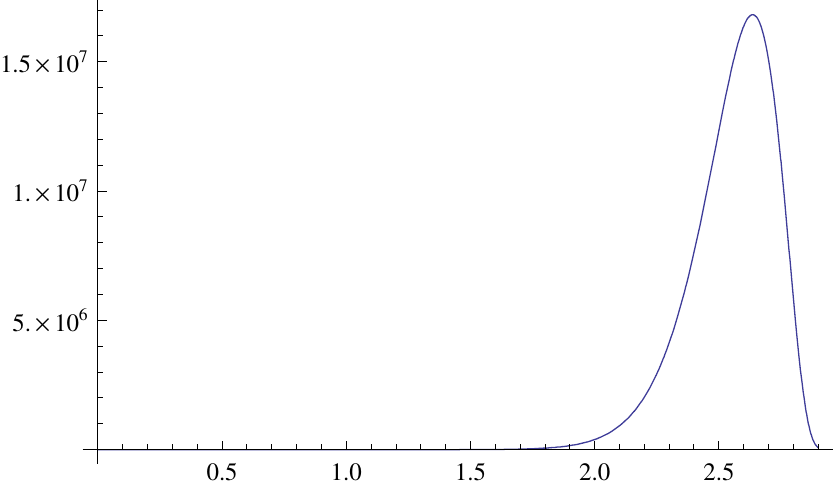}
  \caption{Evolution of the expression \eqref{expressionqq} for growing $N$ whereby we chose $R = 3$. }
\end{figure}

\noindent Another element which looks contradictory is the following. In perturbation theory, or for large momenta, only the small area perturbing around $A =0$ is important, while we have just shown that the configurations get concentrated on the boundary of the Gribov region. Perhaps this can be explained by noticing that the Gribov parameter $\gamma^2$ cannot be accessed in perturbation theory. Indeed, as $\gamma^2 \propto \Lambda_\QCD^2 \propto  \e^{ -\frac{1}{g^2}}$, perturbatively, $\gamma = 0$.

\subsubsection{The correct horizon function}
In order to establish renormalizability \cite{Zwanziger:1992qr}, and as an infrared regulator \cite{Zwanziger:1993dh}, the horizon function \eqref{2h1} was refined into the function
\begin{equation}\label{danhorizon}
S_\h = \lim_{\theta \to 0}  \int \d^d x \ h_2 (x) = \lim_{\theta \to 0}  \int \d^d x \int \d^d y   \left( D_\mu^{ac}(x) \gamma^2(x) \right)(\mathcal M^{-1})^{ab}(x,y) \left( D_\mu^{bc} (y) \gamma^2(y) \right) \;,
\end{equation}
whereby $\gamma(z)$ is defined through
\begin{eqnarray}\label{thetadep}
\gamma^2 (z) &=& \e^{\ii \theta z} \gamma^2 \;.
\end{eqnarray}
The $\lim_{\theta \to 0}$ operation corresponds to replacing the space time dependent $\gamma^2(z)$ with the constant Gribov parameter $\gamma^2$, and at $\theta = 0$ we get $D_\mu^{ac}(x) \gamma^2 = f_{abc}gA^b \gamma^2$, and so \eqref{2h1} is indeed recovered.  We note here that the limit, $\lim_{\theta \to 0}$, in expression \eqref{danhorizon} will be taken after an appropriate localization of the horizon function, a point which we shall outline in detail in what follows. \\
\\
In conclusion, by \eqref{SZhsub1}, the non-local action is given by
\begin{equation}\label{nonlocalGZ}
S_{\nl} = S_\YM + S_\gf + S_\h - \gamma^4 \int \d^d x \ d (N^2 -1 ) \;,
\end{equation}
with $S_\YM$ the Yang-Mills action and $S_\gf$ the gauge fixing term, and the horizon condition reads, by \eqref{partiallnZ},   
\beq
\label{partiallnZgamma}
\frac{\p \ln Z}{\p \gamma^2} = 0_+
\eeq
or, from $Z =  \int [d \Phi] \e^{-S_{\nl}}$, where $\int [d\Phi]$ stands for integration over all fields,
\begin{eqnarray}\label{horizoncondition}
\braket{h_2(x)} &=& \gamma^4 d (N^2 -1) - 0_- \,,
\end{eqnarray}
where $h_2(x)$ is defined in \eqref{danhorizon}, and we have made use of translation invariance to eliminate a factor $V = \int \d^dx$.

\subsection{The local GZ action}
\label{localGZ}
In this section, we shall localize the action $S_\nl$ by introducing some extra fields. Looking at the following standard formula for Gaussian integration for a pair of bosonic fields, see expression \eqref{gauss8}
\begin{multline}\label{alg2}
 C {\det}^{-1} A \exp  \int \d^d x \d^d y \   \overline J^a (A^{-1})^{ab}(x,y) J^b(y) \\
 = \int [\d \varphi][\d \overline{\varphi}] \exp\Bigl[ \int \d^d x \d^d y\left( - \overline{\varphi}^a(x) A^{ab} (x,y)\varphi^b(y)\right)
 + \int \d^d x \ \left(\varphi^a \overline J^a(x)  + \overline{\varphi}^a(x) J^a (x)\right) \Bigr] \;,
\end{multline}
we observe that we can get rid of the inverse of the Faddeev-Popov operator in $S_h$, eq.~\eqref{danhorizon}, by introducing new fields. For every index $i$, defined by $\ldots_i^a = \ldots^{ac}_\mu$, we can write,\footnote{The complex conjugate notation for $\varphi$ and $\overline\varphi$ is purely formal, because, if taken literally, the integration \eqref{phiintgeral} over $\varphi$ and $\overline\varphi$ would explode exponentially for positive eigenvalues of $\mathcal M$.  To give a meaning to this integration, we write $\overline\varphi_i^a = i\chi_i^a$, and we take $\varphi_i^a$ and $\chi_i^a$ to be a pair of real independent bose fields, so \eqref{alg2} correctly represents the identity
\beq
\label{convergenceid}
\int d\varphi d\chi \ \exp(i \chi \mathcal M \varphi + i\chi J +\varphi J) = \int d\phi \ \delta(\mathcal M \phi + J) \exp ( \varphi J) = {\det}^{-1} \mathcal M \exp(- J \mathcal M^{-1} J), 
\eeq
for every real, non-singular matrix $\mathcal M$ and real $J$.  We note that $\mathcal M$ need not be hermitian, as happens for $\mathcal M(A)$ when the gauge condition is taken off-shell, $\p \cdot A \neq 0$.  However there is no harm in writing $\overline\varphi$ for $i \chi$.} with $A \to - \mathcal M$, and 
$J_i^a = \overline J_i^a = D_i^a\gamma^2(x) = 
D_\mu^{ac}\gamma^2(x)$,
\begin{eqnarray}
\label{phiintgeral}
\exp\left( - S_h \right) & = & \prod_{i = 1}^{d (N^2 +1)} \det \mathcal M\int [\d \varphi][\d \overline{\varphi}] \ \exp \Biggl( \lim_{\theta\to 0} \Bigl[   \int \d^d x \int \d^d y \ \overline \varphi_i^a (x) \mathcal M^{ab}(x,y)   \varphi^b_i (y)  \nonumber\\
&&  + \int \d^d x \left( D_i^{a} (x) \gamma^2(x) \right) \varphi_i^a(x) +  \left( D_i^a (x) \gamma^2(x) \right) \overline \varphi_i^a(x) \Bigr] \Biggr) \;,
\end{eqnarray}
whereby we have introduced a pair of conjugate bosonic fields $\left(\overline \varphi_\mu^{ac}, \varphi_\mu^{ac}\right)$ $=$ $\left(\overline \varphi_i^{a}, \varphi_i^{a}\right)$.  We can then also lift the determinants $ \det \mathcal M$ into the exponential by introducing a pair of Grassmann fields $\left( \overline \omega_\mu^{ac},\omega_\mu^{ac} \right)$ $=$ $\left( \overline \omega_i^{a},\omega_i^{a} \right)$. Making use of the standard Gaussian formula for Grassmann variables
\begin{multline}\label{alg1}
	C \left(\det A\right) \exp\left(  -\int \d^d x \d^d y \   J_\omega^a(x) (A^{-1})^{ab}(x,y) J_{\overline{\omega}}^b(y) \right) \\
= \int [\d\omega][\d\overline{\omega}] \exp\left[ \int \d^d x \d^d y \ \overline{\omega}^a(x) A^{ab} \omega^b(y) + \int \d^d x \ (J^a_\omega(x) \omega^a(x) + \overline{\omega}^a(x) J_{\overline{\omega}}^a (x)) \right] \;,
\end{multline}
whereby we set the sources $J_\omega^a $ and $J_{\overline{\omega}}^b$ equal to zero, we obtain
\begin{eqnarray*}
\exp\left( - S_h \right) = \prod_{i = 1}^{d (N^2 +1)}  \int [\d\omega][\d\overline{\omega}][\d \varphi][\d \overline{\varphi}]
\exp \Bigl[ \int \d^d x \int \d^d y  \left( \overline \varphi_i^a (x) \mathcal M^{ab}(x,y)   \varphi^b_i (y)\right.\nonumber\\
 \left.- \overline \omega_i^a (x) \mathcal M^{ab}(x,y)   \omega^b_i (y)\right) +  \lim_{\theta\to 0} \int \d^d x \left( D_i^{a} (x) \gamma^2(x) \right) \varphi_i^a(x)  +  \left( D_i^a (x) \gamma^2(x) \right) \overline \varphi_i^a(x) \Bigr] \;.
\end{eqnarray*}
By \eqref{nonlocalGZ}, the new localized action thus becomes
\begin{eqnarray}\label{SGZphys}
S_\GZ &=& S_0' +  S_\gamma \;,
\end{eqnarray}
with
\begin{equation}\label{SGZ}
S_0' = S_\YM + S_\gf   + \int \d^d x\left( \overline \varphi_\mu^{ac} \p_\nu D_\nu^{ab} \varphi_\mu^{bc}  - \overline \omega_\mu^{ac} \p_\nu D_\nu^{ab} \omega_\mu^{bc}   \right) \,,
\end{equation}
and
\begin{eqnarray}\label{xhor1a}
S_\gamma &=& -  \lim_{\theta \to 0} \int \d^d x\left[ \left( D_\mu^{ac} (x) \gamma^2(x) \right) \varphi_\mu^{ac}(x)  + \left( D_\mu^{ac} (x) \gamma^2(x) \right) \overline \varphi_\mu^{ac}(x) + \gamma^4d(N^2-1)  \right]\nonumber   \\
&=&  \lim_{\theta\to 0} \int \d^d x \; \left[  \gamma^2(x) D_\mu^{ca} \left(\varphi_\mu^{ac}(x) + \overline \varphi_\mu^{ac}(x)\right) - \gamma^4 d(N^2-1) \right]  \nonumber\\
&=&  \int \d^d x  \; \left[ \gamma^2 D_\mu^{ca} \left(\varphi_\mu^{ac}(x) + \overline \varphi_\mu^{ac}(x)\right) - \gamma^4 d (N^2-1) \right]\;. \label{bb1}
\end{eqnarray}
Notice that, as already remarked, the limit $\theta\to 0 $,  in equation \eqref{bb1} has been performed after localization. As one can see from \eqref{thetadep}, taking this limit is equivalent to setting $\gamma^2(x)$ equal to the constant $\gamma^2$. At the level of the classical action, total derivatives may be neglected, $S_\gamma$ becomes
\begin{eqnarray}\label{h1bis}
S_\gamma &=&  \int \d^d x \left[ \gamma^2 g  f^{abc}A_\mu^a \left( \varphi_\mu^{bc} +  \overline \varphi_\mu^{bc} \right) - \gamma^4 d (N^2-1) \right] \,.
\end{eqnarray}
Notice here that starting from the first horizon function $h_1(x)$ given in \eqref{2h1} and undertaking the same procedure, we would end up with exactly the same action $S_\gamma$. This can be understood as we have neglected the total derivatives. Although at the classical level the actions derived from $h_1(x)$ and $h_2(x)$ are the same, at the quantum level they may be different.\\
\\
We use the relation between the local action $S_\GZ$ and the nonlocal action $S_{\nl}$
\begin{equation}
 Z =    \int [\d A] [\d b] [\d c][ \d \overline{c}] \e^{- S_\nl} = \int [\d A ][\d b][ \d c ][\d\overline{c}][ \d \varphi][ \d\overline{\varphi}][ \d \omega][ \d \overline{\omega}]\e^{-S_\GZ} \;.
\end{equation}
to obtain from the horizon condition $\frac{\p \ln Z}{\p \gamma^2} = 0_+$
the local form
\begin{eqnarray}\label{horizonconditionlocal}
- \braket{g f^{abc} A^a_{\mu} ( \varphi^{bc}_{\mu} +  \overline{\varphi}^{bc}_{\mu})}   + 2 \gamma^2 d (N^2 -1)   = 0_+  \;,
\end{eqnarray}
where we have again used translation invariance to eliminate a volume factor $V = \int \d^d x$.
Alternately we can write the horizon condition as
\begin{eqnarray}\label{gapgamma}
- \frac{\p \Gamma}{\p \gamma^2} &=& 0_+\;,
\end{eqnarray}
where the ``free energy," $- \Gamma$, is defined by
\begin{eqnarray}
\e^{-\Gamma} &=& \int [\d\Phi] \e^{-S_\GZ} \;,
\end{eqnarray}
and $\int [\d\Phi]$ stands for the integration over all the fields.\\
\\
There exists a freedom of redefinition of fields and we shall perform a shift of the field $\omega^a_i$, 
\begin{equation}\label{shift}
\omega^a_i (x) \to  \omega^a_i (x) + \int \d^d z (\mathcal M^{-1})^{ad} (x,z) g f_{dk\ell} \p_\mu [ (D_\mu^{ke} c^e) (z) \varphi_i^\ell (z)]\;,
\end{equation}
so that the action becomes
\begin{eqnarray}\label{SGZecht}
S_\GZ &=& S_0 +  S_{\gamma} \,,
\end{eqnarray}
where $S_0'$ has been replaced by $S_0$
\begin{eqnarray}\label{S0}
S_0 &=& S_0' + \int \d^d x \left( - g f^{abc} \p_\mu \overline \omega_i^a    (D_\mu^{bd} c^d)  \varphi_i^c \right)\,,
\end{eqnarray}
and an integration by parts on $\p_\mu$ has been performed.  For reasons having to do with BRST symmetry, this shift makes the action $S_\GZ$ renormalizable \cite{Zwanziger:1992qr}, and in section \ref{renorma}, we shall present the proof that it is. We would like to stress that this is far from trivial, especially since no new parameter is needed to take into account vacuum divergences, which would lead to a modification of the vacuum term $- \int \d^dx \ d(N^2-1)$. In addition, the algebraic formalism employed in the next section also gives a clean and simple argument why the extra term appearing in equation \eqref{S0} is necessary, without invoking the non-local shift \eqref{shift}.

  Another way to see that the term $- g f^{abc} \p_\mu \overline \omega_i^a    (D_\mu^{bd} c^d)  \varphi_i^c$ is allowed is to observe that it gives the action, considered in its dependence on the variables $(\overline c, \overline{\omega})$ and $(c, \omega)$, the structure of a triangular matrix, with terms in $\overline c c$, $\overline\omega \omega$, and $\overline\omega c$, but no term in $\overline c \omega$.  It is easy to see by drawing diagrams that, because of this triangular structure, any term in the action that involves $\overline\omega c$ --- whatever its precise form may be --- contributes only to diagrams with an entering $\overline c$ and an exiting $\omega$.  Thus the $\overline\omega c$ terms in the action do not affect correlators of pure gluons (and quarks), and we have the pleasant freedom to choose them at our convenience.

\subsection{The gluon and the ghost propagator}
Now that we have the local GZ action at our disposal, we can easily calculate the gluon and ghost propagator, at lowest order. We shall show that we obtain the same results as Gribov obtained, see section \ref{gribovgluonghost}.

\subsubsection{The gluon propagator}
To calculate the tree level gluon propagator, we only need the free part of the action $S_\GZ$,
\begin{eqnarray}
    S_\GZ^0 &=& \int \d^d x \Bigl[ \frac{1}{4} \left( \p_{\mu} A_{\nu}^a - \p_{\nu}A_{\mu}^a\right)^2 + \frac{1}{2\alpha} \left( \p_{\mu} A^a_{\mu} \right)^2 +\overline{\varphi}^{ab}_{\mu} \p^2 \varphi^{ab}_{\mu} \nonumber\\
 &&  \hspace{1cm} + \gamma^2 gf^{abc}A^a_{\mu} (\varphi_{\mu}^{bc} + \overline{\varphi}^{bc}_{\mu} )  +\ldots \Bigr]\;,
\end{eqnarray}
where the limit $\alpha \to 0$ is understood in order to recover the Landau gauge.  The $\ldots$ stands for the constant term $-d (N^2 -1) \gamma^4$ and other terms in the ghost- and $\omega, \overline{\omega}$-fields irrelevant for the calculation of the gluon propagator.  We take $\gamma^2 g$ to be of order $g^0$, which is equivalent to the rescaling $\gamma^2 \to \gamma^2/g$.  Next, we integrate out the $\varphi$- and $\overline{\varphi}$-fields. As we are only interested in the gluon propagator, we simply use the equations of motion, $\frac{\p S_\GZ^0 }{\p \overline{\varphi}_{\mu}^{bc}} = 0$ and $\frac{\p S_\GZ^0 }{\p\varphi_{\mu}^{bc}} = 0$, which give
\begin{eqnarray}
    \varphi^{bc}_{\mu} = \overline{\varphi}^{bc}_{\mu} =  \frac{1}{-\p^2}\gamma^2 g f^{abc} A^a_{\mu} \;.
\end{eqnarray}
We use this result to rewrite $ S_\GZ^0$,
\begin{eqnarray}
 S_\GZ^0 &=& \; \int \d^d x \; \left[ \frac{1}{4} \left(\p_{\mu} A_{\nu}^a - \p_{\nu}A_{\mu}^a \right)^2+ \frac{1}{2\alpha} \left(\p_{\mu} A^a_{\mu} \right)^2  + \gamma^4 g^2 f^{abc} A_{\mu}^a \frac{1}{\p^2 } f^{dbc} A^d_{\mu} \right. \nonumber\\
&& \left. \hspace{1cm}- 2 \gamma^4 g(f^{abc}A^a_{\mu} \frac{1}{\p^2} g f^{dbc} A_{\mu}^d ) + \ldots \right] \nonumber\\
&=&  \; \int \d^dx \; \left[ \frac{1}{4} \left( \p_{\mu} A_{\nu}^a - \p_{\nu}A_{\mu}^a \right)^2+ \frac{1}{2\alpha} \left( \p_{\mu} A^a_{\mu} \right)^2  - N \gamma^4 g^2 A_{\mu}^a \frac{1}{\p^2 } A^a_{\mu} + \ldots \right]\;,
\end{eqnarray}
where the last step relies on the relation \eqref{liestructure}. We continue rewriting $ S_\GZ^0$ so we can easily read off the gluon propagator
\begin{eqnarray}
  S_\GZ^0&=&  \; \int \d^d x \; \left[ \frac{1}{2} A^a_{\mu} \Delta^{ab}_{\mu\nu} A^b_{\nu} + \ldots \right] \;, \nonumber\\
\Delta^{ab}_{\mu\nu} &=&\left[ \left(-\p^2 - \frac{2 g^2 N \gamma^4}{\p^2} \right) \delta_{\mu\nu} - \p_{\mu}\p_{\nu} \left(\frac{1}{\alpha} - 1\right) \right]\delta^{ab} \;.
\end{eqnarray}
The gluon propagator can be determined by taking the inverse of $\Delta^{ab}_{\mu\nu}$ and converting it to momentum space. Doing so, we find the following expression
\begin{eqnarray}
  \; \Braket{ A^a_{\mu}(p) A^b_{\nu}(k)} &=& \delta(p+k) (2\pi)^d \underbrace{ \frac{p^2}{p^4 + 2g^2 N \gamma^4}}_{\mathcal{D}(p^2)}\left[\delta_{\mu\nu} - \frac{p_{\mu}p_{\nu}}{p^2} \right]\delta^{ab} \;,\nonumber\\
\end{eqnarray}
which is exactly the same expression as Gribov found, see equation \eqref{gluonprop}. We can already observe that this expression is suppressed in the infrared region,
while displaying complex poles at $p^{2}=\pm i{\hat \gamma}^{2}$, where $\hat\gamma^4 \equiv 2 g^2 N \gamma^4$ is taken to be of order $g^0$. This structure does not allow us to attach the usual particle meaning to the gluon propagator, invalidating the
interpretation of gluons as excitations of the physical spectrum. In other words, gluons cannot be considered as part of the physical spectrum. In this sense, they are confined by the Gribov horizon, whose presence is encoded in the explicit dependence of the propagator on the Gribov parameter $\gamma$.

\subsubsection{The ghost propagator}
\begin{figure}[H]
\begin{center}
\includegraphics[width=8cm]{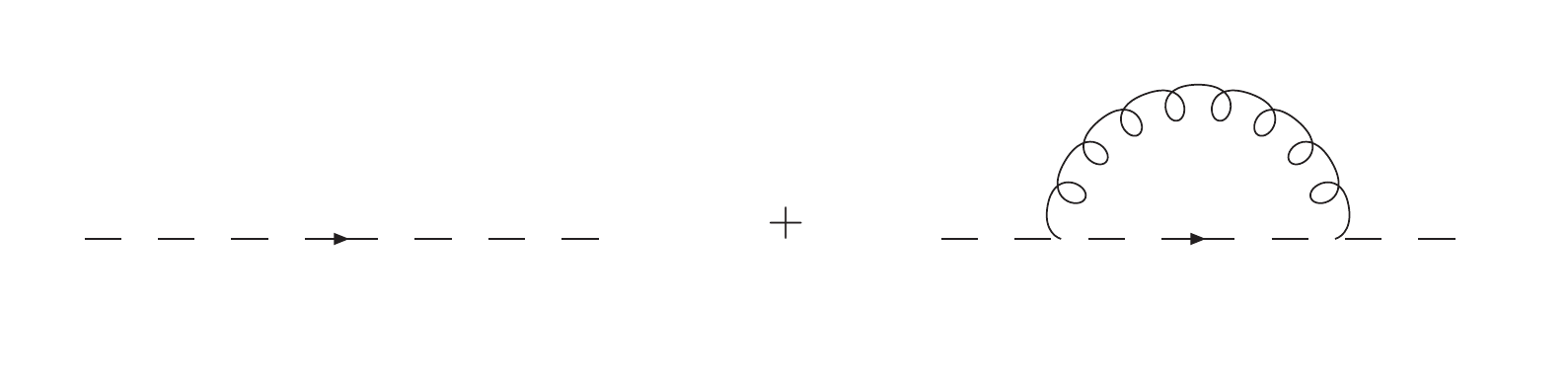}
\caption{The one loop corrected ghost propagator.}\label{2ghostprop}
\end{center}
\end{figure}

We shall calculate the ghost propagator to one-loop order, see figure \ref{2ghostprop}, taking into account the horizon condition \eqref{gapgamma}.

\noindent In momentum space, the ghost propagator is given by
\begin{eqnarray}
\Braket{c^a(p) \overline{c}^b(k)} &=&  \delta^{ab} (2\pi)^d \delta(k-p) \mathcal{G}(k^2)\;,
\end{eqnarray}
whereby
\begin{eqnarray}\label{ghostlabel}
\mathcal{G}(k^2) &=& \frac{1}{k^2} + \frac{1}{k^2} \left[g^2 \frac{N}{N^2 - 1} \int \frac{\d^d
q}{(2\pi)^4} \frac{(k-q)_{\mu} k_{\nu}}{(k-q)^2}
\frac{q^2}{q^4 + 2g^2 N \gamma^4} \right] P_{\mu \nu}(q) \frac{1}{k^2} \nonumber\\
&=& \frac{1}{k^2} (1+ \sigma(k^2)) +
\mathcal{O}(g^4)\;,
\end{eqnarray}
with
\begin{eqnarray}\label{ghi}
\sigma(k^2) &=& Ng^2 \frac{1}{k^2}\int \frac{\d^d q}{(2\pi)^d} \frac{(k-q)_{\mu} k_{\nu}}{(k-q)^2} \frac{q^2}{q^4 + 2g^2 N \gamma^4} \left(\delta_{\mu\nu} - \frac{q_\mu q_\nu}{q^2} \right)\nonumber\\
&=&  Ng^2\frac{k_\mu k_\nu}{k^2}\int \frac{\d^d q}{(2\pi)^d} \frac{1}{(k-q)^2} \frac{q^2}{q^4 + 2g^2 N \gamma^4} \left(\delta_{\mu\nu} - \frac{q_\mu q_\nu}{q^2} \right)\;,
\end{eqnarray}
which is identical to expression \eqref{ghostpropagator}, apart from the redefinition of $\gamma$.  Resumming the one-particle reducible diagrams gives
\beq
\label{1pr}
\mathcal G(k^2) = \frac{1}{k^2} \frac{1}{1 - \sigma(k^2)}.
\eeq 
We are again interested in the low momentum behavior and therefore calculate $\sigma(0)$,
\begin{eqnarray}\label{sigmanul}
\sigma(0) &=&  Ng^2 \frac{k_\mu k_\nu}{k^2} \delta_{\mu\nu} \frac{d-1}{d}\int \frac{\d^d q}{(2\pi)^d} \frac{1}{q^2} \frac{q^2}{q^4 + 2g^2 N \gamma^4}\nonumber\\
&=&  Ng^2  \frac{d-1}{d}\int \frac{\d^d q}{(2\pi)^d}  \frac{1}{q^4 + 2g^2 N \gamma^4} \;.
\end{eqnarray}
Notice the dimensional regularization that is necessary to make this integral well defined. \\
\\
To calculate it, we invoke the gap equation \eqref{gapgamma}. Firstly, we calculate the effective action. The one loop effective action $\Gamma_\gamma^{(1)}$ is obtained from the quadratic part of our action $S_\GZ^0$
\begin{equation}
\e^{-\Gamma_\gamma ^{(1)} }=\int [\d\Phi] \e^{-S_\GZ^0}\;,
\end{equation}
with
\begin{eqnarray*}
 S_\GZ^0 &=&   \; \int \d^dx \; \left[ \frac{1}{4} \left( \p_{\mu} A_{\nu}^a - \p_{\nu}A_{\mu}^a \right)^2+ \frac{1}{2\alpha} \left( \p_{\mu} A^a_{\mu} \right)^2  - N \gamma^4 g^2 A_{\mu}^a \frac{1}{\p^2 } A^a_{\mu}  -d(N^2 - 1) \gamma^4 + \ldots \right]\;.
\end{eqnarray*}
Notice that we need to keep the constant term $-d(N^2 - 1) \gamma^4$ as it enters the horizon condition. A~straightforward calculation gives the one loop effective action in $d$ dimensions,
\begin{eqnarray}
\Gamma_\gamma^{(1)} &=& -d(N^{2}-1)\gamma^{4} +\frac{(N^{2}-1)}{2}\left( d-1\right) \int \frac{\d^{d}q}{\left(
2\pi \right) ^{d}} \ln \frac{q^4  + 2 g^2 N \gamma^4}{ q^2} \;.
\end{eqnarray}
Now we can apply the gap equation \eqref{gapgamma},
\begin{equation}
\frac{\p  \Gamma_\gamma^{(1)}  }  {\p \gamma^2} = - 2 \gamma^2 d(N^{2}-1) + 2g^2 N (N^2 -1)\gamma^2 (d-1)\int \frac{\d^{d}q}{\left(2\pi \right) ^d}  \frac{1}{q^4  + 2 g^2 N \gamma^4} ~=~ 0\;,
\end{equation}
or thus
\begin{eqnarray}
1 &=&   g^2 N \frac{d-1}{d}\int \frac{\d^{d}q}{\left(2\pi \right) ^{d}}  \frac{1}{q^4  + 2 g^2 N \gamma^4}\;,
\end{eqnarray}
which exactly expresses
\begin{eqnarray}\label{sigmagelijkaan1}
\sigma(0) = 1\;,
\end{eqnarray}
see expression \eqref{sigmanul}. By \eqref{1pr} this means that the ghost propagator is enhanced, just as we expected from the semi-classical calculation of Gribov. Moreover, this result has been explicitly checked up to two loops, see \cite{Ford:2009ar,Gracey:2005cx}.  A more detailed calculation to this order yields
\beq
\mathcal G(k^2) \sim \frac{1}{(k^2)^2},
\eeq
as originally found by Gribov.

\section{Algebraic renormalization of the GZ action\label{renorma}}

We shall now prove that the GZ action is renormalizable to all orders by using \textit{algebraic renormalization}, see \cite{Piguet1995,Vandersickel:2011zc}. In \cite{Zwanziger:1989mf} and \cite{Zwanziger:1992qr}, it was first shown that the GZ action was renormalizable.  The Ward identities in the form given below were first given in \cite{Maggiore:1993wq}.  A first algebraic proof was given in \cite{Sobreiro:2004us}, and made complete in \cite{Dudal:2010fq}. A recent alternative proof can be found in \cite{Capri:2010hb}.  While the proof of renormalizability is technical, the underlying reason is simple: the term $S_\gamma$,  given below, that breaks BRST symmetry is soft, of dimension 2.

\subsection{The starting action and BRST}
We start with the action
\begin{eqnarray}\label{GZstart}
S_\GZ &=& S_\YM + S_\gf + S_{0} +  S_{\gamma} \,,
\end{eqnarray}
with
\begin{eqnarray}
\label{SzeroSgamma}
S_{0}&=& \int \d^d x \left[ \overline \varphi_i^a \p_\mu \left( D_\mu^{ab} \varphi^b_i \right)  - \overline \omega_i^a \p_\mu \left( D_\mu^{ab} \omega_i^b \right) - g f^{abc} \p_\mu \overline \omega_i^a   \left( D_\mu^{bd} c^d \right)  \varphi_i^c \right] \nonumber \;, \\
S_{\gamma}&=& -\gamma ^{2}g\int\d^{d}x\left( f^{abc}A_{\mu }^{a}\varphi _{\mu }^{bc} +f^{abc}A_{\mu}^{a}\overline{\varphi }_{\mu }^{bc} + \frac{d}{g}\left(N^{2}-1\right) \gamma^{2} \right) \;.
\end{eqnarray}
We recall that we have simplified the notation of the additional fields $\left( \overline \varphi_\mu^{ac},\varphi_\mu^{ac},\overline \omega_\mu^{ac},\omega_\mu^{ac}\right) $ in $S_0$ as $S_0$ displays a symmetry with respect to the composite index $i=\left( \mu,c\right)$. Therefore, we have set
\begin{equation}
\left( \overline \varphi_\mu^{ac},\varphi_\mu^{ac},\overline \omega_\mu^{ac},\omega_\mu^{ac}\right) =\left( \overline \varphi_i^a,\varphi_i^a,\overline \omega_i^a,\omega_i^a \right)\,.
\end{equation}
The BRST variations \eqref{BRST} can be logically extended for all the fields,
\begin{align}\label{BRST1}
sA_{\mu }^{a} &=-\left( D_{\mu }c\right) ^{a}\,, & sc^{a} &=\frac{1}{2}gf^{abc}c^{b}c^{c}\,,   \nonumber \\
s\overline{c}^{a} &=b^{a}\,,&   sb^{a}&=0\,,  \nonumber \\
s\varphi _{i}^{a} &=\omega _{i}^{a}\,,&s\omega _{i}^{a}&=0\,,\nonumber \\
s\overline{\omega}_{i}^{a} &=\overline{\varphi }_{i}^{a}\,,& s \overline{\varphi }_{i}^{a}&=0\,,
\end{align}
and remains nilpotent, $s^2 = 0$.
We note that $S_0$ is BRST-exact,
\beq
S_0 = s \int \d^d x \ \overline \omega_i^a \p_\mu \left( D_\mu^{ab} \varphi^b_i \right),
\eeq
and the action of $s$ on the field $A$ hidden in $D_\mu^{ab} = D_\mu^{ab}(A)$ explains the origin of the third term of $S_0$, eq.~\eqref{SzeroSgamma}.  Consequently the term  $S_0$ is an allowed gauge-fixing term within the BRST approach \cite{Zwanziger:1992qr} that has no effect whatsover on the expectation-value of gauge-invariant quantities.  However in the GZ action, $S_\GZ = S_0 + S_\gamma$, the $\gamma$-dependent term, $S_\gamma$, breaks this BRST symmetry \cite{Zwanziger:1989mf,Dudal:2008sp},
\begin{equation}\label{breaking}
s S_\GZ ~=~ s  S_{\gamma} ~=~- g \gamma^2 \int \d^d x f^{abc} \left[ A^a_{\mu} \omega^{bc}_\mu -
 \left(D_{\mu}^{am} c^m\right)\left( \overline{\varphi}^{bc}_\mu + \varphi^{bc}_{\mu}\right)  \right]\,.
\end{equation}
However the breaking is soft because the mass dimension of the integrand here, and also in $S_\gamma$, is 2, instead of 4 (in 4 dimensions).  In section \ref{BRSTbreaking}, we shall elaborate on the meaning of this BRST breaking.
\vspace{.5cm}

[As shown below \eqref{S0}, we have the freedom to modify the action by adding any term in $\overline\omega c$.  This allows us to write the starting action in the alternate form,
\beq
\widehat S_\GZ = S_\YM + S_\gf + S_{0} + \widehat S_{\gamma} \,,
\eeq
where
\beq
\label{alternateaction}
\widehat S_{\gamma} \equiv -\gamma ^{2}g\int\d^{d}x\left( f^{abc}A_{\mu }^{a}\varphi _{\mu }^{bc} + sf^{abc}A_{\mu}^{a}\overline{\omega}_{\mu }^{bc} + \frac{d}{g}\left(N^{2}-1\right) \gamma^{2} \right).
\eeq  
For we have
\beqa
\label{relatewidehat}
\widehat S_{\gamma} & = & -\gamma ^{2}g\int\d^{d}x\left( f^{abc}A_{\mu }^{a}\varphi _{\mu }^{bc} + f^{abc}(A_{\mu}^{a}\overline{\varphi}_{\mu }^{bc} + D_{\mu}^{ad}c^d\overline{\omega}_{\mu }^{bc}) + \frac{d}{g}\left(N^{2}-1\right) \gamma^{2} \right), 
\nonumber  \\
&= & S_\gamma -\gamma ^{2}g \int\d^{d}x f^{abc} D_{\mu}^{ad}c^d\overline{\omega}_{\mu }^{bc},
\eeqa 
which indeed differs from $S_\gamma$, eq.\ \eqref{SzeroSgamma}, by a term in $\overline\omega c$.  Note that the second term in $\widehat S_\gamma$ is $s$-exact.  This will simplify the renormalization somewhat, as we shall note parenthetically along the way.]  
\vspace{0.5cm}

In order to discuss the renormalizability of $S_\GZ$ [or $\widehat S_\GZ$], we should treat the breaking as a composite operator to be introduced into the action by means of a suitable set of external sources. This procedure can be done in a BRST invariant way, by embedding $S_\GZ$ into a larger action, namely
\begin{eqnarray}\label{brstinvariant}
\Sigma_\GZ &=& S_{\YM} + S_{\gf} + S_0 + S_\s \,,
\end{eqnarray}
where
\begin{eqnarray}\label{previous}
S_\s &=& s\int \d^d x \left( -U_\mu^{ai} D_\mu^{ab} \varphi_i^b - V_\mu^{ai} D_{\mu}^{ab} \overline \omega_i^{b} - U_\mu^{ai} V_\mu^{ai}  + T_\mu^{a i} g f_{abc} D^{bd}_\mu c^d \overline \omega^c_i \right)\nonumber\\
&=& \int \d^d x \left( -M_\mu^{ai}  D_\mu^{ab} \varphi_i^b - gf^{abc} U_\mu^{ai}   D^{bd}_\mu c^d  \varphi_i^c   + U_\mu^{ai}  D_\mu^{ab} \omega_i^b - N_\mu^{ai}  D_\mu^{ab} \overline \omega_i^b - V_\mu^{ai}  D_\mu^{ab} \overline \varphi_i^b \right. \nonumber\\
&&\left.+ gf^{abc} V_\mu^{ai} D_\mu^{bd} c^d \overline \omega_i^c - M_\mu^{ai} V_\mu^{ai}+U_\mu^{ai} N_\mu^{ai}  + R_\mu^{ai} g f^{abc} D_\mu^{bd} c^d \overline \omega^c_i  + T_\mu^{ai} g f_{abc} D^{bd}_\mu c^d \overline \varphi^c_i\right) \,. \nonumber\\
\end{eqnarray}
We have introduced 3 new doublets ($U_\mu^{ai}$, $M_\mu^{ai}$), ($V_\mu^{ai}$, $N_\mu^{ai}$) and ($T_\mu^{ai}$, $R_\mu^{ai}$) with the following BRST transformations,
and
\begin{align}\label{BRST2}
sU_{\mu }^{ai} &= M_{\mu }^{ai}\,, & sM_{\mu }^{ai}&=0\,,  \nonumber \\
sV_{\mu }^{ai} &= N_{\mu }^{ai}\,, & sN_{\mu }^{ai}&=0\,,\nonumber \\
sT_{\mu }^{ai} &= R_{\mu }^{ai}\,, & sR_{\mu }^{ai}&=0\;.
\end{align}
We have therefore restored the broken BRST at the expense of introducing new sources. However, we do not want to alter our original theory $S_\GZ$. Therefore, at the end, we have to set the sources equal to the following ``physical" values:
\begin{eqnarray}\label{physlimit}
&& \left. U_\mu^{ai}\right|_{\phys} = \left. N_\mu^{ai}\right|_{\phys} = \left. T_\mu^{ai}\right|_{\phys} = 0 \,, \nonumber\\
&& \left. M_{\mu \nu }^{ab}\right|_{\phys}= \left.V_{\mu \nu}^{ab}\right|_{\phys}=  -\left.R_{\mu \nu}^{ab}\right|_{\phys} = \gamma^2 \delta ^{ab}\delta _{\mu \nu } \,,
\end{eqnarray}
which yields
\beq
S_\GZ = \left. \Sigma_\GZ\right|_{\phys}.
\eeq
The doublet ($T_\mu^{ai}$, $R_\mu^{ai}$) was introduced  in \cite{Dudal:2010fq}  and provides for the first time a correct algebraic renormalization of the action $S_{\rm GZ}$ given in \eqref{GZstart}.  This happens because the terms $gf^{abc} V_\mu^{ai} D_\mu^{bd} c^d \overline \omega_i^c$ and $ R_\mu^{ai} g f^{abc} D_\mu^{bd} c^d \overline \omega^c_i $ cancel at the physical value.
\vspace{.5cm}

[Alternatively we may embed $\widehat S_\GZ$ into a larger action, 
\begin{eqnarray}\label{brstinvariant}
\widehat \Sigma_\GZ &=& S_{\YM} + S_{\gf} + S_0 + \widehat S_\s \,,
\end{eqnarray}
where
\beq
\widehat S_\s \equiv S_\s|_{R = T = 0}
\eeq
depends on only the two doublets, $(U, M)$ and $(V, N)$.  The original action $\widehat S_\GZ$ is regained at the physical values of the sources $U, \ M, \ V, \ N$,
\beq
\widehat S_\GZ = \left. \widehat \Sigma_\GZ \right|_{U_\phys, \ M_\phys, \ V_\phys, \ N_\phys}.
\eeq
We shall provide an algebraic renormalization of $\widehat S_\GZ$.]

\subsection{The Ward identities}
Following the procedure of algebraic renormalization outlined in \cite{Piguet1995}, we should try to find all possible Ward identities. Before doing this, in order to be able to write the Slavnov-Taylor identity, we first have to couple all nonlinear BRST transformations to a new source. Looking at \eqref{BRST1}, we see that only $A_\mu^a$ and $c^a$ transform nonlinearly under the BRST $s$. Therefore, we add the following term to the action $\Sigma_\GZ $,
\begin{eqnarray}\label{ext}
S_{\mathrm{ext}}&=&\int \d^d x\left( -K_{\mu }^{a}\left( D_{\mu }c\right) ^{a}+\frac{1}{2}gL^{a}f^{abc}c^{b}c^{c}\right) \;,
\end{eqnarray}
with $K_{\mu }^{a}$ and $L^a$ two new sources which shall be put to zero at the end,
\begin{eqnarray}\label{physlimit2}
\left. K_{\mu }^{a}\right|_{\phys} =\left. L^{a}\right|_{\phys}  = 0\;.&
\end{eqnarray}
These sources are invariant under the BRST transformation,
\begin{align}
s K_{\mu }^{a} &=0\;, & s L^{a} &= 0\;.
\end{align}
The new action is therefore given by
\begin{eqnarray}\label{enlarged}
\Sigma'_\GZ &=& \Sigma_\GZ + S_{\mathrm{ext}} \;.
\end{eqnarray}
The next step is now to find all the Ward identities obeyed by the action $\Sigma'_\GZ$. We have listed all the identities below:

\begin{table}
\caption{Quantum numbers of the fields.}
\label{2tabel1}
\begin{center}
\begin{tabular}{|c|c|c|c|c|c|c|c|c|}
\hline
& $A_{\mu }^{a}$ & $c^{a}$ & $\overline{c}^{a}$ & $b^{a}$ & $\varphi_{i}^{a} $ & $\overline{\varphi }_{i}^{a}$ &                $\omega _{i}^{a}$ & $\overline{\omega }_{i}^{a}$   \\
\hline
\textrm{dimension} & $1$ & $0$ &$2$ & $2$ & $1$ & $1$ & $1$ & $1$ \\
$\mathrm{ghost\; number}$ & $0$ & $1$ & $-1$ & $0$ & $0$ & $0$ & $1$ & $-1$ \\
$Q_{f}\textrm{-charge}$ & $0$ & $0$ & $0$ & $0$ & $1$ & $-1$& $1$ & $-1$ \\
\hline
\end{tabular}
\end{center}
\end{table}

\begin{table}
\caption{Quantum numbers of the sources.}
\begin{center}
\label{2tabel2}
\begin{tabular}{|c|c|c|c|c|c|c|c|c|}\hline
        $U_{\mu}^{ai}$&$M_{\mu }^{ai}$&$N_{\mu }^{ai}$&$V_{\mu }^{ai}$& $R_{\mu }^{ai}$  &  $T_{\mu }^{ai}$ &$K_{\mu }^{a}$&$L^{a}$  \\
\hline
         $2$ & $2$ & $2$ &$2$ & 2&2  & $3$ & $4$  \\
         $-1$& $0$ & $1$ & $0$ & 0& -1 & $-1$ & $-2$  \\
         $-1$ & $-1$ & $1$ & $1$ &1&1& $0$ & $0$  \\
\hline
\end{tabular}
\end{center}
\end{table}
\label{pagewardidentities}
\begin{enumerate}
\item The Slavnov-Taylor identity is given by
\begin{equation}\label{slavnov}
\mathcal{S}(\Sigma'_\GZ )=0\;,
\end{equation}
with
\begin{multline*}
\mathcal{S}(\Sigma'_\GZ ) =\int \d^d x\left( \frac{\delta \Sigma'_\GZ
}{\delta K_{\mu }^{a}}\frac{\delta \Sigma'_\GZ }{\delta A_{\mu
}^{a}}+\frac{\delta \Sigma'_\GZ }{\delta L^{a}}\frac{\delta \Sigma'_\GZ
}{\delta c^{a}} \right. \nonumber\\
\left. +b^{a}\frac{\delta \Sigma'_\GZ }{\delta \overline{c}^{a}}+\overline{\varphi }_{i}^{a}\frac{\delta \Sigma'_\GZ }{\delta \overline{\omega }_{i}^{a}}+\omega _{i}^{a}\frac{\delta \Sigma'_\GZ }{\delta \varphi _{i}^{a}} +M_{\mu }^{ai}\frac{\delta \Sigma'_\GZ}{\delta U_{\mu}^{ai}}+N_{\mu }^{ai}\frac{\delta \Sigma'_\GZ }{\delta V_{\mu }^{ai}} + R_{\mu }^{ai}\frac{\delta \Sigma'_\GZ }{\delta T_{\mu }^{ai}}\right) \;.
\end{multline*}

\item The $U(f)$ invariance is given by
\begin{equation}\label{ward1}
U_{ij} \Sigma'_\GZ =0\;,
\end{equation}
\begin{multline}
U_{ij}=\int \d^dx\Bigl( \varphi_{i}^{a}\frac{\delta }{\delta \varphi _{j}^{a}}-\overline{\varphi}_{j}^{a}\frac{\delta }{\delta \overline{\varphi}_{i}^{a}}+\omega _{i}^{a}\frac{\delta }{\delta \omega _{j}^{a}}-\overline{\omega }_{j}^{a}\frac{\delta }{\delta \overline{\omega }_{i}^{a}} \\
-  M^{aj}_{\mu} \frac{\delta}{\delta M^{ai}_{\mu}} -U^{aj}_{\mu}\frac{\delta}{\delta U^{ai}_{\mu}} + N^{ai}_{\mu}\frac{\delta}{\delta N^{aj}_{\mu}}
  +V^{ai}_{\mu}\frac{\delta}{\delta V^{aj}_{\mu}}    +  R^{aj}_{\mu}\frac{\delta}{\delta R^{ai}_{\mu}} + T^{aj}_{\mu}\frac{\delta}{\delta T^{ai}_{\mu}} \Bigr)  \;. \nonumber
\end{multline}
By means of the diagonal operator $Q_{f}=U_{ii}$, the
$i$-valued fields and sources turn out to possess an additional quantum number.
One can find all quantum numbers in Table \ref{2tabel1} and Table \ref{2tabel2}.

\item The Landau gauge condition reads
\begin{eqnarray}\label{gaugeward}
\frac{\delta \Sigma'_\GZ }{\delta b^{a}}&=&\partial_\mu A_\mu^{a}\;.
\end{eqnarray}

\item The antighost equation yields
\begin{eqnarray}
\frac{\delta \Sigma'_\GZ }{\delta \overline{c}^{a}}+\partial _{\mu}\frac{\delta \Sigma'_\GZ }{\delta K_{\mu }^{a}}&=&0\;.
\end{eqnarray}

\item The linearly broken local constraints yield
\begin{eqnarray}
\frac{\delta \Sigma'_\GZ }{\delta \overline{\varphi }^{a}_i}+\partial _{\mu }\frac{\delta \Sigma'_\GZ }{\delta M_{\mu }^{ai}} + g f_{dba}    T^{d i}_\mu \frac{\delta \Sigma'_\GZ }{\delta K_{\mu }^{b i}} &=&gf^{abc}A_{\mu }^{b}V_{\mu}^{ci} \;, \nonumber\\
\frac{\delta \Sigma'_\GZ }{\delta \omega ^{a}_i}+\partial _{\mu}\frac{\delta \Sigma'_\GZ }{\delta N_{\mu}^{ai}}-gf^{abc}\overline{\omega }^{b}_i\frac{\delta \Sigma'_\GZ }{\delta b^{c}}&=&gf^{abc}A_{\mu }^{b}U_{\mu }^{ci} \;.
\end{eqnarray}

\item The exact $\mathcal{R}_{ij}$ symmetry reads
\begin{equation}
\mathcal{R}_{ij}\Sigma'_\GZ =0\;,
\end{equation}
with
\begin{multline}\label{rij}
\mathcal{R}_{ij} = \int \d^dx\left( \varphi _{i}^{a}\frac{\delta}{\delta\omega _{j}^{a}}-\overline{\omega }_{j}^{a}\frac{\delta }{\delta \overline{\varphi }_{i}^{a}}+V_{\mu }^{ai}\frac{\delta }{\delta N_{\mu}^{aj}}-U_{\mu }^{aj}\frac{\delta }{\delta M_{\mu }^{ai}} + T^{a i}_\mu \frac{\delta }{\delta R_{\mu }^{aj}}  \right) \;.
\end{multline}

\item The integrated Ward identity is given by
\begin{equation}
\mathcal{F}_i \ \Sigma'_\GZ \equiv \int \d^d x \left( c^a \frac{ \delta \Sigma'_\GZ }{ \delta \omega^{ a}_i} + \overline \omega^{a}_i \frac{ \delta \Sigma'_\GZ }{ \delta  \overline c^a } + U^{a i}_\mu \frac{ \delta \Sigma'_\GZ }{ \delta  K^a_\mu }  \right) = 0\;.
\end{equation}

\item Due to the presence of the sources $T_{\mu }^{ai}$ and $R_{\mu }^{ai}$, the Ghost-Ward identity is broken, an identity which is important in ordinary Yang-Mills theory. However, it shall turn out that this is not a problem for the renormalization procedure being undertaken.
\vspace{.5cm}

[The alternate action,
\beq
\widehat\Sigma'_\GZ \equiv \widehat\Sigma_\GZ + S_\ext \ ,
\eeq
satisfies the above 7 Ward identities with $R = T = 0$.  In addition it satisfies the Ghost-Ward identity which is characteristic of Faddeev-Popov theory in the Landau gauge,
\begin{equation}\label{GW}
\mathcal{G}^{a} \ \widehat\Sigma'_\GZ = \Delta _{\mathrm{cl}}^{a}\,,
\end{equation}
with
\begin{eqnarray}
\mathcal{G}^{a} &=&\int \d^dx\left[ \frac{\delta }{\delta c^{a}}+gf^{abc} \left( \overline{c}^{b}\frac{\delta }{\delta b^{c}} + \varphi_i^{b}\frac{\delta }{\delta \omega_i^{c}} + \overline\omega_i^{b}\frac{\delta }{\delta \overline\varphi_i^{c}} + V_\mu^{ib}\frac{\delta }{\delta N_\mu^{ic}} + U_\mu^{ib}\frac{\delta }{\delta M_\mu^{ic}} \right) \right] \,, \nonumber \\
\
\end{eqnarray}
and
\begin{equation}
\Delta _{\mathrm{cl}}^{a}=g\int \d^{4}xf^{abc}\left( K_{\mu}^{b}A_{\mu }^{c} - L^{b}c^{c}\right) \;,
\end{equation}
where all fermionic derivatives are left derivatives.  This Ghost-Ward identity, which is not satisfied by $S_\GZ$, simplifies the proof of renormalizability of $\widehat S_\GZ$.]
\vspace{.5cm}
\end{enumerate}

\subsection{The counter-term}
The next step in the algebraic renormalization is to translate all these symmetries into constraints on the counter-term $\Sigma_\GZ^c$, which is an integrated polynomial in the fields and sources of dimension four and with ghost number zero. The classical action $\Sigma_\GZ'$ changes under quantum corrections according to
\begin{eqnarray}\label{deformed}
    \Sigma_\GZ' \rightarrow \Sigma_\GZ' + h \Sigma_\GZ^c\,,
\end{eqnarray}
whereby $h$ is the perturbation parameter. Demanding that the perturbed action $(\Sigma_\GZ' + h \Sigma_\GZ^c)$ fulfills the same set of Ward identities obeyed by $\Sigma_\GZ'$, see \cite{Piguet1995}, it follows that the counterterm $\Sigma_\GZ^c$ is constrained by the following identities.
\vspace{.4cm}

[In the alternate formulation we have
\begin{eqnarray}\label{deformed}
    \widehat\Sigma_\GZ' \rightarrow \widehat\Sigma_\GZ' + h \widehat\Sigma_\GZ^c\,.
\end{eqnarray}
The 7 identities listed below that are satisfied by the counter-term $\Sigma_\GZ^c$  are also satisfied by $\widehat\Sigma_\GZ^c$, with $R = T = 0$.  As in Faddeev-Popov theory in the Landau gauge, it also satisfies the integrated Ghost-Ward identity,
\begin{equation}\label{GWc}
\mathcal{G}^{a} \ \widehat\Sigma_\GZ^c = 0 \,,
\end{equation}
which is not satisfied by $\Sigma_\GZ^c$.]\\
\begin{enumerate}
\item The linearized Slavnov-Taylor identity yields
\begin{equation}
\mathcal{B}\Sigma_\GZ^{c}=0\;,
\end{equation}
with $\mathcal{B}$ the nilpotent linearized Slavnov-Taylor operator,
\begin{multline}
\mathcal{B}=\int \d^{4}x\Bigl( \frac{\delta \Sigma_\GZ'}{\delta K_{\mu }^{a}}\frac{\delta }{\delta A_{\mu }^{a}}+\frac{\delta \Sigma_\GZ' }{\delta A_{\mu }^{a}}\frac{\delta }{\delta K_{\mu }^{a}}+\frac{\delta \Sigma_\GZ' }{\delta L^{a}}\frac{\delta }{\delta c^{a}}+\frac{\delta\Sigma_\GZ' }{\delta c^{a}}\frac{\delta }{\delta L^{a}}+b^{a}\frac{\delta }{\delta \overline{c}^{a}}\\
+\overline{\varphi}_{i}^{a}\frac{\delta }{\delta \overline{\omega }_{i}^{a}}+\omega_{i}^{a}\frac{\delta }{\delta \varphi_{i}^{a}}
+M_{\mu }^{ai}\frac{\delta }{\delta U_{\mu }^{ai}} + N_{\mu }^{ai}\frac{\delta }{\delta V_{\mu }^{ai}} + R_{\mu }^{ai}\frac{\delta }{\delta T_{\mu }^{ai}}  \Bigr) \,,
\end{multline}
and
\begin{equation}
\mathcal{B}^2=0\;.
\end{equation}

\item The $U(f)$ invariance reads
\begin{eqnarray}
U_{ij} \Sigma_\GZ^{c} &=&0 \;.
\end{eqnarray}

\item The Landau gauge condition
\begin{eqnarray}\label{2lgc}
\frac{\delta \Sigma_\GZ^{c}}{\delta b^{a}}&=&0\,.
\end{eqnarray}

\item The antighost equation
\begin{eqnarray}\label{2age}
\frac{\delta \Sigma_\GZ^{c}}{\delta \overline c^{a}}+\p_\mu\frac{\delta \Sigma_\GZ^{c}}{\delta K_{\mu}^a} &=&0 \,.
\end{eqnarray}

\item The linearly broken local constraints yield
\begin{eqnarray}
\left( \frac{\delta  }{\delta \overline{\varphi }^{a}_i}+\partial _{\mu }\frac{\delta}{\delta M_{\mu }^{ai}} +\partial _{\mu }\frac{\delta }{\delta M_{\mu }^{ai}} + g f_{abc}    T^{b i}_\mu \frac{\delta }{\delta K_{\mu }^{c i}}\right) \Sigma_\GZ^{c} &=&0 \;,\nonumber\\
\left( \frac{\delta }{\delta \omega ^{a}_i}+\partial _{\mu}\frac{\delta }{\delta N_{\mu}^{ai}}-gf^{abc}\overline{\omega }^{b}_i \frac{\delta }{\delta b^{c}} \right) \Sigma_\GZ^{c}&=&0 \;.
\end{eqnarray}

\item The exact $\mathcal{R}_{ij}$ symmetry reads
\begin{equation}
\mathcal{R}_{ij}\Sigma_\GZ^{c}=0\;,
\end{equation}
with  $\mathcal{R}_{ij}$ given in \eqref{rij}.

\item Finally, the integrated Ward identity becomes
\begin{equation}
\int \d^d x \left( c^a \frac{ \delta \Sigma_\GZ^{c}}{ \delta \omega^{ a}_i} + \overline \omega^{a}_i \frac{ \delta \Sigma_\GZ^{c}}{ \delta  \overline c^a } + U^{a i}_\mu \frac{ \delta \Sigma_\GZ^{c}}{ \delta  K^a_\mu }  \right) = 0 \;.
\end{equation}
\end{enumerate}
Now we can write down the most general counterterm $\Sigma_\GZ^{c}$ of $d=4$, which obeys the linearized Slavnov-Taylor identity, has ghost number zero, and vanishing $Q_f$ number,
\begin{eqnarray}\label{counterterm}
\Sigma^c_\GZ &=& a_0 S_{\YM} + \mathcal{B} \int \d^d \!x\,   \biggl\{  a_{1} K_{\mu}^{a} A_{\mu}^{a} + a_2 \partial _{\mu} \overline{c}^{a} A_{\mu}^{a}+a_3 \,L^{a}c^{a}
+a_4 U_{\mu}^{ai}\,\partial _{\mu }\varphi _{i}^{a} +a_5 \,V_{\mu}^{ai}\,\partial _{\mu }\overline{\omega }_{i}^{a}\nonumber\\
&& + a_6 \overline{\omega}_{i}^{a} \partial^{2} \varphi_{i}^{a} + a_7 U_{\mu}^{ai}V_{\mu}^{ai} + a_8 gf^{abc}U_{\mu}^{ai}\varphi_{i}^{b}A_{\mu}^{c}+a_9 gf^{abc}V_{\mu}^{ai}\overline{\omega }_{i}^{b}A_{\mu }^{c}\nonumber \\
&& +a_{10}gf^{abc}\overline{\omega }_{i}^{a}A_{\mu }^{c}\,\partial _{\mu }\varphi _{i}^{b}+a_{11}gf^{abc}\overline{\omega }_{i}^{a}(\partial _{\mu }A_{\mu}^{c})\varphi _{i}^{b} +b_1 R_{\mu}^{ai} U_{\mu}^{ai} +b_2 T_{\mu }^{ai} M_{\mu }^{ai} \nonumber\\
&&+ b_3 g f_{abc} R_{\mu }^{ai} \overline{\omega }_{i}^{b} A_{\mu}^{c} + b_4 g f_{abc} T_{\mu}^{ai} \overline{\varphi }_{i}^{b} A_{\mu}^{c} + b_5 R_{\mu}^{ai} \p_\mu \overline{\omega }_{i}^{a}  + b_6 T_{\mu}^{ai} \p_\mu \overline{\varphi }_{i}^{a} \biggr\} \;,
\end{eqnarray}
with $a_0, \ldots, a_{11}$ and $b_1, \ldots, b_6$ arbitrary parameters.  Now we can unleash the constraints on the counterterm. Firstly, although the ghost Ward identity \eqref{GW} is broken, we know that this is not so in the standard Yang-Mills case. Therefore, we can already set $a_3=0$ as this term is not allowed in the counterterm of the standard Yang-Mills action, which is a special case of the action we are studying.\footnote{In particular, since we will always assume the use of a mass independent renormalization scheme, we may compute $a_3$ with all external mass scales (= sources) equal to zero. Said otherwise, $a_3$ is completely determined by the dynamics of the original Yang-Mills action, in which case it is known to vanish to all orders.}
\vspace{.1cm}

[In the alternate formulation we have an identical expansion for $\widehat\Sigma_\GZ^c$, but with $b_1 = \ldots = b_6 = 0$.  The integrated Ghost-Ward identity, $\mathcal G \ \widehat\Sigma_\GZ^c = 0$, yields $a_3 = 0$ as in Faddeev-Popov theory in the Landau gauge.  This simplifies the above argument that $a_3 = 0$.]\\
\\
Secondly, due to the Landau gauge condition \eqref{2lgc} and the antighost equation \eqref{2age} we find,
\begin{eqnarray}
a_1 &=& a_2\;.
\end{eqnarray}
Next, the linearly broken constraints (5) give the following relations
\begin{align}
 a_1 &= -a_8  = - a_{9} = a_{10} = a_{11} = -b_3 = b_4\;, \nonumber\\
  a_4 &= a_5 = -a_6 = a_7\;, \quad b_1 =b_2 = b_5 = b_6 = 0 \;.
\end{align}
The $R_{ij}$ symmetry does not give any new information, while the integrated Ward identity relates the two previous strings of parameters:
\begin{multline}
 a_1 = -a_8  = - a_{9} = a_{10} = a_{11} = -b_3 = b_4  \equiv     a_4 = a_5 = -a_6 = a_7 \;.
\end{multline}
Taking all this information together, we obtain the following counterterm
\begin{multline}\label{final}
\Sigma^c= a_{0}S_{YM}  + a_{1}\int \d^dx\Biggl(  A_{\mu}^{a}\frac{ \delta S_{YM}}{\delta A_{\mu }^{a}}  + \p_\mu \overline{c}^a \p_\mu c^a  + K_{\mu }^{a}\partial _{\mu }c^{a}  + M_\mu^{a i} \p_\mu \varphi^{a}_i -  U_\mu^{a i} \p_\mu \omega^{a}_i \\
+ N_\mu^{a i} \p_\mu \overline{\omega}_i^{a} +  V_\mu^{a i}\p_\mu \overline{\varphi}^{a}_i   +  \p_\mu \overline{\varphi}^{a}_i \p_\mu \varphi^{a}_i +  \p_\mu \omega^{a}_i \p_\mu \overline{\omega}^{a}_i + V_\mu^{a i} M_\mu^{a i} - U_\mu^{a i}N_\mu^{a i}  - g f_{abc} U_\mu^{ai} \varphi^{b}_i \p_\mu c^c \\
- g f_{abc} V_\mu^{ai} \overline{\omega}^{b}_i \p_\mu c^c - g f_{abc} \p_{\mu} \overline{\omega}^a_i \varphi^{b}_i  \p_\mu c^c   - g f_{abc} R_\mu^{ai} \p_\mu c^b \overline \omega_i^c + g f_{abc} T_\mu^{ai} \p_\mu c^b \overline \varphi_i^c \Biggr) \;.
\end{multline}

\subsection{The renormalization factors}
As a final step, we have to show that the counterterm \eqref{final} can be reabsorbed by means of a multiplicative renormalization of the fields and sources.  We set $\phi = (A^a_\mu, c^a, \overline c^a, b^a, \varphi^a_i, \overline \varphi^a_i, \omega^a_i, \overline \omega^a_i)$ and $\Phi = (K^a_\mu, L^a, M^{ai}_\mu,  N^{ai}_\mu, V^{ai}_\mu, U^{ai}_\mu , R^{ai}_\mu, T^{ai}_\mu)$, and we define the renormalization constants
\begin{eqnarray}
g_0 &=& Z_gg \nonumber\\
\phi_0 &=& Z^{1/2}_\phi \phi \nonumber\\
 \Phi_0 &=& Z_\Phi \Phi ,
\end{eqnarray}
If we try to absorb the counterterm into the original action, we easily find,
\begin{eqnarray}\label{Z1}
Z_{g} &=&1-h \frac{a_0}{2}\,,  \nonumber \\
Z_{A}^{1/2} &=&1+h \left( \frac{a_0}{2}+a_{1}\right) \,,
\end{eqnarray}
and
\begin{eqnarray}\label{Z2}
Z_{\overline{c}}^{1/2} &=& Z_{c}^{1/2} = Z_A^{-1/4} Z_g^{-1/2} = 1-h \frac{a_{1}}{2}\,, \nonumber \\
Z_{b}&=&Z_{A}^{-1}\,, \nonumber\\
Z_{K }&=&Z_{c}^{1/2}\,,  \nonumber\\
Z_{L} &=&Z_{A}^{1/2}\,.
\end{eqnarray}
The results \eqref{Z1} and \eqref{Z2} are already known from the renormalization of the original Yang-Mills action in the Landau gauge. Further, we also obtain
{\allowdisplaybreaks \begin{eqnarray}\label{Z3}
Z_{\varphi}^{1/2} &=& Z_{\overline \varphi}^{1/2} = Z_g^{-1/2} Z_A^{-1/4} = 1 - h \frac{a_1}{2}\,, \nonumber\\
Z_\omega^{1/2} &=& Z_A^{-1/2} \,,\nonumber\\
Z_{\overline \omega}^{1/2} &=& Z_g^{-1} \,,\nonumber\\
Z_M &=& 1- h\frac{a_1}{2} = Z_g^{-1/2} Z_A^{-1/4}\,, \nonumber\\
Z_N &=& Z_A^{-1/2} \,, \nonumber\\
Z_U &=& 1 + h \frac{a_0}{2} = Z_g^{-1} \,, \nonumber\\
Z_V &=& 1- h \frac{a_1}{2} = Z_g^{-1/2}Z_A^{-1/4} \,,  \nonumber\\
Z_T &=& 1+h \frac{a_0}{2} = Z_g^{-1}  \,,  \nonumber\\
Z_R &=& 1- h \frac{a_1}{2} = Z_g^{-1/2}Z_A^{-1/4}\;.
\end{eqnarray}}

\noindent This concludes the proof of the renormalizability of the action \eqref{GZstart} which is the physical limit of $\Sigma_\GZ' $. Notice that in the physical limit \eqref{physlimit}, we have that
\begin{eqnarray}\label{Zgamma}
Z_{\gamma^2} &=& Z_g^{-1/2} Z_A^{-1/4}\;.
\end{eqnarray}

\section{Relation between Gribov no-pole condition and the GZ action}
To end this chapter, we would like to point out that the equivalence between the no-pole condition and Zwanziger's horizon condition has been checked up to third order in the gauge fields. All the details of the calculation can be found in \cite{Gomez:2009tj}.

\chapter{Features of the GZ action\label{sec2}}
In this section we shall elaborate on the GZ action by discussing a variety of topics. Firstly, we shall calculate all the propagators of this action and discuss the transversality of the gluon propagator. Secondly, we devote some effort to scrutinizing the BRST breaking of the GZ action and its consequences. We shall also demonstrate how one can restore this broken BRST by the introduction of additional fields. Next, we shall briefly touch the hermiticity of the GZ action and we shall also devote some words on the form of the horizon function in relation to renormalizability. Finally, we shall consider the Kugo-Ojima criterion in relation to the GZ action.

\section{The propagators of the GZ action}
Firstly, we shall calculate all the propagators of the GZ action, which is useful when doing loop calculations with the GZ action.  We start by taking only the quadratic part $S_{\RGZ}$ of the action $S_{\GZ}$ into account
\begin{multline}
S_{\RGZ} ~=~ \int \d^4 x \Bigl[ \frac{1}{4} (\p_\mu A_\nu^a - \p_\nu A_\mu^a)^2 + b^a \p_\mu A_\mu^a + \overline c^a \p_\mu^2 c^a  + \overline \varphi^a_i \p_\mu^2 \varphi^a_i - \overline \omega^a_i \p_\mu^2 \omega^a_i -\gamma^2 g f^{abc} A_\mu^a \varphi^{bc}_\mu\\ -\gamma^2 g f^{abc} A_\mu^a \overline \varphi^{bc}_\mu \Bigr]\;.
\end{multline}
We see three different parts appear:
\begin{multline}\label{3parts}
S_{\RGZ} = \int \d^4 x \Bigl[  \frac{A_\mu^a}{2} ( - \p^2 \delta^{\mu\nu} + \p_\mu \p_\nu) \delta^{ab} A_\nu^b   + \frac{1}{2} b^a \p_\mu A_\mu^a - \frac{1}{2} \p_\mu b^a A_\mu^a  +   \overline \varphi^{ab}_\mu \p^2  \varphi^{ab}_\mu  \\
 - \gamma^2 g f^{abc} A_\mu^a  (\varphi^{bc}_\mu  + \overline \varphi^{bc}_\mu) \Bigr]
+ \int \d^4 x \Bigl[  \overline \omega^a_i (-\p^2) \omega^a_i \Bigr] + \int \d^4 x \Bigl[ \overline c^a \p_\mu^2 c^a \Bigr]\;.
\end{multline}

\subsection*{The $\overline \omega \omega$ propagator}
The goal is to calculate the propagator
\begin{eqnarray}
\Braket{\widetilde{ \overline \omega}^a_\mu(p) \widetilde \omega^b_\nu(k)} \;.
\end{eqnarray}
As we are working in Euclidean space, the path integral is given by
\begin{eqnarray}
P &=& \int [\d \Phi] \e^{-S_\GZ}\;,
\end{eqnarray}
with $[\d \Phi]$ the integration over all the fields. In order to calculate the propagator in momentum space we can now employ formula \eqref{ghostapp}
\begin{align}
	I &= \int [\d\omega][\d\overline{\omega}] \exp\left[ \int \d^d x \d^d y \ \overline{\omega}_\mu(x) A_{\mu\nu} (x,y)\omega_\nu(y) + \int \d^d x \ (J^{\mu}_\omega(x) \omega_\mu(x) + \overline{\omega}_\nu(x) J_{\overline{\omega}}^\nu (x)) \right]  \nonumber \\
	&=C \det A \exp\left[  -\int \d^d x \d^d y \   J_\omega^\mu(x) A^{-1}_{\mu\nu}(x,y) J_{\overline{\omega}}^\nu(y) \right] \;,
\end{align}
where in our case:
\begin{eqnarray}
A(x,y) = \delta(x-y) (\p^2 ) \delta^{\mu\nu} \;, \nonumber\\
A^{-1}(x,y) = \delta(x-y)  \frac{1}{\p^2 }\delta^{\mu\nu} \;.
\end{eqnarray}
Now going to Fourierspace,
\begin{align}
	I &= \int [\d\omega][\d\overline{\omega}] \exp\left[ \int \frac{\d^d p}{(2\pi)^d}   \widetilde{\overline{\omega}}(-p)(-p^2 ) \widetilde{\omega}(p) + \int \frac{\d^d p}{(2\pi)^d}\ ( \widetilde {J}_\omega(-p) \widetilde{\omega}(p) + \widetilde{\overline{\omega}}(p) \widetilde{J}_{\overline{\omega}} (-p)) \right]  \nonumber \\
	&=C \det A \exp \left[ -\int \frac{\d^d p}{(2\pi)^d}    \widetilde{J}_\omega(-p)  \frac{1}{p^2} \widetilde{J}_{\overline{\omega}}(p) \right] \;,
\end{align}
we can calculate the propagator in Fourierspace:
\begin{equation}
\left. - (2\pi)^8 \frac{\delta }{\delta \widetilde J_{\overline{\omega}}(-p)} \frac{\delta }{\delta \widetilde J_\omega(-k)}I \right|_{J = 0} = \Braket{\widetilde{ \overline \omega}^{ab}_\mu(p) \widetilde \omega^{cd}_\nu(k)} ~=~ \delta^{ac}\delta^{bd} \delta^{\mu\nu} \frac{-1}{p^2 }\delta(p+ k) (2\pi)^4 \;.
\end{equation}

\subsection*{The ghost propagator}
Completely analogously, we find
\begin{eqnarray}
  \Braket{ \widetilde{ \overline c}^a(k) \widetilde c^b(p) } ~=~ \delta^{ab} \frac{1}{p^2} (2\pi)^4 \delta(p+ k) \;.
\end{eqnarray}

\subsection*{The mixed operators}
For the final part, we shall use the following general formula,
\begin{eqnarray}\label{general}
	I(A, J) &=& \int [\d \varphi] \exp \left[ - \frac{1}{2} \int \d^d x \d^d y\  \varphi(x) A(x,y) \varphi(y) +   \int \d^d x \ \varphi (x) J(x) \right] \nonumber \\
	&=& C (\det A)^{-1/2} \exp \left[ \frac{1}{2} \int \d^dx \d^dy\ J(x) A^{-1}(x,y) J(y) \right] \;.
\end{eqnarray}
However, as we see in the action \eqref{3parts} we need to rewrite the complex conjugate Bose fields $\overline \varphi$ and $\varphi$ in terms of real fields. Therefore, we shall introduce the real fields $U$ and $V$
\begin{eqnarray}
\varphi_\mu^{ab} &=& \frac{V_\mu^{ab} + i U_\mu^{ab} }{2}  \;, \nonumber\\
\overline \varphi_\mu^{ab} &=& \frac{V_\mu^{ab} - i U_\mu^{ab} }{2} \;.
\end{eqnarray}
We can thus rewrite the relevant part of the action as
\begin{multline}\label{boz}
S_{\RGZ}' = \int \d^4 x \Bigl[  \frac{A_\mu^a}{2} ( - \p^2 \delta^{\mu\nu} + \p_\mu \p_\nu) \delta^{ab} A_\nu^b   + \frac{1}{2} b^a \p_\mu A_\mu^a - \frac{1}{2} \p_\mu b^a A_\mu^a  + \\
 \frac{1}{4} \left( V^{ab}_\mu \p^2 V^{ab}_\mu + U^{ab}_\mu \p^2  U^{ab}_\mu \right) - \frac{1}{2}\gamma^2 g f^{abc} A_\mu^a  V^{bc}_\mu  - \frac{1}{2}\gamma^2 g f^{abc} A_\mu^a  V^{bc}_\mu \Bigr]\;.
\end{multline}
We  rewrite this in matrixform as
\begin{multline*}
\exp[-S_\RGZ'] =\exp [- \frac{1}{2} \int \d^4 x \underbrace{ \begin{bmatrix}
 A^a_{\mu}(x) & b^m(x) &  V^{k \ell }_{\alpha}(x) &  U^{k \ell }_{\alpha}(x)
\end{bmatrix}}_{X} \nonumber\\
\times
\underbrace{\begin{bmatrix} - \p^2  P_{\mu\nu}  \delta^{ab}  &  - \p_\mu \delta^{an}  &  - \gamma^2 g   f^{aij}  \delta_{\mu\kappa} & 0 \\ \\
  \p_\mu \delta^{bm} &0&0&0 \\ \\
   - \gamma^2 g   f^{bk\ell }  \delta_{\alpha\nu} &0& \frac{1}{2} \p^2   & 0 \\
   && \delta^{\alpha \kappa} \delta^{ki} \delta^{\ell j} & \\
   0 & 0&0 &  \frac{1}{2} \p^2   \\
   &&& \delta^{\beta \lambda }\delta^{sp} \delta^{tq}
 \end{bmatrix}}_{A}
\begin{bmatrix}
A^b_{\nu}(x) \\ \\ b^n(x) \\ \\ V^{ij}_{\kappa}(x) \\ \\ U^{pq}_{\lambda}(x)
\end{bmatrix}] \;,
\end{multline*}
where
\begin{eqnarray}
P_{\mu\nu} = \delta_{\mu\nu} - \frac{\p_{\mu}\p_{\nu}}{\p^2}, \hspace{1cm}   L_{\mu\nu} =  \frac{\p_{\mu}\p_{\nu}}{\p^2}\;,
\end{eqnarray}
are the respective transverse and longitudinal projectors. Notice that we have rewritten \eqref{boz} in a symmetric way. Now we can apply the general formula \eqref{general}, meaning that we have to find the inverse of $A$.
\begin{multline*}
\begin{bmatrix} - \p^2  P_{\mu\nu}  \delta^{ab}  &  - \p_\mu \delta^{an}  &  - \gamma^2 g   f^{aij}  \delta_{\mu\kappa} & 0 \\ \\
  \p_\nu\delta^{mb} &0&0&0 \\ \\
   - \gamma^2 g   f^{bk\ell }  \delta_{\alpha\nu} &0& \frac{1}{2} \p^2  & 0 \\
   && \delta^{\alpha \kappa} \delta^{ki} \delta^{\ell j} & \\
   0 & 0&0 &  \frac{1}{2} \p^2   \\
   &&& \delta^{\beta \lambda }\delta^{sp} \delta^{tq}
 \end{bmatrix}
 \underbrace{
 \begin{bmatrix}
 A^{b\ c}_{\nu\ \tau} &  B^{b\ o}_{\nu } & C_{\nu\ \omega}^{b\ xy} & D^{b\ gh}_{\nu\ \chi} \\ \\
 E^{n\ c}_{\hphantom{f\ } \tau}   &  F^{n \ o} & G_{\hphantom{f\ } \omega}^{ n\ xy} & H^{n \ gh}_{\hphantom{f\ } \chi} \\ \\
 I^{ij\ c}_{\kappa\ \tau}  &J_{\kappa }^{ij\ o} & K_{\kappa \ \omega}^{ij \ xy} & L^{ij \ gh}_{\kappa \ \chi} \\ \\
 M^{pq \ c}_{\lambda \ \tau} & N^{pq\ o}_{\lambda} & O^{pq \ xy}_{\lambda \ \omega} & P^{pq \ gh}_{\lambda \ \chi}
 \end{bmatrix}}_{A^{-1}}
 \\ =
  \begin{bmatrix}
 \delta^{ac} \delta_{\mu \tau} & 0 & 0 & 0\\ \\0 & \delta^{mo}& 0&0 \\ \\
 0&0&  \delta^{kx} \delta^{\ell y} \delta_{\alpha \omega} & 0 \\ \\ 0&0& 0& \delta^{sg} \delta^{th} \delta_{\beta \chi}
 \end{bmatrix} \;.
\end{multline*}
\normalsize After some calculation we find for $A^{-1}(x) $:
\begin{eqnarray*}
 \begin{bmatrix}
 \delta^{bc} \left[ \frac{-\p^2}{\p^4  + 2 N g^2 \gamma^4}  P_{\nu \tau} \right] &   \frac{\p_\nu}{\p^2} \delta^{bo}  & f^{bxy} P_{\nu \omega} \frac{ -2 g \gamma^2}{ \p^4  + 2 g^2 N \gamma^4}    &0  \\  \\
  \p_\tau \delta^{nc} \frac{-1}{\p^2} & \delta^{no} \frac{ 2 g^2 N \gamma^4}{ \p^4} &  f^{nxy} \p_\omega \frac{- 2 g \gamma^2}{ \p^4} &  0 \\ \\
  P_{\kappa \tau} f^{ijc} \frac{ -2 g \gamma^2}{ \p^4  + 2 g^2 N \gamma^4}  & \p_\kappa f^{ijo} \frac{- 2 g \gamma^2}{ \p^4}  & f^{ijr} f^{xyr} P_{\kappa \omega} \frac{4 g^2 \gamma^4}{ (- \p^2)( \p^4   + 2 g^2 N \gamma^4 )} &  0 \\
  & & + \frac{-2}{-\p^2 } \delta^{ix} \delta^{jy} \delta_{\kappa \omega} &\\ \\
  0 & 0& 0& \frac{-2}{-\p^2 } \delta^{pg} \delta^{qh} \delta_{\lambda \chi}
 \end{bmatrix}\;.
\end{eqnarray*} \normalsize
For the propagators, we need to go to Fourierspace,
\begin{multline}\label{generalfourierspace}
\int [\d \varphi] \exp \left[ - \frac{1}{2} \int \d^d x \  X(x) A(x) X^T(x) +   \int \frac{\d^d p}{(2\pi)^d}  \ \widetilde{X} (-p) \widetilde{J}(p) \right] \\
	= C (\det A)^{-1/2} \exp \frac{1}{2} \int \frac{\d^d p}{(2\pi)^d}  \widetilde J^T( -p ) A^{-1}(p)  \widetilde J(p)\;,
\end{multline}
with
\begin{eqnarray}
\widetilde{J}^T &=& \begin{bmatrix}J_A & J_b& J_V &J_U  \end{bmatrix} \;,
\end{eqnarray}
and $A^{-1}(p)$ given by
\begin{eqnarray*}
 \begin{bmatrix}
 \delta^{bc} \left[ \frac{p^2 }{p^4  + 2 N g^2 \gamma^4}  P_{\nu \tau} \right] &   \frac{ - \ii p_\nu }{p^2} \delta^{bo}  & f^{bxy} P_{\nu \omega} \frac{ -2 g \gamma^2}{ p^4  + 2 g^2 N \gamma^4}    &0  \\  \\
   \ii p_\tau \delta^{nc} \frac{1}{p^2} & \delta^{no} \frac{ 2 g^2 N \gamma^4}{ p^4} &  f^{nxy}\ii p_\omega \frac{- 2 g \gamma^2}{ p^4} &  0 \\ \\
  P_{\kappa \tau} f^{ijc} \frac{ -2 g \gamma^2}{ p^4  + 2 g^2 N \gamma^4}  & \p_\kappa f^{ijo} \frac{- 2 g \gamma^2}{ p^4}  & f^{ijr} f^{xyr} P_{\kappa \omega} \frac{4 g^2 \gamma^4}{  p^2( p^4 + 2 g^2 N \gamma^4 )} &  0 \\
  & & + \frac{-2}{ p^2 } \delta^{ix} \delta^{jy} \delta_{\kappa \omega} &\\ \\
  0 & 0& 0& \frac{-2}{p^2} \delta^{pg} \delta^{qh} \delta_{\lambda \chi}
 \end{bmatrix}\;.
\end{eqnarray*}
We now have all the ingredients to calculate the propagators.

\subsubsection*{$AA$-propagator}
We have for example,
\begin{align*}
 \frac{\delta}{\delta \widetilde J_{A_\mu^a }(-p)} \frac{\delta}{\delta \widetilde  J_{A_\nu^b }(-k)  } I &= \frac{1}{(2\pi)^{8}} \Braket{ \widetilde A_\mu^a (p) \widetilde A_\nu^b (k)} \\
 &=  \frac{1}{(2\pi)^4}\delta^{ab} \delta(k+ p) \left[ \frac{ p^2 }{p^4 + 2 N g^2 \gamma^4}  P_{\mu \nu} \right]\;,
\end{align*}
or equivalently
\begin{eqnarray}
\Braket{ \widetilde A_\mu^a (p)  \widetilde A_\nu^b (k)}&=&   \frac{ p^2 }{p^4 + \lambda^4}  P_{\mu \nu} \delta^{ab} \delta(k+ p) (2\pi)^4 \;,
\end{eqnarray}
where we have defined
\begin{eqnarray}\label{deflambda}
\lambda^4  &=& 2 N g^2 \gamma^4 \;.
\end{eqnarray}

\subsubsection*{$Ab$-propagator }
Next,
\begin{eqnarray}
 \frac{\delta}{\delta \widetilde J_{A_{\mu}^a}(-p) } \frac{\delta}{\delta \widetilde J_{b^b}(-k) } I &=& \frac{1}{(2\pi)^{8}} \Braket{  \widetilde A_\mu^{a}  (p) \widetilde b^c (k)} ~=~ - \ii \frac{p_\mu}{p^2} \delta^{ab}  \frac{ \delta(p+k)}{(2\pi)^4} \;,
\end{eqnarray}
or thus
\begin{eqnarray}
\Braket{\widetilde A_\mu^a(p) \widetilde b^b(k)} &=& - \ii \frac{p_\mu}{ p^2} \delta^{ab} \delta(p+k) (2\pi)^4 \;.
\end{eqnarray}

\subsubsection*{$bb$-propagator }
\begin{eqnarray}
\Braket{b^a(p) b^b(k)} &=&   \delta^{ab} \frac{ \lambda^4}{p^4}\delta(p+k)(2\pi)^4 \;.
\end{eqnarray}

\subsubsection*{The propagators with $U$ and $V$}
In an analogue fashion, we find
\begin{eqnarray}
\Braket{ \widetilde A^a_{\mu}(p) \widetilde{ V}^{bc}_{\nu}(k)} &=&  f^{abc}  \frac{- 2g \gamma^2}{p^4   + \lambda^4}  P_{\mu \nu}(p) (2\pi)^4 \delta(p+k) \;, \nonumber\\
 \Braket{ \widetilde b^a(p) \widetilde{ V}^{bc}_{\nu}(k)} &=& f^{abc} \ii p_\nu \frac{- 2 g \gamma^2}{ p^2 (p^2}  (2\pi)^4 \delta(p+k)  \;,\nonumber\\
  \Braket{\widetilde{ V}^{ab}_{\mu}(p) \widetilde{ V}^{cd}_{\nu}(k)} &=& \left(  f^{abr} f^{cdr} P_{\mu \nu} \frac{4 g^2 \gamma^4}{  p^2( p^4  + 2 g^2 N \gamma^4 )} +  \frac{-2}{ p^2} \delta^{ac} \delta^{bd} \delta_{\mu \nu}  \right)    (2\pi)^4 \delta(p+k) \;, \nonumber\\
   \Braket{\widetilde{ U}^{ab}_{\mu}(p) \widetilde{ U}^{cd}_{\nu}(k)} &=&  \frac{-2}{p^2 } \delta^{ac} \delta^{bd} \delta_{\mu \nu}    (2\pi)^4 \delta(p+k) \;, \nonumber\\
\Braket{ \widetilde A^a_{\mu}(p) \widetilde{ U}^{bc}_{\nu}(k)} &=&  \Braket{ \widetilde b^a(p) \widetilde{ U}^{bc}_{\nu}(k)} ~=~ \Braket{ \widetilde{ V}^{ab}_{\nu} (p) \widetilde{ U}^{cd}_{\nu}(k)} ~=~0 \;,
\end{eqnarray}
which can be rewritten in terms of $\varphi$ and $\overline \varphi$ again,
\begin{eqnarray}
\Braket{ \widetilde A^a_{\mu}(p) \widetilde{ \varphi}^{bc}_{\nu}(k)} &=& \Braket{ \widetilde A^a_{\mu}(p) \widetilde{\overline \varphi}^{bc}_{\nu}(k)} ~=~ f^{abc}  \frac{- g \gamma^2}{p^4  + \lambda^4}  P_{\mu \nu}(p) (2\pi)^4 \delta(p+k) \;, \nonumber\\
 \Braket{ \widetilde b^a(p) \widetilde{ \varphi}^{bc}_{\nu}(k)} &=&  \Braket{ \widetilde b^a(p) \widetilde{ \overline \varphi}^{bc}_{\nu}(k)} ~=~  f^{abc} \ii p_\nu \frac{-  g \gamma^2}{ p^4}  (2\pi)^4 \delta(p+k) \;, \nonumber\\
  \Braket{\widetilde{ \varphi}^{ab}_{\mu}(p) \widetilde{ \overline \varphi }^{cd}_{\nu}(k)} &=& \left(  f^{abr} f^{cdr} P_{\mu \nu} \frac{ g^2 \gamma^4}{  p^2( p^4 + 2 g^2 N \gamma^4 )} +  \frac{-1}{ p^2 } \delta^{ac} \delta^{bd} \delta_{\mu \nu}  \right)    (2\pi)^4 \delta(p+k) \;, \nonumber\\
   \Braket{\widetilde{ \varphi}^{ab}_{\mu}(p) \widetilde{ \varphi }^{cd}_{\nu}(k)} &=&  \Braket{\widetilde{\overline \varphi}^{ab}_{\mu}(p) \widetilde{\overline \varphi }^{cd}_{\nu}(k)} ~=~ f^{abr} f^{cdr} P_{\mu \nu} \frac{ g^2 \gamma^4}{ p^2( p^4 + 2 g^2 N \gamma^4 )}   (2\pi)^4 \delta(p+k) \;. \nonumber\\
\end{eqnarray}

\noindent In summary, we have the following large set of propagators in the theory:
\begin{eqnarray}\label{summarypropGZ}
\Braket{\widetilde{ \overline \omega}^{ab}_\mu(k) \widetilde \omega^{cd}_\nu(p)} &=& \delta^{ac}\delta^{bd} \delta^{\mu\nu} \frac{-1}{p^2 } \delta(p+k) (2\pi)^4 \;, \nonumber\\
\Braket{ \widetilde{ \overline c}^a(k) \widetilde c^b(p) } &=& \delta^{ab} \frac{1}{p^2} \delta(p+k) (2\pi)^4 \;,\nonumber\\
\Braket{ \widetilde A_\mu^a (p)  \widetilde A_\nu^b (k)}&=&   \frac{ p^2 }{p^4  + \lambda^4}  P_{\mu \nu} \delta^{ab} \delta(k+ p) (2\pi)^4\;, \nonumber\\
\Braket{\widetilde A_\mu^a(p) \widetilde b^b(k)} &=& - \ii \frac{p_\mu}{ p^2} \delta^{ab} \delta(p+k) (2\pi)^4 \;, \nonumber\\
\Braket{b^a(p) b^b(k)} &=&   \delta^{ab} \frac{ \lambda^4}{p^4 }\delta(p+k)(2\pi)^4 \;, \nonumber\\
\Braket{ \widetilde A^a_{\mu}(p) \widetilde{ \varphi}^{bc}_{\nu}(k)} &=& \Braket{ \widetilde A^a_{\mu}(p) \widetilde{\overline \varphi}^{bc}_{\nu}(k)} ~=~ f^{abc}  \frac{- g \gamma^2}{p^4   + \lambda^4}  P_{\mu \nu}(p) (2\pi)^4 \delta(p+k) \;, \nonumber\\
 \Braket{ \widetilde b^a(p) \widetilde{ \varphi}^{bc}_{\nu}(k)} &=&  \Braket{ \widetilde b^a(p) \widetilde{ \overline \varphi}^{bc}_{\nu}(k)} ~=~  f^{abc} \ii p_\nu \frac{-  g \gamma^2}{ p^4}  (2\pi)^4 \delta(p+k) \;, \nonumber\\
  \Braket{\widetilde{ \varphi}^{ab}_{\mu}(p) \widetilde{ \overline \varphi }^{cd}_{\nu}(k)} &=& \left(  f^{abr} f^{cdr} P_{\mu \nu} \frac{ g^2 \gamma^4}{ p^2( p^4  + 2 g^2 N \gamma^4 )} +  \frac{-1}{ p^2 } \delta^{ac} \delta^{bd} \delta_{\mu \nu}  \right)    (2\pi)^4 \delta(p+k)  \;,\nonumber\\
   \Braket{\widetilde{ \varphi}^{ab}_{\mu}(p) \widetilde{ \varphi }^{cd}_{\nu}(k)} &=&  \Braket{\widetilde{\overline \varphi}^{ab}_{\mu}(p) \widetilde{\overline \varphi }^{cd}_{\nu}(k)} ~=~ f^{abr} f^{cdr} P_{\mu \nu} \frac{ g^2 \gamma^4}{ p^2( p^4  + 2 g^2 N \gamma^4 )}   (2\pi)^4 \delta(p+k) \;.\nonumber\\
\end{eqnarray}

\section{The transversality of the gluon propagator}
One might wonder whether the gluon propagator still remains transverse in the presence of the Gribov horizon, this can however be proven easily as done in \cite{Dudal:2008sp}. We start with the following Ward identity, as given in expression \eqref{gaugeward}
\begin{equation}\label{gp1}
    \frac{\delta\Gamma}{\delta b^a}=\p_\mu A_\mu^a\;.
\end{equation}
whereby $\Gamma$ is the quantum effective action. Introducing sources $I^a(J_\mu^a)$ for the fields $b^a (A_\mu^a)$ and performing the Legendre transformation, this identity translates into
\begin{equation}\label{gp2}
    I^a=\p_\mu\frac{\delta Z^c}{\delta    J_\mu^a}\;.
\end{equation}
whereby $Z_c$ is the generator of the connected Green functions. Subsequently, acting with $\frac{\delta}{\delta J_\mu^b}$ on this expression, and by setting all sources equal to zero, we retrieve
\begin{equation}\label{gp3}
    0=\p_\mu^x\left.\frac{\delta^2 Z^c}{\delta J_\mu^a(x) \delta    J_\mu^b(y)}\right|_{I,J=0}=\p_\mu^x\braket{A_\mu^a(x)A_\nu^b(y)}\;.
\end{equation}
As the gluon propagator is in fact the connected two-point function, we have proven the transversality of the gluon propagator.

\section{The hermiticity of the GZ action}
Let us also shortly comment on the hermiticity of the GZ action. We recall that the action is given by equation \eqref{GZstart}
\begin{eqnarray}
S_\GZ &=& S_{0} +  S_{\gamma} \,,
\end{eqnarray}
with
\begin{eqnarray}
S_{0}&=&S_\YM + S_\gf + \int \d^d x \left( \overline \varphi_i^a \p_\mu \left( D_\mu^{ab} \varphi^b_i \right)  - \overline \omega_i^a \p_\mu \left( D_\mu^{ab} \omega_i^b \right) - g f^{abc} \p_\mu \overline \omega_i^a    D_\mu^{bd} c^d  \varphi_i^c \right) \nonumber \;, \\
S_{\gamma}&=& -\gamma ^{2}g\int\d^{d}x\left( f^{abc}A_{\mu }^{a}\varphi _{\mu }^{bc} +f^{abc}A_{\mu}^{a}\overline{\varphi }_{\mu }^{bc} + \frac{d}{g}\left(N^{2}-1\right) \gamma^{2} \right) \;.
\end{eqnarray}
If we define
\begin{align}
\varphi^\dagger &= \overline \varphi \;,   & \overline \varphi^\dagger &=  \varphi \;,  & \omega^\dagger &=  \omega \;, & \overline \omega^\dagger &= \overline \omega \;,
\end{align}
we see that the GZ action is almost Hermitian, provided that the gauge condition $\p \cdot A = 0$ is treated on-shell,\footnote{The gauge condition may be treated on-shell provided the $b$-field is shifted appropriately under hermitian conjugation.} up to the term $- g f^{abc} \p_\mu \overline \omega_i^a D_\mu^{bd} c^d  \varphi_i^c$. However, we recall that this term was introduced for renormalization reasons by the shift \eqref{shift}. Therefore, returning to the action before the shift, the GZ action is Hermitian. Moreover, for practical purposes, the term $- g f^{abc} \p_\mu \overline \omega_i^a D_\mu^{bd} c^d  \varphi_i^c$ is almost redundant as it cannot couple to any Feynman diagram without external $\overline c$ and $\omega$ legs, as discussed in the paragraph below \eqref{S0}. Therefore, we can conclude that in practice, the GZ action is Hermitian.  [Similar remarks hold also for $\widehat S_\GZ$.]  One comment which one can make is that the kinetic term $\overline \varphi \p^2 \varphi$ in the GZ action, seems to have to wrong sign, as does the term $\overline \varphi_i^a \p_\mu \left( D_\mu^{ab} \varphi^b_i \right)$, when the Faddeev-Popov operator is positive.  This has already been noted above, and explained in \eqref{convergenceid}. \\
\\

\section{The soft breaking of the BRST symmetry\label{BRSTbreaking}}
\subsection{The breaking}
We recall here that the GZ action \eqref{GZstart} is not invariant under the BRST transformation \eqref{BRST1}. Indeed, in equation \eqref{breaking} it was shown that
\begin{eqnarray}\label{deltabrst}
   \Delta_{\gamma} &\equiv& sS = sS_\gamma= -g \gamma^2 \int \d^4 x f^{abc} \left[ A^a_{\mu} \omega^{bc}_\mu -
 \left(D_{\mu}^{am} c^m\right)\left( \overline{\varphi}^{bc}_\mu + \varphi^{bc}_{\mu}\right)  \right] \;.
\end{eqnarray}
or alternatively
\begin{eqnarray}\label{deltabrst}
   \widehat \Delta_{\gamma} &\equiv& s \widehat S = s \widehat S_\gamma= -g \gamma^2 \int \d^4 x f^{abc} \left( A^a_{\mu} \omega^{bc}_\mu -
 \left(D_{\mu}^{am} c^m\right) \varphi^{bc}_{\mu}  \right) \;.
\end{eqnarray}
One sees that it is exactly the presence of the Gribov parameter $\gamma$ which prevents the action from being invariant under the BRST symmetry. Indeed, when $\gamma = 0$, we can integrate out the fields $\overline \varphi, \varphi, \overline \omega, \omega$ and we are left with the original Yang-Mills theory. Therefore, this breaking is clearly due to the introduction of the horizon into the Yang-Mills action. Nevertheless, this fact does not prevent the use of the Slavnov-Taylor identity to prove the renormalizability of the theory. Since the breaking $\Delta_{\gamma}$ is soft, i.e.~it is of dimension two in the fields, it can be neglected in the deep ultraviolet, where we recover the usual notion of exact BRST invariance as well as of BRST cohomology for defining the physical subspace, see section \ref{globalmark}. However, in the nonperturbative infrared region, the breaking term cannot be neglected and the BRST invariance is lost.  Breaking of BRST invariance may be regarded as a consequence of the fact that the non-perturbative gauge fixing introduced above has been done in the Landau gauge, and has not been carried out in other covariant gauges.  In fact it has been shown that if the GZ action is modified by $S_\GZ \to S_\GZ + {\alpha \over 2} b^2$, then the expectation-value of gauge invariant quantities becomes dependent on the would-be gauge parameter $\alpha$ \cite{Lavrov:2011, Lavrov:2012}.  According to the Maggiore-Schaden construction \cite{Maggiore:1993wq} that will be described shortly, the soft BRST breaking introduced here may alternatively be regarded as a spontaneous symmetry breaking.

\subsection{The BRST breaking as a tool to prove that the Gribov parameter is a physical parameter}
The breaking term \eqref{deltabrst} has an interesting consequence as it allows one to give a simple algebraic proof of the fact that the Gribov parameter $\gamma$ is a physical parameter of the theory \cite{Dudal:2008sp}, and that as such it can enter the explicit expression of gauge invariant objects: e.g.~the correlation functions $\braket{F^{2}(x)F^{2}(y)} $ or the vacuum condensate $\braket{F^2}$.\\
\\
We can demonstrate this as follows, see \cite{Dudal:2008sp}. Taking the derivative of both sides of equation \eqref{deltabrst} with respect to $\gamma^{2}$ one gets,
\begin{eqnarray}
s\frac{\partial S}{\partial \gamma ^{2}}&=& \frac{1}{\gamma^2} \Delta _{\gamma } = -g  \int \d^4 x f^{abc} \left( A^a_{\mu}
\omega^{bc}_\mu - \left(D_{\mu}^{am} c^m\right)\left( \overline{\varphi}^{bc}_\mu + \varphi^{bc}_{\mu}\right)  \right)\;.
\end{eqnarray}
As we have seen, the BRST operator $s$ which is defined in equation \eqref{BRST1} is nilpotent, thus we have that $\frac{\partial S}{\partial \gamma ^{2}}$ cannot be cast in the form of a BRST exact variation, namely
\begin{eqnarray}
\frac{\partial S}{\partial \gamma ^{2}}\neq s\widetilde{\Delta}_{\gamma }\;,
\end{eqnarray}
for some local integrated dimension two quantity $\widehat{\Delta}_{\gamma }$. Otherwise $s\frac{\partial S}{\partial \gamma ^{2}}$ would have to be equal to zero.  To show indeed that $\gamma^2$ is a physical parameter, we assume for a moment the contrary
\begin{eqnarray}\label{d22}
sS_{\gamma }&=&0\;,
\end{eqnarray}
whereby there is no breaking term $\Delta _{\gamma }$. Since $S_{\gamma }$ depends on the auxiliary fields $\bigl(\overline{\varphi }_{\mu }^{ac}$, $\varphi _{\mu} ^{ac}$, $\overline{\omega }_{\mu}^{ac}$, $\omega _{\mu }^{ac}\bigr)$ which constitute a set of BRST doublets, we know that,
\begin{equation}  \label{d23}
S_{\gamma }=s\widetilde{S}_{\gamma }\;.
\end{equation}
due to the doublet theorem, see appendix \ref{doublettheorem},i.e.~$S_{\gamma }$ is $s$-exact. Subsequently, taking the derivative of both sides of expression \eqref{d23} with respect to $\gamma ^{2}$, one obtains
\begin{equation}\label{d24}
\frac{\partial S_{\gamma }}{\partial \gamma ^{2}}=s\frac{\partial \widetilde{S}_{\gamma }}{\partial \gamma ^{2}}\;,
\end{equation}
which implies that $\gamma^{2}$ is an unphysical parameter. Indeed, for a gauge invariant quantity $\mathcal G$, we would have that
\begin{equation}
\frac{\p \braket{\mathcal{G}}}{\p\gamma^2} = \frac{\delta}{ \delta \gamma^2} \int [\d \phi] \mathcal G\ \e^{-S_\GZ} \sim \int [\d \phi] \mathcal G\ s \widetilde{S}_{\gamma }\e^{-S_\GZ} \sim \int [\d \phi]s\left( \mathcal G\  \widetilde{S}_{\gamma } \right) \e^{-S_\GZ} \sim  \Braket{s(\ldots)} = 0
\end{equation}
and correlation functions of gauge invariant operators would be completely independent from $\gamma^2$. However, when $sS_{\gamma } \not= 0$, due to the presence of the soft breaking term $\Delta _{\gamma }$, we have that $\frac{\p \braket{\mathcal{G}}}{\p\gamma^2} \not= 0$. Therefore, the existence of the breaking $\Delta _{\gamma }$ could be an ingredient to introduce a nonperturbative mass gap in a local and renormalizable way.

\subsection{The Maggiore-Schaden construction and spontaneous breaking of BRST symmetry}
\label{MSconstruction}

Let us also discuss the following paper  \cite{Maggiore:1993wq}. The authors of this paper proposed to interpret the BRST breaking as a kind of \textit{spontaneous} symmetry breaking.  For the benefit of the reader, we shall firstly explain the Maggiore-Schaden construction, and then clarify its status.

\subsubsection{The Maggiore-Schaden construction explained}
In \cite{Maggiore:1993wq}, the first step was to add the following BRST exact term to the Yang-Mills action $S_\YM$:
\begin{eqnarray}\label{mg1}
    S_1 &=& s \int \d^4 x \left( \overline{c}^{a} \p_{\mu}A^a_{\mu} + \overline{\omega}_{\mu}^{ac} \p_{\nu} D_{\nu}^{ab} \varphi^{bc}_{\mu} \right)
    \;,
\end{eqnarray}
with $s$, the usual nilpotent BRST  operator as defined in \eqref{BRST1}. The first term is the usual gauge fixing $S_\gf$, while, the second term introduces the new fields $\overline{\omega}_{\mu}^{ac}$, $\omega_{\mu}^{ac}$, $\overline{\varphi}_{\mu}^{ac}$ and $\varphi_{\mu}^{ac}$. As $s$ is nilpotent, the action $S_\YM + S_1$ does not break the BRST symmetry. Moreover, we recall that, see expression \eqref{BRST1}
\begin{align}
s\varphi _{i}^{a} &=\omega _{i}^{a}\,,&s\omega _{i}^{a}&=0\,,\nonumber \\
s\overline{\omega}_{i}^{a} &=\overline{\varphi }_{i}^{a}\,,& s \overline{\varphi }_{i}^{a}&=0\,.
\end{align}
Therefore, as the nilpotent symmetry defines two doublets $(\varphi,\omega)$ and $(\overline{\varphi},\overline{\omega})$, and by appendix \ref{doublettheorem}, we can exclude these fields from the physical subspace. The second step was to shift a number a fields in the following way
\begin{eqnarray}
\label{primedfields}
\varphi^{ab}_{\mu} &=& \varphi^{\prime ab}_{\mu} + \gamma^2 \delta^{ab} x_{\mu} \;, \nonumber\\
\overline{\varphi}^{ab}_{\mu} &=& \overline{ \varphi}^{\prime ab}_{\mu} + \gamma^2 \delta^{ab} x_{\mu}\;, \nonumber\\
\overline{c}^{ a} &=& \overline{c}^{\prime a} + g \gamma^2 f^{abc} \overline{\omega}^{bc}_{\mu} x_{\mu} \;,\nonumber\\
b^{a} &=& b^{ \prime a} + g \gamma^2 f^{abc} \overline{\varphi}^{bc}_{\mu} x_{\mu}\;,
\end{eqnarray}
where the primed fields $(\varphi^{\prime ab}_{\mu}, \overline{\varphi}^{\prime ab}_{\mu}, \overline{c}^{\prime a}, b^{ \prime a}) $ are new fields which all have vanishing vacuum expectation value and which respect translation invariance. The BRST operator $s$ acts on these new variables according to
\begin{align}
\label{sacts}
s \;\overline{c}^{\prime a} &=b^{\prime a}\;,&   s b^{\prime a}&=0\;,  \nonumber\\
s \varphi _{\mu}^{\prime ab} &=\omega _{\mu}^{ab}\;, & s \omega _{\mu}^{ab} &=0\;.
\end{align}
Notice now that we can write $ s \overline{\omega}_{\mu}^{ab}$ as follows,
\begin{equation}
\label{xdependents}
s \overline{\omega}_{\mu}^{ab} = \overline{\varphi}^{\prime ab}_{\mu} + \gamma^2 \delta^{ab} x_\mu;    \ \ \ \ \ \ \ \ \ \ \ \   \ \ \ \ \ \ \ \  \ \ \ \ \ \ \ \ \ \ \
s \overline{\varphi }^{\prime ab}_\mu =0\; ,  \ \ \ \ \ \ \
\end{equation}
and thus, this BRST variation becomes $x$-dependent. Moreover, also the vacuum expectation value becomes $x$-dependent, i.e.~
\begin{equation} \label{mg4}
\Braket{ s \; \overline{\omega}_{\mu}^{ab}} = \gamma^2 \delta^{ab} x_\mu \;.
\end{equation}
Because the expectation value of an $s$-exact quantity is broken, and because the action is $s$-exact, it is concluded in \cite{Maggiore:1993wq} that the BRST symmetry $s$ is spontaneously broken.
As a third step, the action $S_\YM + S_1$ is rewritten in the primed variables. One finds for $S_1$
\begin{eqnarray}
    S_1 &=&  s \int \d^4 x \left( \overline{c}^{\prime a} \p_{\mu}A^a_{\mu} + \overline{\omega}_{\mu}^{ac} \p_{\nu} D_{\nu}^{ab} \varphi^{\prime bc}_{\mu}
   - g \gamma^2 \overline{\omega}^{ab}_{\mu} f_{abc} A^c_{\mu}  \right) \;.
\end{eqnarray}
After working out the BRST variation $s$,
\begin{align}\label{first}
    S_1 =& \int \d^4 x \left[ b^{\prime a} \p_{\mu} A^a_{\mu} + \overline{c}^{\prime a} \p_{\mu} \left(D_{\mu}^{ab} c^b \right) \right] + \int \d^4 x \left[ \overline{\varphi}^{\prime ac}_{\mu} \p_\nu D_{\nu}^{ab} \varphi^{ \prime bc}_{\mu} + \gamma^2  x_{\mu}   \p_\nu D_{\nu}^{ab} \varphi^{\prime ba}_{\mu} \right.\nonumber\\
   & \left. +  \overline{\omega}^{ac}_{\mu} \p_{\nu} \left(g  f^{akb} D^{kd}_{\nu} c^d \varphi^{\prime bc}_{\mu}  \right) - \overline{\omega}^{ac}_{\mu} \p_{\nu} D^{ab}_{\nu} \omega^{bc}_{\mu} \right]  + \int \d^4 x\left[ -g \gamma^2 \overline{\varphi}^{\prime ab}_{\mu} f_{abc} A^c_{\mu} - g \gamma^2 \overline{\omega}^{ab}_{\mu} f_{abc} D^{cd}_{\mu} c^d \right]\;,
\end{align}
one finds that the action $S_\YM + S_1$ is very similar to the GZ action, up to three parts. Firstly, the constant part $4\gamma^4 (N^2 -1)$ is missing, therefore a BRST invariant part $ S_2 = - \gamma^2 s\int \d^4 x \p_{\mu} \overline{\omega}^{aa}_{\mu}$ is added to the action. This gives,
\begin{eqnarray}
S_2 = - \int \d^4 x  \p_\mu \overline \varphi^{aa}_\mu- \int \d^4 x \gamma^4 \delta^{aa} \p_\mu x_\mu = -\int \d^4 x 4 \gamma^4 (N^2 - 1)
\end{eqnarray}
where we have neglected the first term as it is a total derivative of a field that respects translation invariance.  Secondly, the last term on the first line of expression \eqref{first}, i.e.~$ \gamma^2  x_{\mu}   \p_\nu D_{\nu}^{ab} \varphi^{\prime ba}_{\mu}$ needs to be deformed. If we naively assume that we may perform a partial integration and neglect the surface terms we find
\begin{equation}\label{partial}
\int \d^4 x \gamma^2  x_{\mu}   \p_\nu D_{\nu}^{ab} \varphi^{\prime ba}_{\mu}  = - \int \d^4 x \gamma^2 D_{\mu}^{ab} \varphi^{\prime ba}_{\mu}  =   -  \gamma^2 g f^{abc} \int \d^4 x A_{\mu}^a  \varphi^{\prime bc}_{\mu}
\end{equation}
Thirdly, comparing with expression \eqref{GZstart}, we see that there still is an extra term present which precisely yields the alternate action $\widehat S_\GZ$, eq.\ \eqref{relatewidehat}:
\beqa
\label{sec}
S_\YM + S_1 + S_2 & = & S_\GZ -  \; g \gamma^2 f^{abc} \int \d^4x \; \overline{\omega}_{\mu}^{ab} D_{\mu}^{cd} c^d 
\nonumber \\
& = & \widehat S_\GZ \;.
\eeqa
\\
In conclusion, we can say that the action $\widehat S_\GZ$ (which differs from $S_\GZ$ by a physically irrelevant term) is obtained from an exact $s$-variation

\subsection{Local version of Maggiore-Schaden construction}
In \cite{Dudal:2008sp} this construction was re-examined and some comments were made.  The main point of criticism is assumptions that were made concerning the partial integration with neglect of surface terms at infinity.
The main point of criticism are the assumptions which have been made concerning the partial integration. Looking at the partial integration \eqref{partial}, one sees that the surface terms are neglected. Normally, this is not a problem as we always deal with fields which vanish at infinity. However, here, the surface term has the following form
\begin{equation}\label{surface}
\left[ \gamma^2  x_{\mu}  D_{\nu}^{ab} \varphi^{\prime ba}_{\mu} \right]^\infty
\end{equation}
The presence of the  $x_\mu$ spoils the argument that one can discard this surface term. In fact, to be correct, one would have to impose extra conditions on the fields to justify the dropping of the surface terms, i.e.~the fields need to drop faster to zero than usually assumed. The partial integration is really needed to obtain the similarity with the GZ action.\\
\\
Keeping these criticisms in mind, we present a version of the Maggiore-Schaden construction that is based on local quantities, so partial integration may be done without surface terms at infinity, following a method proposed in \cite{zwanziger:2010x}.  We shall consider the BRST symmetry of a local lagrangian density and derive a conserved Noether BRST-current and corresponding Ward identities.\\
\\
  Consider the local Lagrangian density
\begin{eqnarray}\label{actionbrst2}
    \widehat{\mathcal L } & \equiv & \mathcal L_\YM + s \left( - \p_{\mu} \overline{c}^{\prime a} A^a_{\mu} -  \p_{\nu} \overline{\omega}_{\mu}^{ac}  D_{\nu}^{ab} \varphi^{\prime bc}_{\mu}
    - \gamma^2 D_{\mu}^{ab} \overline{\omega}^{ba}_{\mu} \right) \nonumber\\
      &=& \mathcal L_\YM - \p_{\mu} b^{\prime a} A^a_{\mu} - \p_{\mu} \overline{c}^{\prime a} D_{\mu}^{ab} c^b - \p_\nu \overline{\varphi}^{\prime ac}_{\mu}  D_{\nu}^{ab} \varphi^{\prime bc}_{\mu} - \gamma^2   D_{\mu}^{ab} \varphi^{\prime ba}_{\mu} \nonumber\\
     && - \left( \p_{\nu} \overline{\omega}^{ac}_{\mu} \right)  g  f^{akb} D^{kd}_{\nu} c^d \varphi^{\prime bc}_{\mu}  + \left( \p_{\nu} \overline{\omega}^{ac}_{\mu} \right)  D^{ab}_{\nu} \omega^{bc}_{\mu}   \nonumber\\
   && - g \gamma^2 \overline{\omega}^{ab}_{\mu} f_{abc} D^{cd}_{\mu} c^d - \gamma^2 D_{\mu}^{ab} \overline{\varphi}^{\prime ba}_{\mu} - \gamma^4 d(N^2-1) \;.
\end{eqnarray}    
where $s$ acts formally as in \eqref{sacts} and \eqref{xdependents}.  We shall discuss the status of $s$ shortly.  Although the  BRST operator $s$ produces an $x$-dependent term $s\overline\omega_\mu^{ab} = \overline\varphi_\mu^{ab} + \gamma^2 x_\mu \delta^{ab}$, the Lagrangian density $\widehat{\mathcal L}$ is nevertheless translation-invariant because $\overline\omega_\mu^{ab}$ appears in the first line only in the combination $\p_\nu\overline\omega_\mu^{ab}$ or $\overline\omega_\mu^{ab}f_{abc}$.  Because $s$ is nilpotent, $s^2 = 0$, this Lagrangian density is $s$-invariant,
\beq
s \widehat{\mathcal L} = 0 \;.
\eeq
\\
We now consider a local infinitesimal operator that acts on all fields $\Phi_a$ according to
\beq
\label{localoperator}
\delta_\epsilon \Phi_a(x) \equiv \epsilon(x) \ s\Phi_a(x) \;.
\eeq
To control all partial integrations, we stipulate --- and this is the main point --- that the otherwise arbitrary function $\epsilon(x)$ vanishes outside a small region.  To be concrete, suppose that the fields satisfy periodic boundary conditions, $\Phi_a(x + L) = \Phi_a(x)$, where $L_\mu$ are the edges of the volume, as happens at finite temperature.  Then the {\em global} BRST operator $s$, satsifying $s\overline\omega_\mu^{ab}(x) = \overline\varphi_\mu^{ab}(x) + \gamma^2 x_\mu \delta^{ab}$, is not well defined because $x_\mu$ is not periodic.  This substantiates the criticism of \cite{Dudal:2008sp} regarding the status of the global BRST operator $s$ for, with periodic boundary conditions, the global operator $s$ does not exist.  However the {\em local} operator $\delta_\epsilon = \epsilon(x) s$ is well defined, provided we stipulate that the support of $\epsilon(x)$ is restricted to a region contained within $- L_\mu/2 < x_\mu < L_\mu/2$, and this will be sufficient for our purposes.  (In practice we will choose $\epsilon(x) = \eta \delta(x - z)$, which is concentrated at the point $x = z$.  Here $\eta$ is an infinitesimal constant.)  With this restriction, the right hand side of \eqref{localoperator} extends unambiguously to a function that is periodic in $x_\mu$ with period $L_\mu$.  Moreover Noether's theorem holds,
\beqa
\delta_\epsilon \widehat{\mathcal L}(x) & = & \epsilon(x) s \Phi_a \frac{\p \widehat{\mathcal L}}{\p \Phi_a} + \p_\mu[ \epsilon(x) s \Phi_a ] \frac{\p \widehat{\mathcal L}}{\p \p_\mu \Phi_a}
\nonumber \\
& = & \epsilon(x) s\widehat{\mathcal L}(x) + \p_\mu\epsilon(x) \ s \Phi_a \ \frac{\p \widehat{\mathcal L}}{\p \p_\mu\Phi_a}
\nonumber \\
& = & \p_\mu\epsilon(x) \ j_\mu(x),
\eeqa 
and yields the Noether current of the would-be BRST symmetry,
\beq
j_\mu \equiv s \Phi_a \ \frac{\p \widehat{\mathcal L}}{\p \p_\mu\Phi_a} \;.
\eeq
The BRST-current $j_\mu(x)$ inherits an explicit $x$-dependence from the transformation $s\overline\omega_\mu^{ab} = \overline\varphi_\mu^{\prime ab} + \gamma^2 x_\mu \delta^{ab}$ given by
\beqa
j_\mu & = & u_\mu + \frac{\p \widehat{\mathcal L}} {\p \p_\mu \overline\omega_\nu^{ac}} \gamma^2 x_\nu \delta^{ac}
\nonumber \\
& = & u_\mu + t_{\mu \nu} \ x_\nu \;,
\eeqa
where
\beqa
u_\mu & = & - D_\nu^{ab}c^b F_{\mu \nu}^a + (1/2) g f^{ade}c^d c^e \left( \p_\mu \overline c'^a -g f^{abc} \p_\mu \overline\omega_\nu^{bk} \varphi_\nu^{\prime ck} + \gamma^2 g f^{cab} \overline\omega_\mu^{bc} \right)
\nonumber \\
&& - b^{\prime a} D_\mu^{ad} c^d - \omega_\nu^{ab} \left( \p_\mu \overline\varphi_\nu^{\prime ab} + \gamma^2 \delta_{\mu \nu} \delta^{ab} \right) + \overline\varphi_\nu^{\prime ab} \left( D^{ac}_{\mu} \omega^{cb}_{\nu} -  g  f^{akc} D^{kd}_{\mu} c^d \varphi^{\prime cb}_{\nu} \right) \; 
\eeqa
and
\beq
t_{\mu \nu} = \gamma^2( D^{ab}_{\mu} \omega^{ba}_{\nu} -  g  f^{akb} D^{kd}_{\mu} c^d \varphi^{\prime ba}_{\nu} ) \; 
\eeq
have no explicit $x$-dependence.  With periodic boundary conditions $j_\mu(x)$ is, strictly speaking, not well defined, because of this explicit $x$-dependence.  However the product  $\p_\mu\epsilon(x) \ j_\mu(x)$ is well-defined, and in our manipulations we shall always keep the well-defined product involving $\epsilon(x)$ until, at the end of the day, $\epsilon(x)$ will drop out and we obtain identities 
satisfied by the quantities $u_\mu(x)$ and $t_{\mu \nu}(x)$ 
 that are manifestly consistent with translation invariance and periodicity.\\
\\
The renormalizable action $\widehat S_\GZ$ we shall consider is the integral of the Lagrangian density 
\beq
\widehat S_\GZ = \int d^dx \ \widehat{\mathcal L} \; ,
\eeq
which is formally equivalent to $S_{GZ}$.  Because $\epsilon(x)$ vanishes outside a small region, a partial integration is justified in calculating the variation of the action and we obtain
\beqa
\delta_\epsilon \widehat S_\GZ & = & - \int d^dx \ \epsilon(x) \ \p_\mu j_\mu
\nonumber    \\
&=& - \int d^dx \ \epsilon(x) \ ( \p_\mu u_\mu + t_{\mu \mu} + \p_\mu t_{\mu \nu} \ x_\nu) \;.
\eeqa
\\
In classical physics the action is stationary under any infinitesimal variation along the classical path, so $\delta_\epsilon \widehat S_\GZ = 0$ for any $\epsilon(x)$ and, choosing $\epsilon(x) = \eta \delta(x-z)$, we obtain, classically $\p_\mu j_\mu(z) = 0$
for $-L_\mu/2 < z_\mu < L_\mu/2$.  One may also show that, classically, the tensor $t_{\mu \nu}$ is separately conserved $\p_\mu t_{\mu \nu} = 0$, for it is the conserved Noether current of a symmetry generator $p_\nu$ of the Lagrangian density
\beq
\label{psymmetry}
a_\nu p_\nu\widehat{\mathcal L} = 0
\eeq
where $a_\nu p_\nu$ is a translation of $\overline\omega_\mu^{cd}$ by a constant, $a_\nu p_\nu \overline\omega_\mu^{cd} = \delta^{cd} a_\mu$, and $p_\nu \Phi_a = 0$, for $\Phi_a \neq \overline\omega_\mu^{cd}$.  Thus, classically, we obtain two identities that are translationally invariant and consistent with periodicity, $\p_\mu t_{\mu \nu} = 0$, and $\p_\mu u_\mu = - t_{\mu \mu}$.  We note that $t_{\mu \nu}$ is $s$-exact,
\beq
t_{\mu \nu} = s \gamma^2 D_\mu^{ab} \varphi_\nu^{ba},
\eeq
so $u_\mu$ is classically a conserved current, modulo an $s$-exact piece, $\p_\mu u_\mu = - s \gamma^2 D_\mu^{ab} \varphi_\mu^{ba}$. 
\\
\\
In quantum field theory a conserved current leads to a Ward identity.  To obtain it we start from the expectation value
\beq
\Braket{O} = N \int d\Phi \ O \ \exp(- \widehat S_\GZ) \;,
\eeq
and make the infinitesimal change of variable $\Phi_a' = \Phi_a + \delta_\epsilon \Phi_a$, which gives the identity
\beq
0 = \Braket{ \delta_\epsilon O - \delta_\epsilon \widehat S_\GZ \ O }
\eeq
or
\beqa
\label{ccidentity}
\Braket{\delta_\epsilon O} & = & - \int d^d x \ \epsilon(x) \  \Braket{ \p_\mu j_\mu(x) \ O }
\nonumber   \\
& = & - \int d^dx \ \epsilon(x) \ \Braket{ ( \p_\mu u_\mu + t_{\mu \mu} + \p_\mu t_{\mu \nu} \ x_\nu) \ O } \; .
\eeqa
To summarize so far: although the global BRST symmetry is not well defined when periodic boundary conditions are imposed, the corresponding local transformation involving the factor $\epsilon(x)$ is well defined, and the Noether BRST-current of the Lagrangian density $\widehat{\mathcal L}$ is conserved classically, and the usual Ward identity holds for the Noether BRST-current provided that $\epsilon(x)$ vanishes outside a sufficiently small region.\\
\\
As an application, we apply the Ward identity \eqref{ccidentity} for the case $O = \overline\omega_\nu^{cd}(y)$.  On the left-hand side of \eqref{ccidentity}, the $s$-exact quantity acquires a non-zero expectation value
\beqa
\Braket{ \delta_\epsilon \overline\omega_\nu^{cd}(y)} & = &
\epsilon(y)\Braket{ s \overline\omega_\nu^{cd}(y)}
\nonumber  \\
& = & \epsilon(y) \Braket{\overline\varphi_\nu^{\prime cd}(y) + \gamma^2 y_\nu \delta^{cd}} 
\nonumber \\
& = & \gamma^2 \epsilon(y) y_\nu \delta^{cd},
\eeqa
and we obtain
\beq
\epsilon(y) \gamma^2 y_\nu \delta^{cd} = - \int d^d x \ \epsilon(x) [F_\nu^{cd}(x-y) + G_{\lambda \nu}^{cd}(x-y) x_\lambda ] \   \; ,
\eeq
where
\beq
F_\nu^{cd}(x-y) \equiv \Braket{ ( \p_\mu u_\mu + t_{\mu \mu})(x)  \ \overline\omega_\nu^{cd}(y) }
\eeq
\beq
G_{\lambda \nu}^{cd}(x-y) \equiv \Braket{\p_\mu t_{\mu \lambda}(x) \ \overline\omega_\nu^{cd}(y) } \; .
\eeq
We specialize to $\epsilon(y) = \eta \delta(y-z)$ and find
\beq
\delta(z-y) \gamma^2 z_\nu \delta^{cd} = - F_\nu^{cd}(z-y) - G_{\lambda \nu}^{cd}(z-y) z_\lambda   \; ,
\eeq
or, with $x = z-y$,
\beq
\delta(x) \gamma^2 z_\nu \delta^{cd} = - F_\nu^{cd}(x) - G_{\lambda \nu}^{cd}(x) z_\lambda   \; .
\eeq
This holds for all $x$ and $z$.  The explicit dependence on $z$ is consistent on both sides of this equation, and  we obtain two identities,
\beqa
\label{Gdelta}
G_{\mu \nu}^{cd}(x) & = & 
- \delta(x) \gamma^2 \delta_{\mu \nu} \delta^{cd}
\nonumber \\
F_\nu^{cd}(x) & = & 0\; ,
\eeqa
both of which are consistent with translation invariance and periodic boundary conditions.
The identity involving $G_{\mu \nu}^{cd}(x)$ could have been derived using the Goldstone theorem for the symmetry \eqref{psymmetry}, for which $t_{\mu \nu}$ is ($ - \gamma^2$ times) the Noether current, and which is spontaneously broken
\beq
\Braket{ p_\mu \overline\omega_\nu^{ad}} = \delta_{\mu \nu} \delta^{cd} \; .
\eeq 
\\

To solve the identity satisfied by $G_{\lambda \nu}^{cd}(x)$, we introduce the quantity
\beqa
\Gamma_{\mu \lambda \nu}^{cd}(x-y) & \equiv & \Braket{ t_{\mu \lambda}(x) \ \overline\omega_\nu^{cd}(y) } \; ,
\nonumber  \\
& = & \gamma^2 \Braket{ \left( D^{ab}_{\mu} \omega^{ba}_{\lambda} -  g  f^{akb} D^{kd}_{\mu} c^d \varphi^{\prime ba}_{\lambda} \right) \hspace{-.1cm}(x) \ \overline\omega_\nu^{cd}(y) } \; ,
\eeqa
so 
\beq
G_{\lambda \nu}^{cd}(x) = \p_\mu \Gamma_{\mu \lambda \nu}^{cd}(x) \; ,
\eeq
and the identity satisfied by $G_{\lambda \nu}^{cd}(x)$ reads
\beq
 \p_\mu \Gamma_{\mu \lambda \nu}^{cd}(x) = - \gamma^2 \delta(x) \delta_{\lambda \nu} \delta^{cd} \; .
\eeq
We take fourier transforms and obtain
\beq
 ik_\mu \widetilde \Gamma_{\mu \lambda \nu}^{cd}(k) = - \gamma^2 \delta_{\lambda \nu} \delta^{cd} \; .
\eeq
We decompose $\widetilde \Gamma_{\mu \lambda \nu}^{cd}(k)$ into invariants,
\beq
\widetilde \Gamma_{\mu \lambda \nu}^{cd}(k) = \delta^{cd} ( k_\mu \delta_{\lambda \nu} \Gamma_1 + k_\lambda \delta_{\mu \nu} \Gamma_2 + k_\nu \delta_{\lambda \mu} \Gamma_3 + k_\mu k_\lambda k_\nu \Gamma_4) \;,
\eeq
and the last identity yields
\beqa
i k^2 \Gamma_1 = - \gamma^2   
\nonumber \\
\Gamma_2 + \Gamma_3 + k^2 \Gamma_4 = 0 \; .
\eeqa
This is easily solved for $\Gamma_1$ and $\Gamma_4$ with the result,
\beq
\label{goldstonepole}
\widetilde \Gamma_{\mu \lambda \nu}^{cd}(k) = \delta^{cd} \left[ \gamma^2 ik_\mu \frac{ \delta_{\lambda \nu}}{k^2} + k_\lambda \left( \delta_{\mu \nu} - \frac{k_\mu k_\nu}{k^2} \right) \Gamma_2 + k_\nu \left( \delta_{\lambda \mu} - \frac{k_\lambda k_\mu}{k^2} \right) \Gamma_3 \right] \;.
\eeq
The pole term in $\frac{\gamma^2}{k^2}$ is a Goldstone pole, and arises from the symmetry-breaking $\Braket{s \overline\omega_\nu^{cd}} \neq 0$.  Since the BRST current is fermionic, the Goldstone particle is a fermion.  However because of the explicit $x$-dependence, the pattern of symmetry-breaking is not the usual one, and from the Ward identity for the BRST-current $j_\mu$ we obtain two identities corresponding to the two quantities $u_\mu$ and $t_{\mu \nu}$.\\
\\
The formula we have obtained holds to all orders in perturbation theory.  However we may check it at tree-level.  To zeroth order in $g$ we have
\beq
\Gamma_{\mu \lambda \nu}^{cd}(x-y)\Big|_0 = \gamma^2 \Braket{ \p_{\mu} \omega^{aa}_{\lambda} \hspace{-.1cm}(x) \ \overline\omega_\nu^{cd}(y) }_0 \; ,
\eeq
which has the fourier transform
\beq
\widetilde \Gamma_{\mu \lambda \nu}^{cd}(k)\Big|_0 = \delta^{cd}  ik_\mu \gamma^2 \frac{ \delta_{\lambda \nu}}{k^2} \; .
\eeq
This is consistent with the exact result, and shows that there are no radiative corrections to the first term in \eqref{goldstonepole}.\\
\\
The identity $F_\nu^{cd} = 0$ may be treated similarly.  It reads
\beq
\Braket{ \p_\mu u_\mu(x) \ \overline\omega_\nu^{cd}(y)} = - \Braket{ t_{\mu \mu}(x) \ \overline\omega_\nu^{cd}(y)}.
\eeq  
To solve it we observe that the right hand side is the quantity $-\Gamma_{\mu \mu \nu}^{cd}(x-y)$, which we have just considered, and we define the propagator
\beq
\Lambda_{\mu \nu}^{cd}(x-y) \equiv \Braket{ u_\mu(x) \ \overline\omega_\nu^{cd}(y)} \; .
\eeq
The the last identity now reads
\beq
\p_\mu \Lambda_{\mu \nu}^{cd}(x-y) = - \Gamma_{\mu \mu \nu}^{cd}(x-y) \; ,
\eeq
or, in terms of fourier transforms,
\beqa
i k_\mu \widetilde\Lambda_{\mu \nu}^{cd}(k) & = &  - \widetilde\Gamma_{\mu \mu \nu}^{cd}(k)
\nonumber \\
& = & - \delta^{cd} k_\nu \left[ i \gamma^2  \frac{1}{k^2} + (d-1) \Gamma_2  \right] \; ,
\eeqa
where we have used our previous solution \eqref{goldstonepole}.  We decompose $\widetilde\Lambda_{\mu \nu}^{cd}(k)$ into invariants
\beq
\widetilde\Lambda_{\mu \nu}^{cd}(k) = \delta^{cd} \left( \delta_{\mu \nu} \Lambda_1 + k_\mu k_\nu \Lambda_2 \right)
\eeq
and obtain
\beq
i(\Lambda_1 + k^2 \Lambda_2) = - \left[ i \gamma^2  \frac{1}{k^2} + (d-1) \Gamma_2  \right] .
\eeq
This is easily solved for $\Lambda_1$ with the result
\beq
\widetilde\Lambda_{\mu \nu}^{cd}(k) = \delta^{cd} \left[ \delta_{\mu \nu} \left( - \gamma^2  \frac{1}{k^2} + i(d-1) \Gamma_2  \right) -(\delta_{\mu \nu}k^2 - k_\mu k_\nu ) \Lambda_2 \right] \; .
\eeq
This formula holds to all orders of perturbation theory and, as before, we may verify that the pole term is obtained in zeroth order of perturbation theory using 
\beq
\Lambda_{\mu \nu}^{cd}(x-y)|_0 = - \gamma^2 \Braket{ \omega_\mu^{aa}(x) \ \overline\omega_\nu^{cd}(y) } \Big|_0 \; .
\eeq
\\
To conclude, we have seen that with periodic boundary conditions, the would-be BRST operator $s$ is not globally well defined because of an explicit $x$-dependence.  However the local infinitesimal variation $\delta_\epsilon = \epsilon(x) s$ is well-defined provided that $\epsilon(x)$ is non-zero only in a small region, and this is sufficient for the existence of a classically conserved Noether BRST-current and corresponding Ward identity.  A Goldstone pole results from the non-zero expectation-value of the $s$-exact quantity $s \overline\omega_\nu^{ab}$.  These are characteristic features normally associated with spontaneous symmetry breaking.  However the pattern of symmetry breaking is modified, and there are two independent Ward identities, corresponding to the two quantities $u_\mu$ and $t_{\mu \nu}$, both of which are consistent with translation invariance and periodicity in $x$.

\subsubsection{Other symmetries}

The Lagrangian density $\widehat{\mathcal L}$, eq.~\eqref{actionbrst2}, formally possess all the symmetries of
\beq
\widehat{\mathcal L}_0 \equiv \widehat{\mathcal L}|_{\gamma = 0},
\eeq
that it inherits from the formal change of variable \eqref{primedfields}, for we have
\beq
\widehat{\mathcal L}_0(\varphi, \overline\varphi, \overline c, b) = \widehat{\mathcal L}(\varphi', \overline\varphi', \overline c', b').
\eeq
Among these symmetries is the BRST symmetry that was discussed in the preceding paragraph.  The other symmetries of $\widehat{\mathcal L}_0$ may be treated in the same way: the symmetries of $\widehat{\mathcal L}_0$ become  would-be symmetries of $\widehat{\mathcal L}$ that are not defined globally.  But for each such symmetry there is a conserved Noether current of $\widehat{\mathcal L}$ and corresponding Ward identity, each with its Goldstone boson or fermion.\footnote{In \cite{zwanziger:2010x} Goldstone bosons and fermions were shown to exist at the critical point $\frac{\p \Gamma}{\p \gamma} = 0$ selected by the horizon condition.  In the present discussion we have not used this condition explicitly, and it appears that we have found Goldstone bosons and fermions for any non-zero value of $\gamma$.  However it may be that a solution with  non-zero $\gamma$ may exist only if the horizon condition is imposed.  We are grateful to Martin Schaden for a discussion on this point.}  These will be discussed elsewhere.



\section{Restoring the BRST}
\subsection{Adapting the BRST symmetry $s$ is not possible}
One can ask whether it might be possible to modify the BRST operator, i.e. $s\rightarrow s_m$, in such a way that the new operator $s_m$ would be still nilpotent, while defining an exact symmetry of the action, $s_mS=0$. In \cite{Dudal:2008sp} a simple argument was presented, discarding such a possibility. Firstly, we recall that the BRST transformation \eqref{BRST1} defines an exact symmetry of the action when $\gamma =0$ as this corresponds to the physical situation in which the restriction to the Gribov region has not been implemented. Therefore, one should search for possible modifications of the BRST operator which depends on $\gamma$, namely
\begin{eqnarray}
s_m&=&s+s_\gamma  \;,
\end{eqnarray}
whereby
\begin{eqnarray}
s_\gamma&=&\gamma\textrm{-dependent}\;\textrm{terms}\;,
\label{mstwee}
\end{eqnarray}
so as to guarantee a smooth limit when $\gamma$ is set to zero. However, taking into account the fact that $\gamma $ has mass dimension one, that all auxiliary fields $\left( \overline{\varphi }_{\mu }^{ac},\varphi _{\mu }^{ac},\overline{\omega }_{\mu}^{ac},\omega _{\mu }^{ac}\right)$ have dimension one too, and that the BRST operator $s$ does not alter the dimension of the fields\footnote{It is understood that the dimensions are assigned to the fields $A^{a}_{\mu}$, $b^a$, $c^a$, ${\bar c}^a$ as in Table~\ref{2tabel1} of sec.~\ref{renorma}. It is apparent that the BRST operator $s$ does not alter the dimension of the fields.}, it does not seem possible to introduce extra $\gamma$-dependent terms in the BRST transformation of the fields $\left( \overline{\varphi }_{\mu }^{ac},\varphi _{\mu }^{ac},\overline{\omega }_{\mu }^{ac},\omega _{\mu }^{ac}\right) $ while preserving locality, Lorentz covariance as well as color group structure.  However this argument does not exclude the local Maggiore-Schaden construction just introduced.

\subsection{Restoring the BRST by the introduction of new fields}
However there is another way to restore the BRST \cite{Dudal:2010hj} following the idea that was initiated in \cite{Sorella:2009vt} and \cite{Kondo:2009qz}. In these papers it was shown that the broken BRST symmetry $s$ can be rewritten as a non-local symmetry, being not nilpotent in \cite{Sorella:2009vt} and nilpotent in \cite{Kondo:2009qz}. In a subsequent paper, one succeeded in localizing the non-local BRST symmetry of \cite{Sorella:2009vt} by the introduction of new fields.  Let us summarize here these results.

\subsubsection{The non-local BRST symmetry $s'$}
First we review the results of \cite{Sorella:2009vt}, and write down a non-local BRST symmetry which is obeyed by the GZ action. We start from the standard GZ action \eqref{GZstart}, set $g\gamma^2=\theta^2$, and drop the vacuum term for brevity because it does not influence any variation of the action,
\begin{multline}\label{GZb1}
 S_{\GZ} = S_\YM +  \int \d^d x\,\left( b^a \p_\mu A_\mu^a +\overline c^a \p_\mu D_\mu^{ab} c^b \right) + \int \d^d x\left( \overline \varphi_\mu^{ac} \p_\nu D_\nu^{ab} \varphi_\mu^{bc} \right. \\ \left. - \overline \omega_\mu^{ac} \p_\nu D_\nu^{ab} \omega_\mu^{bc}  - g f^{abc} \p_\mu \overline \omega_\nu^{ae} D_\mu^{bd} c^d  \varphi_\nu^{ce}  -\theta^{2}  f^{abc}A_\mu^a \left( \varphi_\mu^{bc} +  \overline \varphi_\mu^{bc}\right)\right)\,.
\end{multline}
Following \cite{Sorella:2009vt}, we first drop the final term of the above expression, i.e.~$g f^{abc} \p_\mu \overline \omega_\nu^{ae} D_\mu^{bd} c^d \varphi_\nu^{ce}$, so that there is no $\overline\omega c$ term in the action. This has no influence on the soft breaking of the action because the $s$-variation of the action we now consider,
\begin{eqnarray}
S'_{\GZ} &=& S_\YM +  \int \d^d x\,\left( b^a \p_\mu A_\mu^a +\overline c^a \p_\mu D_\mu^{ab} c^b \right)  +\nonumber\\&&+ \int \d^d x\left( \overline \varphi_\mu^{ac} \p_\nu D_\nu^{ab} \varphi_\mu^{bc}  - \overline \omega_\mu^{ac} \p_\nu D_\nu^{ab} \omega_\mu^{bc}  - \theta^{2}  f^{abc}A_\mu^a \left( \varphi_\mu^{bc} +  \overline \varphi_\mu^{bc}\right)\right)\,,
\end{eqnarray}
can be written as
\begin{eqnarray}\label{GZb4}
s  S'_{\GZ} &=& \Delta_\gamma -  \int \d^d x\ s \left( g f_{akb} (D_\nu c)^k \p_\nu \overline \omega_\mu^{ac}  \varphi_\mu^{bc}   \right) \\ \nonumber
&=& \int \d^d x\Bigl( - \theta^2c^k D_\mu^{ka}f^{abc}\left(\varphi_\mu^{bc} +  \overline \varphi_\mu^{bc}\right) - \theta^2f^{abc}A_\mu^a\omega_\mu^{bc}
    +gf^{abc}(D_\nu^{bp}c^p)\left(\p_\nu\overline\varphi_\mu^{ae}\varphi_\mu^{ce}-\p_\nu\overline\omega_\mu^{ae}\omega_\mu^{ce}\right)\Bigr)
\end{eqnarray}
where $\Delta_\gamma$ can be found in equation \eqref{deltabrst}. Therefore, we can indeed safely leave out the term $g f^{abc} \p_\mu \overline \omega_\nu^{ae} D_\mu^{bd} c^d \varphi_\nu^{ce}$. After the localization however, this term shall be picked up again. According to \cite{Sorella:2009vt}, the positivity of the Faddeev-Popov operator inside the Gribov region allows to rewrite \eqref{GZb4} as
\begin{align}\label{GZb6}
s S'_{\GZ}  &= \int \d^d x\left( c^a D_\nu^{ab}\Lambda_\nu^b + \theta^2 f^{abc} A_\mu^a \omega_\mu^{bc}\right)\nonumber\\
&=\int \d^d x\left((D_\nu^{ma}\Lambda_\nu^a)[(\p_\nu D_\nu)^{-1}]^{mc}\frac{\delta}{\delta \overline c^c} \hat S_{\GZ} + \theta^2f^{abc}A_\mu^a[(\p_\nu D_\nu)^{-1}]^{bm}\frac{\delta}{\delta \overline\omega_\mu^{mc}} \hat S_{\GZ}\right)\,,
\end{align}
with
\begin{equation}
    \Lambda_\nu^a= - \theta^2 f^{abc}(\varphi_\nu^{bc}+\overline\varphi_\nu^{bc})-gf^{bap}\left(\p_\nu\overline\varphi_\mu^{bc}\varphi_\mu^{pc}-\p_\nu\overline\omega_\mu^{bc}\omega_\mu^{pc}\right)\,.
\end{equation}
From \eqref{GZb6}, we can now read off a new nonlocal BRST symmetry, $s'\hat S_{\GZ}=0$, generated by
\begin{eqnarray}\label{GZb7}
s'A_{\mu }^{a} &=&-D_{\mu }^{ab} c^b\,, \quad s'c^{a} ~=~\frac{1}{2}gf^{abc}c^{b}c^{c}\,, \quad s'\overline{c}^{a} ~=~b^{a}-(D_\nu^{kc}\Lambda_\nu^c)[(\p_\nu D_\nu)^{-1}]^{ka}\,, \quad   s'b^{a}~=~0\,,  \nonumber \\
s'\varphi _{\mu}^{ac} &=&\omega_{\mu}^{ac}\,,\quad s'\omega_{\mu}^{ac}~=~0\,, \quad s'\overline{\omega}_{\mu}^{ac} ~=~\overline{\varphi }_{\mu}^{ac}+\theta^2 [( D_\nu \p_\nu )^{-1}]^{ap} f^{pqc}A_\mu^q\,,\quad s' \overline{\varphi }_{\mu}^{ac}~=~0\,.
\end{eqnarray}
We repeat that this symmetry $s'$ is not nilpotent, $s'^2\neq0$, see also \cite{Sorella:2009vt}.

\subsubsection{The local BRST symmetry $s_\theta$}
In \cite{Dudal:2010hj}, this non-local BRST symmetry $s'$ was localized into $s_\theta$. This was done by adding an auxiliary part to the GZ action,
\begin{equation}
S_\GZ^\loc = S_\GZ + S_\aux
\end{equation}
with $S_\GZ$ given in \eqref{GZstart} and $S_\aux$ given by
\begin{eqnarray}
\label{Saux}
S_\aux  &=&\int \d^dx\left(\Omega^a \p_\mu D_\mu^{ab}\overline \Omega^b  + \alpha^a \p_\mu D_\mu^{ab}\overline \alpha^b + gf^{abc}(\p_\mu \alpha^a)(D_\mu^{bd}c^d)\overline \Omega^c-\overline\alpha^a D_\nu^{ab} \overline\Lambda_\nu^b+\overline\Omega^a s(D_\nu^{ab} \overline\Lambda_\nu^b) \right. \nonumber\\
&& \left.+ \Psi_\nu^{ac}\p_\mu D_\mu^{ab}\overline \Psi_\nu^{bc}+\beta_\nu^{ac}\p_\mu D_\mu^{ab}\overline \beta_\nu^{bc} +gf^{abc} (\p_\mu\beta_{\nu}^{ae})(D_\mu^{bd}c^d)\overline \Psi_\nu^{ce}-f^{abc}A_\mu^a\overline{\beta}_\mu^{bc}-f^{abc}\overline{\Psi}_\mu^{bc}D_\mu^{ad}c^d\right)\nonumber\\
 &&+\int \d^4x\left( R^a \p_\mu D^{ab}_\mu \overline R^b +  Q^a \p_\mu D_\mu^{ab}\overline Q^b + g f^{abc}\p_\mu Q^a  D_\mu^{bd} c^d \overline R^c - \overline Q^d \kappa^d + \overline R^d s (\kappa^d) \right)\;,
\end{eqnarray}
where
\beqa
\overline\Lambda_\nu^a & \equiv & f^{abc}(\varphi_\nu^{bc} + \overline\varphi_\nu^{bc}) 
\nonumber \\ 
\kappa^d & \equiv & D_\mu^{bd}( g f^{abc} \p_\mu \beta_\nu^{ae} \varphi_\nu^{ce} )\;.
\eeqa
Here we have introduced of a new set of doublets $(\alpha^a, \Omega^a )$, $(\overline\Omega, \overline\alpha^a)$, $(\beta_\mu^{ab},\Psi_\mu^{ab})$, $(\overline\Psi_\mu^{ab}, \overline\beta_\mu^{ab})$, $(Q^a, R^a)$, $(\overline R^a, \overline Q^a )$ i.e.~
\begin{align}\label{GZb9}
s \alpha^a &= \Omega^a\,, &   s\Omega^a &=0\,, & s \overline\Omega^a &= \overline\alpha^a\,, &   s\overline\alpha^a &= 0\,,\nonumber\\
s \beta_\mu^{ab} &=\Psi_\mu^{ab}\,, & s\Psi_\mu^{ab}&=0\,, & s \overline\Psi_\mu^{ab} &=\overline\beta_\mu^{ab}\,, &  s\overline\beta_\mu^{ab} &=0\,, \nonumber\\
s Q^a &=  R^a\,, & s R^a &=0 \,, & s \overline R^a &= \overline Q^a\,,  &       s \overline Q^a &=0 \;.
\end{align}
$\alpha^a$, $\overline{\alpha}^a$, $\beta_\mu^{ab}$, $\overline\beta_\mu^{ab}$, $Q^a$ and $\overline Q^a$ are bosonic fields, while the other fields $\Omega^a$, $\overline{\Omega}^a$, $\Psi_\mu^{ab}$, $\overline\Psi_\mu^{ab}$, $R^a$ and $\overline R^a$ are anti-commuting fields. It can be checked that the action $S_\GZ^\loc$ is invariant under the following BRST transformation $s_\theta$
\begin{align}\label{GZb18}
s_\theta A_{\mu }^{a} &=-\left( D_{\mu }c\right) ^{a}\,, &  s_\theta c^{a} &=\frac{1}{2}gf^{abc}c^{b}c^{c}\,,& s_\theta \overline{c}^{a} &=b^{a}-\theta^2\alpha^a -\theta^2 Q^a  \,,&  s_\theta b^{a}&=0\,,  \nonumber \\
s_\theta \varphi _{\mu}^{ac} &=\omega_{\mu}^{ac}\,,& s_\theta\omega_{\mu}^{ac}&=0\,,& s_\theta \overline{\omega}_{\mu}^{ac} &=\overline{\varphi }_{\mu}^{ac}+\theta^2\beta_\mu^{bc}\,,& s_\theta \overline{\varphi }_{\mu}^{ac}&=0\,,\nonumber\\
 s_\theta \alpha^a &=\Omega^a\,, &  s_\theta\Omega^a &=0\,,& s_\theta \overline\Omega^a &=\overline\alpha^a\,, &  s_\theta\overline\alpha^a &=\theta^2 c^a\,,\nonumber\\
s_\theta \beta_\mu^{ab} &=\Psi_\mu^{ab}\,,&   s_\theta\Psi_\mu^{ab}&=0\,,& s_\theta \overline\Psi_\mu^{ab} &=\overline\beta_\mu^{ab}\,,&  s_\theta\overline\beta_\mu^{ab}&=\theta^2 \omega_\mu^{ab}\,,\nonumber\\
s_\theta Q^a &=  R^a \,, & s_\theta R^a &=0 \,,& s_\theta \overline R^a &= \overline Q^a\,,& s_\theta \overline Q^a &= \theta^2 c^a \,.
\end{align}
Again, this symmetry $s_\theta$ is not nilpotent, $s_\theta^2 \not=0$. However, one can check that $s_\theta^4 = 0$. \\
\\
Let us summarize some properties of the action $S_\GZ^\loc$.
\begin{itemize}
\item Firstly, we need to ask what happens if we set $\theta^2 =0$, or equivalently, $\gamma^2 = 0$. In this case, we are not considering the restriction to the Gribov horizon anymore and we are back in the standard Yang-Mills framework. In order for the theory to be meaningful, we would expect that $\left. S_\GZ^\loc \right|_{\theta^2 =0}$ is equivalent with the ordinary Yang-Mills action $S_\YM + S_\gf$. This is indeed the case, as is proven in \cite{Dudal:2010hj}.
\item Secondly, one should also prove that the GZ action and the new action $S_\GZ^\loc$ are equivalent. In other words, for $\phi\in\bigl\{A_\mu^a$, $b^a$, $\overline{c}^a$, $c^a$, $\varphi_\mu^{ab}$, $\overline\varphi_\mu^{ab}$, $\omega_\mu^{ab}$, $\overline\omega_\mu^{ab}\bigr\}$, we require the following identification
\begin{eqnarray}\label{GZb32}
  \Braket{\phi(x_1)\ldots\phi(x_n)}_{\loc}=  \int [\d\Phi]_{\loc} \phi(x_1)\ldots\phi(x_n) \e^{-S_{\GZ}^{\loc}} =   \Braket{\phi(x_1)\ldots\phi(x_n)}_{\GZ}\,.
\end{eqnarray}
This was also proven in detail in \cite{Dudal:2010hj}.
\item Thirdly, in order for the action $S_\GZ^\loc$ to be meaningful at the quantum level, one should prove that this action is renormalizable. However this has not yet been proven. In \cite{Dudal:2010hj} many other symmetries were given which could be useful for the renormalization of the action. It is worth mentioning that $s_\theta$ not being nilpotent, is not necessarily a problem to prove  renormalizability.  From the result $s_\theta^4 = 0$, it should be possible to construct a nilpotent symmetry, see \cite{Fucito:1997xm}.
\end{itemize}

\subsection{Two different phases of the restored BRST symmetry}

We investigate the BRST operators $s'$ and $s_\theta$ just defined.  They are closely related.  Indeed, in \eqref{GZb7}, we have $s'\overline{\omega}_{\mu}^{ac} ~=~\overline{\varphi }_{\mu}^{ac}+\theta^2 [( D_\nu \p_\nu )^{-1}]^{ap} f^{pqc}A_\mu^q$, while in \eqref{GZb18}, $s_\theta$ acts according to $s_\theta \overline \omega_\mu^{ac} = \overline\varphi_\mu^{ac} + \theta^2 \beta_\mu^{ac}$, where $\beta_\nu^{ac}$ satisfies the equation of motion
\beq
\label{betaeq}
D_\mu^{ab} \p_\mu \beta_\nu^{bc} = f^{abc} A_\nu^b,
\eeq
obtained by (left) differentiating \eqref{Saux} with respect to $\overline\beta_\nu^{ac}$.  This equation has the formal solution
\beq
\label{formalsol}
\beta_\nu^{ac} = ( D_\mu \p_\mu )^{-1}]^{ap} f^{pqc}A_\nu^q,
\eeq
and with it, the action of $s_\theta$ agrees with the action of $s'$.

We shall arrive at the unexpected conclusion that there are two different solutions $\beta_\nu^{(1)bc}$ and $\beta_\nu^{(2)bc}$ to equation \eqref{betaeq} which provide two different phases of $s'$ and $s_\theta$.

{\it Solution 1:} For simplicity we suppose that $A_\mu$ is transverse, $\p_\mu A_\mu = 0$, so the Faddeev-Popov operator $D_\mu(A) \p_\mu$ is hermitian.  We quantize in a finite Euclidean periodic box of edge $L$, so the spectrum of $D_\mu(A) \p_\mu$ is real and discrete, and there is a complete set of eigenvectors,
\beq
D_\mu^{ab}(A) \p_\mu \psi_n^b = \lambda_n \psi_n^{a}.
\eeq
For generic $A$ not precisely on the Gribov horizon, as we shall suppose, the null space of $D_\mu(A) \p_\mu$ consists of constant functions $\p_\mu \psi = 0$.  To obtain solution 1, we use the eigenfunction expansion of the inverse operator,
\beq
[( D_\mu \p_\mu )^{-1}]^{ab}(x, y) = {\sum}'_n \frac{\psi_n^a(x) \psi_n^b(y)}{\lambda_n},
\eeq
where the prime means sum over non-zero eigenvalues.  Solution 1 is thus
\beqa
\beta_\nu^{(1)ac}(x) & = & {\sum}_n^\prime \psi_n^a(x) \ \beta_{n \nu}^c
\nonumber \\
\beta_{n \nu}^c & = & \frac{1}{\lambda_n} \int \d^dy \ \psi_n^b(y) f^{bqc} A_\nu^q(y)\;.
\eeqa
As an explicit example of solution 1 in SU (2) gauge theory, take $A_\nu^b = c \cos(kx_1) \delta_{\nu 2} \delta^{b3}$, where $k = 2 \pi m/L$, $m$ is a non-zero integer, and $x_1$ is the first componant of $x_\mu$, which is transverse, $\p_\mu A_\mu^b = c \delta^{b3} \p_2 \cos(kx_1) = 0$.  It is easy to verify that $\psi^a = \cos(kx_1) u^a$ , where $u^a$ is any constant color vector, is an eigenfunction.  Indeed we have $A_\mu^c \p_\mu \cos(kx_1) = c \cos(kx_1) \delta^{c3} \p_2 \cos(kx_1) = 0$, so $D_\mu^{ab} \p_\mu \psi^b = \p^2 \psi^a = - k^2 \psi^a$.  For this choice of $A_\nu^b$, the eigenfunction expansion of $\beta_\nu^{ac}$ is given by $\beta_\nu^{ac} = (-k^2)^{-1} c \cos(kx_1) f^{a3c} \delta_{\nu 2} = (-k^2)^{-1} f^{abc} A_\nu^b $.

{\it Solution 2:}  The simple identity,
\beq
D_\mu^{pb}(A) \p_\mu \ \delta^{bc}  x_\nu = g f^{pqc} A_\mu^q,
\eeq
reveals unexpectedly that a second solution of \eqref{betaeq}, for any $A_\nu^b(x)$, is given by\footnote{The horizon function $H(A) = - \int \d^d x g f^{ba\ell} A_\mu^a (x) [( D_\mu \p_\mu)^{-1})^{\ell m} g f^{bkm} A^k_\mu ](x)$  was derived in \eqref{horizonfunction} by solving the eigenvalue problem (by summing the perturbation series).  Consequently solution 1, the eigenfunction expansion, should be used to evaluate $H(A)$.}
\beq
\label{npsolution}
\beta_\nu^{(2)ac} = g^{-1} x_\nu \delta^{ac}.
\eeq
Remarkably, it is independent of $A_\mu^b$.  Solutions 1 and 2 are different, as one sees by comparison with the explicit example of solution 1.  Note that a perturbative expansion yields solution 1, whereas solution 2 is non-perturbative.  

With solution 2, the BRST transformations \eqref{GZb7}, $s'\overline{\omega}_{\mu}^{ac} ~=~\overline{\varphi }_{\mu}^{ac}+\theta^2 [( D_\nu \p_\nu )^{-1}]^{ap} f^{pqc}A_\mu^q$ $s\overline{\omega}_{\mu}^{ac}$ and $s_\theta \overline \omega_\mu^{ac} = \overline\varphi_\mu^{ac} + \theta^2 \beta_\mu^{ac}$ coincide with the BRST transformation of Maggiore and Schaden, $s\overline{\omega}_{\mu}^{ac} ~=~\overline\varphi_{\mu}^{\prime ac}+\gamma^2 x_\mu \delta^{ac}$, after obvious relabeling, and with $\theta^2 = g \gamma^2$.  One can reject solution 2 by imposing periodic boundary conditions, in which case $x_\mu$ is not well defined.  However even with periodic boundary conditions, the local infinitesimal variation $\delta = \epsilon(x) s'$, or $\delta = \epsilon(x) s_\theta$, remains well-defined for solution 2 provided that $\epsilon(x)$ vanishes outside a sufficiently small region.  As shown in sec.~\ref{MSconstruction}, consideration of this variation is sufficient to establish the existence of a conserved Noether current and corresponding Ward identity, and the presence of the Goldstone pole normally associated with a spontaneously broken symmetry.  On an infinite volume, the system may choose the second solution \eqref{npsolution}, in which case we have
\beqa
\Braket{s_\theta \overline\omega_\mu^{ab} } & = & \Braket{\overline\varphi_\mu^{ab} + \theta^2 \beta_\mu^{ab}} 
\nonumber \\ 
& = & \theta^2 g^{-1} x_\mu \delta^{ab}.
\eeqa
In this case $s_\theta$ is spontaneously broken because the vacuum state is not $s_\theta$-invariant,
\beq
\label{vaccumNI}
s_\theta \Psi_0 \neq 0\; .
\eeq

In conclusion we note that the action $S_{\GZ}^\loc = S_\GZ + S_\aux$, is local and possesses a localized BRST operator $s_\theta$. Its field equation possess two solutions.  Solution 1 preserves BRST symmetry, whereas solution 2 spontaneously breaks this symmetry.  At this point we do not know which solution the system will choose.  If the correlators of the system described by the action $S_\GZ^\loc$ coincide with the correlators described by $\widehat S_\GZ$ (for the fields appearing in $\widehat S_\GZ$), then according to the calculation of sec. \ref{MSconstruction}, BRST will in fact be spontaneously broken.  However $S_\GZ^\loc$ appears to have two phases: one in which BRST is preserved and one in which it is spontaneously broken.

\subsubsection{Interpreting the restored BRST symmetry}

BRST symmetry is thought to encode the geometrical character of a quantum gauge field theory.  It is therefore encouraging that the local action $S_\GZ^\loc$ possesses a BRST symmetry --- as does $\widehat S_\GZ$ in the local sense discussed in sec. \ref{MSconstruction} --- even though this theory is presumably only approximate (because it was derived by integrating over the Gribov region instead of the fundamental modular region).\footnote{The $s$-invariance of $\widehat S_\GZ$ for any value of $\gamma$ is in fact more than is needed for physics. Indeed the theory defined by $\widehat S_\GZ$ need not be a gauge theory unless the horizon condition holds, so it would be sufficient for physics if the action were BRST-symmetric only for that special value of $\gamma$ determined by the horizon condition.  Instead $\widehat S_\GZ$ is BRST-invariant  for every value of $\gamma$.}  If BRST is spontaneously broken, as indicated by the calculation of sec.~\ref{MSconstruction}, the vacuum itself is not BRST-invariant, $s \Psi_0 \neq 0$, eq.~\eqref{vaccumNI}, and we cannot identify the positive metric subspace by the usual condition $s \Psi = 0$.  Here we enter into an unfamiliar realm in which it is not known how to define a positive-metric physical subspace.  However, to the authors' knowledge, proof of positivity of the subspace $s \Psi = 0$ relies heavily on the existence of an asymptotic field for every fundamental field, including the gluon and quark fields.  But the very meaning of confinement is that the asymptotic gluon and quark fields do not exist in the confined phase of~QCD.  Thus at present there is a general lack of understanding of how to construct a positive-metric physical subspace in the confined phase.  Making a virtue of necessity, we are tempted to interpret the difficulty we have encountered, of finding the positive-metric physical subspace when BRST is spontaneously broken, as a welcome signal that gluons and quarks are not in the physical subspace of the phase we have found.  This suggests a scenario in which spontaneous breaking of BRST symmetry serves as an order parameter of the confined phase.  In this interpretation, the two solutions of the equations of motion of $S_\GZ^\loc$ that we have found, in which BRST is either preserved or spontaneously broken, correspond respectively to the deconfined or confined phases of QCD.

\section{The GZ action and its relation to the Kugo-Ojima confinement criterion}
\subsection{Introduction: the Kugo-Ojima criterion}
Some important results concerning a possible origin of confinement, were given in \cite{Kugo:1979gm,Kugo:1995km}. In these papers, confinement was related to the enhancement of the ghost propagator.\\
\\
Let us explain this a bit more in detail. The starting point of the analysis of Kugo and Ojima is a well defined nilpotent (BRST) symmetry operator $Q_B$ and a ghost charge. Their analysis yields two results. Firstly they showed that with a well defined nilpotent symmetry, all unphysical states, see section \ref{globalmark}, form so-called quartets \cite{Kugo:1979gm} that decouple from the physical spectrum. In this way, only physical states, which are closed under the symmetry $Q_B$ but not exact, survive. In this way they proved that the longitudinal and temporal gauge polarization, the ghost and the antighost fields can be excluded from the physical spectrum. In fact this idea is very general, and can be applied whenever a system admits a nilpotent symmetry $s$.\\
\\
Secondly, for the Faddeev-Popov action, they also showed the following. Using the equation of motion for the gluon field, the conserved global color current can be written as
\begin{equation}
J_\mu^a = \p_\mu F^a_{\mu\nu} + \{ Q_B , D_\mu^{ab} \overline c^b \} \;,
\end{equation}
and so the color charge, which is the integrated zero component of $J_\mu^a$, is given by
\begin{equation}
Q^a = \int \d^3 x \left( \p_i F^a_{0i} + \{ Q_B, D_0^{ab} \overline c^b \} \right) \;.
\end{equation}
Thus there are two criteria which need to be satisfied in order to have color confinement, namely $Q^a = 0$ for all physical states. The first is that the gluon propagator should not have massless poles, so the first term of the last expression vanishes because it is the integral of a derivative\footnote{If the gluon propagator has massless poles, the first term is ill-defined.}. The second criterion is that $\{ Q_B, D_0^{ab} \overline c^b \} $ should be well defined, which is the case when
\begin{equation}\label{KO0}
 u(0)=-1\;,
\end{equation}
with $u(p^2)$ defined through the following Green function\footnote{Strictly speaking, the KO analysis is done in Minkowski space. We shall however, as any functional or lattice approach, consider the corresponding operator in Euclidean space.},
\begin{equation}\label{KO1}
    \int \d^dx \e^{\ii px}\Braket{D_\mu^{ad}c^d(x) D_\nu^{be}\overline c^e(0)}_{\FP}= \left(\left(\delta_{\mu\nu}-\frac{p_\mu p_\nu}{p^2}\right)u(p^2)-\frac{p_\mu p_\nu}{p^2}\right)\delta^{ab}\;.
\end{equation}
$\braket{O}_{\FP}$ stands for the expectation value taken with the Faddeev-Popov action. If the two criteria are met, then $Q^a$ is well defined and we have $\bra{\varphi}Q^a \ket{\psi}_\phys = 0$, so color confinement is guaranteed.\\
\\
The second criterion can be connected to the ghost propagator. Indeed, in \cite{Kugo:1995km}, it was shown that one can parameterize the ghost propagator, defined as follows
\begin{eqnarray}\label{KO5}
\Braket{c^a(-p) \overline{c}^b(p)}&=\delta^{ab} G (p^2)\;,
\end{eqnarray}
in terms of\footnote{Actually, in \cite{Kugo:1995km}, another notation $v(p^2)$ has been used instead of $w(p^2)$, the relation being  $ w(p^2) = p^2v(p^2)$.}
\begin{eqnarray}\label{KO6}
G (p^2) &=& \frac{1}{p^2(1 + u(p^2) + w (p^2))}\;.
\end{eqnarray}
This relation was also discussed in \cite{Zwanziger:1992qr,Kondo:2009wk,Kondo:2009ug,Aguilar:2009nf,Boucaud:2009sd}. It is usually assumed that\footnote{$w (p^2) = 0$ has been checked up to two loops, see \cite{Gracey:2005cx}.} $w (p^2) = 0$, so that $u = -1$ implies an enhanced ghost propagator. Notice that this scenario is exactly predicted by the GZ framework.

\subsection{Remarks on the KO criteria}
Two comments should be made concerning the argument of Kugo and Ojima. First of all, in the KO framework \cite{Kugo:1979gm,Kugo:1995km}, the existence of a globally well-defined BRST charge is assumed.  This assumption has now been substantiated by the construction of a BRST operator in lattice gauge theory which does have non-perturbative validity \cite{vonSmekal:2008}.  On the other hand, the derivation of the two criteria was done by employing the usual Faddeev-Popov gauge-fixed action, which does not take into account Gribov copies.\\
\\
Moreover the KO argument does not really hold for the GZ action, as (1) the GZ action breaks the usual BRST symmetry, either explicitly, see equation \eqref{deltabrst}, or spontaneously, as in the Maggiore-Schaden construction discussed above, and (2) the GZ action does take into account Gribov copies.\\
\\
Another point which should be mentioned is that the Kugo and Ojima criterion \eqref{KO0} is sometimes useful as a starting point in looking for a solution to the Dyson-Schwinger or functional renormalization-group equations, see e.g. \cite{Fischer:2008uz}. 

\subsection{Imposing u(0) = -1 as a boundary condition in the Faddeev-Popov action}
In \cite{Dudal:2009xh}, the following was tried. The authors imposed the constraint $u(0)=-1$  directly into the Faddeev-Popov theory, by appropriately modifying the measure one starts from. By using thermodynamic arguments, as iat the end of sec.~\ref{non-localGZaction}, it was concluded that one precisely finds the GZ action! The boundary condition $u(0) = -1$ is exactly encoded in the horizon condition for the $\gamma$, see expression \eqref{horizoncondition}. The details can be found in \cite{Dudal:2009xh,Vandersickel:2011zc}.

\subsection{Conclusion}
We can thus conclude:
\begin{itemize}
\item Imposing a boundary condition into a theory can have serious consequences. Here for example, we have seen that imposing the boundary condition $u(0) = -1$ in the Faddeev-Popov measure with BRST symmetry, leads us to the GZ action without BRST symmetry. Therefore, imposing a boundary condition can change the symmetry content of a theory.
\item The relevance of the KO criterion in the GZ formalism is not clear because the KO analysis was done in the Faddeev-Popov formalism. Due to the breaking of the BRST symmetry in the GZ action, one cannot simply redo the KO analysis in the GZ framework.
\end{itemize}

\section{The relation of the GZ action to the lattice data\label{sec3}}
\subsection{The lattice data}\label{lattice.data}
 It would be interesting to verify the GZ theory. As we are working with a quenched Yang-Mills theory, we cannot use experimental data, but we can compare our analytical results with lattice calculations. Two particular quantities have been tested in great detail, namely the ghost and the gluon propagator, as they are believed to play an important role in confinement scenarios. The first calculations of the gluon propagator in 4 dimensions in the Landau gauge were already carried out in 1987, see e.g.\cite{Mandula:1987rh} on very small lattices of the order of $4^3\times 8$, and of the ghost propagator around 1996, see \cite{Suman:1995zg} on lattices of the order of $32^4$. Ever since, many papers appeared on the subject. Let us review them briefly.

\subsubsection{The $4d$ and $3d$ studies}
Let us start with the gluon and the ghost propagator in 4 and 3 dimensions. There have been many investigations of them, see \cite{Bernard:1993tz,Marenzoni:1993td,Marenzoni:1994ap,Nakamura:1995sf,Gutbrod:1996sq,Aiso:1997au,Cucchieri:1999sz,Leinweber:1998uu,Cucchieri:2001tw,Cucchieri:2003di,Bloch:2003sk,Furui:2003jr,
Cucchieri:2004mf,Silva:2004bv,Sternbeck:2005tk,Furui:2006py,Boucaud:2006pc,Sternbeck:2006cg,Oliveira:2008uf,Cucchieri:2007md,Bogolubsky:2007ud,Cucchieri:2007rg,Oliveira:2007tr,Dudal:2010tf,Bogolubsky:2009dc,Cucchieri:2010xr,Bornyakov:2009ug,Maas:2008ri} for a selection of some papers.  A useful overview of the lattice results can be found in \cite{Cucchieri:2010xr}, where the results are examined in the light of the GZ framework.\\
\\
Concerning the gluon propagator\footnote{The propagators were investigated both in $SU(2)$ as in $SU(3)$; for a comparison between these two settings, see \cite{Cucchieri:2007zm,Oliveira:2009nn}.} in $4d$ in the infrared, it quickly became clear that perturbative behavior was not seen, and that the gluon propagator is not enhanced at small momentum.  Instead a massive behavior was seen \cite{Aiso:1997au,Leinweber:1998uu,Bloch:2003sk,Silva:2004bv,Furui:2006py}. Further studies confirmed this behavior; however, in order to reach the deep infrared, larger lattices were required \cite{Cucchieri:2006xi,Oliveira:2008uf}. Therefore, one can say that until 2007 not so much was known about the gluon propagator in 4d, although the hope was that it would vanish at zero momentum in order to agree with the GZ framework. We must say, many studies already reported a finite gluon propagator at zero momentum \cite{Bloch:2003sk,Boucaud:2006pc} although the volumes were not large enough to reach the far infrared \cite{Sternbeck:2006cg}.  Finally, in 2007, the papers \cite{Cucchieri:2007md,Bogolubsky:2007ud} appeared, with simulations on huge lattices\footnote{It should be mentioned that the lattice spacing is also very large, in the range of $0.18 - 0.22$ fm. }. From these papers, it was found that the gluon propagator in 4d did \textit{not vanish} at zero momentum! After these papers, more papers appeared \cite{Cucchieri:2007rg,Bogolubsky:2009dc,Bornyakov:2009ug,Dudal:2010tf}, which arrived at the same conclusion, i.e.~the gluon propagator seems to attain a finite value at zero momentum. Some papers were still in agreement with a vanishing gluon propagator, see e.g.~\cite{Oliveira:2007tr}.\\
\\
Fewer studies of the gluon propagator have been performed in $3d$. The first one was in 1999 by Cucchieri \cite{Cucchieri:1999sz}, where a gluon propagator was reported which is infrared suppressed. The deep infrared was not accessible at that time. Later papers confirmed this suppressed behavior, \cite{Cucchieri:2001tw,Cucchieri:2003di,Cucchieri:2006tf}. It was however not clear what happened at zero momentum, but at least in $3d$ the gluon propagator was not in contradiction with a vanishing gluon propagator \cite{Cucchieri:2003di}. Positivity violation was already clearly seen, see \cite{Cucchieri:2004mf}. The hope was that at larger lattices, a turnover would be seen so the gluon propagator would vanish at zero momentum. Then the paper \cite{Cucchieri:2007md} appeared, with simulations on huge lattices, where a finite gluon propagator at $p = 0$ in $3d$ was reported.  Paper \cite{Cucchieri:2007rg} supported this result.  The upshot in $3d$ is seen in \cite{Cucchieri:2010xr} where on lattices of size $320^3$ the gluon propagator exhibits a clear {\em turnover} in the infrared namely, $D(p)$ has a maximum below which it {\em decreases} as $p$ decreases, and approaches a finite value $D(0) \neq 0$ from above.  The {\em decrease} of $D(p)$ for $p \to 0$, has no other explanation than the proximity of the Gribov horizon in infrared directions, in accordance with the Gribov scenario, but the {\em finite value} of $D(0) \neq 0$ contradicts perturbative calculations with the GZ action.  By contrast, in $4d$ on $128^4$ lattices \cite{Cucchieri:2010xr} there is something like a shoulder at very low momenta as the finite value of $D(0)$ is approached.\\
\\
The ghost propagator was more complicated to study on the lattice because one has to access the inverse of the Faddeev-Popov operator. In $4d$, papers \cite{Suman:1995zg,Cucchieri:1997dx,Furui:2003jr,Gattnar:2004bf,Sternbeck:2005tk} showed a behavior between the zeroth-order perturbative behavior $1/p^2$ and the $1/p^4$ singularity predicted by the GZ framework. Many problems however needed to be overcome \cite{Cucchieri:2010xr}, including finite-volume effects, discretization effects and Gribov-copy effects, which could influence the results obtained to far. The hope was that at larger volumes indeed the enhancement of the ghost propagator would be more clearly seen. The breakthrough came in 2007, whereby in the same papers \cite{Cucchieri:2007md,Bogolubsky:2007ud,Bogolubsky:2009dc} as for the gluon propagator, the ghost propagator was investigated at large lattices up to $128^4$. The result was very surprising, as the enhancement completely disappeared and the $1/p^2$ behavior was recovered in the deep infrared. \\
\\
The ghost propagator in $3d$ was even less studied \cite{Cucchieri:2006tf,Cucchieri:2007md,Cucchieri:2008fc}. The main conclusion is however again the same. The larger the volumes, the less enhancement of the ghost propagator is seen. Finally a consensus was reached, and one finds again the $1/p^2$ behavior at very large lattices.\\
\\
From these lattice results one may conclude that something is missing in the current GZ framework. Much research has been done on this problem, see e.g.~papers \cite{Dudal:2007cw,Dudal:2008rm,Dudal:2008sp,Dudal:2008xd,Dudal:2011gd}, where a dynamically refined GZ action was proposed in 3 and 4 dimensions.

\subsubsection{The $2d$ studies}
$2d$ appeared to be different from the $3d$ and $4d$ case. Not so many studies were carried out in 2d, but they are all consistent \cite{Maas:2007uv,Cucchieri:2007rg,Cucchieri:2011um}. The gluon propagator goes to zero at zero momentum; this was even checked with lattices up to $2560^2$ \cite{Cucchieri:2011um}, while the ghost propagator displays enhanced behavior.  Therefore, the lattice data in $2d$ are at least qualitatively in agreement with the GZ framework. However, two comments are in order. Firstly, the ghost propagator does not exactly display enhancement like $1/p^4$, as the following enhanced behavior was reported: consistent with $\lim_{p^2 \to 0} \mathcal G(p^2) \sim \frac{1}{p^2} \frac{1}{p^{2/5}} $, see \cite{Maas:2007uv}, and in agreement with Schwinger-Dyson calculations \cite{Zwanziger:2001kw}. Secondly, a qualitative comparison has not been done so far, and would be needed before drawing any further conclusions.\\
\\
Let us also mention that there has been done some work on the GZ action in two dimensions, whereby it was shown that the dynamical refinement is not possible \cite{Dudal:2008xd}, therefore meaning that the GZ framework would still be valid in $2d$. This qualitatively agrees with the lattice data so far.

\subsection{Reflection positivity}\label{reflection.positivity}

It was pointed in \cite{Zwanziger:1991xx} that if the gluon propagator $D(k)$ vanishes at $k = 0$, then reflection positivity would be violated.  In fact it would be maximally violated, as we now explain.  

Reflection positivity and the other axioms of Euclidean quantum field theory imply that $D(k)$ satisfies the K\"{a}llen-Lehmann representation,
\beq
D(k) = \int_0^\infty dm^2 {\rho(m^2) \over k^2 + m^2},
\eeq
where $\rho(m^2) \geq 0$ is a positive weight.  This implies that the gluon propagator in $x$-space,
\beq
\widetilde D(x) = \int_0^\infty dm^2 \ \rho(m^2) \ \widetilde D_m(x),
\eeq
is positive for all $x$, $\widetilde D(x) > 0$, because the free propagator of a particle of mass $m$ is positive for all $x$, as one sees from
\beq
\widetilde D_m(x) = 2^{-1} \int_0^\infty d\alpha \ (2\pi \alpha)^{-d/2} \exp[2^{-1}(m^2 \alpha + x^2\alpha^{-1})] > 0.
\eeq
If the zero-momentum propagator vanishes,
\beq
D(0) = \int d^dx \ \widetilde D(x) = 0,
\eeq
and if $\widetilde D(x)$ is positive (or zero) for all $x$, then the gluon propagator vanishes identically,  $\widetilde D(x) = 0$ for all $x$, which is false.  Thus, if the gluon propagator vanishes at $k = 0$, then reflection positivity is violated.  In fact it is maximally violated in the sense that if $D(0) = 0$, then the gluon correlator $ \widetilde D(x) $ is positive and negative in equal measure.  The Gribov propagator $D(k) = k^2/[(k^2)^2 + m^4]$ maximally violates positivity, because $D(0) = 0$, and the K\"{a}llen-Lehmann representation is violated by having poles at imaginary mass-square, $k^2 = \pm i m^2$.\\
\\
As discussed in the preceding section, numerical investigation indicates that $D(0)$ vanishes in Euclidean dimension 2, whereas it is finite, $D(0) > 0$, in Euclidean dimension 3 and 4, so reflection positivity is maximally violated for $d = 2$, but it is not maximally violated for $d = 3$ and $d = 4$.  Nevertheless, numerical investigation does show clear violation of reflection positivity in Euclidean dimension 3 \cite{Cucchieri:2004mf} and in dimension 4, for both quenched and unquenched cases \cite{Bowman:2007hd}.  This evidence of violation of reflection positivity in dimension $d = 2, 3$ and $4$ stands in contrast to perturbation theory, where BRST symmetry assures that the perturbative gluon spectrum is physical and unitary.  Violation of reflection positivity is a signal that the gluon is not a physical particle, in accordance with the confinement scenario.

\section{Color confinement}
\label{color.confinement}

  In this section we present an exact bound that results from the restriction of the functional integral to the Gribov region which implies that the color degree of freedom is  confined \cite{Zwanziger:1991xx, Zwanziger:1991xy}.

\subsection{Exact bounds on free energy}

  Let the free energy $W(J)$ be defined by
\beq
\exp[W(J)] = \int_\Omega dA \ \rho(A) \exp(J, \ A),
\eeq
where only transverse configurations, $\p_\mu A_\mu = 0$, that lie inside the Gribov region $\Omega$, are integrated over, and the source term is given by $(J, A) \equiv \int d^Dx \ J_\mu^a(x) A_\mu^a(x)$.  Since $A$ is identically transverse, only the transverse part of the source $J$ contributes, and we impose that it is also identically transverse, $\p_\mu J_\mu = 0$. The result is quite general, for we shall suppose only that $\rho(A)$ is a non-negative weight, $\rho(A) \geq 0$, that it is normalized, $\int dA \ \rho(A) = 1$, with support restricted to the Gribov region, $\rho(A) = 0$ for $A \notin \Omega$.  Since the fundamental modular region $\Lambda$ is contained in $\Omega$, the result holds in particular if the integral is restricted to $\Lambda$.  \\
\\
We specialize the source to a single fourier component,
\beq
J_\mu^a(x) = H_\mu^a \cos(k x_1),
\eeq
where we have aligned the $1$-axis along $k$, and transversality of $J_\mu$ reads $H_1^a = 0$, so the source term is given by
\beq
(J, A) = \int d^Dx \ H_i^a \cos(k x_1) A_i^a(x).
\eeq
where $i = 2,. ...D$.  It may be interpreted by analogy with magnetic spin systems, where $A_\mu^a(x)$ is the analog of a spin variable, but with a color index $a$, and $H_\mu^a$ is the analog of an external magnetic field, also with a color index, that is modulated by a plane wave $\cos(k x_1)$.  (The external color-magnetic field $H_\mu^a$ should not be confused with the Yang-Mills color-magnetic field tensor $F_{i j}^a$.)  In the limit $k \to 0$, the source becomes a constant color-magnetic field.  The free energy $W(J)$ now depends only on the parameters $k$ and $H_i^a$, and we write
\beq
W(J) = W_k(H).
\eeq
This specialized free energy is sufficient to calculate the gluon propagator,
\beq
\label{gluonprop}
\delta^{ab} \delta_{ij} \ D(k) = 2 \ { \p^2 w_k(0) \over \p H_i^a \p H_j^b },
\eeq   
where $k$ is aligned along the 1-axis, and $i, j = 2, ... D$.  Here $w_k(H)$ is the free energy per unit volume,
\beq
w_k(H) = {W_k(H) \over V},
\eeq
where $V$ is the Euclidean volume.\\
\\
	An exact bound for $W_k(H)$ on a finite lattice was given in \cite{Zwanziger:1991xx, Zwanziger:1991xy} which holds for any numerical gauge fixing to Landau gauge, with support restricted to the Gribov region $\Omega$, which is to say, with support on relative or absolute minima of the gauge fixing functional.  In the continuum limit and at large Euclidean volume $V$, the lattice bound implies the bound in $D$ Euclidean dimensions,
\beq
\label{LIcontbound}
w_{k}(H) \leq (2 D)^{1/2} k |H|,
\eeq
where $|H|^2 = \sum_{\mu, b} (H_\mu^b)^2$.\\
\\
	More recently, a stricter continuum bound for $w_k(H)$ at finite $H$ was obtained~\cite{Zwanziger:2011a, Zwanziger:2011b}, that also holds for any numerical gauge fixing  with support inside the Gribov region $\Omega$, 
\beq
\label{newbound}
w_k(H) \leq 2^{-1/2} k \ {\rm tr}[ ( H^a H^a )^{1/2}].
\eeq
Here $H^a H^a$ is the matrix with elements $\sum_a H_i^a H_j^a$.  It has positive eigenvalues, and the positive square root is understood.  This bound is in fact optimal for a probability distribution $\rho(A)$ of which it is known only that its support lies inside the Gribov region.  Either bound yields in the zero-momentum limit
\beq
\label{0momentum} 
w_{0}(H) = \lim_{k \to 0} w_{k}(H) = 0.
\eeq\\
These bounds imply that the magnetization vanishes in the limit $k \to 0$.  To show this we write $H_\mu^a = h e_\mu^a$, where $h$ represents the magnitude of the external magnetic field in the direction of the fixed unit vector $e_\mu^a$, with $\sum_{\mu a} (e_\mu^a)^2 = 1$, and we write $w_k(H) = w_k(h)$.  The magnetization is given by
\beq
m_k(h) = { \p w_k(h) \over \p h } = \langle \ V^{-1}\int d^Dx \ \cos(k x_1) e_i^a \  A_i^a(x) \ \rangle_h.
\eeq
It vanishes at zero magnetic field, $m_k(0) = 0$, if there is no spontaneous magnetization, as we suppose, and increases monotonically, 
${ \p m_k(h) \over \p h} \geq 0$ as follows from
\beq
{ \p m_k(h) \over \p h } = { \p^2 w_k(h) \over \p h^2 } = (1/2) D(h) \geq 0, 
\eeq
which is positive because the gluon propagator $D(h)$ is positive at any $h$.  The free energy may be expressed in terms of the magnetization and the bound reads,
\beq
{w(h) \over h} = {\int_0^h dh' \ m_k(h') \over h} \leq 2^{-1/2} k \ {\rm tr} [ (e^a e^a)^{1/2} ]
\eeq
by \eqref{newbound}.  Because $m_k(h)$ increases monotonically, this bound implies that $m_k(h)$ approaches a finite limit, $\lim_{h \to \infty}m_k(h) = m_k(\infty)$ which satisfies
\beq
m_k(\infty) = \lim_{h \to \infty}{\int_0^h dh' \ m_k(h') \over h} \leq 2^{-1/2} k \ {\rm tr} [ (e^a e^a)^{1/2} ].
\eeq  
Thus the magnetization at finite $h$ satisfies the bounds,
\beq
\label{magnetizationbound}
0 \leq m_k(h) \leq m_k(\infty) \leq 2^{-1/2} k \ {\rm tr} [ (e^a e^a)^{1/2} ].
\eeq
We arrive at the remarkable conclusion that in the limit of constant external magnetic field, $k \to 0$, the color magnetization per unit volume vanishes, {\it no matter how strong the external magnetic field} $h$,
\beq
\lim_{k \to 0} m_k(h) = \langle \ V^{-1}\int_V d^Dx \ e_i^a \  A_i^a(x) \ \exp[ \  h \int_V d^Dx \ e_i^a \  A_i^a(x) \ ] \ \rangle = 0.
\eeq
Thus the system does not respond to a constant external color-magnetic field.  In this precise sense, the color degree of freedom may be said to be confined.  This results from the proximity of the Gribov horizon in infrared directions.\\
\\	
If $w_k(H)$ were analytic in $H$ in the limit $k \to 0$, the bound, $\lim_{k \to 0} w_k(H) = 0$, would imply that all derivatives of the generating function $w_0(H)$ vanish, including in particular the gluon propagator $D(k)$ (\ref{gluonprop}) at $k = 0$.  In \cite{Zwanziger:1991xx, Zwanziger:1991xy} it was assumed that $w_k(H)$ is analytic in $H$ at $k = 0$, and it was concluded that the gluon propagator at $k = 0$ vanishes, $D(0) = 0.$  However, as reported in the preceding section, although this agrees with lattice data in Eucldean dimension 2, it disagrees with recent lattice data which indicate a finite value, $D(0) \neq 0$, in Euclidean dimensions 3 and 4.  If this is true, then $w_k(H)$ must become non-analytic in $H$ in the limit $k \to 0$.  Two models were recently exhibited \cite{Zwanziger:2011a, Zwanziger:2010b} for which the free energy $w_k(H)$ vanishes in the limit $\lim_{k \to 0} w_k(H) = 0$ for all $H$ --- and thus exhibit confinement of color --- but for which the gluon propagator $D(k)$ is nevertheless finite at $k = 0$, in agreement with the numerical data in 3 and 4 Euclidean dimensions.

\subsection{Simple model}

	We exhibit here a similar model, defined by
\beq
\label{model}
w_{k,{\rm mod}}(H) =   {\rm tr} \Big[ \Big(g^2(k) + {k^2 H^a H^a \over 2} \Big)^{1/2} - g(k) \Big],
\eeq
where $g(k) \geq 0$ is a function at our disposal, and $H^a H^a$ is the matrix with elements $H_i^a H_j^a$, for $i, j = 2,... \ D$.   This model possesses the following  features of the exact theory \cite{Zwanziger:2011a}:
(i) It satisfies $w_{k,{\rm mod}}(0) = 0$, which is necessary at $H = 0$ for a normalized probability distribution $\int dA \ \rho(A) = 1$.
(ii) It satisfies the bound \eqref{newbound},  
\beq
w_{k, {\rm mod}}(H)  \leq 2^{-1/2} k \ {\rm tr} [ ( H^a H^a )^{1/2} ] .
\eeq
(iii) The magnetization
\beq
m_{k,{\rm mod}}(h) = {\p w_{k,{\rm mod}}(h) \over \p h} = {\rm tr} \Big[ {k^2 h e^a e^a \over 2} \Big(g^2(k) + {k^2 h^2 e^a e^a \over 2} \Big)^{-1/2}  \Big],
\eeq
where $H_i^a = h e_i^a$, satisfies the bound \eqref{magnetizationbound}. \\
(iv) The gluon propagator 
\beq
\label{gluprop}
(1/2)D(k, h) = {\rm tr} \Big[ {k^2 g^2(k) e^a e^a \over 2} \Big(g^2(k) + {k^2 h^2 e^a e^a \over 2} \Big)^{-3/2}  \Big]  \geq 0.
\eeq
is positive at all $h$, as required for it to be a covariance.  This assures that the magnetization is monotonically increasing,
\beq
(1/2)D(k, h) = {\p m_{k,{\rm mod}}(h) \over \p h} \geq 0.
\eeq \\
	In this model, the free energy $w_{k, {\rm mod}}(H)$ and magnetization $m_{k, {\rm mod}}(H)$ both vanish in the limit $k \to 0$ for all $h$, in agreement with the exact bounds given above, and consistent with confinement of color.  If $w_{k,  {\rm mod}}(H)$ were analytic in $H$ at $k = 0$, then the gluon propagator $(1/2)D(k, h) = {\p^2 w_{k,{\rm mod}}(h) \over \p h^2}$ would vanish at $k = 0$ for all $h$.  However from \eqref{gluprop} we find for the gluon propagator, $D(k) = D(k, h = 0)$,
\beq
\label{glupropa}
D(k) = {k^2  \over g(k)}.
\eeq
Whether or not it vanishes at $k = 0$ depends on the behavior at $k = 0$ of the function at our disposal, $g(k)$.  If $g(k) = m^2 k^2$ at low momentum, where $m$ is some mass, then the model yields at low momentum
\beq
\lim_{k \to 0}D(k) = {1 \over m^2},
\eeq  	
which is finite, and agrees with numerical results in 3 and 4 Euclidean dimensions.\footnote{ The first two terms in the expansion of (\ref{model}) in powers of $H$ are given by
$w_{k, {\rm mod}}(H) = { k^2 \over 8 g(k) } H_i^a H_i^a + g(k) O[ k^4 H^4 / g^4(k) ]$.  For $g(k) = m^2 k^2$, which makes the first term finite at $k = 0$, the coefficient of the $H^4$ term is of order $1/k^2$, which diverges as $k \to 0$, as do all higher order coefficients.  This exhibits the non-analyticity of $w_{k, {\rm mod}}(H)$ in $H$ at $k = 0$.}  On the other hand, if $g(k)$ does not vanish as rapidly as $k^2$ at $k = 0$, then we get 
$\lim_{k \to 0}D(k) = 0 $, which agrees with the numerical data in 2 Euclidean dimensions.  Thus we have a model for which the free energy $w_k(h)$ vanishes at $k = 0$ for all $h$, in agreement with the exact bound provided by the Gribov horizon.  Nevertheless, the gluon propagator $D(k)$ may remain finite at $h = 0$.

\chapter{Overview of various approaches to the Gribov problem}
\label{overview.of.approaches}

In this section we present a critique of the present approach and some related analytic approaches, particularly stochastic quantization, with a view toward lessons learned, open problems, and directions for future research.\\  
\\
\section{GZ action}
In the present review we have gathered in one place the derivation and properties of the GZ action, eq.~\eqref{SGZphys}, which otherwise are widely scattered in the literature.  This action, in Landau gauge, incorporates a cut-off of the functional integral outside the Gribov region $\Omega$ which is done to avoid counting Gribov copies that are related by ``large" gauge transformations.  It must be said at the outset that, as explained in sect.~\ref{discussFMR}, there is an approximation involved in the derivation of the GZ action, in which the fundamental modular region $\Lambda$ is replaced by the Gribov region $\Omega$ (although it has been conjectured that they may give the same expectation values \cite{Zwanziger:1993dh, Greensite:2004ke}).  The GZ action is of interest nevertheless because it is local and renormalizable even though the cut-off is non-perturbative.  Its non-perturbative character is manifested by the appearance in it of the Gribov mass $\hat\gamma$.\footnote{$\hat\gamma$ is not an independent parameter; its value is fixed in terms of $\Lambda_{\rm QCD}$ by the horizon condition ${\p \Gamma \over \p \hat\gamma} = 0$, where $\Gamma$ is the quantum effective action.}\\
\\
If one calculates with the GZ action instead of the Faddeev-Popov action, and expands in powers of the coupling constant $g$, one obtains an alternative, perturbatively renormalizable series, in which the zeroth-order gluon propagator is the Gribov propagator, $D_0(k) = {k^2 \over (k^2)^2 + {\hat\gamma}^4}$.  Its poles occur at the unphysical locations $k^2 = \pm i \hat\gamma^2$.  This does not correspond to a physical particle but is appropriate for a propagator of gluons that are confined.  This would clearly be a non-perturbative phenomenon in conventional Faddeev-Popov perturbation theory, but it appears already in zeroth order in the alternative perturbative expansion provided by the GZ action.  In this alternative perturbation series, the gluon propagator $D(k)$ vanishes at $k = 0$, $D(0) = 0$.  As discussed in sect. \ref{reflection.positivity}, this corresponds to maximal violation of reflection positivity, so the K\"{a}llen-Lehmann representation is maximally violated, and this is manifested concretely in the unphysical poles of the gluon propagator.  Lattice data also show clear violation of reflection positivity \cite{Cucchieri:2004mf, Bowman:2007hd}, and we take unphysical singularities of correlators of gauge-non-invariant fields to be a correct description of confinement in continuum QCD.\\
\\
Limitations of the GZ approach in perturbation calculations are mentioned in the Introduction.  They may possibly be remedied by non-perturbative calculations.

 \section{Refined GZ action}
In response to this challenge, allowable condensates which preserve renormalizability, such as $A^2$, have been treated in a mean-field approach, by introducing them into the GZ-action.  This results in what is known as the refined GZ action (RGZ) that has been studied extensively \cite{Dudal:2007cw,Dudal:2008sp,Dudal:2008rm,Dudal:2008xd,Dudal:2011gd}.  By this method, satisfactory agreement with lattice data for the gluon and ghost propagators has been achieved, and the glue-ball spectrum has been calculated.  A separate review article would be required to provide an overview of these developments.      \\
\\  
\section{Coulomb gauge}
The present review has been devoted to non-perturbative QCD dynamics in the Euclidean Landau gauge.  However the Coulomb gauge also requires a cut-off to avoid Gribov copies, as Gribov pointed out in his original article \cite{Gribov:1977wm}.  The cut-off in the Coulomb gauge has also been implemented by a local action, along the lines of the GZ action in Landau gauge  \cite{Zwanziger:2006sc}.  QCD calculations in the Coulomb gauge have also been done by variational calculation in the Hamiltonian formulation \cite{Reinhardt:2011ab, Szczepaniak:2001ss}.  An exact bound exists in the Coulomb gauge, that goes by the slogan ``no confinement without Coulomb confinement."  Thus if the (gauge-invariant) Wilson potential $V(r)$ is confining, then the temporal gluon propagator in Coulomb gauge, $D_{00}(r, t)$, has an instantaneous part, $V_{\rm coul}(r) \delta(t)$, that is also confining, \cite{Zwanziger:2002hd}.  The interest of this bound is that the Wilson potential, obtained from gauge-invariant Wilson loop $P\exp(i \oint A_\mu dx^\mu),$ involves $n$-point functions of all orders $n$, whereas the gluon propagator $D_{00}(r, t)$ is a 2-point function.  A proof of renormalizability in the Coulomb gauge remains on open challenge.\\
\\
\section{Dyson-Schwinger equation} \label{DSequations}
The Dyson-Schwinger (DS) equation is perhaps the most highly explored analytic approach to non-perturbatve calculations in QCD \cite{Fischer:2008uz, Alkofer:2000wg, Zwanziger:2001kw, Lerche:2002ep, Pawlowski:2003hq, Alkofer:2003jj, Boucaud:2008ky}.  It is similar in spirit to the functional renormalization-group equation \cite{Fischer:2008uz, Pawlowski:2003hq}, and has also been developed in the pinch technique formalism \cite{Binosi:2009qm, Aguilar:2009nf}.  Most DS calculations have been done with the Faddeev-Popov action and, because they are non-perturbative, the question arises whether Gribov  copies are treated correctly.  To address this question, we must review the derivation of the DS equations.  The functional DS equation, which expresses the tower of the DS equations for all correlators, results from the identity,
\beq
\label{nodalregion}
0 = \int_R dA \ {\delta \over \delta A_\mu^b} \det[M(A)] \exp[- S_{\rm YM}(A)/\hbar + (J, A)/\hbar],
\eeq
which states that the integral of a derivative vanishes.  Here $M(A) = - \p_\mu D_\mu(A)$ is the Faddeev-Popov operator and, for Landau gauge, the integral extends over transverse configurations $\p \cdot A = 0$, and we have introduced $\hbar$ to keep track of the loop order in a diagram.  This identity is correct, provided that the integrand vanishes on the boundary $\p R$ of the region of integration $R$ in (transverse) $A$-space.  The region of integration $R$ is usually assumed to be all of A-space.  However the identity also holds if $R$ is the Gribov region, $R = \Omega$, because the Faddeev-Popov determinant, $\det[M(A)]$ vanishes on its boundary $\p \Omega$, which is the first Gribov horizon.  Indeed, because the Faddeev-Popov determinant is the infinite product of eigenvalues $\det[M(A)] = \prod_n \lambda_n(A)$, the $n$-th Gribov horizon, defined by $\lambda_n(A) = 0$, is a {\em nodal surface} of the integrand for every integer $n$, and the identity holds if the boundary $\p R$ coincides with any one of these nodal surfaces.  From the last identity the functional DS equation for the generating functional of correlators,
\beq
Z_R(J) \equiv \int_R dA \ \det[M(A)] \exp[- S_{\rm YM}(A)/\hbar + (J, A)/\hbar],
\eeq
follows,
\beq
\label{DSequatins}
\left[ J_\mu^b(x) - {\delta S_{\rm eff} \over \delta A_\mu^b(x)} \left( \hbar {\delta \over \delta J} \right) \right] Z_R(J) = 0,
\eeq
where  $S_{\rm eff}(A) \equiv S_{\rm FP}(A) - \hbar \ln \det M(A)$.\footnote{In practice Faddeev-Popov ghosts are introduced, but the argument still stands.}   With $Z_R(J) = \exp[W_R(J)/\hbar]$, where $W_R(J)$ is the generating functional of connected correlators,  the last equation reads
\beq
J_\mu^b(x) = {\delta S_{\rm eff} \over \delta A_\mu^b(x)}\left( x, { \delta W_R \over \delta J  } + \hbar { \delta \over \delta J } \right).
\eeq
Next we change variable from $W_R(J)$ to the quantum effective action $\Gamma_R(A)$ by the Legendre transformation,
\beq
\label{Legendre}
A(J) = {\delta W_R(J) \over \delta J}; \ \ \ \ \ \ \ \ \ \Gamma_R(A) = (J, \ A) - W_R(J), \ \ \ \ \ \ \ J(A) = {\delta \Gamma_R(A) \over \delta A}.
\eeq
This yields the functional DS equation for the quantum effective action $\Gamma_R(A)$,
\beq
\label{DSGamma}
{\delta \Gamma_R \over \delta A_\mu^b(x)} = {\delta S_{\rm eff} \over \delta A_\mu^b(x)} \left( x; \ A + \hbar {\cal D} { \delta \over \delta A } \right),
\eeq
where
$\hbar {\cal D}_{\mu \nu}^{cd}(x, y; \ A)$ is the gluon propagator in the presence of the source $A$, and $\cal D(A)$ is expressed in terms of $\Gamma_R(A)$ by
\beq
\label{glupropwsource}
({\cal D} ^{-1})_{\mu \nu}^{cd}(x, y; A) = {\delta^2 \Gamma_R \over \delta A_\mu^c(x) \delta A_\nu^d(y)}.
\eeq
The DS equations for the one-particle irreducible correlators are obtained by expanding $\Gamma_R(A)$ in powers of $A$, and equating like coefficients.\\
\\
Observe now that the functional DS equation depends only on $S_{\rm eff}$ and is {\em independent} of which region $R$, bounded by a nodal surface $\p R$, is chosen, or if all of space is chosen in \eqref{nodalregion}.  Consequently, for each region $R$ that is bounded by a nodal surface --- and there are an infinite number of nodal surfaces --- there is a different solution $Z_R(J)$ of the DS equation for the generating functional $Z(J)$, and moreover, because the equation for $Z(J)$ is linear, a linear combination of solutions $Z_R$ is also a solution for $Z(J)$.  Each different solution $Z_R(J)$ gives a different solution $\Gamma_R(A)$, because $\Gamma_R(A)$ is obtained from $Z_R(J)$ by an invertible change of variable.  Thus there is an enormous ambiguity in the solutions to the DS equation in Faddeev-Popov theory \cite{Zwanziger:2001kw}.  {\em One} of these solutions belongs to the choice $R = \Omega$, which corresponds to a cut-off at the Gribov horizon $\p \Omega$.  The good news is that the DS equations for Faddeev-Popov theory do have a solution that corresponds to a cut-off at the Gribov horizon.\footnote{Alternatively, it has been argued \cite{Hirschfeld:1978yq} that integrating over all of $A$-space gives the correct answer by systematic cancellation of Gribov copies because the signed Faddeev-Popov determinent gives the signed intersection number of the gauge orbit with the gauge-fixing surface, which is a topological invariant.  The question remains: how to select this particular solution?}  Taking the optimistic view, if we impose enough additional conditions, which result for example from the positivity of the integrand $\det M \exp(-S)$ in the Gribov region $\Omega$, we may resolve the ambiguity and select this particular solution.  The bad news is that, in practice, we know little about what additional conditions would be necessary to resolve this ambiguity completely.  In view of this, it is perhaps not surprising that two different solutions have been found to the DS equations in Faddeev-Popov theory, namely ``scaling" and ``decoupling" solutions  \cite{Fischer:2008uz}.\\
\\
The DS equations for the propagators have also been derived from the GZ action, and its infrared behavior was investigated, with two solutions being found \cite{Huber:2009}.  Subsequently one of these solutions could be eliminated \cite{Huber:2010}.  Interestingly, the surviving solution has precisely the same infrared critical exponents as in Faddeev-Popov theory in the Landau gauge \cite{Zwanziger:2001kw, Lerche:2002ep}.

\section{Stochastic quantization}
\label{stochastic.quantization}
An avenue for non-perturbative calculations which is insufficiently explored is stochastic quantization \cite{Parisi:1980ys} with stochastic gauge fixing \cite{Zwanziger:1981kg}.  This method elegantly by-passes the problem of Gribov copies, and is geometrically unobjectionable, whereas the GZ action relies on an approximate solution of the Gribov problem in which the fundamental modular region $\Lambda$ is replaced by the Gribov region $\Omega$, as discussed in sect.~\ref{thegribovproblem}.  We discuss briefly the advantages and challenges presented by stochastic quantization, beginning with a brief introduction to this approach in case it is not familiar to the reader.\\
\\

\subsection{Background}
Stochastic quantization is conveniently defined by a quantum field theoretic version of the time-dependent Fokker-Planck equation
\beq
 {\p P({\bf x}, t) \over \p t} = {\bf \nabla}_i \  \left[ \hbar {\bf \nabla}_i - K_i({\bf x}) \right] P({\bf x}, t).
 \eeq  
This equation determines the probability distribution $P({\bf x}, t)$ of Brownian particles subject to a drift force $K_i({\bf x})$.  It results from the local flow equation $\dot{P} = - {\bf \nabla} \cdot {\bf j}$, where the current, ${\bf j} = - \hbar{\bf \nabla} P + {\bf K} P$, consists of a diffusion or fluctuation term, where $\hbar$ is the diffusion constant, and a drift term.  If the drift force is conservative, ${\bf K} = - {\bf \nabla} V({\bf x})$, the probability distribution $P({\bf x}, t)$ relaxes to the Boltzmann distribution $\lim_{t \to \infty}P({\bf x},t) = N\exp[-V({\bf x})/ \hbar ]$, for any initial distribution, provided that $\exp[-V({\bf x})/\hbar]$ is normalizable.\\
\\  
Consider now the functional version of this equation, 
 \beq
{\p P(A, t) \over \p t} =  \int d^dx \ { \delta \over \delta A_\mu^b(x) } \left[ \hbar {\delta \over \delta A_\mu^b(x) } - K_\mu^b(x; \ A) \right] P(A, t),
\eeq
where $P(A, t)$ is a time-dependent probability distribution in Euclidean $A$-space, $t$ is an artificial ``fifth time" that corresponds to machine time in Monte Carlo simulations of lattice gauge theory, $K_\mu^b(x; \ A)$ is a ``drift force" in $A$-space, and $x_\mu$ is a Euclidean position-vector.  If the drift force is conservative, $K_\mu^b(x; \ A) = - { \delta S(A) \over \delta A_\mu^b(x) }$, where the Euclidean action $S(A)$ is the analog of the potential energy $V({\bf x})$, then $P(A,t)$ relaxes to the Euclidean probability distribution, $P_{\rm eq}(A) \equiv \lim_{t \to \infty}P(A, t) = N \exp[- S(A)/\hbar]$, for any initial probability distribution, provided $\exp[-S(A)/\hbar]$ is normalizable.  Euclidean quantum field theory is recovered by calculating expectation-values from the equilibrium distribution
\beq
\label{expecteqprob}
\langle O \rangle = \int dA \ O(A) P_{\rm eq}(A).
\eeq
In a gauge theory, the (would-be) equilibrium distribution $\exp[-S_{\rm YM}(A) / \hbar]$, where $S_{\rm YM}(A)$ is the Euclidean Yang-Mills action, is not in fact normalizable, and it appears that we are back with the original problem.\footnote{In the original version of stochastic quantization \cite{Parisi:1980ys}, this was dealt with by taking the limit of the expectation-values of gauge-invariant observables,
\beq
\langle O \rangle = \lim_{t \to \infty} \int dA \ O(A) P(A, t)
\eeq}
However in gauge theory, we are interested only in gauge-invariant observables.  For these one may modify the drift force by introducing a gauge-fixing ``force" that has the form of an infinitesimal gauge transformation,
\beq
K_\mu^b(x) \to - { \delta S_{\rm YM} \over \delta A_\mu^b(x) } + D_\mu^{ab} \omega^b,
\eeq
because this has no effect on the expectation-value of gauge-invariant observables \cite{Zwanziger:1981kg}.  The only scalar with the correct engineering dimension and color dependence is given by $\omega^b = a^{-1} \p \cdot A^b$, and we take for the drift force,
\beq
\label{driftforce}
K_\mu^b(x; A) = - { \delta S_{\rm YM} \over \delta A_\mu^b(x) } + a^{-1} D_\mu^{ab} \p \cdot A^b,
\eeq
where $a$ is a gauge parameter.  The second term is a globally restoring force for $a > 0$, in the sense that it always points back toward the origin in $A$-space, $(A, K_{\rm gf}) \leq 0$, for we have
\beq
a^{-1} (A_\mu, D_\mu \p \cdot A) =  - a^{-1} (\p \cdot A,  \p \cdot A) \leq 0.
\eeq
Expectation-values are calculated from the equilibrium distribution $P_{\rm eq}(A) = \lim_{t \to \infty} P(A, t)$ determined by the Fokker-Planck equation with drift force \eqref{driftforce}.       \\
\\
This method of gauge fixing, that relies on a {\em gauge-fixing force that is an infinitesimal gauge transformation}, is a geometrically correct procedure that bypasses the problem of Gribov copies.  It is not available in an action formalism because the gauge-fixing force is non-conservative, and cannot be written as the gradient of some gauge-fixing action,
\beq
K_{\rm gf} \neq - {\delta S_{\rm gf} \over \delta A_\mu^b}.
\eeq\\
\\
Note that this is a property of non-Abelian gauge groups. For the $U(1)$ gauge group, we have $D_\mu(A) = \p_\mu$, and the gauge-fixing force is conservative, $K_{{\rm gf} \ \mu} = a^{-1} \p_\mu \p \cdot A = - {\delta S_{\rm gf} \over \delta A_\mu}$, where $S_{\rm gf} = (2a)^{-1} \int d^dx \ (\p \cdot A)^2$.

\subsection{Time-dependent approach}

Calculations may be done using either the time-dependent or time-independent Fokker-Planck equation.  In the time-dependent approach, the solution $P(A, t)$ is expressed as a path integral over paths $A_\mu^b(x, t)$, analogous to an integral over Brownian motion paths with a drift force.  Because of the resemblance of the Fokker-Planck to the Schr\"{o}dinger equation, this path integral formally resembles the Feynman path integral in quantum mechanics, but with a local action that is 5-dimensional (and should not be confused with the 4-dimensional Euclidean action).  It has been shown by BRST methods that this method is renormalizable \cite{ZinnJustin:1987ux, Chan:1985kf, Baulieu:1999wz, Baulieu:2000bgz}.  However calculations are challenging in this approach.  For example, in a perturbative expansion, the zeroth-order gluon propagator is given by $[ k_5^2 + (k_\mu^2)^2]^{-1}$, where $\mu = 1.,,,4$.  To our knowledge the time-dependent Fokker-Planck  has not been used in non-perturbative calculations.\\
\\

\subsection{Time-independent approach}

The equilibrium probability distribution $P_{\rm eq} = \lim_{t \to \infty}P(A, t)$, which is used to calculate expectation values of gauge-invariant observables, satisfies the time-independent Fokker-Planck equation,
\beq
\label{TIFokkerPlanck}
0 =  \int d^dx \ {\delta \over \delta A_\mu^b(x) } \left[ \hbar {\delta \over \delta A_\mu^b(x) } - K_\mu^b(x; \ A) \right] P_{\rm eq}(A),
\eeq
Because the force $K_\mu^b$ is not conservative, we cannot give a closed-form solution for the equilibrium distribution $P_{\rm eq}(A)$.  The basis of the time-independent approach is that this equation may be converted by a change of variables to a useful functional equation for the correlators, that serves as a substitute to the functional DS equation in a theory based on a local action \cite{Zwanziger:2002ea}.  Before doing so, we note that the solution $P_{\rm eq}(A)$ to this equation (if it exists) has {\em no nodal surfaces}, and is {\em unique and positive} up to normalization.  This is shown in Appendix C.  In contrast, the Faddeev-Popov weight $ \det [M(A)] \exp[- S_{\rm YM}(A)/\hbar]$ has many nodal surfaces, and this leads to a great ambiguity in the solution of the functional DS equation in Faddeev-Popov theory, as we have seen.\\
\\ 
To express the time-independent Fokker-Planck equation as an equation for the correlators, we multiply by $\exp[(J, \ A)/\hbar]$ and integrate over all $A$.  After integration by parts we obtain
\beq
0 = \int d^dx \ J_\mu^b(x) \left[ J_\mu^b(x) + K_\mu\left( x;  \hbar { \delta \over \delta J } \right) \right] Z(J),
\eeq
where
\beq
Z(J) = \int dA \ \exp[(J, \ A)/\hbar] \ P_{\rm eq}(A)
\eeq
is the generating functional of the correlators.  [If we replace $J$ by $iJ$, then $Z(iJ) = \widetilde{P}_{\rm eq}(J)$ is the functional fourier transform of $P_{\rm eq}(A)$.]  With the change of variable $Z(J) = \exp[ W(J)/\hbar]$, where $W(J)$ is the generating functional of connected correlators,  this reads
\beq
0 = \int d^dx \ J_\mu^b(x) \left[ J_\mu^b(x) + K_\mu\left( x, { \delta W \over \delta J } + \hbar { \delta \over \delta J } \right) \right].
\eeq
Finally we make the change of variable from $W(J)$ to the quantum effective action $\Gamma(A)$ by the Legendre transformation \eqref{Legendre}, which expresses the time-independent Fokker-Planck equation as a functional equation for the quantum effective action,
\beq
\label{TIforgamma}
0 = \int d^dx \ {\delta \Gamma \over \delta A_\mu^b(x)} \left[ {\delta \Gamma \over \delta A_\mu^b(x)} + K_\mu^b \left( x; \ A + \hbar {\cal D} { \delta \over \delta A } \right) \right],
\eeq
where
$\hbar {\cal D}_{\mu \nu}^{cd}(x, y; \ A)$ is the gluon propagator in the presence of the source $A$, and is expressed in terms of $\Gamma(A)$ in \eqref{glupropwsource}, and $K_\mu^b = - {\delta S_{\rm YM} \over \delta A_\mu^b(x)} + {K_{\rm gf}}_\mu^b$.  The term $ - {\delta S_{\rm YM} \over \delta A_\mu^b(x)}\left( x; \ A + \hbar {\cal D} { \delta \over \delta A } \right)$ occurs in the familiar functional DS equation in Faddeev-Popov theory \eqref{DSGamma}, and we shall not write it out here, while the second term, which results from stochastic gauge fixing, is simpler and is given by
\beq
{K_{\rm gf \ }}_\mu^b\left( x; \ A + \hbar {\cal D} { \delta \over \delta A } \right) = a^{-1} D_\mu^{bd}(A) \p \cdot A^d(x) + a^{-1} \hbar g f^ {bcd} \p_\nu^{(y)}{\cal D}_{\mu \nu}^{cd}(x, y; \ A)|_{y = x}.
\eeq
\\  
Although the time-independent Fokker-Planck equation in the form \eqref{TIforgamma} appears to have a quite different structure from the DS equation \eqref{DSGamma} in Faddeev-Popov theory, it may be used, like the DS equation in Faddeev-Popov theory, to generate a perturbation series for the correlators or, more ambitiously, to obtain a non-perturbative solution \cite{Zwanziger:2002ea}.  In a perturbative expansion, the parameter $\hbar$ counts the number of independent loops.  This method has the advantage that the solution to equation \eqref{TIforgamma} is unique.  This is true because the solution to the time-independent Fokker-Planck equation \eqref{TIFokkerPlanck}, is unique (up to normalization) as shown in Appendix C, and because the change of variable from \eqref{TIFokkerPlanck} to \eqref{TIforgamma} is invertible.\footnote{The uniqueness of the solution of the exact equation \eqref{TIforgamma} does not necessarily imply the uniqueness of the solution by a truncation of \eqref{TIforgamma}.}  In contrast, the DS equation in Faddeev-Popov theory has many different solutions because of the existence of an infinite number of nodal surfaces of the Faddeev-Popov determinant $\det M(A)$, as explained above.\\
\\
In \cite{Zwanziger:2001kw} the Landau-gauge limit, $a \to 0$, of this equation was written as the DS equation in Faddeev-Popov theory plus a correction term.  With neglect of the correction term, that is, from the DS equation in Faddeev-Popov theory, the infrared critical exponent of the gluon propagator was  found to be $D(k) \sim (k^2)^{2/5}$, $(k^2)^{0.2952}$ and $(k^2)^{0.1906}$ in Euclidean dimension $d = 2, 3,$ and $4$ respectively \cite{Zwanziger:2001kw}.  The result for $d = 4$ agrees with the independent calculation of \cite{Lerche:2002ep}.  An improved calculation, based directly on the time-independent Fokker-Planck equation \eqref{TIforgamma}, yielded instead $D(k) \sim (k^2)^{0.043}$ in Euclidean dimension $d = 4$ \cite{Zwanziger:2002ea}, which vanishes at $k = 0$, but very weakly.  The smallness of the exponent is notable because the vanishing of $D(k)$ at $k = 0$ was imposed by hand (rightly or wrongly) as a boundary condition.  The result, $D(k) \sim (k^2)^{0.043}$ is in reasonable agreement with the (subsequent) lattice calculations on very large lattices, \cite{Cucchieri:2007md} and other references cited above, which favor $ D(k) \approx (k^2)^0$ for $d = 3$ and 4, taking into account that there is an unestimated truncation error in the analytic calculation from \eqref{TIforgamma}, and numerical uncertainty in  the lattice calculation. \\
\\ 
In summary, use of the time-independent Fokker-Planck equation \eqref{TIforgamma} has two great advantages of principle.  First the Gribov problem is by-passed by gauge fixing with the gauge-fixing force that is an infinitesimal gauge transformation.  Second, the solution to \eqref{TIforgamma} is unique.  The price paid is that the time-independent stochastic dynamics involves a non-conservative (but local) drift force, and  this exceeds the bounds of conventional quantum field theory that is formulated in terms of a local action.  So if we wish to pursue this approach, we must abandon the comfortable {\it terra firma} of conventional quantum field theory, and embark on an unexplored sea.  New techniques would have to be developed to establish renormalizability, Ward identities, and other needed properties.\\
\\
Assuming these results can be established, let us speculate on the characteristics of the time-independent stochastic dynamics defined by eq.~\eqref{TIforgamma}.  First we should expect that the singularities of correlators of gauge-non-invariant fields will be unphysical, as  happens with the GZ action.  This is practically assured because (1) the probability distribution gets concentrated inside the Gribov region in the Landau gauge limit, $a \to 0$, of \eqref{TIforgamma} \cite{Zwanziger:2001kw}, as it does with the GZ action, and (2) violation of reflection positivity in the gluon propagator has been confirmed by numerical simulations in lattice gauge theory.  There is also a simple a priori reason to expect unphysical singularities to result from \eqref{TIforgamma}: because of the non-conservative drift force, the rules of calculation deviate from the conventional Feynman rules, so the singularities that they determine should also be different.  If the singularities are unphysical, as seems to be assured, then the corresponding particles are confined.  If so, one needs a mechanism which assures that the singularities of the correlators of gauge-invariant objects, such as glue-balls or hadrons, should remain physical.  For this too there is a simple argument.  The correlators of gauge-invariant objects are independent of the gauge parameter $a^{-1}$, and thus exist in the limit $a^{-1} \to 0$.  But $a^{-1}$ is the coefficient of the gauge-fixing drift force, so in the limit $a^{-1} \to 0$, the non-conservative drift force weakens, and the drift force approaches the physical limit, $\lim_{a^{-1} \to 0}K_\mu^b = - {\delta S_{\rm YM} \over \delta A_\mu^b}$.  Thus, for the correlators of gauge-invariant objects, we expect to recover the physical singularity structure associated with a local action.  Note also that the gauge-fixing force $K_{gf}$ provides a restoring force that is tangent to gauge orbits and, as this force weakens in the limit $a^{-1} \to 0$, the probability escapes to infinity along gauge-orbit directions, and we may expect the correlators of gauge-non-invariant objects to vanish in this limit, as happens in (non-gauge-fixed) lattice gauge theory.  In this scenario in the version of continuum gauge theory provided by time-independent stochastic quantization, confinement of gauge non-invariant fields is an almost kinematic consequence of non-Abelian gauge invariance.\\
\\

\section*{Acknowledgements}
N.~Vandersickel is supported by the Research-Foundation Flanders (FWO Vlaanderen). We wish like to thank Attilio Cucchieri, David Dudal, Orlando Oliveira and Silvio P. Sorella for comments and improvements of this manuscript.  Laurent Baulieu, Jutho Haegeman and Axel Maas are acknowledged for useful discussions.

\appendix

\chapter{Some formulae and extra calculations}
\section{Gaussian integrals}
\subsection{Gaussian integral for scalar variables}
\begin{eqnarray}\label{gauss1}
	I(A, J) &=& \int [\d \varphi] \exp \left[ - \frac{1}{2} \int \d^d x \d^d y\  \varphi(x) A(x,y) \varphi(y) +   \int \d^d x \ \varphi (x) J(x) \right] \nonumber \\
	&=& C (\det A)^{-1/2} \exp \frac{1}{2} \int \d^dx \d^dy\ J(x) A^{-1}(x,y) J(y),
\end{eqnarray}
with $C$ an infinite constant, which, in practice, can always be omitted.

\subsection{Gaussian integral for complex conjugated scalar variables}
\begin{eqnarray}\label{gauss8}
	I(A, J) &=& \int [\d \varphi] [\d \overline \varphi] \exp \left[ -  \int \d^d x \d^d y\  \overline \varphi(x) A(x,y) \varphi(y) +   \int \d^d x \left( \overline \varphi(x)  J_{\overline \varphi}(x) + \varphi (x) J_\varphi(x) \right)\right] \nonumber \\
	&=& C (\det A)^{-1} \exp  \int \d^d x \d^d y\ J_\varphi(x) A^{-1}(x,y) J_{\overline \varphi}(y)\;,
\end{eqnarray}
again with $C$ an infinite constant.

\subsection{Gaussian integral for Grassmann variables}
\begin{eqnarray}\label{ghostapp}
	I(A,\eta,\bar{\eta}) &=& \int [\d\theta][\d\bar{\theta}] \exp\left[ \int \d^d x \d^d y \ \bar{\theta}(x) A(x,y) \theta(y) + \int \d^dx \ (\bar{\eta}(x) \theta(x) + \bar{\theta}(x) \eta(x)) \right]  \nonumber \\
	&=&C \det A \exp  -\int \d^d x \d^d y \   \bar{\eta}(x) A^{-1}(x,y) \eta(y),
\end{eqnarray}
again with $C$ an infinite constant.

\section{$D_\mu (A) \omega = 0$ is a gauge invariant equation.\label{appgaugeinv}}
To prove that $D_\mu (A) \omega = 0$ is a gauge invariant equation, we can write from expression \eqref{covariantderivativeadjoint},
\begin{equation}
D_\mu \omega = \p_\mu \omega - \ii g A_\mu \omega + \ii g \omega A_\mu \;.
\end{equation}
Now performing a $SU(N)$ transformation, we know that $D_\mu \omega$ is in the adjoint representation by definition,
\begin{eqnarray}
D_\mu \omega  = 0 \to U D_\mu \omega U^\dagger = 0 \;,
\end{eqnarray}
so working out this equation we find,
\begin{eqnarray}
U D_\mu \omega U^\dagger &=& U \p_\mu \Omega U^\dagger - \ii g U A_\mu U^\dagger U \omega U^\dagger + \ii g U \omega U^\dagger U A_\mu U^\dagger \nonumber\\
&=& \p_\mu \omega' - (\p_\mu U) \omega U^\dagger  - U  \omega (\p_\mu U^\dagger ) - \ii g (A_\mu' + \frac{\ii}{g} \p_\mu U U^\dagger ) \omega ' + \ii g \omega '(A_\mu' + \frac{\ii}{g} \p_\mu U U^\dagger ) \nonumber\\
&=& D_\mu' \omega' - \p_\mu U \omega  U^\dagger - U \omega \p_\mu U^\dagger + \p_\mu U U^\dagger (U \omega U^\dagger) - (U \omega U^\dagger) \p_\mu U U^\dagger \nonumber\\
&=&  D_\mu' \omega'\;,
\end{eqnarray}
whereby we made use of equation \eqref{notinf} and the simple formula
\begin{eqnarray}
U U^\dagger = 1 &\Rightarrow& \p_\mu U U^\dagger +  U \p_\mu U^\dagger = 0 \;.
\end{eqnarray}
We have thus indeed proven that $D_\mu (A) \omega = 0$ is a gauge invariant equation.

\section{$\sigma$ decreases with increasing $k^2$. \label{sigma}}
We shall prove that the following function
\begin{eqnarray*}
f(k,A) &=& \frac{k_\mu  k_\nu}{k^2} \int\frac{ \d^d q}{(2 \pi)^2} f(q^2)  \frac{ 1 }{(k-q)^2} P_{\mu\nu} = \int\frac{ \d^4 q}{(2 \pi)^2} f(q^2)  \frac{ 1 }{(k-q)^2}  \left( 1 - \frac{k_\mu k_\nu}{k^2} \frac{q_\mu q_\nu}{q^2} \right)\;,
\end{eqnarray*}
decreases with increasing $k^2$. Let us prove this in 2 dimensions for simplicity. We assume $k = (k_x, k_y)$ to be oriented along the $x$ axis. Using polar coordinates, we obtain
\begin{eqnarray}
f(k,A) &=&  \int_0^{\infty}\frac{ \d q}{(2 \pi)^2} q f(q^2)  \int_{0}^{2\pi} \d \theta \frac{ 1 - \cos^2 \theta }{k^2 + q^2 - k q \cos \theta }  \nonumber\\
&=&  \int_0^{\infty}\frac{ \d q}{(2 \pi)^2} q f(q^2)  \left( \theta(q^2 - k^2) \frac{\pi}{k^2} + \theta(k^2 - q^2) \frac{\pi}{k^2} \right) \;,
\end{eqnarray}
whereby we have used the following Poisson-like $\theta$-integral which can be easily calculated using a contour integration,
\begin{equation}\label{contourint}
    \int_{0}^{2\pi}\d\theta
\frac{1-\cos^2\theta}{k^2+q^2-2qk\cos\theta}=\left\{\begin{array}{c}
                                                    \frac{\pi}{q^ 2}\qquad\mbox{if\;} k^2\leq q^2 \\
                                                    \frac{\pi}{k^ 2}\qquad\mbox{if\;} q^2\leq k^2
                                                  \end{array}\right.\;.
\end{equation}
Now deriving $f(k,A)$ w.r.t.~$k^2$ and using the property of the $\theta$ function: $\frac{\p}{\p x} \theta(x-y) = \delta(x-y)$, we find
\begin{eqnarray}
\frac{\p}{\p k^2} f(k,A) &=& -\int_0^{\infty}\frac{ \d q}{(2 \pi)^2} q f(q^2)  \theta(k^2 - q^2) \frac{\pi}{k^4} = - \theta(k)\frac{\pi}{k^4} \int_0^{k}\frac{ \d q}{(2 \pi)^2} q f(q^2) \;,
\end{eqnarray}
and thus $f(k,A)$ is a decreasing function for increasing $k^2$.

\section{Determinant of $K_{\mu\nu}$ \label{sigma2}}
We calculate the determinant of
\begin{eqnarray}
K_{\mu \nu}^{ab} (k) &=&\delta^{ab} \left( \underbrace{\beta\frac{1}{V} \frac{2}{d} \frac{N g^2}{N^2 - 1}}_{\lambda} \delta_{\mu\nu} \frac{1}{k^2} + \delta_{\mu\nu} k^2 + \left(\frac{1}{\alpha} - 1 \right)k_\mu k_\nu \right)\;.
\end{eqnarray}
We can write
\begin{eqnarray}\label{B4}
\left(\det  K_{\mu \nu}^{ab} (k) \right)^{-1/2} &=& \e^{-\frac{1}{2} \ln \det  K_{\mu \nu}^{ab}} = \e^{-\frac{1}{2}\Tr \ln   K_{\mu \nu}^{ab}}\;.
\end{eqnarray}
Therefore, we need to determine
\begin{eqnarray}
\Tr \ln   K_{\mu \nu}^{ab} 
&=& (N^2 - 1) \Tr \ln \left( \delta_{\mu\kappa} \left(\frac{\lambda}{k^2} + k^2 \right) \left( \delta_{\kappa\nu} +  \frac{1}{\frac{\lambda}{k^2} + k^2}\left(\frac{1}{\alpha} - 1 \right)k_\kappa k_\nu \right) \right) \nonumber\\
&=& (N^2 - 1) \left[ \Tr \ln \left( \delta_{\mu\nu} \left(\frac{\lambda}{k^2} + k^2 \right)\right) + \Tr \ln \left( \delta_{\mu\nu} +  \frac{k^2}{\lambda + k^4}\left(\frac{1}{\alpha} - 1 \right)k_\mu k_\nu \right)   \right] \nonumber\\
&=& (N^2 - 1)\Biggl[ d \sum_k \ln \frac{k ^4 + \lambda}{k ^2}\nonumber\\
 && + \Tr \left( \frac{k^2}{\lambda + k^4}\left(\frac{1}{\alpha} - 1 \right)k_\mu k_\nu + \left( \frac{k^2}{\lambda + k^4}\left(\frac{1}{\alpha} - 1\right) \right)^2 k_\mu k_\kappa k_\kappa k_\nu\right)  \Biggr]\;,
\end{eqnarray}
whereby we used $\ln (1 + x) = x - \frac{x^2}{2} + \ldots$. We can now take the trace of the diagonal elements of the second term, and again use $x - \frac{x^2}{2} + \ldots = \ln (1 + x)$. We obtain,
\begin{eqnarray}
\Tr \ln   K_{\mu \nu}^{ab} &=& (N^2 - 1)\left[ d \sum_k \ln \frac{k ^4 + \lambda}{k ^2}  + \sum_k \ln \left( 1 +  \frac{k^2}{\lambda + k^4}\left(\frac{1}{\alpha} - 1 \right) k^2 \right)\right] \nonumber\\
&=& (N^2 - 1)\left[ d \sum_k \ln \frac{k ^4 + \lambda}{k ^2} - \sum_k \ln \frac{k^4 + \lambda}{k^2} + \sum_k \ln \left( \frac{\lambda}{k^2} + \frac{k^2}{\alpha}\right) \right]\;.
\end{eqnarray}
By working out the last term, we see that it is proportional to $\alpha$,
\begin{eqnarray}
\sum_k \ln \left( \frac{\lambda}{k^2} + \frac{k^2}{\alpha}\right) &=&  \sum_k \ln \left( \frac{k^4}{\alpha} + \lambda \right)  - \sum_k \ln k^2 \nonumber\\
&=& V \int \frac{\d^d k}{(2\pi)^d} \ln \left( \frac{k^2}{\sqrt{\alpha}} + \ii \sqrt \lambda \right)+ V \int \frac{\d^d k}{(2\pi)^d} \ln \left( \frac{k^2}{\sqrt{\alpha}} - \ii \sqrt \lambda \right) \nonumber\\
&\sim& \alpha^{d/4} \;,
\end{eqnarray}
whereby $\int \d^q q \ln q^2$ is zero in dimensional regularization. Therefore, in the limit $\alpha \to 0$, becomes zero.  In conclusion, we find
\begin{align}\label{B8}
\left(\det  K_{\mu \nu}^{ab} (k) \right)^{-1/2}  &=\exp  \left[(N^2 - 1) \frac{(d-1)}{2}V \int \frac{\d^d k}{(2\pi)^d} \ln \frac{k^4 + \lambda}{k^2} \right] \nonumber\\
&=\exp  \left[(N^2 - 1) \frac{(d-1)}{2}V  \int \frac{\d^d k}{(2\pi)^d} \ln \left( k^2 + \frac{1}{V} \frac{2}{d} \frac{\beta N g^2}{N^2 - 1}\frac{1}{k^2} \right) \right]\;.
\end{align}

\chapter{Cohomologies and the doublet theorem}
\section{Cohomology}
Suppose $\delta$ is a nilpotent operator, $\delta^2 = 0$. The cohomology of $\delta$ is given by the solutions of the equation
\begin{eqnarray}\label{298}
\delta \Delta &=& 0\;,
\end{eqnarray}
which cannot be written in the form
\begin{eqnarray}\label{coho}
 \Delta = \delta \Omega \;.
\end{eqnarray}
A quantity $\Delta$ obeying equation \eqref{298} is called \textit{closed}, while a quantity of the form \eqref{coho} is called \textit{exact}. The cohomology of $\Delta$ is thus identified by quantities which are closed but not exact. More precisely, a non trivial quantity $\Delta$ is always defined up to the addition of an arbitrary exact part, i.e.~one speaks of cohomology classes. In fact, take now two closed quantities $\Delta_1$ and $\Delta_2$. These quantities belong to the same cohomology class if
\begin{eqnarray}
\Delta_1 - \Delta_2 &=& \delta (\ldots) \;,
\end{eqnarray}
i.e.~when $\Delta_1$ and $\Delta_2$ differ by an exact part.\\
\\
In this way one can always write $\Delta$ obeying \eqref{298} as a sum of a trivial part and a non trivial part.
\begin{eqnarray}\label{siex}
\Delta &=& \Delta_{\mathrm{n.triv}} + \underbrace{ \delta (\ldots)}_{\Delta_{\mathrm{triv}}}\;,
\end{eqnarray}
whereby $ \Delta_{\mathrm{n.triv}}$ does not contain parts that can be written as $\delta (\ldots)$. In quantum field theory, these non trivial parts shall be the most interesting parts, as they will be related to the renormalization of the physical parameters of the theory.

\section{Doublet theorem}\label{doublettheorem}
Now there is a very important theorem which shall be very useful later on \cite{Piguet1995}.  Suppose our theory contains a pair of fields, sources or parameters $(u_i,v_i)$ which form a doublet:
\begin{align}\label{uomzetten}
\delta u_i &= v_i & \delta v_i &= 0\;,
\end{align}
whereby the subscript $i$ is a certain index (e.g.~color). We assume $u_i$ to be commuting, while $v_i$ is an anticommuting quantity. Then we can prove that $u_i$ and $v_i$ shall never enter the non trivial part of the cohomology of $\delta$.\\
\\
The proof is as follows. We introduce two operators $\hat{P}$ and $\hat{A}$
\begin{eqnarray}
\hat{P} &=& \int \d x \left(  u_i \frac{\partial}{\partial u_i} + v_i \frac{\partial}{\partial v_i} \right )\nonumber \\
\hat{A} &=&  \int \d x\ \left( u_i \frac{\partial}{\partial v_i}   \right)\;,
\end{eqnarray}
Functionally, we write for the nilpotent operator $\delta$
\begin{eqnarray}
\delta &=& v_i \frac{\p }{\p u_i}\;,
\end{eqnarray}
so we obtain
\begin{align*}
\delta\  \hat{A}   = \int \d x\left( v_i \frac{\partial}{\partial v_i} + v_j u_i \frac{\partial}{\partial u_j}\frac{\partial}{\partial v_i} \right), \nonumber\\
 \hat{A}\ \delta \  = \int \d x \left( u_i \frac{\partial}{\partial u_i} -   u_i v_j \frac{\partial}{\partial v_i} \frac{\partial}{\partial u_j} \right)\;,
\end{align*}
and thus
\begin{equation} \label{antic2}
\{ \delta, \hat{A} \} = \hat{P}.
\end{equation}
Analogously we also have
\begin{eqnarray*}
 \hat{P}  \delta &=&  \int \d x \left(  u_i v_j \frac{\partial}{\partial u_i} \frac{\partial}{\partial u_j}  + v_i \frac{\partial}{\partial u_i} - v_i v_j \frac{\partial}{\partial v_i} \frac{\partial}{\partial u_j}  \right) \nonumber\\
\delta \ \hat{P}  &=&\int \d x \left(  v_j \frac{\partial}{\partial u_j} +  v_j u_i  \frac{\partial}{\partial u_j} \frac{\partial}{\partial u_i} + v_j v_i \frac{\partial}{\partial u_j} \frac{\partial}{\partial v_i} \right)\;,
\end{eqnarray*}
and thus
\begin{equation}\label{antic}
[\delta \ , \hat{P}] = 0.
\end{equation}
As $\widehat P$ is a counting operator for the total number of $u_i$ and $v_i$, we can expand\footnote{We assume $\Delta$ to be a polynomial in $u_i$ and $v_i$. } $\Delta$, see expression \eqref{siex}, in eigenvectors of $\widehat P$,
\begin{equation}
\Delta = \sum_{n \geq 0} \Delta_n\;,
\end{equation}
whereby $\widehat P  \Delta_n = n \Delta_n$ and $n$ represents the total number of $u_i$ and $v_i$ in $\Delta_n$. Now from the cohomology condition \eqref{298} and the commutation relation \eqref{antic}, we find that
\begin{equation}
0 = \sum_{n \geq 0} n \delta \Delta_n \quad \Rightarrow \quad \sum_{n \geq 1} n \delta \Delta_n = 0 \;.
\end{equation}
Looking at expression \eqref{uomzetten}, we easily obtain that 
\begin{equation}
 \delta \Delta_n = 0  \quad \forall n \geq 1 \;.
\end{equation}
Finally, using this property and invoking expression \eqref{antic2}, we obtain
\begin{eqnarray}
\Delta &=&  \Delta_0 + \sum_{n \geq 1} \frac{1}{n}  \hat{P} \Delta_n \nonumber\\
  &=&  \Delta_0  + \sum_{n \geq 1} \frac{1}{n} \delta \hat{A} \Delta_n \nonumber\\
  &=& \Delta_0 + \delta (\ldots) \;.
\end{eqnarray}
In conclusion, as $\delta^2 = 0$
\begin{equation}
\delta  \Delta = \delta \Delta_0 \;,
\end{equation}
whereby $ \Delta_0$ is independent of the doublet $(u_i,\ v_i)$. The quantities $u_i$ and $v_i$ shall thus never enter the non trivial part of the cohomology.

\chapter{Positivity and uniqueness of solution of time-independent Fokker-Planck equation}

\section{Positivity}

The proof is given in a finite number of dimensions, but that should be a good guide.  We use discrete notation and write $A_i$ instead of $A_\mu^b(x)$, so the time-independent Fokker-Planck equation reads,

\beq
 0 =  \p_i  [ -  \p_i + K_i(A) ] F(A),
 \eeq
where $\p_i \equiv {\p \over \p A_i}$.    Suppose that this equation possesses a solution $F(A)$ that is positive somewhere and negative elsewhere.  Call $R$ the region where $F(A)$ is negative, so $F(A) < 0$ for $A$ in the interior of $R$, it is positive outside, $F(A) > 0$ for $A \notin R$, and the boundary $\p R$ of the region $R$ is a nodal surface, $F(A) = 0$ for $A \in \p R$.  Now integrate the equation over the region $R$,
\beq
 0 = \int_R dA \  \p_i  [ -  \p_i + K_i(A) ] F(A),
 \eeq
 where $dA$ is the volume element in $A$-space,
 and use Gauss's law to obtain
 \beq
 0 = \int_{\p R} d\Sigma_i \  [ -  \p_i + K_i(A) ] F(A),
 \eeq
 where $d\Sigma_i$ is the surface element oriented along the outward normal $n_i$ on $\p R$.  By hypothesis $F(A)$ vanishes on the boundary $\p R$, which gives
\beq
 0 = \int_{\p R} d\Sigma_i \  \p_i  F(A).
 \eeq
Moreover $F(A)$ is negative inside $R$ and positive outside, so $F(A)$ increases as one leaves the region $R$, and the gradient along $d\Sigma_i$ is positive,
$d\Sigma_i \  \p_i  F(A) > 0$, which is a contradiction.  Consequently the solution $F(A)$ is either positive everywhere or negative everywhere, and $F(A)$ cannot have any nodal surface.  We may normalize $F(A)$ so it is positive everywhere $F(A) > 0$, with $\int dA \ F(A) = 1$.

\section{Uniqueness}

Now suppose that there are two linearly independent solutions to the time-independent Fokker-Planck equation, $F_1(A)$ and $F_2(A),$ and that each is normalized $\int dA \ F_1(A) = \int dA \ F_2(A) = 1$.  Because the time-independent Fokker-Planck equation is linear, the difference $F(A) = F_1(A) - F_2(A)$ is also a solution, and it satisfies $\int dA \ F(A) = 0$.  However we have just proven that $F(A)$ is either positive everywhere or negative everywhere, or zero, and we conclude $F(A) = F_1(A) - F_2(A) = 0$.  Thus if the time-independent  Fokker-Planck equation possesses a non-zero solution $F_1(A)$, this solution is positive and unique up to normalization.  This result in no way depends upon whether the drift force $K_i(A)$ is conservative.  However when $K_i(A)$ is non-conservative, the equilibrium current does not vanish $j_{{\rm eq} \ i} \equiv - \p_i P_{\rm eq} + K_i P_{\rm eq} \neq 0$, for if $j_{{\rm eq} \ i} = 0$ then the drift force is given by $K_i = \p_i \ln P_{\rm eq}$, and is conservative.


\providecommand{\href}[2]{#2}\begingroup\raggedright\endgroup

\end{document}